\documentstyle[aaspp4,11pt,flushrt]{article}
\begin{document}


\def\lya{Ly$\alpha$~}
\def\ljeans{\lambda_{\rm J}}
\def\ie{i.e.~}
\def\kjeans{k_{\rm J}}
\def\mjeans{M_{\rm J}}
\def\beq{\begin{equation}}
\def\eeq{\end{equation}}
\def\lapprox{\la}
\def\Msun{$M_\odot$}
\def\lyasource{{\dot N_\alpha}}
\def\beqa{\begin{eqnarray}}
\def\eeqa{\end{eqnarray}}
\def\xb{{\bf x}}
\def\yb{{\bf y}}
\def\rb{{\bf r}}
\def\vb{{\bf v}}
\def\ub{{\bf u}}
\def\kb{{\bf k}}
\def\Omm{{\Omega_m}}
\def\Ommz{{\Omega_m^{\,z}}}
\def\Omr{{\Omega_r}}
\def\Omk{{\Omega_k}}
\def\Oml{{\Omega_{\Lambda}}}
\def\nb{\bar{n}}
\def\etal{et al.\ }
\def\Ng{N_\gamma}
\def\zr{z_{\rm reion}}
\def\HI{\rm H\,I}
\def\cN{c_{\rm N}}
\def\kB{k_{\rm B}}
\def\Ni{N_{\rm ion}}
\def\fg{f_{\rm gas}}
\def\fe{f_{\rm eject}}
\def\fin{f_{\rm int}}
\def\fw{f_{\rm wind}}
\def\NGST{{\it NGST}}


\title{\bf In the Beginning: The First Sources of Light\\ and the
Reionization of the Universe}

\author{Rennan Barkana \footnote{Present address: Canadian Institute
for Theoretical Astrophysics, 60 St. George Street \#1201A, Toronto,
Ontario, M5S 3H8, CANADA}} \affil{Institute for Advanced Study, Olden
Lane, Princeton, NJ 08540}

\author{Abraham Loeb} 
\affil{Department of Astronomy, Harvard University, 60 Garden St.,
Cambridge, MA 02138}


\begin{abstract}

The formation of the first stars and quasars marks the transformation
of the universe from its smooth initial state to its clumpy current
state. In popular cosmological models, the first sources of light
began to form at a redshift $z=30$ and reionized most of the hydrogen
in the universe by $z=7$. Current observations are at the threshold of
probing the hydrogen reionization epoch. The study of high-redshift
sources is likely to attract major attention in observational and
theoretical cosmology over the next decade.
\end{abstract}


\newpage
\tableofcontents
\newpage


\section{\bf Preface: The Frontier of Small-Scale Structure}
\label{sec1}

The detection of cosmic microwave background (CMB) anisotropies
(Bennett et al.\  1996; de Bernardis et al.\  2000; Hanany et al.\  2000)
confirmed the notion that the present large-scale structure in the
universe originated from small-amplitude density fluctuations at early
times.
Due to the natural instability of gravity, regions that were denser than
average collapsed and formed bound objects, first on small spatial
scales and later on larger and larger scales. The present-day
abundance of bound objects, such as galaxies and X-ray clusters, can be
explained based on an appropriate extrapolation of the detected
anisotropies to smaller scales.
Existing observations with the {\it Hubble Space Telescope}\, (e.g.,
Steidel et al.\ 1996; Madau et al.\ 1996; Chen et al.\ 1999; Clements
et al.\ 1999) and ground-based telescopes (Lowenthal et al.\ 1997; Dey
et al.\ 1999; Hu et al.\ 1998, 1999; Spinrad et al.\ 1999; Steidel et
al.\ 1999), have constrained the evolution of galaxies and their
stellar content at $z\la 6$.  However, in the bottom-up hierarchy of
the popular Cold Dark Matter (CDM) cosmologies, galaxies were
assembled out of building blocks of smaller mass. The elementary
building blocks, i.e., the first gaseous objects to form, acquired a
total mass of order the Jeans mass ($\sim 10^4 M_\odot$), below which
gas pressure opposed gravity and prevented collapse (Couchman \& Rees
1986; Haiman \& Loeb 1997; Ostriker \& Gnedin 1996). In variants of
the standard CDM model, these basic building blocks first formed at
$z\sim 15$--$30$.

\begin{figure}[htbp]
\includegraphics{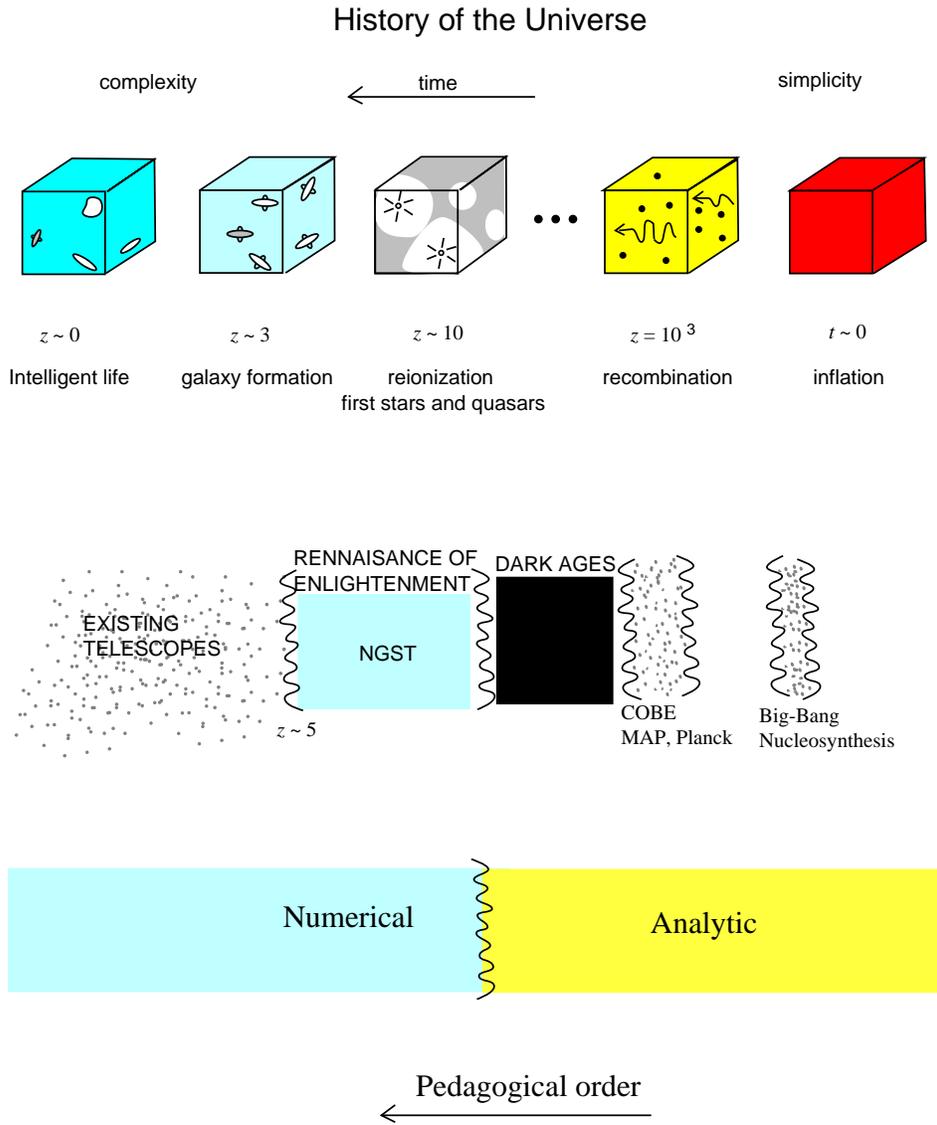}
\vspace{6.2in}
\caption{Milestones in the evolution of the universe from simplicity
to complexity. The ``end of the dark ages'' bridges between the
recombination epoch probed by microwave anisotropy experiments ($z\sim
10^3$) and the horizon of current observations ($z\sim 5$--6).}
\label{fig1a}
\end{figure}

An important qualitative outcome of the microwave anisotropy data is
the confirmation that the universe started out simple. It was by and
large homogeneous and isotropic with small fluctuations that can be
described by linear perturbation analysis. The current universe is
clumpy and complicated.  Hence, the arrow of time in cosmic history
also describes the progression from simplicity to complexity (see
Figure~\ref{fig1a}). While the conditions in the early universe can be
summarized on a single sheet of paper, the mere description of the
physical and biological structures found in the present-day universe
cannot be captured by thousands of books in our libraries.  The
formation of the first bound objects marks the central milestone in
the transition from simplicity to complexity.  Pedagogically, it would
seem only natural to attempt to understand this epoch before we try to
explain the present-day universe. Historically, however, most of the
astronomical literature focused on the local universe and has only
been shifting recently to the early universe. This violation of
the pedagogical rule was forced upon us by the limited state of our
technology; observation of earlier cosmic times requires detection of
distant sources, which is feasible only with large telescopes and
highly-sensitive instrumentation.

\begin{figure}[htbp]
\includegraphics{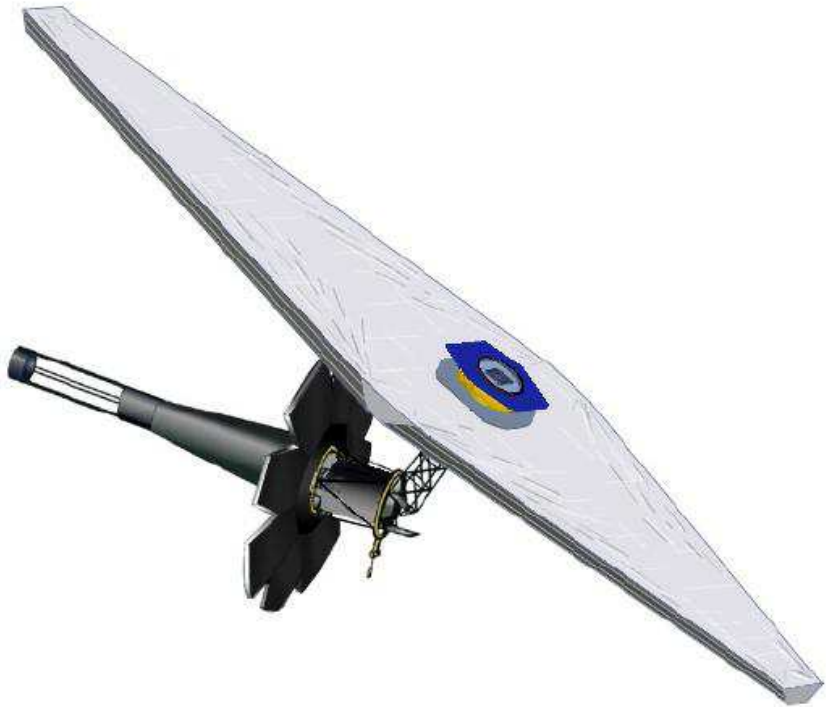}
\vspace{5.5in}
\caption{Artist's illustration of one of the current designs (GSFC) of
the Next Generation Space Telescope. More details about the telescope
can be found at http://ngst.gsfc.nasa.gov/}
\label{fig1b}
\end{figure}

For these reasons, advances in technology are likely to make the high
redshift universe an important frontier of cosmology over the coming
decade. This effort will involve large (30 meter) ground-based
telescopes and will culminate in the launch of the successor to the
{\it Hubble Space Telescope},\, called {\it Next Generation Space
Telescope} (\NGST\,). Figure~\ref{fig1b} shows an artist's illustration
of this telescope which is currently planned for launch in 2009.
\NGST\, will image the first sources of light that formed in the
universe. With its exceptional sub-nJy ($1~{\rm nJy}=10^{-32}{\rm
erg~cm^{-2}~s^{-1}~Hz^{-1}}$) sensitivity in the 1--3.5$\mu$m infrared
regime, \NGST\, is ideally suited for probing optical-UV emission from
sources at redshifts $\ga 10$, just when popular Cold Dark Matter
models for structure formation predict the first baryonic objects to
have collapsed.

The study of the formation of the first generation of sources at early
cosmic times (high redshifts) holds the key to constraining the
power-spectrum of density fluctuations on small scales. Previous
research in cosmology has been dominated by studies of {\it Large Scale
Structure} (LSS); future studies are likely to focus on {\it Small
Scale Structure} (SSS).

The first sources are a direct consequence of the growth of linear
density fluctuations. As such, they emerge from a well-defined set of
initial conditions and the physics of their formation can be followed
precisely by computer simulation. The cosmic initial conditions for
the formation of the first generation of stars are much simpler than
those responsible for star formation in the Galactic interstellar
medium at present. The cosmic conditions are fully specified by the
primordial power spectrum of Gaussian density fluctuations, the mean
density of dark matter, the initial temperature and density of the
cosmic gas, and the primordial composition according to Big-Bang
nucleosynthesis. The chemistry is much simpler in the absence of
metals and the gas dynamics is much simpler in the absence of both
dynamically-significant magnetic fields and feedback from luminous
objects.

The initial mass function of the first stars and black holes is
therefore determined by a simple set of initial conditions (although
subsequent generations of stars are affected by feedback from
photoionization heating and metal enrichment).  While the early
evolution of the seed density fluctuations can be fully described
analytically, the collapse and fragmentation of nonlinear structure
must be simulated numerically.  The first baryonic objects connect the
simple initial state of the universe to its complex current state, and
their study with hydrodynamic simulations (e.g., Abel et al.\ 1998a,
Abel, Bryan, \& Norman 2000; Bromm, Coppi, \& Larson 1999) and with
future telescopes such as \NGST\, offers the key to advancing our
knowledge on the formation physics of stars and massive black holes.

\begin{figure}[htbp]
\includegraphics{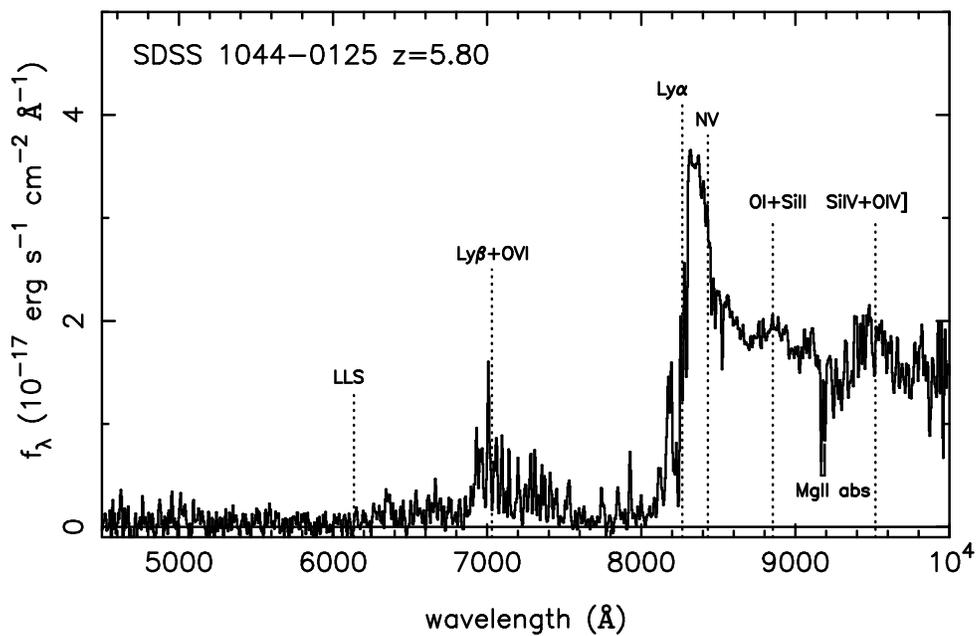}
\vspace{6.2in}
\caption{Optical spectrum of the highest-redshift known quasar at
$z=5.8$, discovered by the Sloan Digital Sky Survey (Fan et
al.\  2000). }
\label{fig1c}
\end{figure}

The {\it first light} from stars and quasars ended the ``dark ages''
\footnote{The use of this term in the cosmological context was coined
by Sir Martin Rees.} of the universe and initiated a ``renaissance of
enlightenment'' in the otherwise fading glow of the microwave
background (see Figure~\ref{fig1a}).  It is easy to see why the mere
conversion of trace amounts of gas into stars or black holes at this
early epoch could have had a dramatic effect on the ionization state
and temperature of the rest of the gas in the universe. Nuclear fusion
releases $\sim 7\times 10^6$ eV per hydrogen atom, and thin-disk
accretion onto a Schwarzschild black hole releases ten times more
energy; however, the ionization of hydrogen requires only 13.6 eV. It
is therefore sufficient to convert a small fraction, $\sim 10^{-5}$ of
the total baryonic mass into stars or black holes in order to ionize
the rest of the universe. (The actual required fraction is higher by
at least an order of magnitude [Bromm, Kudritzky, \& Loeb 2000]
because only some of the emitted photons are above the ionization
threshold of 13.6 eV and because each hydrogen atom recombines more
than once at redshifts $z\ga 7$). Recent calculations of structure
formation in popular CDM cosmologies imply that the universe was
ionized at $z\sim 7$--12 (Haiman \& Loeb 1998, 1999b,c; Gnedin \&
Ostriker 1997; Chiu \& Ostriker 2000; Gnedin 2000a), and has remained
ionized ever since. Current observations are at the threshold of
probing this epoch of reionization, given the fact that galaxies and
quasars at redshifts $\sim 6$ are being discovered (Fan et al.\ 2000;
Stern et al.\ 2000). One of these sources is a bright quasar at
$z=5.8$ whose spectrum is shown in Figure~\ref{fig1c}. The plot
indicates that there is transmitted flux shortward of the Ly$\alpha$
wavelength at the quasar redshift.  The optical depth at these
wavelengths of the uniform cosmic gas in the intergalactic medium is
however (Gunn \& Peterson 1965),
\begin{equation}
\tau_{s}={\pi e^2 f_\alpha \lambda_\alpha n_{\HI}(z_s) \over m_e
cH(z_s)} \approx 6.45\times 10^5 x_{\HI} \left({\Omega_bh\over
0.03}\right)\left({\Omega_m\over 0.3}\right)^{-1/2} \left({1+z_s\over
10}\right)^{3/2} \label{G-P}
\end{equation}
where $H\approx 100h~{\rm
km~s^{-1}~Mpc^{-1}}\Omega_m^{1/2}(1+z_s)^{3/2}$ is the Hubble
parameter at the source redshift $z_s$, $f_\alpha=0.4162$ and
$\lambda_\alpha=1216$\AA\, are the oscillator strength and the
wavelength of the Ly$\alpha$ transition; $n_{\HI}(z_s)$ is the neutral
hydrogen density at the source redshift (assuming primordial
abundances); $\Omega_m$ and $\Omega_b$ are the present-day density
parameters of all matter and of baryons, respectively; and $x_{\HI}$
is the average fraction of neutral hydrogen. In the second equality we
have implicitly considered high redshifts (see
equations~(\ref{highz1}) and (\ref{highz2}) in \S \ref{sec2.1}).
Modeling of the transmitted flux (Fan et al.\ 2000) implies
$\tau_s<0.5$ or $x_{\HI}\la 10^{-6}$, i.e., the low-density gas
throughout the universe is fully ionized at $z=5.8$!  One of the
important challenges for future observations will be to identify {\it
when and how the intergalactic medium was ionized}. Theoretical
calculations (see \S \ref{sec6.3.1}) imply that such observations are
just around the corner.

\begin{figure}[htbp]
\includegraphics{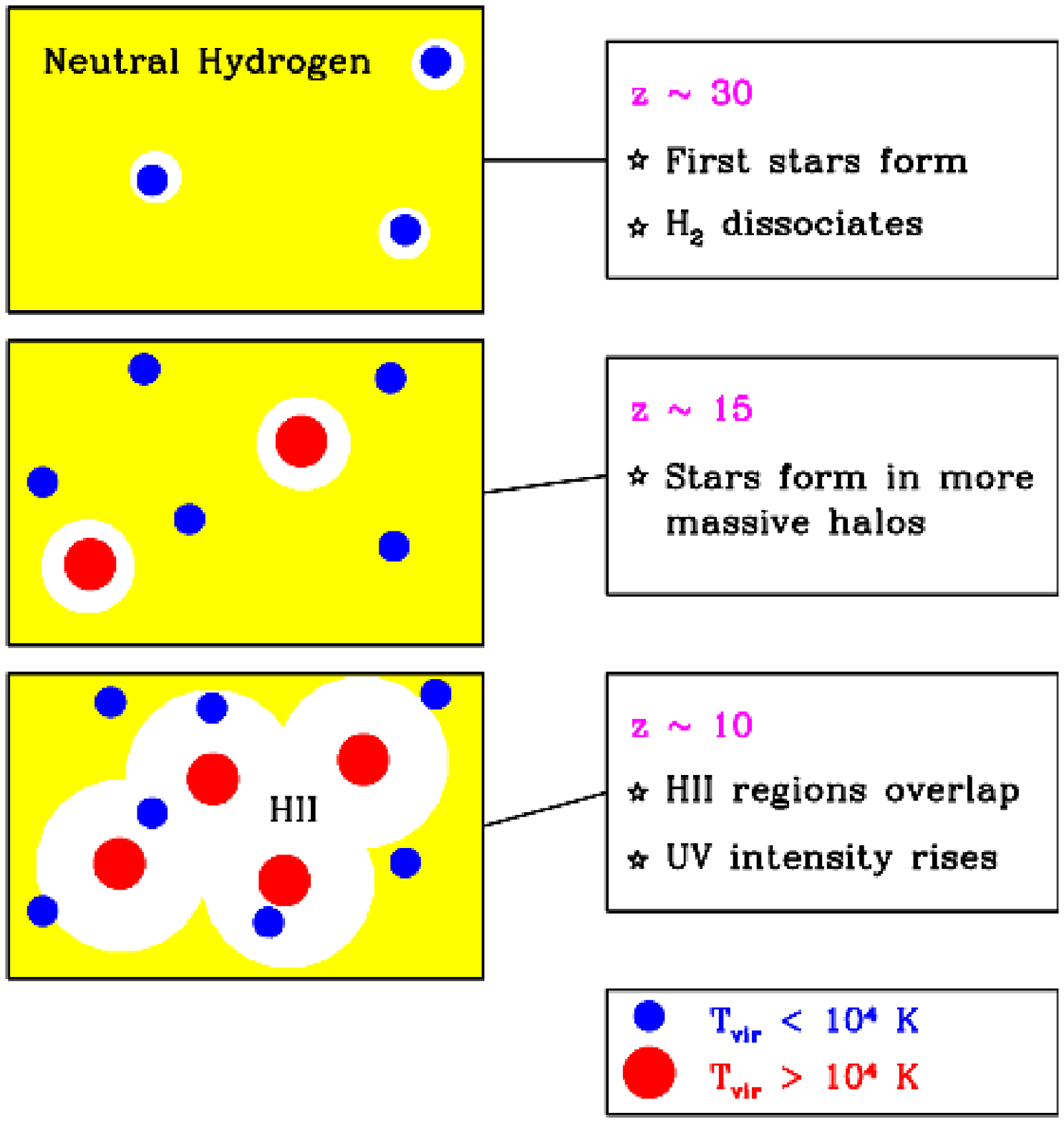}
\vspace{6.2in}
\caption{Stages in the reionization of hydrogen in the intergalactic
medium. }
\label{fig1d}
\end{figure}

Figure~\ref{fig1d} shows schematically the various stages in a
theoretical scenario for the history of hydrogen reionization in the
intergalactic medium. The first gaseous clouds collapse at redshifts
$\sim 20$--$30$ and fragment into stars due to molecular hydrogen
(H$_2$) cooling. However, H$_2$ is fragile and can be easily
dissociated by a small flux of UV radiation.  Hence the bulk of the
radiation that ionized the universe is emitted from galaxies with a
virial temperature $\ga 10^4$ K, where atomic cooling is effective and
allows the gas to fragment (see the end of \S \ref{sec3.3} for an
alternative scenario).

Since recent observations confine the standard set of cosmological
parameters to a relatively narrow range, we assume a $\Lambda$CDM
cosmology with a particular standard set of parameters in the
quantitative results in this review. For the contributions to the
energy density, we assume ratios relative to the critical density of
$\Omm=0.3$, $\Oml=0.7$, and $\Omega_b=0.045$, for matter, vacuum
(cosmological constant), and baryons, respectively. We also assume a
Hubble constant $H_0=100\,h\mbox{ km s}^{-1}\mbox{Mpc}^{-1}$ with
$h=0.7$, and a primordial scale invariant ($n=1$) power spectrum with
$\sigma_8=0.9$, where $\sigma_8$ is the root-mean-square amplitude of
mass fluctuations in spheres of radius $8\ h^{-1}$ Mpc. These
parameter values are based primarily on the following observational
results: CMB temperature anisotropy measurements on large scales
(Bennett et al.\ 1996) and on the scale of $\sim 1^\circ$ (Lange et
al.\ 2000; Balbi et al.\ 2000); the abundance of galaxy clusters
locally (Viana \& Liddle 1999; Pen 1998; Eke, Cole, \& Frenk 1996) and
as a function of redshift (Bahcall \& Fan 1998; Eke, Cole, Frenk, \&
Henry 1998); the baryon density inferred from big bang nucleosynthesis
(see the review by Tytler et al.\ 2000); distance measurements used to
derive the Hubble constant (Mould et al.\ 2000; Jha et al.\ 1999;
Tonry et al.\ 1997; but see Theureau et al.\ 1997; Parodi et al.\
2000); and indications of cosmic acceleration from distances based on
type Ia supernovae (Perlmutter et al.\ 1999; Riess et al.\ 1998).

This review summarizes recent theoretical advances in understanding
the physics of the first generation of cosmic structures. Although the
literature on this subject extends all the way back to the sixties
(Saslaw \& Zipoy 1967, Peebles \& Dicke 1968, Hirasawa 1969, Matsuda
et al.\ 1969, Hutchins 1976, Silk 1983, Palla et al.\ 1983, Lepp \&
Shull 1984, Couchman 1985, Couchman \& Rees 1986, Lahav 1986), this
review focuses on the progress made over the past decade in the modern
context of CDM cosmologies.



\section{\bf Hierarchical Formation of Cold Dark Matter Halos}
\label{sec2}

\subsection{The Expanding Universe}
\label{sec2.1}

The modern physical description of the universe as a whole can be
traced back to Einstein, who argued theoretically for the so-called
``cosmological principle'': that the distribution of matter and energy
must be homogeneous and isotropic on the largest scales. Today
isotropy is well established (see the review by Wu, Lahav, \& Rees
1999) for the distribution of faint radio sources, optically-selected
galaxies, the X-ray background, and most importantly the cosmic
microwave background (henceforth, CMB; see, e.g., Bennett et al.\
1996). The constraints on homogeneity are less strict, but a
cosmological model in which the universe is isotropic but
significantly inhomogeneous in spherical shells around our special
location is also excluded (Goodman 1995).

In General Relativity, the metric for a space which is spatially
homogeneous and isotropic is the Robertson-Walker metric, which can be
written in the form \beq \label{RW}
ds^2=dt^2-a^2(t)\left[\frac{dR^2}{1-k\,R^2}+R^2
\left(d\theta^2+\sin^2\theta\,d\phi^2\right)\right]\ , \eeq where
$a(t)$ is the cosmic scale factor which describes expansion in time,
and $(R,\theta,\phi)$ are spherical comoving coordinates. The constant
$k$ determines the geometry of the metric; it is positive in a closed
universe, zero in a flat universe, and negative in an open
universe. Observers at rest remain at rest, at fixed
$(R,\theta,\phi)$, with their physical separation increasing with time
in proportion to $a(t)$. A given observer sees a nearby observer at
physical distance $D$ receding at the Hubble velocity $H(t)D$, where
the Hubble constant at time $t$ is $H(t)=d\ln a(t)/dt$. Light emitted by
a source at time $t$ is observed at $t=0$ with a redshift
$z=1/a(t)-1$, where we set $a(t=0) \equiv 1$.

The Einstein field equations of General Relativity yield the Friedmann
equation (e.g., Weinberg 1972; Kolb \& Turner 1990) \beq
H^2(t)=\frac{8 \pi G}{3}\rho-\frac{k}{a^2}\ ,\eeq which relates the
expansion of the universe to its matter-energy content. For each
component of the energy density $\rho$, with an equation of state
$p=p(\rho)$, the density $\rho$ varies with $a(t)$ according to the
equation of energy conservation \beq d (\rho R^3)=-p d(R^3)\ . \eeq
With the critical density \beq \rho_C(t) \equiv \frac{3 H^2(t)}{8 \pi
G} \eeq defined as the density needed for $k=0$, we define the ratio
of the total density to the critical density as \beq \Omega \equiv
\frac{\rho}{\rho_C}\ . \eeq With $\Omm$, $\Oml$, and $\Omr$ denoting
the present contributions to $\Omega$ from matter (including cold dark
matter as well as a contribution $\Omega_b$ from baryons), vacuum
density (cosmological constant), and radiation, respectively, the
Friedmann equation becomes \beq \frac{H(t)}{H_0}= \left[ \frac{\Omm}
{a^3}+ \Oml+ \frac{\Omr}{a^4}+ \frac{\Omk}{a^2}\right]^{1/2}\ , \eeq
where we define $H_0$ and $\Omega_0=\Omm+\Oml+\Omr$ to be the present
values of $H$ and $\Omega$, respectively, and we let \beq \Omk \equiv
-\frac{k}{H_0^2}=1-\Omega_0. \eeq In the particularly simple
Einstein-de Sitter model ($\Omm=1$, $\Oml=\Omr=\Omk=0$), the scale
factor varies as $a(t) \propto t^{2/3}$. Even models with non-zero
$\Oml$ or $\Omk$ approach the Einstein-de Sitter behavior at high
redshifts, i.e., when \beq (1+z)\gg {\rm max} \left [(1-\Omega_m
-\Omega_\Lambda) / \Omega_m, (\Omega_\Lambda/\Omega_m)^{1/3}\right]
\label{highz1} \eeq 
(as long as $\Omr$ can be neglected). The Friedmann equation implies
that models with $\Omk=0$ converge to the Einstein-de Sitter limit
faster than do open models. E.g., for $\Omega_m=0.3$ and
$\Omega_\Lambda=0.7$ equation~(\ref{highz1}) corresponds to the
condition $z\gg 1.3$, which is easily satisfied by the reionization
redshift. In this high-$z$ regime, $H(t) \approx 2/(3t)$, and the age
of the universe is
\begin{equation}
t\approx {2\over 3\, H_0 \sqrt{\Omega_m}} \left(1+z\right)^{-3/2} =
5.38 \times 10^8 \left(\frac{1+z}{10}\right)^{-3/2}{\rm yr}\ , 
\label{highz2}
\end{equation}
where in the last expression we assumed our standard cosmological
parameters (see the end of \S \ref{sec1}).

In the standard hot Big Bang model, the universe is initially hot and
the energy density is dominated by radiation. The transition to matter
domination occurs at $z \sim 10^4$, but the universe remains hot
enough that the gas is ionized, and electron-photon scattering
effectively couples the matter and radiation. At $z \sim 1200$ the
temperature drops below $\sim 3300$ K and protons and electrons
recombine to form neutral hydrogen. The photons then decouple and
travel freely until the present, when they are observed as the CMB.

\subsection{Linear Gravitational Growth}
\label{sec2.2}

Observations of the CMB (e.g., Bennett et al.\ 1996) show that the
universe at recombination was extremely uniform, but with spatial
fluctuations in the energy density and gravitational potential of
roughly one part in $10^5$. Such small fluctuations, generated in the
early universe, grow over time due to gravitational instability, and
eventually lead to the formation of galaxies and the large-scale
structure observed in the present universe.

As in the previous section, we distinguish between fixed and comoving
coordinates. Using vector notation, the fixed coordinate ${\bf r}$
corresponds to a comoving position $\xb=\rb/a$. In a homogeneous
universe with density $\rho$, we describe the cosmological expansion
in terms of an ideal pressure-less fluid of particles each of which is
at fixed $\xb$, expanding with the Hubble flow $\vb=H(t) \rb$ where
$\vb=d\rb/dt$. Onto this uniform expansion we impose small
perturbations, given by a relative density perturbation \beq
\delta(\xb)=\frac{\rho(\rb)}{\bar{\rho}}-1\ , \eeq where the mean
fluid density is $\bar{\rho}$, with a corresponding peculiar velocity
$\ub \equiv \vb - H \rb$. Then the fluid is described by the
continuity and Euler equations in comoving coordinates (Peebles 1980,
1993): \beqa \frac{\partial \delta}{\partial t}+\frac{1}{a}{\bf
\nabla} \cdot \left[(1+\delta) \ub\right] &=& 0 \\ \frac{\partial
\ub}{\partial t}+H\ub+\frac{1}{a}(\ub \cdot {\bf \nabla})
\ub&=&-\frac{1}{a}{\bf \nabla}\phi\ .  \eeqa The potential $\phi$ is
given by the Poisson equation, in terms of the density perturbation:
\beq \nabla^2\phi=4 \pi G \bar{\rho} a^2 \delta\ . \eeq This fluid
description is valid for describing the evolution of collisionless
cold dark matter particles until different particle streams cross.
This
``shell-crossing'' typically occurs only after perturbations have
grown to become non-linear, and at that point the individual particle
trajectories must in general be followed. Similarly, baryons can be
described as a pressure-less fluid as long as their temperature is
negligibly small, but non-linear collapse leads to the formation of
shocks in the gas.

For small perturbations $\delta \ll 1$, the fluid equations can be
linearized and combined to yield
\beq \frac{\partial^2\delta}{\partial t^2}+2 H
\frac{\partial\delta}{\partial t}=4 \pi G \bar{\rho} \delta\ . \eeq
This linear equation has in general two independent solutions, only
one of which grows with time. Starting with random initial conditions,
this ``growing mode'' comes to dominate the density evolution. Thus,
until it becomes non-linear, the density perturbation maintains its
shape in comoving coordinates and grows in proportion to a growth
factor $D(t)$. The growth factor is in general given by (Peebles 1980)
\beq
D(t) \propto \frac{\left(\Oml a^3+\Omk
a+\Omm\right)^{1/2}}{a^{3/2}}\int^a \frac{a^{3/2}\,
da}{\left(\Oml  a^3+\Omk a+\Omm\right)^{3/2}}\ ,
\eeq
where we neglect $\Omr$ when considering halos forming at $z \ll
10^4$. In the Einstein-de Sitter model (or, at high redshift, in other
models as well) the growth factor is simply proportional to $a(t)$.

The spatial form of the initial density fluctuations can be described
in Fourier space, in terms of Fourier components \beq \delta_\kb =
\int d^3x\, \delta(\xb) e^{-i \kb \cdot \xb}\ .\eeq Here we use the
comoving wavevector $\kb$, whose magnitude $k$ is the comoving
wavenumber which is equal to $2\pi$ divided by the wavelength. The
Fourier description is particularly simple for fluctuations generated
by inflation (e.g., Kolb \& Turner 1990). Inflation generates
perturbations given by a Gaussian random field, in which different
$\kb$-modes are statistically independent, each with a random
phase. The statistical properties of the fluctuations are determined
by the variance of the different $\kb$-modes, and the variance is
described in terms of the power spectrum $P(k)$ as follows: \beq
\left<\delta_{\kb} \delta_{{\bf k'}}^{*}\right>=\left(2 \pi\right)^3
P(k)\, \delta^{(3)} \left(\kb-{\bf k'}\right)\ , \eeq where
$\delta^{(3)}$ is the three-dimensional Dirac delta function.

In standard models, inflation produces a primordial power-law spectrum
$P(k) \propto k^n$ with $n \sim 1$. Perturbation growth in the
radiation-dominated and then matter-dominated universe results in a
modified final power spectrum, characterized by a turnover at a scale
of order the horizon $cH^{-1}$ at matter-radiation equality, and a
small-scale asymptotic shape of $P(k) \propto k^{n-4}$. On large
scales the power spectrum evolves in proportion to the square of the
growth factor, and this simple evolution is termed linear
evolution. On small scales, the power spectrum changes shape due to
the additional non-linear gravitational growth of perturbations,
yielding the full, non-linear power spectrum. The overall amplitude of
the power spectrum is not specified by current models of inflation,
and it is usually set observationally using the CMB temperature
fluctuations or local measures of large-scale structure.

Since density fluctuations may exist on all scales, in order to
determine the formation of objects of a given size or mass it is
useful to consider the statistical distribution of the smoothed
density field.  Using a window function $W(\yb)$ normalized so that
$\int d^3\yb\, W(\yb)=1$, the smoothed density perturbation field,
$\int d^3\yb \delta(\xb+\yb) W(\yb)$, itself follows a Gaussian
distribution with zero mean. For the particular choice of a spherical
top-hat, in which $W=1$ in a sphere of radius $R$ and is zero outside,
the smoothed perturbation field measures the fluctuations in the mass
in spheres of radius $R$. The normalization of the present power
spectrum is often specified by the value of $\sigma_8 \equiv
\sigma(R=8 h^{-1} {\rm Mpc})$. For the top-hat, the smoothed
perturbation field is denoted $\delta_R$ or $\delta_M$, where the mass
$M$ is related to the comoving radius $R$ by $M=4 \pi \rho_m R^3/3$,
in terms of the current mean density of matter $\rho_m$. The variance
$\langle \delta_M \rangle^2$ is \beq \sigma^2(M)= \sigma^2(R)=
\int_0^{\infty}\frac{dk}{2 \pi^2} \,k^2 P(k) \left[\frac{3
j_1(kR)}{kR} \right]^2\ ,\label{eqsigM}\eeq where $j_1(x)=(\sin x-x
\cos x)/x^2$. The function $\sigma(M)$ plays a crucial role in
estimates of the abundance of collapsed objects, as described below.

\subsection{Formation of Nonlinear Objects}
\label{sec2.3}

The small density fluctuations evidenced in the CMB grow over time as
described in the previous subsection, until the perturbation $\delta$
becomes of order unity, and the full non-linear gravitational problem
must be considered. The dynamical collapse of a dark matter halo can
be solved analytically only in cases of particular symmetry. If we
consider a region which is much smaller than the horizon $cH^{-1}$,
then the formation of a halo can be formulated as a problem in
Newtonian gravity, in some cases with minor corrections coming from
General Relativity. The simplest case is that of spherical symmetry,
with an initial ($t=t_i\ll t_0$) top-hat of uniform overdensity
$\delta_i$ inside a sphere of radius $R$. Although this model is
restricted in its direct applicability, the results of spherical
collapse have turned out to be surprisingly useful in understanding
the properties and distribution of halos in models based on cold dark
matter.

The collapse of a spherical top-hat is described by the Newtonian
equation (with a correction for the cosmological constant) \beq
\frac{d^2r}{dt^2}=H_0^2 \Oml\, r-\frac{GM}{r^2}\ , \eeq where $r$ is
the radius in a fixed (not comoving) coordinate frame, $H_0$ is the
present Hubble constant, $M$ is the total mass enclosed within radius
$r$, and the initial velocity field is given by the Hubble flow
$dr/dt=H(t) r$. The enclosed $\delta$ grows initially as
$\delta_L=\delta_i D(t)/D(t_i)$, in accordance with linear theory, but
eventually $\delta$ grows above $\delta_L$. If the mass shell at
radius $r$ is bound (i.e., if its total Newtonian energy is negative)
then it reaches a radius of maximum expansion and subsequently
collapses. At the moment when the top-hat collapses to a point, the
overdensity predicted by linear theory is (Peebles 1980) $\delta_L\, =
1.686$ in the Einstein-de Sitter model, with only a weak dependence on
$\Omm$ and $\Oml$. Thus a top-hat collapses at redshift $z$ if its
linear overdensity extrapolated to the present day (also termed the
critical density of collapse) is \beq \delta_{\rm
crit}(z)=\frac{1.686}{D(z)}\ ,
\label{deltac} \eeq where we set $D(z=0)=1$.

Even a slight violation of the exact symmetry of the initial
perturbation can prevent the top-hat from collapsing to a
point. Instead, the halo reaches a state of virial equilibrium by
violent relaxation (phase mixing). Using the virial theorem $U=-2K$ to
relate the potential energy $U$ to the kinetic energy $K$ in the final
state, the final overdensity relative to the critical density at the
collapse redshift is $\Delta_c=18\pi^2 \simeq 178$ in the Einstein-de
Sitter model, modified in a universe with $\Omm+\Oml=1$ to the fitting
formula (Bryan \& Norman 1998) \beq \Delta_c=18\pi^2+82 d-39 d^2\ ,
\eeq where $d\equiv \Ommz-1$ is evaluated at the collapse redshift, so
that \beq \Ommz=\frac{\Omm (1+z)^3}{\Omm (1+z)^3+\Oml+\Omk (1+z)^2}\ .
\label{Ommz} \eeq

A halo of mass $M$ collapsing at redshift $z$ thus has a (physical)
virial radius \beq r_{\rm vir}=0.784 \left(\frac{M}{10^8\ h^{-1} \
M_{\sun} }\right)^{1/3} \left[\frac{\Omm} {\Ommz}\ \frac{\Delta_c}
{18\pi^2}\right]^{-1/3} \left (\frac{1+z}{10} \right)^{-1}\ h^{-1}\
{\rm kpc}\ , \label{rvir}\eeq and a corresponding circular velocity,
\beq V_c=\left(\frac{G M}{r_{\rm vir}}\right)^{1/2}= 23.4 \left(
\frac{M}{10^8\ h^{-1} \ M_{\sun} }\right)^{1/3} \left[\frac {\Omm}
{\Ommz}\ \frac{\Delta_c} {18\pi^2}\right]^{1/6} \left( \frac{1+z} {10}
\right)^{1/2}\ {\rm km\ s}^{-1}\ . \label{Vceqn} \eeq In these
expressions we have assumed a present Hubble constant written in the
form $H_0=100\, h\mbox{ km s}^{-1}\mbox{Mpc}^{-1}$. We may also define
a virial temperature \beq \label{tvir} T_{\rm vir}=\frac{\mu m_p
V_c^2}{2 \kB}=1.98\times 10^4\ \left(\frac{\mu}{0.6}\right)
\left(\frac{M}{10^8\ h^{-1} \ M_{\sun} }\right)^{2/3} \left[ \frac
{\Omm} {\Ommz}\ \frac{\Delta_c} {18\pi^2}\right]^{1/3}
\left(\frac{1+z}{10}\right)\ {\rm K} \ , \eeq where $\mu$ is the mean
molecular weight and $m_p$ is the proton mass. Note that the value of
$\mu$ depends on the ionization fraction of the gas; $\mu=0.59$ for a
fully ionized primordial gas, $\mu=0.61$ for a gas with ionized
hydrogen but only singly-ionized helium, and $\mu=1.22$ for neutral
primordial gas. The binding energy of the halo is
approximately\footnote{The coefficient of $1/2$ in
equation~(\ref{Ebind}) would be exact for a singular isothermal
sphere, $\rho(r)\propto 1/r^2$.} \beq \label{Ebind} E_b= {1\over 2}
\frac{GM^2}{r_{\rm vir}} = 5.45\times 10^{53} \left(\frac{M}{10^8\
h^{-1} \ M_{\sun} }\right)^{5/3} \left[ \frac {\Omm} {\Ommz}\
\frac{\Delta_c} {18\pi^2}\right]^{1/3} \left(\frac{1+z}{10}\right)
h^{-1}\ {\rm erg}\ . \eeq Note that the binding energy of the baryons
is smaller by a factor equal to the baryon fraction $\Omega_b/\Omm$.

Although spherical collapse captures some of the physics governing the
formation of halos, structure formation in cold dark matter models
proceeds hierarchically. At early times, most of the dark matter is in
low-mass halos, and these halos continuously accrete and merge to form
high-mass halos. Numerical simulations of hierarchical halo formation
indicate a roughly universal spherically-averaged density profile for
the resulting halos (Navarro, Frenk, \& White 1997, hereafter NFW),
though with considerable scatter among different halos (e.g., Bullock
et al.\ 2000). The NFW profile has the form \beq \rho(r)=\frac{3
H_0^2} {8 \pi G} (1+z)^3 \frac{\Omm}{\Ommz} \frac{\delta_c} {\cN x
(1+\cN x)^2}\ , \label{NFW} \eeq where $x=r/r_{\rm vir}$, and the
characteristic density $\delta_c$ is related to the concentration
parameter $\cN$ by \beq \delta_c=\frac{\Delta_c}{3} \frac{\cN^3}
{\ln(1+\cN)-\cN/(1+\cN)} \ . \eeq The concentration parameter itself
depends on the halo mass $M$, at a given redshift $z$. We note that
the dense, cuspy halo profile predicted by CDM models is not apparent
in the mass distribution derived from measurements of the rotation
curves of dwarf galaxies (e.g., de Blok \& McGaugh 1997; Salucci \&
Burkert 2000), although observational and modeling uncertainties may
preclude a firm conclusion at present (van den Bosch \etal 2000;
Swaters, Madore, \& Trewhella 2000).

\subsection{The Abundance of Dark Matter Halos}
\label{sec2.4}

In addition to characterizing the properties of individual halos, a
critical prediction of any theory of structure formation is the
abundance of halos, i.e., the number density of halos as a function of
mass, at any redshift. This prediction is an important step toward
inferring the abundances of galaxies and galaxy clusters. While the
number density of halos can be measured for particular cosmologies in
numerical simulations, an analytic model helps us gain physical
understanding and can be used to explore the dependence of abundances
on all the cosmological parameters. 

A simple analytic model which successfully matches most of the
numerical simulations was developed by Press \& Schechter (1974). The
model is based on the ideas of a Gaussian random field of density
perturbations, linear gravitational growth, and spherical collapse. To
determine the abundance of halos at a redshift $z$, we use $\delta_M$,
the density field smoothed on a mass scale $M$, as defined in \S
\ref{sec2.2}. Although the model is based on the initial conditions,
it is usually expressed in terms of redshift-zero quantities. Thus, we
use the linearly-extrapolated density field, i.e., the initial density
field at high redshift extrapolated to the present by simple
multiplication by the relative growth factor (see \S
\ref{sec2.2}). Similarly, in this section the 'present power spectrum'
refers to the initial power spectrum, linearly-extrapolated to the
present without including non-linear evolution. Since $\delta_M$ is
distributed as a Gaussian variable with zero mean and standard
deviation $\sigma(M)$ [which depends only on the present power
spectrum, see equation~(\ref{eqsigM})], the probability that
$\delta_M$ is greater than some $\delta$ equals \beq
\int_{\delta}^{\infty}d\delta_M \frac{1}{\sqrt{2 \pi}\, \sigma(M)}
\exp \left[- \frac{\delta_M^2} {2 \,\sigma^2(M)}\right]={1\over 2}
{\rm erfc}\left(\frac{\delta} {\sqrt{2} \,\sigma } \right)\
. \label{PS1} \eeq The fundamental ansatz is to identify this
probability with the fraction of dark matter particles which are part
of collapsed halos of mass greater than $M$, at redshift $z$. There
are two additional ingredients: First, the value used for $\delta$ is
$\delta_{\rm crit}(z)$ given in equation~(\ref{deltac}), which is the
critical density of collapse found for a spherical top-hat
(extrapolated to the present since $\sigma(M)$ is calculated using the
present power spectrum); and second, the fraction of dark matter in
halos above $M$ is multiplied by an additional factor of 2 in order to
ensure that every particle ends up as part of some halo with
$M>0$. Thus, the final formula for the mass fraction in halos above
$M$ at redshift $z$ is \beq
\label{PSerfc} F(>M | z)={\rm erfc}\left(\frac{\delta_{\rm crit}(z)} 
{\sqrt{2}\,\sigma } \right)\ . \eeq

This ad-hoc factor of 2 is necessary, since otherwise only positive
fluctuations of $\delta_M$ would be included. Bond et al.\ (1991)
found an alternate derivation of this correction factor, using a
different ansatz. In their derivation, the factor of 2 has a more
satisfactory origin, namely the so-called ``cloud-in-cloud'' problem:
For a given mass $M$, even if $\delta_M$ is smaller than $\delta_{\rm
crit}(z)$, it is possible that the corresponding region lies inside a
region of some larger mass $M_L>M$, with $\delta_{M_L}>\delta_{\rm
crit}(z)$. In this case the original region should be counted as
belonging to a halo of mass $M_L$. Thus, the fraction of particles
which are part of collapsed halos of mass greater than $M$ is larger
than the expression given in equation~(\ref{PS1}). Bond et al.\ showed
that, under certain assumptions, the additional contribution results
precisely in a factor of 2 correction.

Differentiating the fraction of dark matter in halos above $M$ yields
the mass distribution. Letting $dn$ be the comoving number density of
halos of mass between $M$ and $M+dM$, we have \beq\frac{dn}{dM}=
\sqrt{\frac{2}{\pi}}\, \frac{\rho_m}{M}\, \frac{-d(\ln \sigma)}{dM}
\,\nu_c\, e^{-\nu_c^2/2}\ , \eeq where $\nu_c=\delta_{\rm crit}(z)/
\sigma(M)$ is the number of standard deviations which the critical
collapse overdensity represents on mass scale $M$. Thus, the abundance
of halos depends on the two functions $\sigma(M)$ and $\delta_{\rm
crit} (z)$, each of which depends on the energy content of the
universe and the values of the other cosmological parameters. We
illustrate the abundance of halos for our standard choice of the
$\Lambda$CDM model with $\Omm=0.3$ (see the end of \S \ref{sec1}).

\begin{figure}[htbp]
\epsscale{0.7}
\plotone{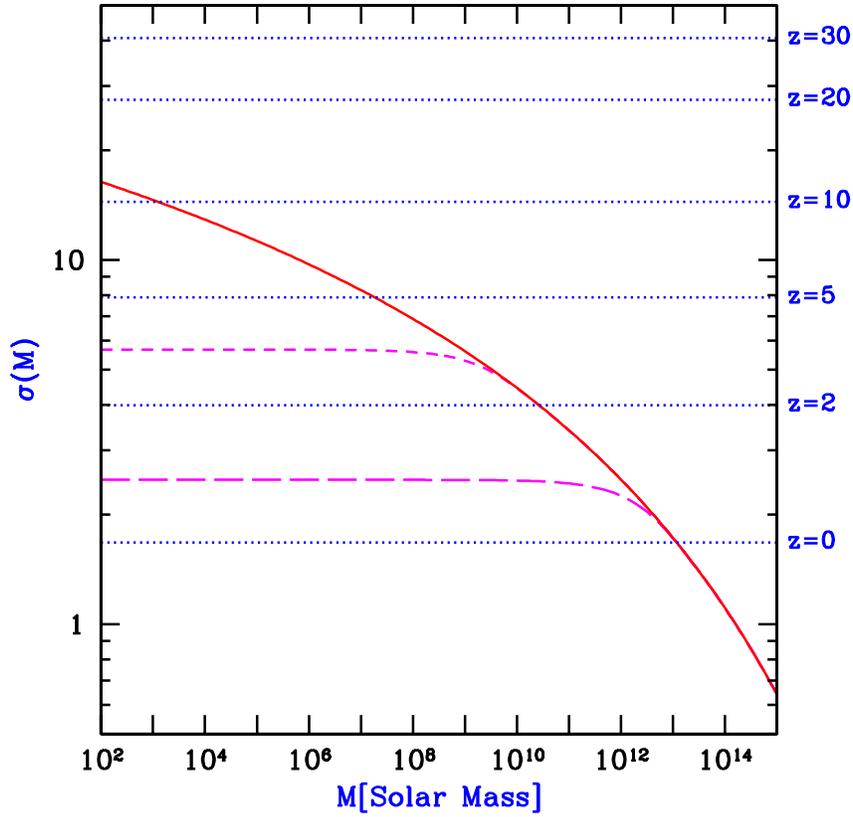}
\caption{Mass fluctuations and collapse thresholds in cold dark matter
models. The horizontal dotted lines show the value of the extrapolated
collapse overdensity $\delta_{\rm crit}(z)$ at the indicated
redshifts. Also shown is the value of $\sigma(M)$ for the cosmological
parameters given in the text (solid curve), as well as $\sigma(M)$ for
a power spectrum with a cutoff below a mass $M=1.7\times 10^8
M_{\sun}$ (short-dashed curve), or $M=1.7\times 10^{11} M_{\sun}$
(long-dashed curve). The intersection of the horizontal lines with the
other curves indicate, at each redshift $z$, the mass scale (for each
model) at which a $1-\sigma$ fluctuation is just collapsing at $z$
(see the discussion in the text).}
\label{fig2a}
\end{figure}

Figure~\ref{fig2a} shows $\sigma(M)$ and $\delta_{\rm crit}(z)$, with
the input power spectrum computed from Eisenstein \& Hu (1999). The
solid line is $\sigma(M)$ for the cold dark matter model with the
parameters specified above. The horizontal dotted lines show the value
of $\delta_{\rm crit}(z)$ at $z=0, 2, 5, 10, 20$ and 30, as indicated
in the figure. From the intersection of these horizontal lines with
the solid line we infer, e.g., that at $z=5$ a $1-\sigma$ fluctuation
on a mass scale of $2\times 10^7 M_{\sun}$ will collapse. On the other
hand, at $z=5$ collapsing halos require a $2-\sigma$ fluctuation on a
mass scale of $3\times 10^{10} M_{\sun}$, since $\sigma(M)$ on this
mass scale equals about half of $\delta_{\rm crit}(z=5)$. Since at
each redshift a fixed fraction ($31.7\%$) of the total dark matter
mass lies in halos above the $1-\sigma$ mass, Figure~\ref{fig2a} shows
that most of the mass is in small halos at high redshift, but it
continuously shifts toward higher characteristic halo masses at lower
redshift. Note also that $\sigma(M)$ flattens at low masses because of
the changing shape of the power spectrum. Since $\sigma \rightarrow
\infty$ as $M \rightarrow 0$, in the cold dark matter model all the
dark matter is tied up in halos at all redshifts, if sufficiently
low-mass halos are considered.
  
Also shown in Figure~\ref{fig2a} is the effect of cutting off the
power spectrum on small scales. The short-dashed curve corresponds to
the case where the power spectrum is set to zero above a comoving
wavenumber $k=10\, {\rm Mpc}^{-1}$, which corresponds to a mass
$M=1.7\times 10^8 M_{\sun}$. The long-dashed curve corresponds to a
more radical cutoff above $k=1\, {\rm Mpc}^{-1}$, or below
$M=1.7\times 10^{11} M_{\sun}$. A cutoff severely reduces the
abundance of low-mass halos, and the finite value of $\sigma(M=0)$
implies that at all redshifts some fraction of the dark matter does
not fall into halos. At high redshifts where $\delta_{\rm crit}(z) \gg
\sigma(M=0)$, all halos are rare and only a small fraction of the dark
matter lies in halos. In particular, this can affect the abundance of
halos at the time of reionization, and thus the observed limits on
reionization constrain scenarios which include a small-scale cutoff in
the power spectrum (Barkana, Haiman, \& Ostriker 2000).

\begin{figure}[htbp]
\epsscale{0.7}
\plotone{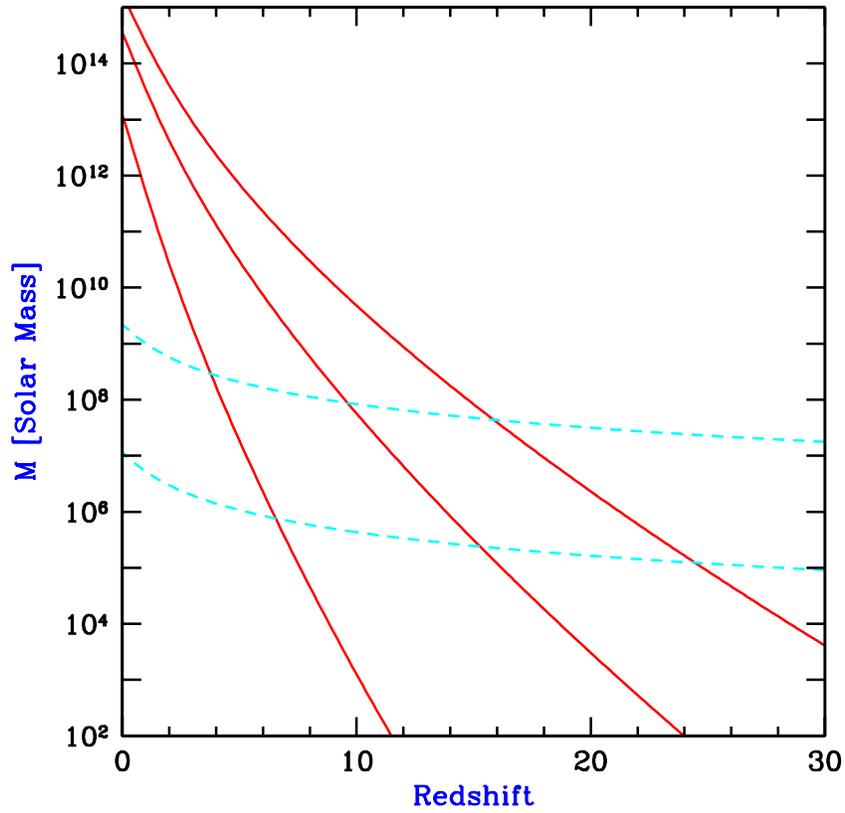}
\caption{Characteristic properties of collapsing halos: Halo mass.
The solid curves show the mass of collapsing halos which correspond to
$1-\sigma$, $2-\sigma$, and $3-\sigma$ fluctuations (in order from
bottom to top). The dashed curves show the mass corresponding to the
minimum temperature required for efficient cooling with primordial
atomic species only (upper curve) or with the addition of molecular
hydrogen (lower curve).}
\label{fig2b}
\end{figure}

\begin{figure}[htbp]
\epsscale{0.7}
\plotone{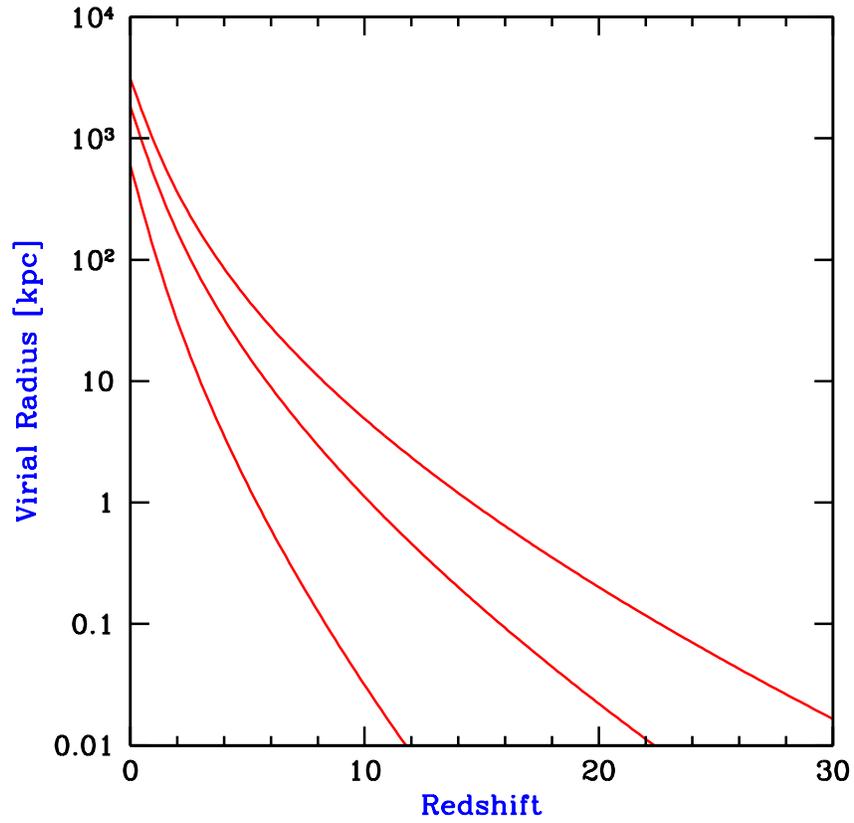}
\caption{Characteristic properties of collapsing halos: Halo virial
radius. The curves show the virial radius of collapsing halos which
correspond to $1-\sigma$, $2-\sigma$, and $3-\sigma$ fluctuations (in
order from bottom to top).}
\label{fig2c}
\end{figure}

\begin{figure}[htbp]
\epsscale{0.7}
\plotone{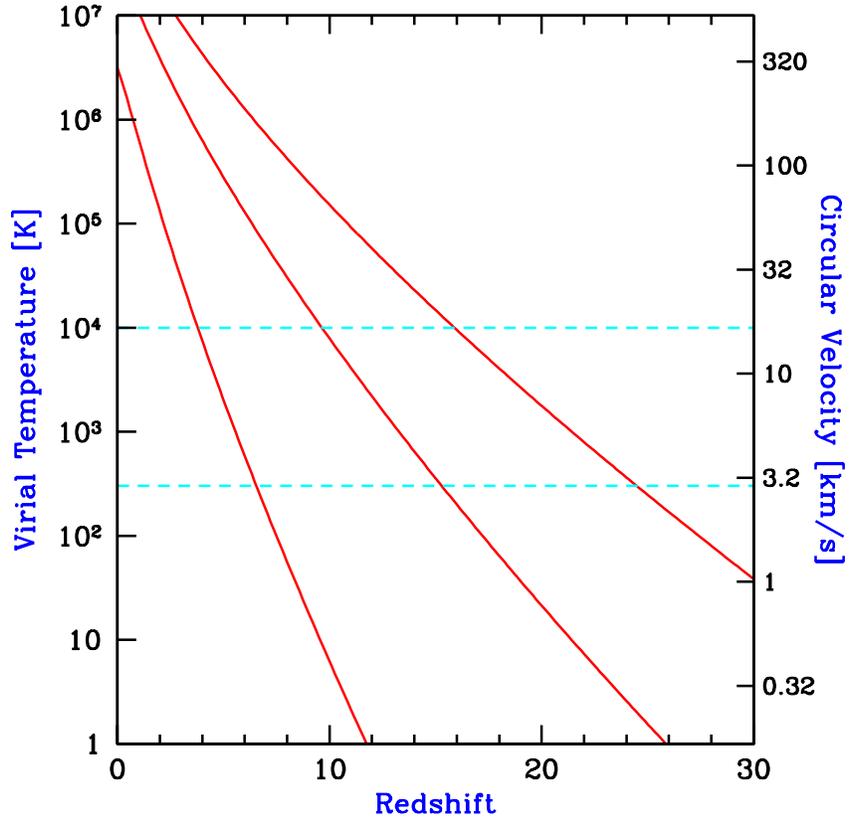}
\caption{Characteristic properties of collapsing halos: Halo virial
temperature and circular velocity. The solid curves show the virial
temperature (or, equivalently, the circular velocity) of collapsing
halos which correspond to $1-\sigma$, $2-\sigma$, and $3-\sigma$
fluctuations (in order from bottom to top). The dashed curves show the
minimum temperature required for efficient cooling with primordial
atomic species only (upper curve) or with the addition of molecular
hydrogen (lower curve).}
\label{fig2d}
\end{figure}

\begin{figure}[htbp]
\epsscale{0.7}
\plotone{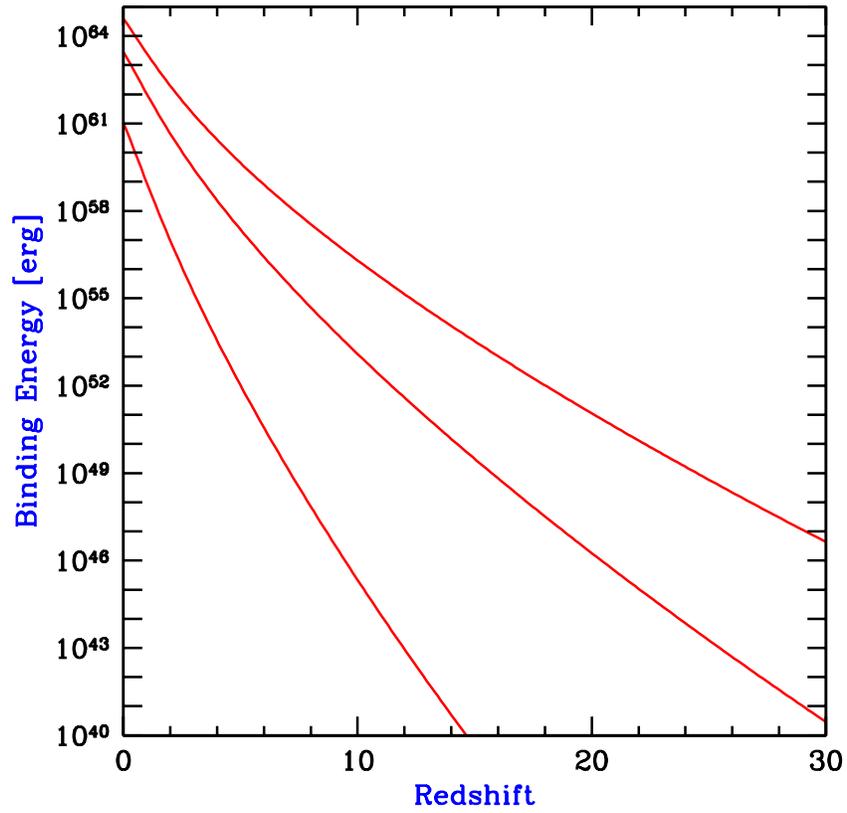}
\caption{Characteristic properties of collapsing halos: Halo binding
energy. The curves show the total binding energy of collapsing halos
which correspond to $1-\sigma$, $2-\sigma$, and $3-\sigma$
fluctuations (in order from bottom to top).}
\label{fig2e}
\end{figure}

In Figures~\ref{fig2b} -- \ref{fig2e} we show explicitly the
properties of collapsing halos which represent $1-\sigma$, $2-\sigma$,
and $3-\sigma$ fluctuations (corresponding in all cases to the curves
in order from bottom to top), as a function of redshift. No cutoff is
applied to the power spectrum. Figure~\ref{fig2b} shows the halo mass,
Figure~\ref{fig2c} the virial radius, Figure~\ref{fig2d} the virial
temperature (with $\mu$ in equation~(\ref{tvir}) set equal to $0.6$,
although low temperature halos contain neutral gas) as well as
circular velocity, and Figure~\ref{fig2e} shows the total binding
energy of these halos. In Figures~\ref{fig2b} and \ref{fig2d}, the
dashed curves indicate the minimum virial temperature required for
efficient cooling (see \S \ref{sec3.3}) with primordial atomic species
only (upper curve) or with the addition of molecular hydrogen (lower
curve). Figure~\ref{fig2e} shows the binding energy of dark matter
halos. The binding energy of the baryons is a factor $\sim
\Omega_b/\Omega_m\sim 15\%$ smaller, if they follow the dark
matter. Except for this constant factor, the figure shows the minimum
amount of energy that needs to be deposited into the gas in order to
unbind it from the potential well of the dark matter. For example, the
hydrodynamic energy released by a single supernovae, $\sim
10^{51}~{\rm erg}$, is sufficient to unbind the gas in all $1-\sigma$
halos at $z\ga 5$ and in all $2-\sigma$ halos at $z\ga 12$.

\begin{figure}[htbp]
\epsscale{0.7}
\plotone{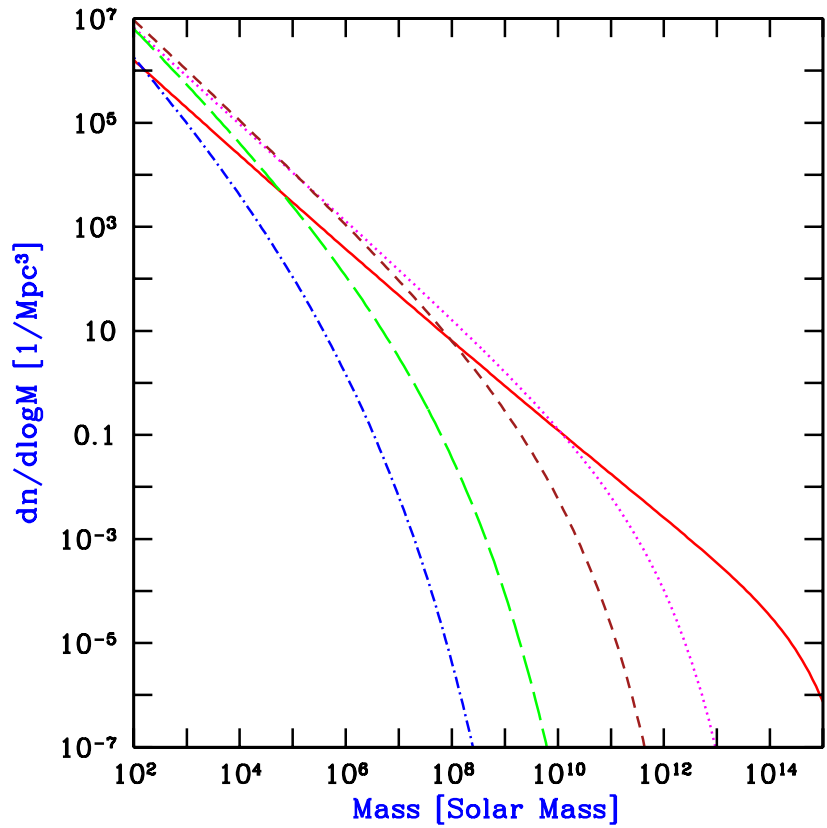}
\caption{Halo mass function at several redshifts: $z=0$ (solid curve),
$z=5$ (dotted curve), $z=10$ (short-dashed curve), $z=20$ (long-dashed
curve), and $z=30$ (dot-dashed curve).}
\label{fig2f}
\end{figure}

At $z=5$, the halo masses which correspond to $1-\sigma$, $2-\sigma$,
and $3-\sigma$ fluctuations are $1.8\times 10^7 M_{\sun}$, $3.0\times
10^{10} M_{\sun}$, and $7.0\times 10^{11} M_{\sun}$, respectively. The
corresponding virial temperatures are $2.0 \times 10^3$ K, $2.8 \times
10^5$ K, and $2.3 \times 10^6$ K. The equivalent circular velocities
are 7.5 ${\rm km\ s}^{-1}$, 88 ${\rm km\ s}^{-1}$, and 250 ${\rm km\
s}^{-1}$. At $z=10$, the $1-\sigma$, $2-\sigma$, and $3-\sigma$
fluctuations correspond to halo masses of $1.3\times 10^3 M_{\sun}$,
$5.7\times 10^7 M_{\sun}$, and $4.8\times 10^9 M_{\sun}$,
respectively. The corresponding virial temperatures are 6.2 K, $7.9
\times 10^3$ K, and $1.5 \times 10^5$ K. The equivalent circular
velocities are 0.41 ${\rm km\ s}^{-1}$, 15 ${\rm km\ s}^{-1}$, and 65
${\rm km\ s}^{-1}$. Atomic cooling is efficient at $T_{\rm vir} \ga
10^4$ K, or a circular velocity $V_c \ga 17\ {\rm km\ s}^{-1}$. This
corresponds to a $1.2-\sigma$ fluctuation and a halo mass of
$2.1\times 10^8 M_{\sun}$ at $z=5$, and a $2.1-\sigma$ fluctuation and
a halo mass of $8.3\times 10^7 M_{\sun}$ at $z=10$. Molecular hydrogen
provides efficient cooling down to $T_{\rm vir} \sim 300$ K, or a
circular velocity $V_c \sim 2.0\ {\rm km\ s}^{-1}$. This corresponds
to a $0.76-\sigma$ fluctuation and a halo mass of $3.5\times 10^5
M_{\sun}$ at $z=5$, and a $1.3-\sigma$ fluctuation and a halo mass of
$1.4\times 10^5 M_{\sun}$ at $z=10$.

In Figure~\ref{fig2f} we show the halo mass function $dn/d\ln(M)$ at
several different redshifts: $z=0$ (solid curve), $z=5$ (dotted
curve), $z=10$ (short-dashed curve), $z=20$ (long-dashed curve), and
$z=30$ (dot-dashed curve). Note that the mass function does not
decrease monotonically with redshift at all masses. At the lowest
masses, the abundance of halos is higher at $z>0$ than at $z=0$.

\section{\bf Gas Infall and Cooling in Dark Matter Halos}
\label{sec3}

\subsection{Cosmological Jeans Mass}
\label{sec3.1}

The Jeans length $\ljeans$ was originally defined (Jeans 1928) in
Newtonian gravity as the critical wavelength that separates
oscillatory and exponentially-growing density perturbations in an
infinite, uniform, and stationary distribution of gas. On scales
$\ell$ smaller than $\ljeans$, the sound crossing time, $\ell/c_s$ is
shorter than the gravitational free-fall time, $(G\rho)^{-1/2}$,
allowing the build-up of a pressure force that counteracts gravity. On
larger scales, the pressure gradient force is too slow to react to a
build-up of the attractive gravitational force.  The Jeans mass is
defined as the mass within a sphere of radius $\ljeans/2$,
$\mjeans=(4\pi/3)\rho(\ljeans/2)^3$.  In a perturbation with a mass
greater than $\mjeans$, the self-gravity cannot be supported by the
pressure gradient, and so the gas is unstable to gravitational
collapse. The Newtonian derivation of the Jeans instability suffers
from a conceptual inconsistency, as the unperturbed gravitational
force of the uniform background must induce bulk motions (compare
Binney \& Tremaine 1987).  However, this inconsistency is remedied
when the analysis is done in an expanding universe.

The perturbative derivation of the Jeans instability criterion can be
carried out in a cosmological setting by considering a sinusoidal
perturbation superposed on a uniformly expanding background.  Here, as
in the Newtonian limit, there is a critical wavelength $\ljeans$ that
separates oscillatory and growing modes.  Although the expansion of
the background slows down the exponential growth of the amplitude to a
power-law growth, the fundamental concept of a minimum mass that can
collapse at any given time remains the same (see, e.g. Kolb \& Turner
1990; Peebles 1993).

We consider a mixture of dark matter and baryons with density
parameters $\Omega_{\rm dm}^{\,z}=\bar{\rho}_{\rm dm}/\rho_{\rm c}$
and $\Omega_{\rm b}^{\,z}=\bar{\rho}_{\rm b}/\rho_{\rm c}$, where
$\bar{\rho}_{\rm dm}$ is the average dark matter density,
$\bar{\rho}_{\rm b}$ is the average baryonic density, $\rho_{\rm c}$
is the critical density, and $\Omega_{\rm dm}^{\,z} + \Omega_{\rm
b}^{\,z} = \Ommz$ is given by equation~(\ref{Ommz}). We also assume
spatial fluctuations in the gas and dark matter densities with the
form of a single spherical Fourier mode on a scale much smaller than
the horizon,
\begin{eqnarray}
\frac{\rho_{\rm dm}(r,t)-\bar{\rho}_{\rm dm}(t)}
{\bar{\rho}_{\rm dm}(t)} & = &
\delta_{\rm dm}(t) \frac{\sin(kr)}{kr}\ , \\
\frac{\rho_{\rm b}(r,t)-\bar{\rho}_{\rm b}(t)}
{\bar{\rho}_{\rm b}(t)} & = &
\delta_{\rm b}(t) \frac{\sin(kr)}{kr}\ ,
\end{eqnarray}
where $\bar{\rho}_{\rm dm}(t)$ and $\bar{\rho}_{\rm b}(t)$ are the
background densities of the dark matter and baryons, $\delta_{\rm
dm}(t)$ and $\delta_{\rm b}(t)$ are the dark matter and baryon
overdensity amplitudes, $r$ is the comoving radial coordinate, and $k$
is the comoving perturbation wavenumber.  We adopt an ideal gas
equation-of-state for the baryons with a specific heat ratio
$\gamma$=$5/3$.  Initially, at time $ t=t_{\rm i}$, the gas
temperature is uniform $ T_{\rm b}(r,t_{\rm i})$=$ T_{\rm i}$, and the
perturbation amplitudes are small $\delta_{\rm dm,i},\delta_{\rm
b,i}\ll 1$.  We define the region inside the first zero of
$\sin(kr)/(kr)$, namely $0<kr<\pi$, as the collapsing ``object''.

The evolution of the temperature of the baryons $T_{\rm b}(r,t)$ in
the linear regime is determined by the coupling of their free
electrons to the Cosmic Microwave Background (CMB) through Compton
scattering, and by the adiabatic expansion of the gas.  Hence, $
T_{\rm b}(r,t)$ is generally somewhere between the CMB temperature,
$T_{\gamma}\propto (1+z)^{-1}$ and the adiabatically-scaled
temperature $T_{\rm ad}\propto (1+z)^{-2}$.  In the limit of tight
coupling to $T_{\gamma}$, the gas temperature remains uniform.  On the
other hand, in the adiabatic limit, the temperature develops a
gradient according to the relation \beq T_{\rm b} \propto
\rho_b^{(\gamma-1)}.  \eeq

The evolution of dark matter overdensity, $\delta_{\rm dm}(t)$, in the
linear regime is described by the equation (see \S 9.3.2 of Kolb \& Turner
1990),
\begin{equation}
       {\ddot{\delta}}_{\rm dm}
 + 2H{\dot \delta}_{\rm dm}
       ={3\over 2}H^2 \left(\Omega_{\rm b} \delta_{\rm b} +
\Omega_{\rm dm}
        \delta_{\rm dm}\right)\label{dm} 
\end{equation} 
whereas the evolution of the overdensity of the baryons, $\delta_{\rm
b}(t)$, is described by
\begin{equation}
        {\ddot{\delta}}_{\rm b}+
       2H{\dot \delta}_{\rm b} ={3\over 2}H^2
\left(\Omega_{\rm b} \delta_{\rm b} +
       \Omega_{\rm dm} \delta_{\rm dm}\right) -\frac{\kB T_{\rm
       i}}{\mu m_p} \left({k\over a}\right)^2
       \left(\frac{a_{\rm i}}{a} \right)^{(1+\beta)} \left(\delta_{\rm
       b}+{2\over 3}\beta [\delta_{\rm b}-\delta_{\rm
       b,i}]\right)\label{b}.
\end{equation}
Here, $H(t)={\dot a}/a$ is the Hubble parameter at a cosmological time
$t$, and $\mu=1.22$ is the mean molecular weight of the neutral
primordial gas in atomic units. The parameter $\beta$ distinguishes
between the two limits for the evolution of the gas temperature. In
the adiabatic limit $\beta=1$, and when the baryon temperature is
uniform and locked to the background radiation, $\beta=0$. The last
term on the right hand side (in square brackets) takes into account
the extra pressure gradient force in $\nabla(\rho_{\rm b} T)=(T \nabla
\rho_{\rm b}+\rho_{\rm b}\nabla T)$, arising from the temperature
gradient which develops in the adiabatic limit. The Jeans wavelength
$\ljeans=2\pi/\kjeans$ is obtained by setting the right-hand side of
equation~(\ref{b}) to zero, and solving for the critical wavenumber
$\kjeans$.  As can be seen from equation~(\ref{b}), the critical
wavelength $\ljeans$ (and therefore the mass $\mjeans$) is in general
time-dependent.  We infer from equation~(\ref{b}) that as time
proceeds, perturbations with increasingly smaller initial wavelengths
stop oscillating and start to grow.

To estimate the Jeans wavelength, we equate the right-hand-side of
equation~(\ref{b}) to zero. We further approximate $\delta_{\rm b}\sim
\delta_{\rm dm}$, and consider sufficiently high redshifts at which
the universe is matter-dominated and flat (equations~(\ref{highz1})
and (\ref{highz2}) in \S \ref{sec2.1}). We also assume $\Omega_{\rm
b}\ll\Omm$, where $\Omega_m=\Omega_{\rm dm}+\Omega_b$ is the total
matter density parameter. Following cosmological recombination at
$z\approx 10^3$, the residual ionization of the cosmic gas keeps its
temperature locked to the CMB temperature (via Compton scattering)
down to a redshift of (p.\ 179 of Peebles 1993)
\begin{equation}
1+z_t\approx 137 (\Omega_b h^2/0.022)^{2/5}\ .
\end{equation}
In the redshift range between recombination and $z_t$, $\beta=0$ and
\beq 
\kjeans\equiv (2\pi/\ljeans)=[2\kB T_{\gamma}(0)/3\mu
m_p]^{-1/2} {\sqrt{\Omega_m}} H_0\ , 
\eeq 
so that the Jeans mass is therefore redshift independent and obtains the
value (for the total mass of baryons and dark matter)
\begin{equation}
M_{\rm J}\equiv {4\pi\over 3} \left({\ljeans\over 2}\right)^3
{\bar\rho}(0)
= 1.35\times 10^5 \left({\Omega_mh^2\over 0.15}\right)^{-1/2} M_\odot\ .
\end{equation}

Based on the similarity of $\mjeans$ to the mass of a globular
cluster, Peebles \& Dicke (1968) suggested that globular clusters form
as the first generation of baryonic objects shortly after cosmological
recombination. Peebles \& Dicke assumed a baryonic universe, with a
nonlinear fluctuation amplitude on small scales at $z\sim 10^3$, a
model which has by now been ruled out. The lack of a dominant mass of
dark matter inside globular clusters (Moore 1996; Heggie \& Hut 1995)
makes it unlikely that they formed through direct cosmological
collapse, and more likely that they resulted from fragmentation during
the process of galaxy formation. Furthermore, globular clusters have
been observed to form in galaxy mergers (e.g., Miller et al.\ 1997).

At $z\la z_t$, the gas temperature declines adiabatically as
$[(1+z)/(1+z_t)]^2$ (i.e., $\beta=1$) and the total Jeans mass obtains
the value,
\begin{equation}
\mjeans= 5.73\times 10^3\left({\Omega_mh^2\over 0.15}\right)^{-1/2}
\left({\Omega_b h^2\over 0.022}\right)^{-3/5}
\left({1+z\over 10}\right)^{3/2}~M_\odot.
\label{eq:m_j}
\end{equation}
Note that we have neglected Compton drag, i.e., the radiation force which
suppresses gravitational growth of structure in the baryon fluid as long as
the electron abundance is sufficiently high to keep the baryons dynamically
coupled to the photons. After cosmological recombination, the net friction
force on the predominantly neutral fluid decreases dramatically, allowing
the baryons to fall into dark matter potential wells, and essentially
erasing the memory of Compton drag by $z \sim 100$ (e.g., \S 5.3.1. of Hu
1995).

It is not clear how the value of the Jeans mass derived above relates
to the mass of collapsed, bound objects. The above analysis is
perturbative (Eqs.~[\ref{dm}] and [\ref{b}] are valid only as long as
$\delta_{\rm b}$ and $\delta_{\rm dm}$ are much smaller than unity),
and thus can only describe the initial phase of the collapse.  As
$\delta_{\rm b}$ and $\delta_{\rm dm}$ grow and become larger than
unity, the density profiles start to evolve and dark matter shells may
cross baryonic shells (Haiman, Thoul, \& Loeb 1996) due to their
different dynamics. Hence the amount of mass enclosed within a given
baryonic shell may increase with time, until eventually the dark
matter pulls the baryons with it and causes their collapse even
for objects below the Jeans mass.


Even within linear theory, the Jeans mass is related only to the
evolution of perturbations at a given time. When the Jeans mass itself
varies with time, the overall suppression of the growth of
perturbations depends on a time-averaged Jeans mass. Gnedin \& Hui
(1998) showed that the correct time-averaged mass is the filtering
mass $M_F=(4 \pi/3)\, \bar{\rho}\, (2 \pi a/k_F)^3$, in terms of the
comoving wavenumber $k_F$ associated with the ``filtering scale''. The
wavenumber $k_F$ is related to the Jeans wavenumber $\kjeans$ by \beq
\frac{1}{k^2_F (t)}=\frac{1}{D(t)} \int_0^t dt' \, a^2(t')
\frac{\ddot{D} (t')+ 2 H(t') \dot{D}(t')} {k_J^2 (t')} \, \int_{t'}^t
\frac{dt''} {a^2(t'')}\ , \eeq where $D(t)$ is the linear growth
factor (\S \ref{sec2.2}).  At high redshift (where $\Ommz \rightarrow
1$), this relation simplifies to (Gnedin 2000b) \beq \frac{1}{k^2_F
(t)}= \frac{3}{a} \int_0^a \frac{d a'}{k_J^2(a')} \left( 1-\sqrt{
\frac{a'} {a}}~ \right)\ . \eeq Then the relationship between the
linear overdensity of the dark matter $\delta_{\rm dm}$ and the linear
overdensity of the baryons $\delta_b$, in the limit of small $k$, can
be written as (Gnedin \& Hui 1998) \beq \frac{\delta_b} {\delta_{\rm
dm}} = 1-\frac{k^2}{k_F^2}+O (k^4)\ . \eeq

Linear theory specifies whether an initial perturbation, characterized
by the parameters $ k$, $\delta_{\rm dm,i}$, $\delta_{\rm b,i}$ and
$t_{\rm i}$, begins to grow.  To determine the minimum mass of
nonlinear baryonic objects resulting from the shell-crossing and
virialization of the dark matter, we must use a different model which
examines the response of the gas to the gravitational potential of a
virialized dark matter halo.

\subsection{Response of Baryons to Nonlinear Dark Matter Potentials}
\label{sec3.2}

The dark matter is assumed to be cold and to dominate gravity, and so
its collapse and virialization proceeds unimpeded by pressure
effects. In order to estimate the minimum mass of baryonic objects, we
must go beyond linear perturbation theory and examine the baryonic
mass that can accrete into the final gravitational potential well
of the dark matter.

For this purpose, we assume that the dark matter had already
virialized and produced a gravitational potential $\phi({\bf r})$ at a
redshift $z_{\rm vir}$ (with $\phi\rightarrow0$ at large distances,
and $\phi<0$ inside the object) and calculate the resulting
overdensity in the gas distribution, ignoring cooling (an assumption
justified by spherical collapse simulations which indicate that
cooling becomes important only after virialization; see Haiman, Thoul,
\& Loeb 1996).

After the gas settles into the dark matter potential well, it
satisfies the hydrostatic equilibrium equation,
\begin{equation}
\nabla p_{\rm b} = -\rho_{\rm b} \nabla \phi
\label{eq:hyd}
\end{equation}
where $p_{\rm b}$ and $\rho_{\rm b}$ are the pressure and mass density
of the gas.  At $z \la 100$ the gas temperature is decoupled from the
CMB, and its pressure evolves adiabatically (ignoring atomic or
molecular cooling),
\begin{equation}
{p_{\rm b}\over {\bar p}_{\rm b}} = \left({\rho_{\rm b}\over {\bar
\rho}_{\rm b}}\right)^{5/3}
\label{eq:adi}
\end{equation}
where a bar denotes the background conditions. We substitute
equation~(\ref{eq:adi}) into~(\ref{eq:hyd}) and get the solution,
\begin{equation}
{\rho_{\rm b}\over {\bar \rho}_{\rm b}}= \left(1- {2\over 5}{\mu
m_p\phi\over {\kB \bar T}}\right)^{3/2}
\end{equation}
where ${\bar T}= {\bar p}_{\rm b} \mu m_p/(\kB {\bar \rho}_{\rm b})$
is the background gas temperature.  If we define $T_{\rm vir}=
-{1\over 3} \mu m_p\phi/\kB$ as the virial temperature for a potential
depth $-\phi$, then the overdensity of the baryons at the
virialization redshift is
\begin{equation}
\delta_{\rm b} = {\rho_{\rm b}\over {\bar \rho}_{\rm b}} - 1 =
\left(1+
{6\over 5}{T_{\rm vir}\over {\bar T}}\right)^{3/2} - 1 .
\label{eq:del}
\end{equation}
This solution is approximate for two reasons: (i) we assumed that the
gas is stationary throughout the entire region and ignored the
transitions to infall and the Hubble expansion at the interface
between the collapsed object and the background intergalactic medium
(henceforth IGM), and (ii) we ignored entropy production at the
virialization shock surrounding the object.  Nevertheless, the result
should provide a better estimate for the minimum mass of collapsed
baryonic objects than the Jeans mass does, since it incorporates the
nonlinear potential of the dark matter.

We may define the threshold for the collapse of baryons by the
criterion that their mean overdensity, $\delta_{\rm b}$, exceeds a
value of 100, amounting to $\ga 50\%$ of the baryons that would
assemble in the absence of gas pressure, according to the spherical
top-hat collapse model (\S \ref{sec2.3}). Equation~(\ref{eq:del}) then
implies that $T_{\rm vir} > 17.2\, {\bar T}$.

As mentioned before, the gas temperature evolves at $z\la 160$
according to the relation ${\bar T}\approx 170 [(1+z) /100]^2\ {\rm
K}$. This implies that baryons are overdense by $\delta_{\rm b} >
100$ only inside halos with a virial temperature $T_{\rm vir}\ga
2.9\times 10^3~[(1+z)/100]^2\ {\rm K}$. Based on the top-hat model (\S
\ref{sec2.3}), this implies a minimum halo mass for baryonic objects
of
\begin{equation}
M_{\rm min}= 5.0 \times 10^3  \left({\Omega_m h^2\over
0.15}\right)^{-1/2} \left({\Omega_b h^2\over 0.022}\right)^{-3/5}
\left({1+z\over 10}\right)^{3/2}~M_\odot,
\label{eq:M_min}
\end{equation}
where we set $\mu=1.22$ and consider sufficiently high redshifts so
that $\Ommz \approx 1$. This minimum mass is coincidentally almost
identical to the naive Jeans mass calculation of linear theory in
equation~(\ref{eq:m_j}) despite the fact that it incorporates shell
crossing by the dark matter, which is not accounted for by linear
theory. Unlike the Jeans mass, the minimum mass depends on the choice
for an overdensity threshold [taken arbitrarily as $\delta_{\rm
b}>100$ in equation~(\ref{eq:M_min})]. To estimate the minimum halo
mass which produces any significant accretion we set, e.g.,
$\delta_{\rm b}=5$, and get a mass which is lower than $M_{\rm min}$
by a factor of 27.

Of course, once the first stars and quasars form they heat the
surrounding IGM by either outflows or radiation. As a result, the
Jeans mass which is relevant for the formation of new objects changes
(Ostriker \& Gnedin 1997; Gnedin 2000a). The most dramatic change
occurs when the IGM is photo-ionized and is consequently heated to a
temperature of $\sim(1$--$2)\times 10^4$ K. As we discuss in \S
\ref{sec6.5}, this heating episode had a dramatic impact on galaxy
formation.

\subsection{Molecular Chemistry, Photo-Dissociation, and Cooling}
\label{sec3.3}

Before metals are produced, the primary molecule which acquires
sufficient abundance to affect the thermal state of the pristine
cosmic gas is molecular hydrogen, H$_2$. The dominant H$_2$ formation
process is
\begin{eqnarray}\label{equ:H2paths}
\rm
H \  \ \ \  + \ \ e^-  \  & \rightarrow & \ \ {\rm H^-} \ \  +  \ \
h\nu,  \\
\rm H^-  \ \ + \ \ H \ \ & \rightarrow & \ \ \rm H_2  \ \ \  +  \ \ 
e^-,
\end{eqnarray}
where free electrons act as catalysts.  The complete set of chemical
reactions leading to the formation of H$_2$ is summarized in Table~1,
together with the associated rate coefficients (see also Haiman,
Thoul, \& Loeb 1996; Abel et al.\ 1997; Galli \& Palla 1998; and the
review by Abel \& Haiman 2000). Table~2 shows the same for deuterium
mediated reactions.  Due to the low gas density, the chemical
reactions are slow and the molecular abundance is far from its value
in chemical equilibrium.  After cosmological recombination the
fractional H$_2$ abundance is small, $\sim 10^{-6}$ relative to
hydrogen by number (Lepp \& Shull 1984; Shapiro, Giroux \& Babul
1994).  At redshifts $z\ll 100$, the gas temperature in most regions
is too low for collisional ionization to be effective, and free
electrons (over and above the residual electron fraction) are mostly
produced through photoionization of neutral hydrogen by UV or X-ray
radiation.

In objects with baryonic masses $\ga 3\times 10^4M_\odot$, gravity
dominates and results in the bottom-up hierarchy of structure
formation characteristic of CDM cosmologies; at lower masses, gas
pressure delays the collapse.  The first objects to collapse are those
at the mass scale that separates these two regimes.  Such objects
reach virial temperatures of several hundred degrees and can fragment
into stars only through cooling by molecular hydrogen (e.g., Abel
1995; Tegmark et al.\ 1997). In other words, there are two independent
minimum mass thresholds for star formation: the Jeans mass (related to
accretion) and the cooling mass. For the very first objects, the
cooling threshold is somewhat higher and sets a lower limit on the
halo mass of $\sim 5 \times 10^4 M_\odot$ at $z \sim 20$.

However, molecular hydrogen (${\rm H_2}$) is fragile and can easily be
photo-dissociated by photons with energies of $11.26$--$13.6$eV, to
which the IGM is transparent even before it is ionized. The
photo-dissociation occurs through a two-step process, first suggested
by Solomon in 1965 (compare Field et al.\ 1966) and later analyzed
quantitatively by Stecher \& Williams (1967).  Haiman, Rees, \& Loeb
(1997) evaluated the average cross-section for this process between
11.26eV and 13.6eV, by summing the oscillator strengths for the Lyman
and Werner bands of ${\rm H_2}$, and obtained a value of $3.71\times
10^{-18}~{\rm cm^2}$.  They showed that the UV flux capable of
dissociating ${\rm H_2}$ throughout the collapsed environments in the
universe is lower by more than two orders of magnitude than the
minimum flux necessary to ionize the universe.  The inevitable
conclusion is that soon after trace amounts of stars form, the
formation of additional stars due to ${\rm H_2}$ cooling is
suppressed.  Further fragmentation is possible only through atomic
line cooling, which is effective in objects with much higher virial
temperatures, $T_{\rm vir}\ga 10^4$K.  Such objects correspond to a
total mass $\ga 10^8 M_\odot [(1+z)/10]^{-3/2}$. Figure~\ref{fig1d}
illustrates this sequence of events by describing two classes of
objects: those with $T_{\rm vir}< 10^{4}$K (small dots) and those with
$T_{\rm vir}> 10^4$K (large dots).  In the first stage (top panel),
some low-mass objects collapse, form stars, and create ionized
hydrogen (\ion{H}{2}) bubbles around them.  Once the UV background
between 11.2--13.6eV reaches a specific critical level, ${\rm H_2}$ is
photo-dissociated throughout the universe and the formation of new
stars is delayed until objects with $T_{\rm vir}\ga 10^4$K collapse
(Haiman, Abel, \& Rees 2000; Ciardi, Ferrara, \& Abel 2000; Ciardi et
al.\ 2000). Machacek, Bryan \& Abel (2000) have confirmed that the
soft UV background can delay the cooling and collapse of low-mass
halos ($\sim 10^6 M_\odot$) based on analytical arguments and
three-dimensional hydrodynamic simulations; they also determined the
halo mass threshold for collapse for a range of UV fluxes. Omukai \&
Nishi (1999; see also Silk 1977) have argued that the
photo-dissociation of H$_2$ could be even more effective due to a
small number of stars embedded within the gas clouds themselves.
  
\begin{figure}[htbp] 
\epsscale{0.7}
\plotone{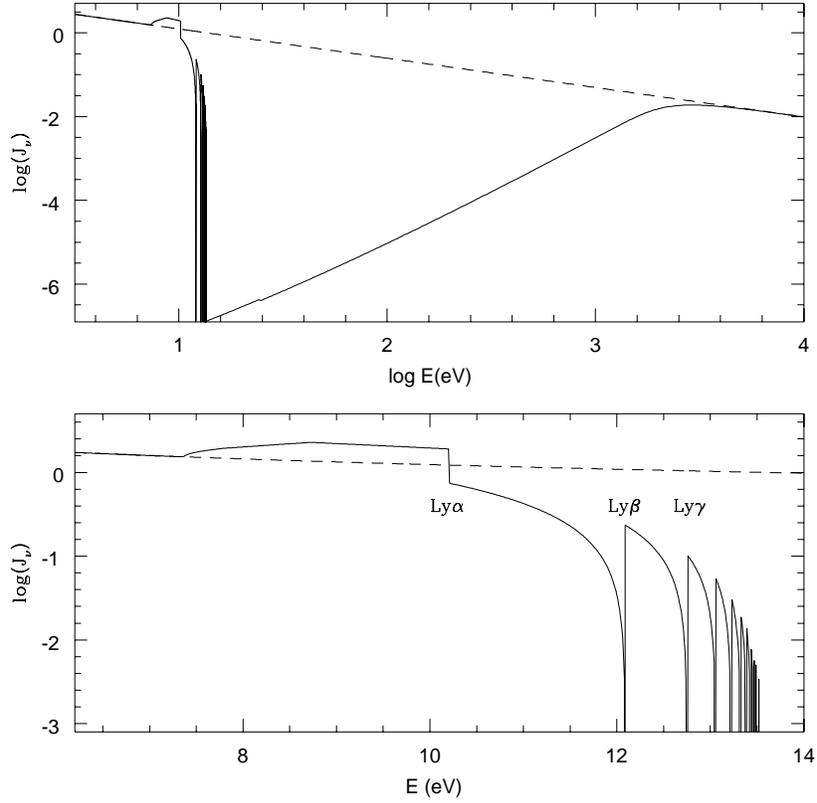}
\caption{The average spectrum during the initial phase of the
reionization epoch.  The upper panel shows that absorption by neutral
hydrogen and helium suppresses the flux above 13.6eV up to the keV
range.  The lower panel shows a close-up of the sawtooth modulation
due to line absorption below 13.6 eV. A constant comoving density of
sources was assumed, with each source emitting a power-law continuum,
which would result in the spectrum shown by the dashed lines if
absorption were not taken into account.}
\label{fig3a}
\end{figure}

When considering the photo-dissociation of H$_2$ before reionization,
it is important to incorporate the {\it processed} spectrum of the UV
background at photon energies below the Lyman limit. Due to the
absorption at the Lyman-series resonances this spectrum obtains the
sawtooth shape shown in Figure~\ref{fig3a}. For any photon energy
above Ly$\alpha$ at a particular redshift, there is a limited redshift
interval beyond which no contribution from sources is possible because
the corresponding photons are absorbed through one of the Lyman-series
resonances along the way. Consider, for example, an energy of 11 eV at
an observed redshift $z=10$. Photons received at this energy would
have to be emitted at the 12.1 eV Ly$\beta$ line from $z=11.1$. Thus,
sources in the redshift interval 10--11.1 could be seen at 11 eV, but
radiation emitted by sources at $z>11.1$ eV would have passed through
the 12.1 eV energy at some intermediate redshift, and would have been
absorbed. Thus, an observer viewing the universe at any photon energy
above Ly$\alpha$ would see sources only out to some horizon, and the
size of that horizon would depend on the photon energy. The number of
contributing sources, and hence the total background flux at each
photon energy, would depend on how far this energy is from the nearest
Lyman resonance. Most of the photons absorbed along the way would be
re-emitted at Ly$\alpha$ and then redshifted to lower energies. The
result is a sawtooth spectrum for the UV background before
reionization, with an enhancement below the Ly$\alpha$ energy (see
Haiman et al.\ 1997 for more details). Unfortunately, the direct
detection of the redshifted sawtooth spectrum as a remnant of the
reionization epoch is not feasible due to the much higher flux
contributed by foreground sources at later cosmic times.

\noindent

\begin{figure}[htbp]
\epsscale{0.7}
\plotone{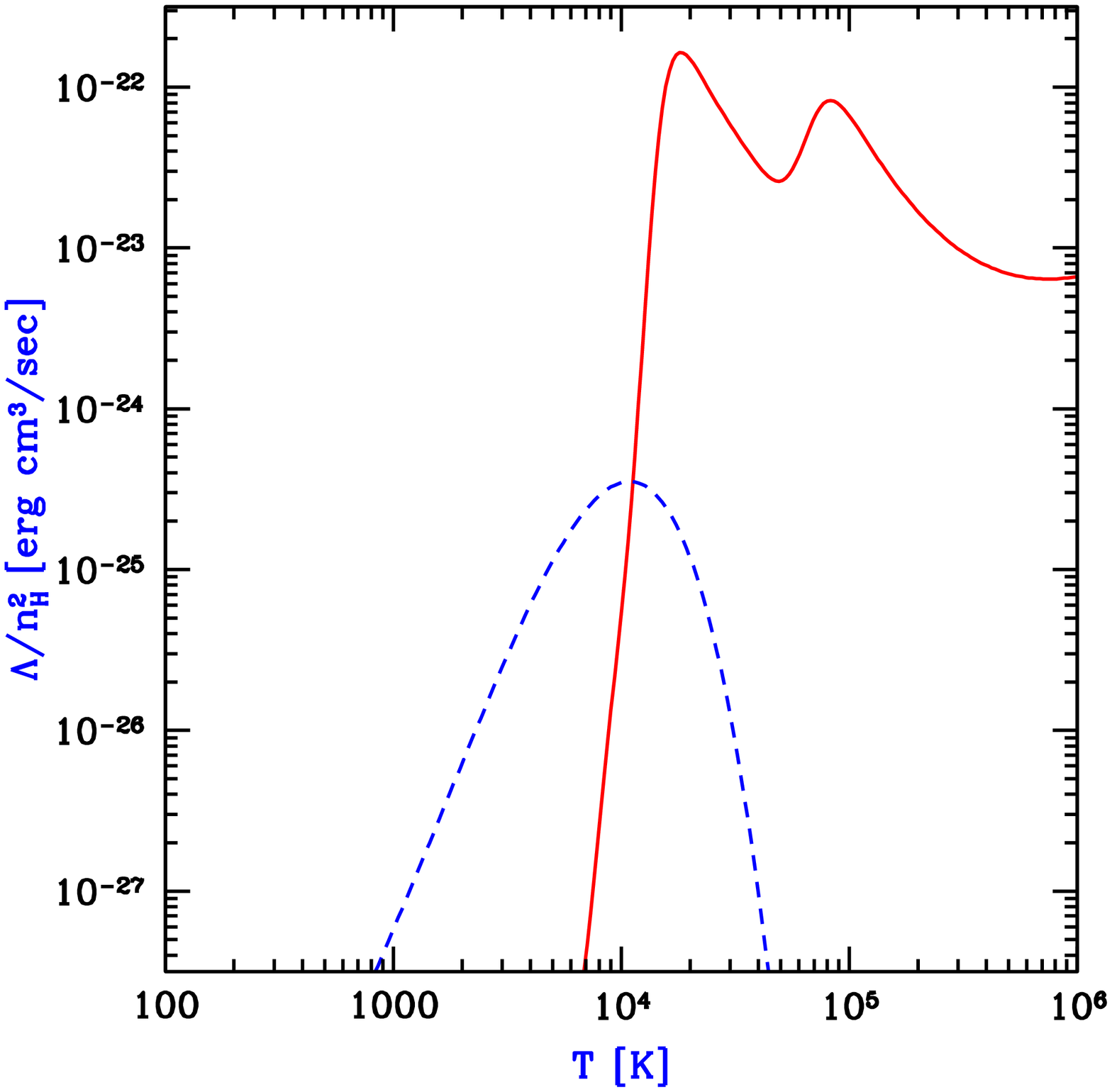}
\caption{Cooling rates as a function of temperature for a primordial
gas composed of atomic hydrogen and helium, as well as molecular
hydrogen, in the absence of any external radiation. We assume a
hydrogen number density $n_H=0.045\ {\rm cm}^{-3}$, corresponding to
the mean density of virialized halos at $z=10$. The plotted quantity
$\Lambda/n_H^2$ is roughly independent of density (unless $n_H \gg 10\
{\rm cm}^{-3}$), where $\Lambda$ is the volume cooling rate (in
erg/sec/cm$^3$). The solid line shows the cooling curve for an atomic
gas, with the characteristic peaks due to collisional excitation of
\ion{H}{1} and \ion{He}{2}. The dashed line (calculated using the code
of Abel available at http://logy.harvard.edu/tabel/PGas/cool.html)
shows the additional contribution of molecular cooling, assuming a
molecular abundance equal to $0.1\%$ of $n_H$.}
\label{cooling}
\end{figure}

\vspace{-0.31in} The radiative feedback on H$_2$ need not be only
negative, however.  In the dense interiors of gas clouds, the
formation rate of H$_2$ could be accelerated through the production of
free electrons by X-rays.  This effect could counteract the
destructive role of H$_2$ photo-dissociation (Haiman, Rees, \& Loeb
1996). Haiman, Abel, \& Rees (2000) have shown that if a significant
($\ga 10\%$) fraction of the early UV background is produced by
massive black holes (mini-quasars) with hard spectra extending to
photon energies $\sim 1\, {\rm keV}$, then the X-rays will catalyze
H$_2$ production and the net radiative feedback will be positive,
allowing low mass objects to fragment into stars. These objects may
greatly alter the topology of reionization (\S \ref{sec6.3}). However,
if such quasars do not exist or if low mass objects are disrupted by
supernova-driven winds (see \S \ref{sec7.2}), then most of the stars
will form inside objects with virial temperatures $\ga 10^4$K, where
atomic cooling dominates.  Figure~\ref{cooling} and Table~3 summarize
the cooling rates as a function of gas temperature in high-redshift,
metal-free objects.

\begin{deluxetable}{rllc} 
 \label{tabH2}
\tablenum{1}
\footnotesize
\tablewidth{13.cm}
\tablecaption{Reaction Rates for Hydrogen Species}
\tablecolumns{4}
\tablehead{
\colhead{} &
\colhead{} &
\colhead{Rate Coefficient} &  
\colhead{} \\ 
\colhead{} &
\colhead{Reaction} &  
\colhead{(cm$^{3}$s$^{-1}$)} &
\colhead{Reference} 
 } 
\startdata
 (1)& H + $e^{-}$ $\rightarrow$ H$^{+}$ + $2e^{-}$ & $5.85\times
10^{-11} 
T^{1/2}\mbox{exp}(-157,809.1/T)(1 + T^{1/2}_{5})^{-1}$ & 1 \nl
 (2)& H$^{+}$ + $e^{-}$ $\rightarrow$ H + $h\nu$ & $8.40\times
10^{-11} 
T^{-1/2}T_{3}^{-0.2}(1 + T^{0.7}_{6})^{-1}$ & 1 \nl
 (3)& H + $e^{-}$ $\rightarrow$ H$^{-}$ + $h\nu$ & See expression in
reference & 2 \nl
 (4)& H + H$^{-}$ $\rightarrow$ H$_{2}$ + $e^{-}$ & $1.30\times
10^{-9}$ 
 & 1 \nl
 (5)& H$^{-}$ + H$^{+}$ $\rightarrow$ 2H & $7.00\times 10^{-7} 
T^{-1/2}$ & 1 \nl
 (6)& H$_{2}$ + $e^{-}$ $\rightarrow$ H + H$^{-}$ & $2.70\times
10^{-8} 
T^{-3/2}\mbox{exp}(-43,000/T)$ & 1 \nl
 (7)& H$_{2}$ + H $\rightarrow$ 3H & See expression in reference & 1
\nl
 (8)& H$_{2}$ + H$^{+}$ $\rightarrow$ H$^{+}_{2}$ + H & $2.40\times
10^{-9} 
\mbox{exp}(-21,200/T)$ & 1 \nl
 (9)& H$_{2}$ + $e^{-}$ $\rightarrow$ 2H + $e^{-}$ & $4.38\times
10^{-10} 
\mbox{exp}(-102,000/T)T^{0.35}$ & 1 \nl
(10)& H$^{-}$ + $e^{-}$ $\rightarrow$ H + $2e^{-}$ & $4.00\times
10^{-12} 
T\,\mbox{exp}(-8750/T)$ & 1 \nl
(11)& H$^{-}$ + H $\rightarrow$ 2H + $e^{-}$ & $5.30\times 10^{-20} 
T\,\mbox{exp}(-8750/T)$ & 1 \nl
(12)& H$^{-}$ + H$^{+}$ $\rightarrow$ H$^{+}_{2}$ + $e^{-}$ & 
See expression in reference & 1 \nl
\enddata
\tablerefs{
(1) Haiman, Thoul, \& Loeb 1996; 
(2) Abel, et al.\ 1997. 
}
\end{deluxetable}



\begin{deluxetable}{rllc} 
 \label{tabD}
\tablenum{2}
\footnotesize
\tablewidth{13.cm}
\tablecaption{Reaction Rates for Deuterium Species}
\tablecolumns{4}
\tablehead{
\colhead{} &
\colhead{} &
\colhead{Rate Coefficient} &  
\colhead{} \\ 
\colhead{} &
\colhead{Reaction} &  
\colhead{(cm$^{3}$s$^{-1}$)} &
\colhead{Reference}
 } 
\startdata
 (1)& D$^{+}$ + $e^{-}$ $\rightarrow$ D + $h\nu$ &  
  $8.40\times 10^{-11} 
T^{-1/2}T_{3}^{-0.2}(1 + T^{0.7}_{6})^{-1}$ & 1 \nl
 (2)& D + H$^{+}$ $\rightarrow$ D$^{+}$ + H  & $3.70\times 10^{-10} 
T^{0.28}\mbox{exp}(-43/T)$ & 3 \nl
 (3)& D$^{+}$ + H $\rightarrow$ D + H$^{+}$ & $3.70\times
10^{-10}T^{0.28}$
& 3 \nl
 (4)& D$^{+}$ + H$_{2}$ $\rightarrow$ H$^{+}$ + HD & $2.10\times
10^{-9}$ 
 & 3 \nl
 (5)& HD + H$^{+}$ $\rightarrow$ H$_{2}$ + D$^{+}$ & $1.00\times
10^{-9} 
\mbox{exp}(-464/T)$ & 3 
\nl
\enddata
\tablerefs{
(1) Haiman, Thoul, \& Loeb 1996; (3) Galli \& Palla 1998.
}
\end{deluxetable}



\begin{deluxetable}{rllc} 
\tablenum{3}
\label{tab3}
\footnotesize
\tablewidth{13.cm}
\tablecaption{Radiative Cooling Processes in the Primordial Gas}
\tablecolumns{4}
\tablehead{
\colhead{} &
\colhead{} &
\colhead{Cooling rate} &  
\colhead{} \\ 
\colhead{} &
\colhead{Cooling due to} &  
\colhead{(erg s$^{-1}$ cm$^{-3}$)} &
\colhead{Reference} \\
 } 
\startdata
 (1)& Molecular hydrogen & See expression in reference
& 1 \nl
 (2)& Deuterium hydride (HD) & See expression in reference
& 2 \nl
 (3)& Atomic H \& He  & See expression in reference
& 3 \nl
 (4)& Compton scattering  & $5.6\times 10^{-36}(1+z)^{4}n_{e}(T-T_{\rm
CMB})$
& 4 \nl
\nl \enddata \tablecomments{ $T$ is the gas temperature in K,
$T_3=T/10^3~{\rm K}$, $T_{5}=T/10^{5}~{\rm K}$, $T_{6}=T/10^{6}~{\rm
K}$, $n_{e}$ is the density of free electrons, $z$ is the redshift, and
$T_{\rm CMB}=2.73\,(1+z)$ K is the temperature of the CMB.}  
\tablerefs{ (1) Galli \& Palla 1998; (2) Flower, Le Bourlot, Pineau
des For\^{e}ts, \& Roueff 2000; (3) Cen 1992; Verner \& Ferland 1996;
Ferland et al.\ 1992; Voronov 1997 (4) Ikeuchi \& Ostriker 1986.  }
\end{deluxetable}

\clearpage


\section{\bf Fragmentation of the First Gaseous Objects}
\label{sec4}

\subsection{Star Formation}
\label{sec4.1}

\subsubsection{Fragmentation into Stars}
\label{sec4.1.1}

As mentioned in the preface, the fragmentation of the first gaseous
objects is a well-posed physics problem with well specified initial
conditions, for a given power-spectrum of primordial density
fluctuations.
This problem is ideally suited for three-dimensional computer
simulations, since it cannot be reliably addressed in idealized 1D or
2D geometries.

Recently, two groups have attempted detailed 3D simulations of the
formation process of the first stars in a halo of $\sim 10^6 M_\odot$
by following the dynamics of both the dark matter and the gas
components, including H$_2$ chemistry and cooling (Deuterium is not
expected to play a significant role; Bromm 2000). Bromm et al.\ (1999)
have used a Smooth Particle Hydrodynamics (SPH) code to simulate the
collapse of a top-hat overdensity with a prescribed solid-body
rotation (corresponding to a spin parameter $\lambda=5\%$) and
additional small perturbations with $P(k)\propto k^{-3}$ added to the
top-hat profile. Abel et al.\ (2000) isolated a high-density filament
out of a larger simulated cosmological volume and followed the
evolution of its density maximum with exceedingly high resolution
using an Adaptive Mesh Refinement (AMR) algorithm.
  
\begin{figure}[htbp] 
\includegraphics{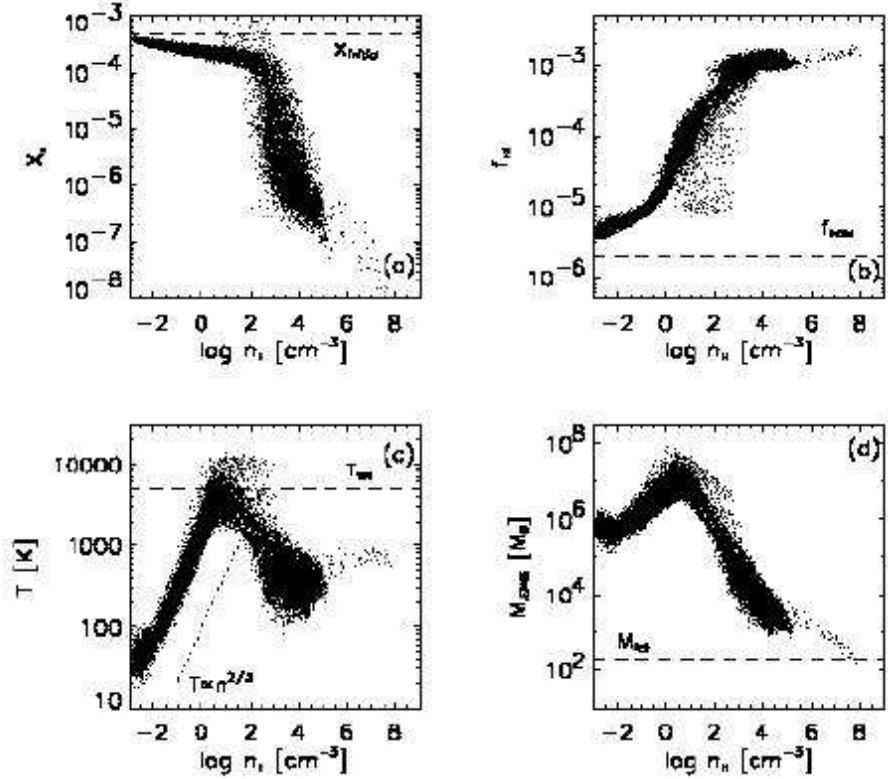}
\vspace{4.2in}
\caption{Numerical results from Bromm et al.\ (1999), showing gas
properties at $z=31.2$ for a collapsing slightly inhomogeneous top-hat
region with a prescribed solid-body rotation. Each point in the figure
is a gas particle in the simulation. {\bf (a)} Free electron fraction
(by number) vs.\ hydrogen number density (in cm$^{-3}$). At densities
exceeding $n\sim 10^{3}$ cm$^{-3}$, recombination is very efficient,
and the gas becomes almost completely neutral.\ {\bf (b)} Molecular
hydrogen fraction vs.\ number density. After a quick initial rise, the
H$_{2}$ fraction approaches the asymptotic value of $f\sim 10^{-3}$,
due to the H$^{-}$ channel.  {\bf (c)} Gas temperature vs.\ number
density. At densities below $\sim 1$ cm$^ {-3}$, the gas temperature
rises because of adiabatic compression until it reaches the virial
value of $T_{vir}\simeq 5000$ K.  At higher densities, cooling due to
H$_{2}$ drives the temperature down again, until the gas settles into
a quasi-hydrostatic state at $T\sim 500$ K and $n\sim 10^{4}$
cm$^{-3}$.  Upon further compression due to accretion and the onset of
gravitational collapse, the gas shows a further modest rise in
temperature. {\bf (d)} Jeans mass (in $M_{\odot}$) vs.\ number
density. The Jeans mass reaches a value of $M_{J}\sim 10^{3}M_{\odot}$
for the quasi-hydrostatic gas in the center of the potential well, and
reaches the resolution limit of the simulation, $M_{\rm res}\simeq 200
M_{\odot}$, for densities close to $n=10^{8}$ cm$^{-3}$.  }
\label{fig4a}
\end{figure}

\begin{figure}[htbp] 
\includegraphics{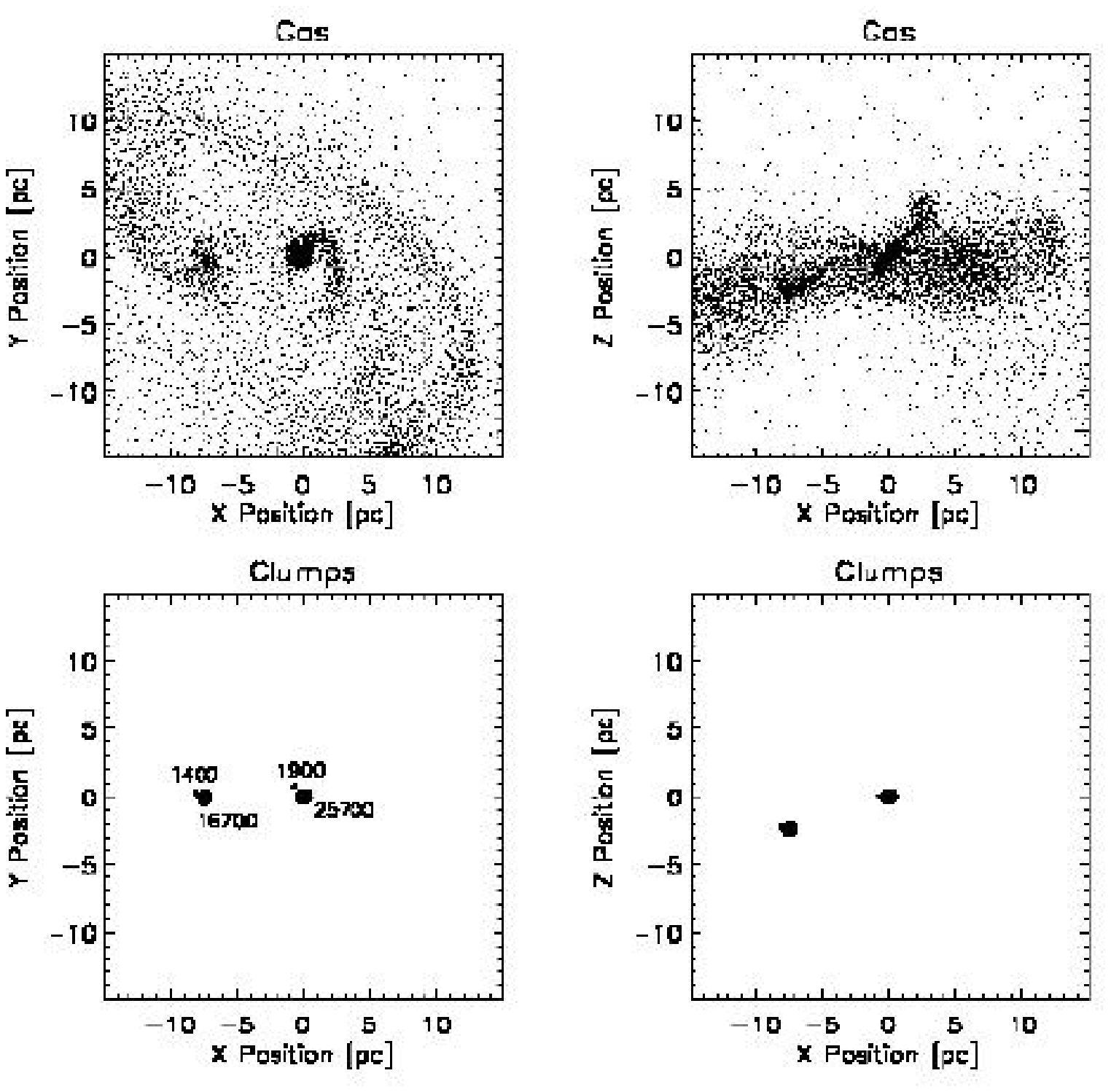}
\vspace{3.7in}
\caption{Gas and clump morphology at $z=28.9$ in the simulation of
Bromm et al.\ (1999).  {\it Top row:} The remaining gas in the diffuse
phase.  {\it Bottom row:} Distribution of clumps. The numbers next to
the dots denote clump mass in units of $M_{\odot}$.  {\it Left
panels:} Face-on view.  {\it Right panels:} Edge-on view.  The length
of the box is 30 pc.  The gas has settled into a flattened
configuration with two dominant clumps of mass close to $20,000
M_{\odot}$. During the subsequent evolution, the clumps survive
without merging, and grow in mass only slightly by accretion of
surrounding gas. }
\label{fig4b}
\end{figure}

The generic results of Bromm et al.\ (1999; see also Bromm 2000) are
illustrated in Figure~\ref{fig4a}. The collapsing region forms a disk
which fragments into many clumps. The clumps have a typical mass $\sim
10^2$--$10^3M_\odot$. This mass scale corresponds to the Jeans mass
for a temperature of $\sim 500$K and the density $\sim 10^4~{\rm
cm^{-3}}$ where the gas lingers because its cooling time is longer
than its collapse time at that point (see Figure~\ref{fig4b}). This
characteristic density is determined by the fact that hydrogen
molecules reach local thermodynamic equilibrium at this density. At
lower densities, each collision leads to an excited state and to
radiative cooling, so the overall cooling rate is proportional to the
collision rate, and the cooling time is inversely proportional to the
gas density. Above the density of $\sim 10^4~{\rm cm^{-3}}$, however,
the relative occupancy of each excited state is fixed at the thermal
equilibrium value (for a given temperature), and the cooling time is
nearly independent of density (e.g., Lepp \& Shull 1983). Each clump
accretes mass slowly until it exceeds the Jeans mass and collapses at
a roughly constant temperature (i.e., isothermally) due to H$_2$
cooling. The clump formation efficiency is high in this simulation due
to the synchronized collapse of the overall top-hat perturbation.

\noindent
  
\noindent
 
\vspace{-0.61in} Bromm (2000, Chapter 7) has simulated the collapse of
one of the above-mentioned clumps with $\sim 1000 M_\odot$ and
demonstrated that it does not tend to fragment into sub-components.
Rather, the clump core of $\sim 100M_\odot$ free-falls towards the
center leaving an extended envelope behind with a roughly isothermal
density profile.  At very high gas densities, three-body reactions
become important in the chemistry of H$_2$.  Omukai \& Nishi (1998)
have included these reactions as well as radiative transfer and
followed the collapse in spherical symmetry up to stellar densities.
Radiation pressure from nuclear burning at the center is unlikely to
reverse the infall as the stellar mass builds up. These calculations
indicate that each clump may end up as a single massive star; however,
it is possible that angular momentum or nuclear burning may eventually
halt the monolithic collapse and lead to further fragmentation.

The Jeans mass (\S \ref{sec3.1}), which is defined based on small
fluctuations in a background of {\it uniform}\/ density, does not
strictly apply in the context of collapsing gas cores. We can instead
use a slightly modified critical mass known as the Bonnor-Ebert mass
(Bonnor 1956; Ebert 1955). For baryons in a background of uniform
density $\rho_b$, perturbations are unstable to gravitational collapse
in a region more massive than the Jeans mass \beq M_J=2.9\,
\frac{1}{\sqrt{\rho_b}}\,\left( \frac{k T} {G \mu m_p} \right)^{3/2}\
. \label{MJb} \eeq Instead of a uniform background, we consider a
spherical, non-singular, isothermal, self-gravitating gas in
hydrostatic equilibrium, i.e., a centrally-concentrated object which
more closely resembles the gas cores found in the above-mentioned
simulations. We consider a finite sphere in equilibrium with an
external pressure. In this case, small fluctuations are unstable and
lead to collapse if the sphere is more massive than the Bonnor-Ebert
mass $M_{\rm BE}$, given by the same expression as equation
(\ref{MJb}) but with a different coefficient (1.2 instead of 2.9) and
with $\rho_b$ denoting in this case the gas (volume) density at the
surface of the sphere.

\begin{figure}[htbp] 
\includegraphics{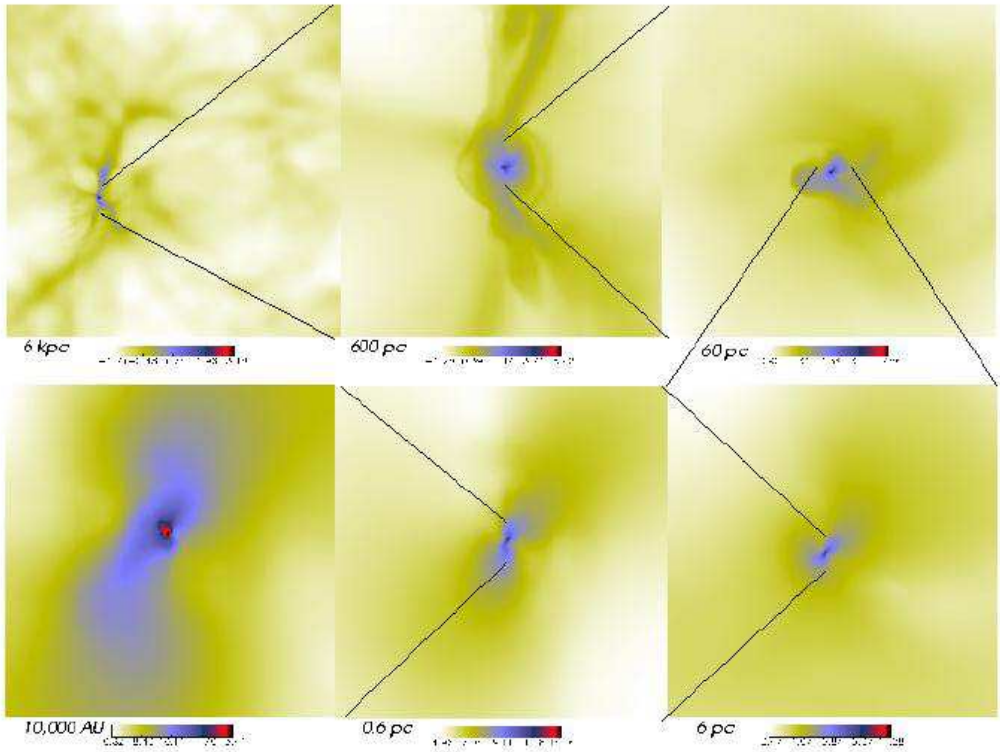}
\vspace{6.2in}
\caption{Zooming in on the core of a star forming region with the {\it
Adaptive Mesh Refinement} simulation of Abel et al.\ (2000). The panels
show different length scales, decreasing clockwise by an order of
magnitude between adjacent panels. Note the large dynamic range of
scales which are being resolved, from 6 kpc (top left panel) down to
10,000 AU (bottom left panel).}
\label{fig4c}
\end{figure}

\begin{figure}[htbp] 
\includegraphics{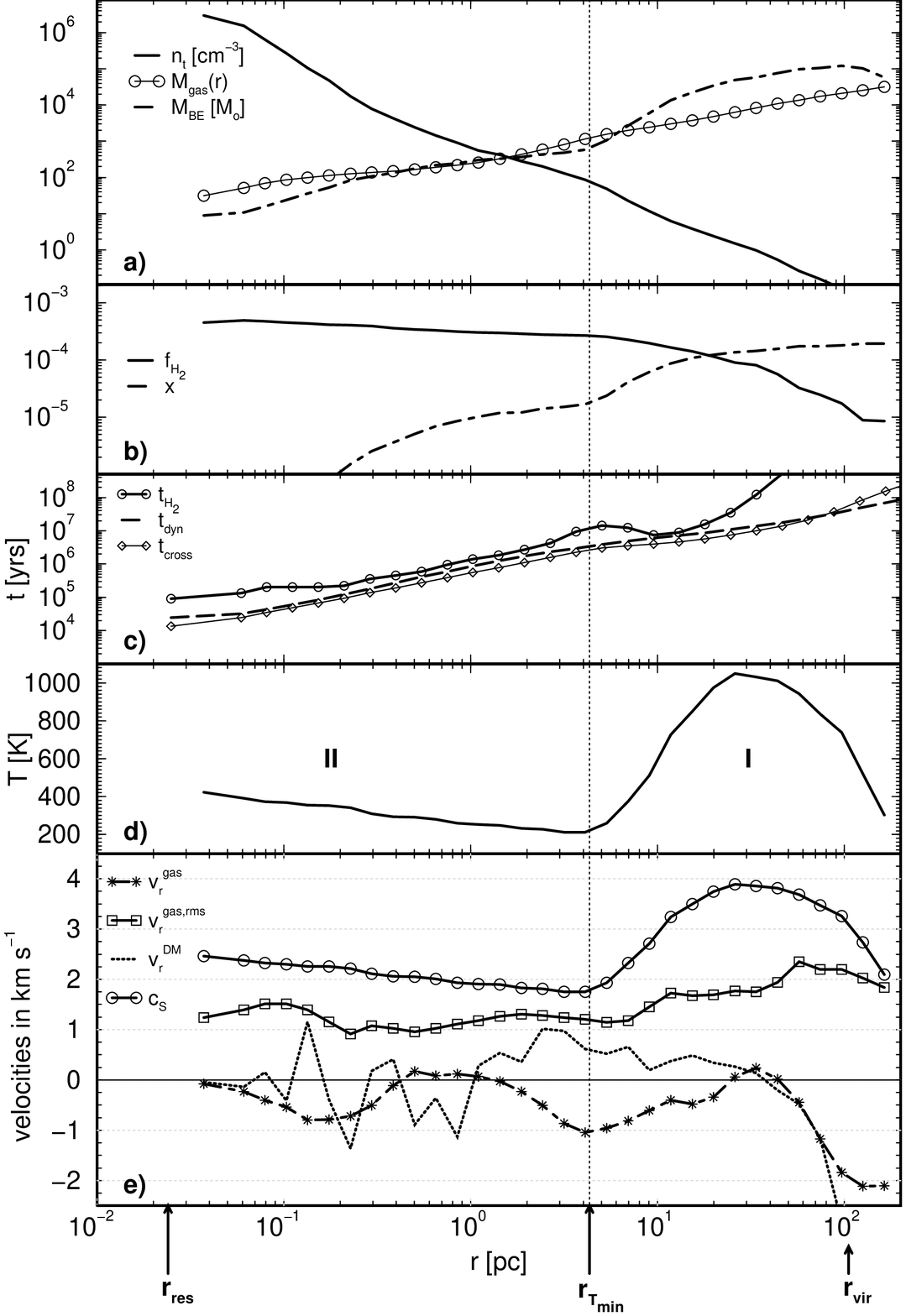}
\vspace{6.1in}
\caption{Gas profiles from the simulation of Abel et al.\ (2000).  The
cell size on the finest grid corresponds to $0.024$ pc, while the
simulation box size corresponds to 6.4 kpc.  Shown are
spherically-averaged mass-weighted profiles around the baryon density
peak shortly before a well defined fragment forms ($z=19.1$). Panel
(a) shows the baryonic number density (solid line), enclosed gas mass
in solar mass (thin solid line with circles), and the local
Bonnor-Ebert mass $M_{\rm BE}$ (dashed line; see text).
Panel (b) plots the molecular hydrogen fraction (by number) $f_{\rm
H_2}$ (solid line) and the free electron fraction $x$ (dashed
line). The H$_2$ cooling time, $t_{\rm H_2}$, the time it takes a
sound wave to travel to the center, $t_{\rm cross}$, and the
free-fall time $t_{\rm ff}=[3\pi/(32G\rho )]^{1/2}$ are given in
panel (c). Panel (d) gives the temperature in K as a function of
radius.  The bottom panel gives the local sound speed, $c_s$ (solid
line with circles), the root-mean-square radial velocities of the dark
matter (dashed line) and the gas (dashed line with asterisks) as well
as the root-mean-square gas velocity (solid line with square
symbols). The vertical dotted line indicates the radius ($\sim 5$ pc)
at which the gas has reached its minimum temperature allowed by H$_2$
cooling. The virial radius of the $5.6\times 10^{6}M_\odot$ halo is
106 pc.  }
\label{fig4d}
\end{figure}

In their simulation, Abel et al.\ (2000) adopted the actual
cosmological density perturbations as initial conditions. The
simulation focused on the density peak of a filament within the IGM,
and evolved it to very high densities (Figure~\ref{fig4c}). Following
the initial collapse of the filament, a clump core formed with $\sim
200M_\odot$, amounting to only $\sim 1\%$ of the virialized gas mass.
Subsequently due to slow cooling, the clump collapsed subsonically in
a state close to hydrostatic equilibrium (see Figure~\ref{fig4d}).
Unlike the idealized top-hat simulation of Bromm et al.\ (2000), the
collapse of the different clumps within the filament is not
synchronized.  Once the first star forms at the center of the first
collapsing clump, it is likely to affect the formation of other stars
in its vicinity.

\noindent
  
\noindent
  
\vspace{-0.61in}
If the clumps in the above simulations end up forming individual very
massive stars, then these stars will likely radiate copious amounts of
ionizing radiation (Carr, Bond, \& Arnett 1984; Tumlinson \& Shull
2000; Bromm et al.\ 2000) and expel strong winds.  Hence, the stars
will have a large effect on their interstellar environment, and
feedback is likely to control the overall star formation efficiency.
This efficiency is likely to be small in galactic potential wells
which have a virial temperature lower than the temperature of
photoionized gas, $\sim 10^4$K. In such potential wells, the gas may
go through only a single generation of star formation, leading to a
``suicidal'' population of massive stars.

The final state in the evolution of these stars is uncertain; but if
their mass loss is not too extensive, then they are likely to end up
as black holes (Bond, Carr, \& Arnett 1984; Fryer, Woosley, \& Heger
2001). The remnants may provide the seeds of quasar black holes
(Larson 1999).  Some of the massive stars may end their lives by
producing $\gamma$-ray bursts. If so then the broad-band afterglows of
these bursts could provide a powerful tool for probing the epoch of
reionization (Lamb \& Reichart 2000; Ciardi \& Loeb 2000).  There is
no better way to end the dark ages than with $\gamma$-ray burst
fireworks.

{\it Where are the first stars or their remnants located today?} The
very first stars formed in rare high-$\sigma$ peaks and hence are
likely to populate the cores of present-day galaxies (White \&
Springel 1999). However, the star clusters which formed in
low-$\sigma$ peaks at later times are expected to behave similarly to
the collisionless dark matter particles and populate galaxy halos
(Loeb 1998).

\subsubsection{Emission Spectrum of Metal-Free Stars}
\label{sec4.1.2}

The evolution of metal-free (Population III) stars is qualitatively
different from that of enriched (Population I and II) stars. In the
absence of the catalysts necessary for the operation of the CNO cycle,
nuclear burning does not proceed in the standard way. At first,
hydrogen burning can only occur via the inefficient PP chain. To
provide the necessary luminosity, the star has to reach very high
central temperatures ($T_{c}\simeq 10^{8.1}$ K). These temperatures
are high enough for the spontaneous turn-on of helium burning via the
triple-$\alpha$ process. After a brief initial period of
triple-$\alpha$ burning, a trace amount of heavy elements
forms. Subsequently, the star follows the CNO cycle. In constructing
main-sequence models, it is customary to assume that a trace mass
fraction of metals ($Z\sim 10^{-9}$) is already present in the star
(El Eid et al.\ 1983; Castellani et al.\ 1983).

\begin{figure}[htbp] 
\includegraphics{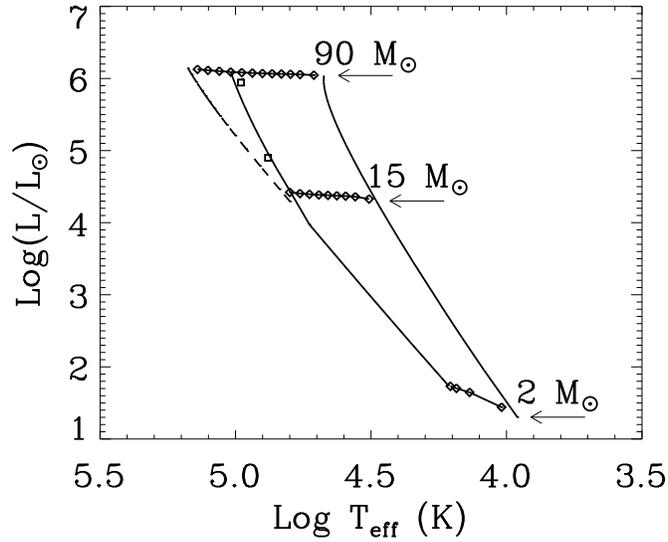}
\vspace{3.3in}
\caption{Luminosity vs.\ effective temperature for zero-age main
sequences stars in the mass range of $2$--$90M_\odot$ (from Tumlinson
\& Shull 2000). The curves show Pop I ($Z_{\odot}$ = 0.02, on the
right) and Pop III stars (on the left) in the mass range 2--90
\Msun. The diamonds mark decades in metallicity in the approach to $Z
= 0$ from 10$^{-2}$ down to 10$^{-5}$ at 2 \Msun, down to 10$^{-10}$
at 15 \Msun, and down to 10$^{-13}$ at 90 \Msun. The dashed line along
the Pop III zero-age main sequence assumes pure H-He composition,
while the solid line (on the left) marks the upper MS with $Z_{\rm C}
= 10^{-10}$ for the $M \geq 15$ \Msun\ models. Squares mark the points
corresponding to pre-enriched evolutionary models from El Eid et al.\
(1983) at 80 \Msun\ and from Castellani et al.\ (1983) at 25 \Msun.
}
\label{fig4e}
\end{figure}

\begin{figure}[htbp] 
\includegraphics{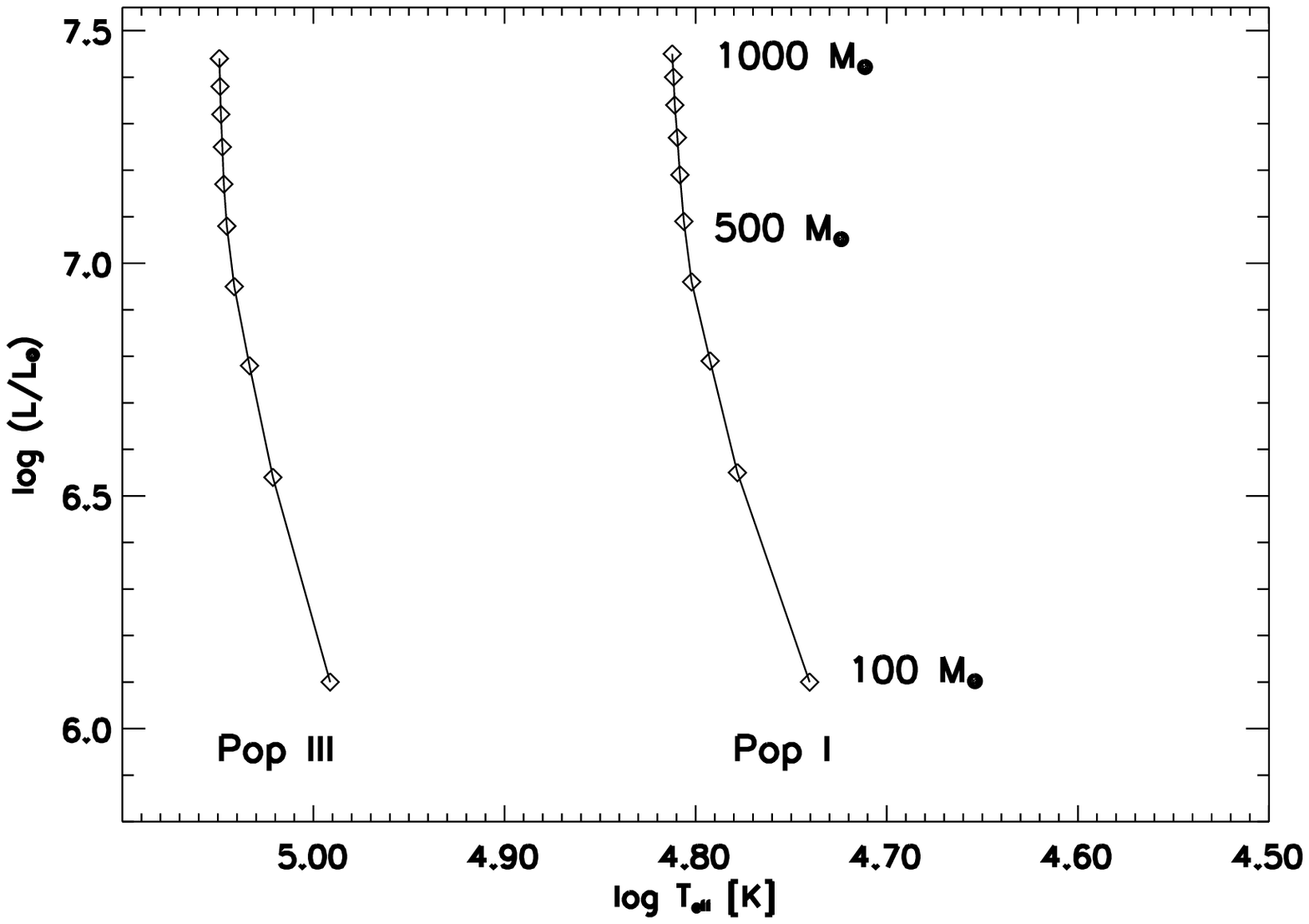}
\vspace{3.3in}
\caption{Same as Figure~\ref{fig4e} but for very massive stars above
$100M_\odot$ (from Bromm et al.\ 2000).  {\it Left solid line:} Pop
III zero-age main sequence (ZAMS).  {\it Right solid line:} Pop I
ZAMS. In each case, stellar luminosity (in $L_{\odot}$) is plotted
vs.\ effective temperature (in K).  {\it Diamond-shaped symbols:}
Stellar masses along the sequence, from $100 M_{\odot}$ (bottom) to
$1000 M_{\odot}$ (top) in increments of $100 M_{\odot}$.  The Pop III
ZAMS is systematically shifted to higher effective temperature, with a
value of $\sim 10^{5}$ K which is approximately independent of
mass. The luminosities, on the other hand, are almost identical in the
two cases.  }
\label{fig4f}
\end{figure}

Figures~\ref{fig4e} and \ref{fig4f} show the luminosity $L$ vs.\
effective temperature $T$ for zero-age main sequence stars in the mass
ranges of $2$--$90M_\odot$ (Figure~\ref{fig4e}) and $100$--$1000
M_\odot$ (Figure~\ref{fig4f}). Note that above $\sim 100M_\odot$ the
effective temperature is roughly constant, $T_{\rm eff}\sim 10^5$K,
implying that the spectrum is independent of the mass distribution of
the stars in this regime (Bromm et al.\ 2000). As is evident from
these Figures (see also Tumlinson \& Shull 2000), both the effective
temperature and the ionizing power of metal-free (Pop III) stars are
substantially larger than those of metal-rich (Pop I) stars.
Metal-free stars with masses $\ga 20M_\odot$ emit between $10^{47}$
and $10^{48}$ \ion{H}{1} and \ion{He}{1} ionizing photons per second
per solar mass of stars, where the lower value applies to stars of
$\sim 20M_\odot$ and the upper value applies to stars of $\ga 100
M_{\odot}$ (see Tumlinson \& Shull 2000 and Bromm et al.\ 2000 for
more details). These massive stars produce $10^4$--$10^5$ ionizing
photons per stellar baryon over a lifetime of $\sim 3\times 10^6$
years [which is much shorter than the age of the universe, equation
(\ref{highz2}) in \S \ref{sec2.1}]. However, this powerful UV emission
is suppressed as soon as the interstellar medium out of which new
stars form is enriched by trace amounts of metals. Even though the
collapsed fraction of baryons is small at the epoch of reionization,
it is likely that most of the stars responsible for the reionization
of the universe formed out of enriched gas.

\noindent
  
\noindent
  
\begin{figure}[htbp] 
\includegraphics{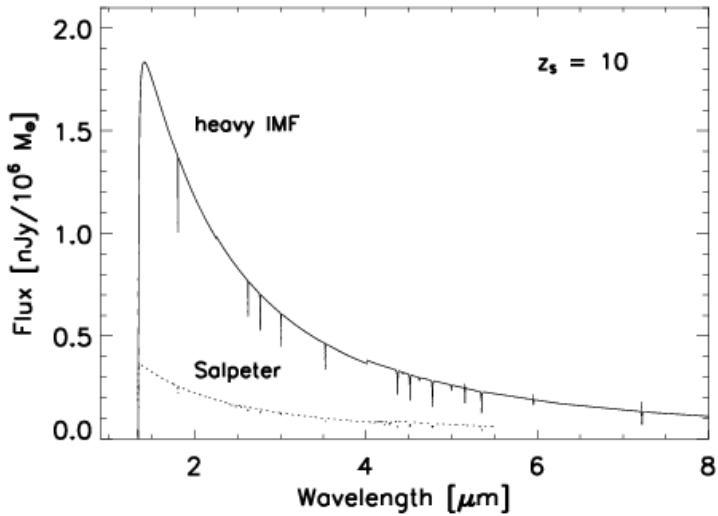}
\vspace{3.3in}
\caption{ Comparison of the predicted flux from a Pop III star cluster
at $z_{s}=10$ for a Salpeter IMF (Tumlinson \& Shull 2000) and a
massive IMF (Bromm et al.\ 2000).  Plotted is the observed flux (in
$\mbox{nJy}$ per $10^{6}M_{\odot}$ of stars) vs.\ observed wavelength
(in $\mu$m) for a flat universe with $\Omega_{\Lambda}=0.7$ and
$h=0.65$.  {\it Solid line:} The case of a heavy IMF.  {\it Dotted
line:} The fiducial case of a standard Salpeter IMF. The cutoff below
$\lambda_{obs} = 1216\mbox{\,\AA \,} (1+z_{s})= 1.34\mu$m is due to
Gunn-Peterson absorption. (The cutoff has been slightly smoothed here
by the damping wing of the Ly$\alpha$ line, with reionization assumed
to occur at $z=7$; see \S \ref{sec9.1.1} for details.) Clearly, for
the same total stellar mass, the observable flux is larger by an order
of magnitude for stars which are biased towards having masses $\ga 100
M_\odot\,$.  }
\label{fig4g}
\end{figure}

\vspace{-0.61in}
{\it Will it be possible to infer the initial mass function (IMF) of
the first stars from spectroscopic observations of the first
galaxies?}  Figure~\ref{fig4g} compares the observed spectrum from a
Salpeter IMF ($dN_\star/dM\propto M^{-2.35}$) and a heavy IMF (with
all stars more massive than $100M_\odot$) for a galaxy at
$z_s=10$. The latter case follows from the assumption that each of the
dense clumps in the simulations described in the previous section ends
up as a single star with no significant fragmentation or mass
loss. The difference between the plotted spectra cannot be confused
with simple reddening due to normal dust. Another distinguishing
feature of the IMF is the expected flux in the hydrogen and helium
recombination lines, such as Ly$\alpha$ and \ion{He}{2} 1640 \AA, from the
interstellar medium surrounding these stars. We discuss this next.

\noindent
  
\subsubsection{Emission of Recombination Lines from the First Galaxies}
\label{sec4.1.3}

The hard UV emission from a star cluster or a quasar at high redshift
is likely reprocessed by the surrounding interstellar medium,
producing very strong recombination lines of hydrogen and helium (Oh
1999; Tumlinson \& Shull 2000; see also Baltz, Gnedin \& Silk
1998). We define $\dot{N}_{\rm ion}$ to be the production rate per
unit stellar mass of ionizing photons by the source. The emitted
luminosity $L_{\rm line}^{\rm em}$ per unit stellar mass in a
particular recombination line is then estimated to be
\begin{equation}
L_{\rm line}^{\rm em} = p_{\rm line}^{\rm em} h\nu \dot{N}_{\rm ion}
(1 - p^{\rm esc}_{\rm cont}) p^{\rm esc}_{\rm line} \mbox{\ \ \ ,}
\end{equation}
where $p_{\rm line}^{\rm em}$ is the probability that a recombination
leads to the emission of a photon in the corresponding line, $\nu$ is
the frequency of the line and $p^{\rm esc}_{\rm cont}$ and $p^{\rm
esc}_{\rm line}$ are the escape probabilities for the ionizing photons
and the line photons, respectively. It is natural to assume that the
stellar cluster is surrounded by a finite \ion{H}{2} region, and hence
that $p^{\rm esc}_{\rm cont}$ is close to zero (Wood \& Loeb 2000;
Ricotti \& Shull 2000).  In addition, $p^{\rm esc}_{\rm line}$ is
likely close to unity in the \ion{H}{2} region, due to the lack of
dust in the ambient metal-free gas. Although the emitted line photons
may be scattered by neutral gas, they diffuse out to the observer and
in the end survive if the gas is dust free. Thus, for simplicity, we
adopt a value of unity for $p^{\rm esc}_{\rm line}$ (two-photon decay
is generally negligible as a way of losing line photons in these
environments).

As a particular example we consider case B recombination which yields
$p_{\rm line}^{\rm em}$ of about 0.65 and 0.47 for the Ly${\alpha}$
and \ion{He}{2} 1640\,\AA \,lines, respectively. These numbers
correspond to an electron temperature of $\sim 3\times 10^4$K and an
electron density of $\sim 10^{2}-10^{3}$ cm$^{-3}$ inside the
\ion{H}{2} region (Storey \& Hummer 1995). For example, we consider
the extreme and most favorable case of metal-free stars all of which
are more massive than $\sim 100M_\odot$. In this case $L_{\rm
line}^{\rm em} = 1.7\times 10^{37}$ and $2.2\times 10^{36}$ erg
s$^{-1}M_{\odot}^{-1}$ for the recombination luminosities of
Ly$\alpha$ and \ion{He}{2} 1640\,\AA\,per stellar mass (Bromm et al.\
2000). A cluster of $10^{6} M_{\odot}$ in such stars would then
produce 4.4 and 0.6 $\times 10^{9}L_{\odot}$ in the Ly$\alpha$ and
\ion{He}{2} 1640\,\AA\,lines. Comparably-high luminosities would be
produced in other recombination lines at longer wavelengths, such as
\ion{He}{2} 4686\,\AA\,and H$\alpha$ (Oh 2000; Oh, Haiman, \& Rees
2000).

The rest-frame equivalent width of the above emission lines measured
against the stellar continuum of the embedded star cluster at the line
wavelengths is given by
\begin{equation}
W_{\lambda} =\left(\frac{L_{\rm line}^{\rm em}}{L_{\lambda}}\right)
\mbox{\ \ \ ,}
\end{equation}
where $L_{\lambda}$ is the spectral luminosity per unit wavelength of
the stars at the line resonance.  The extreme case of metal-free stars
which are more massive than $100M_\odot$ yields a spectral luminosity
per unit frequency $L_{\nu} = 2.7\times 10^{21}$ and $1.8\times
10^{21}$ erg s$^{-1}$ Hz$^{-1}M_{\odot}^{-1}$ at the corresponding
wavelengths (Bromm et al.\ 2000).  Converting to $L_{\lambda}$, this
yields rest-frame equivalent widths of $W_{\lambda}$ = 3100\,\AA\,and
1100\,\AA\,for Ly$\alpha$ and \ion{He}{2} 1640\,\AA, respectively.
These extreme emission equivalent widths are more than an order of
magnitude larger than the expectation for a normal cluster of hot
metal-free stars with the same total mass and a Salpeter IMF under the
same assumptions concerning the escape probabilities and recombination
(Kudritzki et al.\ 2000). The equivalent widths are, of course, larger
by a factor of $(1+z_{s})$ in the observer frame.  Extremely strong
recombination lines, such as Ly$\alpha$ and \ion{He}{2} 1640\,\AA, are
therefore expected to be an additional spectral signature that is
unique to very massive stars in the early universe. The strong
recombination lines from the first luminous objects are potentially
detectable with \NGST\, (Oh, Haiman, \& Rees 2000).

High-redshift objects could also, in principle, be detected through
their cooling radiation. However, a simple estimate of the radiated
energy shows that it is very difficult to detect the corresponding
signal in practice. As it cools, the gas loses much of its
gravitational binding energy, which is of order $k_B T_{\rm vir}$ per
baryon, with the virial temperature given by equation (\ref{tvir}) in
\S \ref{sec2.3}. Some fraction of this energy is then radiated as
Ly$\alpha$ photons. The typical galaxy halos around the reionization
redshift have $T_{\rm vir} \sim 1$ eV, and this must be compared to
the nuclear energy output of 7 MeV per baryon in stellar
interiors. Clearly, for a star formation efficiency of $\ga 1\%$, the
stellar radiation is expected to be far more energetic than the
cooling radiation. Both forms of energy should come out on a
time-scale of order the dynamical time. Thus, even if the cooling
radiation is concentrated in the Ly$\alpha$ line, its detection is
more promising for low redshift objects, while \NGST\, will only be
able to detect this radiation from the rare 4-$\sigma$ halos (with
masses $\ga 10^11 M_{\sun}$) at $z \sim 10$ (Haiman, Spaans, \&
Quataert 2000; Fardal et al.\ 2000).

\subsection{Black Hole Formation}
\label{sec4.2}

Quasars are more effective than stars in ionizing the intergalactic
hydrogen because (i) their emission spectrum is harder, (ii) the
radiative efficiency of accretion flows can be more than an order of
magnitude higher than the radiative efficiency of a star, and (iii)
quasars are brighter, and for a given density distribution in their
host system, the escape fraction of their ionizing photons is higher
than for stars.

Thus, the history of reionization may have been greatly altered by the
existence of massive black holes in the low-mass galaxies that
populate the universe at high redshifts. For this reason, it is
important to understand the formation of massive black holes (i.e.,
black holes with a mass far greater than a stellar mass). The problem
of black hole formation is not a priori more complicated than the
problem of star formation. Surprisingly, however, the amount of
theoretical work on star formation far exceeds that on massive black
hole formation.  One of the reasons is that stars form routinely in
our interstellar neighborhood where much data can be gathered, while
black holes formed mainly in the distant past at great distances from
our telescopes.  As more information is gathered on the high-redshift
universe, this state of affairs may begin to change.

Here we adopt the view that massive black holes form out of gas and not
through the dynamical evolution of dense stellar systems (see Rees 1984 for
a review of the alternatives).  To form a black hole inside a given dark
matter halo, the baryons must cool.  For most objects, this is only
possible with atomic line cooling at virial temperatures $T_{\rm vir}\ga
10^4$K and thus baryonic masses $\ga 10^7 M_\odot [(1+z)/10]^{3/2}$. After
losing their thermal pressure, the cold baryons collapse and form a thin
disk on a dynamical time (Loeb \& Rasio 1994).  The basic question is then
the following: what fraction of the cold baryons is able to sink to the
very center of the potential well and form a massive black hole?  Just as
for star formation, the main barrier in this process is angular momentum.
The centrifugal force opposes radial infall and keeps the gas in disks at a
typical distance which is 6--8 orders of magnitude larger than the
Schwarzschild radius corresponding to the total gas mass. Eisenstein \&
Loeb (1995b) demonstrated that a small fraction of all objects have a
sufficiently low angular momentum that the gas in them inevitably forms a
compact semi-relativistic disk that evolves to a black hole on a short
viscous time-scale. These low-spin systems are born in special cosmological
environments that exert unusually small tidal torques on them during their
cosmological collapse.  As long as the initial cooling time of the gas is
short and its star formation efficiency is low, the gas forms the compact
disk on a free-fall time. In most systems the baryons dominate gravity
inside the scale length of the disk. Therefore, if the baryons in a
low-spin system acquire a spin parameter which is only one sixth of the
typical value, i.e., an initial rotation speed $\sim (16\%\times 0.05)
\times V_c$, then with angular momentum conservation they would reach
rotational support at a radius $r_{\rm disk}$ and circular velocity $V_{\rm
disk}$ such that $V_{\rm disk}\, r_{\rm disk}\sim (16\%\times 0.05)\, V_c\,
r_{\rm vir}$, where $r_{\rm vir}$ is the virial radius and $V_c$ the
circular velocity of the halo.  Using the relations: $(GM_{\rm halo}/r_{\rm
vir})\sim V_c^2$, and $[G (\Omega_b/\Omega_m) M_{\rm halo}/r_{\rm
disk}]\sim V_{\rm disk}^2$, we get $V_{\rm disk}\sim 18 V_c$.  For $T_{\rm
vir}\sim 10^4$K, the dark matter halo has a potential depth corresponding
to a circular velocity of $V_c \sim 17~{\rm km~s^{-1}}$, and the low-spin
disk attains a characteristic rotation velocity of $V_{\rm disk}\sim
300~{\rm km~s^{-1}}$ (sufficient to retain the gas against supernova-driven
winds), a size $\la \, 1$ pc, and a viscous evolution time which is
extremely short compared to the Hubble time.

Low-spin dwarf galaxies populate the universe with a significant
volume density at high redshift; these systems are eventually
incorporated into higher mass galaxies which form later. For example,
a galactic bulge of $\sim 10^{10} M_\odot$ in baryons forms out of
$\sim 10^{3}$ building blocks of $\sim 10^7M_\odot$ each. In order to
seed the growth of a quasar, it is sufficient that only one of these
systems had formed a low-spin disk that produced a black hole
progenitor. Note that if a low-spin object is embedded in an overdense
region that eventually becomes a galactic bulge, then the black hole
progenitor will sink to the center of the bulge by dynamical friction
in less than a Hubble time (for a sufficiently high mass $\ga 10^6
M_{\sun}\,$; p.\ 428 of Binney \& Tremaine) and seed quasar activity.
Based on the phase-space volume accessible to low-spin systems
($\propto j^3$), we expect a fraction $\sim 6^{-3} = 5\times 10^{-3}$
of all the collapsed gas mass in the universe to be associated with
low-spin disks (Eisenstein \& Loeb 1995b).  However, this is a
conservative estimate.  Additional angular momentum loss due to
dynamical friction of gaseous clumps in dark matter halos (Navarro,
Frenk, \& White 1995) or bar instabilities in self-gravitating disks
(Shlosman, Begelman, \& Frank 1990) could only contribute to the black
hole formation process. The popular paradigm that all galaxies harbor
black holes at their center simply {\it postulates} that in all
massive systems, a small fraction of the gas ends up as a black hole,
but does not explain quantitatively why this fraction obtains its
particular small value.  The above scenario offers a possible physical
context for this result.

If the viscous evolution time is shorter than the cooling time and if
the gas entropy is raised by viscous dissipation or shocks to a
sufficiently high value, then the black hole formation process will go
through the phase of a supermassive star (Shapiro \& Teukolsky 1983,
\S 17; see also Zel'dovich \& Novikov 1971). The existence of angular
momentum (Wagoner 1969) tends to stabilize the collapse against the
instability which itself is due to general-relativistic corrections to
the Newtonian potential (Shapiro \& Teukolsky 1983, \S 17.4). However,
shedding of mass and angular momentum along the equatorial plane
eventually leads to collapse (Bisnovati-Kogan, Zel'dovich \& Novikov
1967; Loeb \& Rasio 1994; Baumgarte \& Shapiro 1999a).  Since it is
convectively unstable (Loeb \& Rasio 1994) and supported by radiation
pressure, a supermassive star should radiate close to the Eddington
limit (with modifications due to rotation; see Baumgarte \& Shapiro
1999b) and generate a strong wind, especially if the gas is enriched
with metals. The thermal$+$wind emission associated with the collapse
of a supermassive star should be short-lived and could account for
only a minority of all observed quasars.

After the seed black hole forms, it is continually fed with gas during
mergers. Mihos \& Hernquist (1996) have demonstrated that mergers tend
to deposit large quantities of gas at the centers of the merging
galaxies, a process which could fuel a starburst or a quasar. If both
of the merging galaxies contain black holes at their centers,
dynamical friction will bring the black holes together. The final
spiral-in of the black hole binary depends on the injection of new
stars into orbits which allow them to extract angular momentum from
the binary (Begelman, Blandford, \& Rees 1980). If the orbital radius
of the binary shrinks to a sufficiently small value, gravitational
radiation takes over and leads to coalescence of the two black holes.
This will provide powerful sources for future gravitational wave
detectors (such as the LISA project; see http://lisa.jpl.nasa.gov).

The fact that black holes are found in low-mass galaxies in the local
universe implies that they are likely to exist also at high
redshift. Local examples include the compact ellipticals M32 and NGC
4486B.  In particular, van der Marel et al.\ (1997) infer a black hole
mass of $\sim 3.4\times 10^6M_\odot$ in M32, which is a fraction $\sim
8\times 10^{-3}$ of the stellar mass of the galaxy, $\sim 4\times 10^8
M_\odot$, for a central mass-to-light ratio of $\gamma_V=2$. In NGC
4486B, Kormendy et al.\ (1997) infer a black hole mass of $6\times
10^8 M_\odot$, which is a fraction $\sim 9\%$ of the stellar mass.

Despite the poor current understanding of the black hole formation
process, it is possible to formulate reasonable phenomenological
prescriptions that fit the quasar luminosity function within the
context of popular galaxy formation models. These prescription are
described in \S \ref{sec8.2.2}.


\section{\bf Galaxy Properties}
\label{sec5}

\subsection{Formation and Properties of Galactic Disks}
\label{sec5.1}

The formation of disk galaxies within hierarchical models of structure
formation was first explored by Fall \& Efstathiou (1980). More
recently, the distribution of disk sizes was derived and compared to
observations by Dalcanton, Spergel, \& Summers (1997) and Mo, Mao, \&
White (1998). Although these authors considered a number of detailed
models, we adopt here the simple model of an exponential disk in a
singular isothermal sphere halo. We consider a halo of mass $M$,
virial radius $r_{\rm vir}$, total energy $E$, and angular momentum
$J$, for which the spin parameter is defined as \beq \lambda \equiv J
|E|^{1/2} G^{-1} M^{-5/2}\ . \eeq The spin parameter simply expresses
the halo angular momentum in a dimensionless form. The gas disk is
assumed to collapse to a state of rotational support in the dark
matter halo. If the disk mass is a fraction $m_d$ of the halo mass and
its angular momentum is a fraction $j_d$ of that of the halo, then the
exponential scale radius of the disk is given by (Mo et al.\ 1998)
\beq R_d=\frac{1}{\sqrt{2}} \left(\frac{j_d}{m_d}\right)
\lambda\,r_{\rm vir}\ . \eeq

The observed distribution of disk sizes suggests that the specific
angular momentum of the disk is similar to that of the halo (e.g.,
Dalcanton et al.\ 1997; Mo et al.\ 1998), and so we assume that
$j_d/m_d=1$. Although this result is implied by observed galactic
disks, its origin in the disk formation process is still unclear. The
formation of galactic disks has been investigated in a large number of
numerical simulations (Navarro \& Benz 1991; Evrard, Summers, \& Davis
1994; Navarro, Frenk, \& White 1995; Tissera, Lambas, \& Abadi 1997;
Navarro \& Steinmetz 1997; Elizondo, et al.\ 1999). The overall
conclusion is that the collapsing gas loses angular momentum to the
dark matter halo during mergers, and the disks which form are much
smaller than observed galactic disks. The most widely discussed
solution for this problem is to prevent the gas from collapsing into a
disk by injecting energy through supernova feedback (e.g. Eke,
Efstathiou, \& Wright 1999; Binney, Gerhard, \& Silk 2001; Efstathiou
2000). However, some numerical simulations suggest that feedback may
not adequately suppress the angular momentum losses (Navarro \&
Steinmetz 2000).

With the assumption that $j_d/m_d=1$, the distribution of disk sizes
is then determined by the Press-Schechter halo abundance and by the
distribution of spin parameters [along with equation~(\ref{rvir}) for
$r_{\rm vir}$]. The spin parameter distribution is approximately
independent of mass, environment, and cosmological parameters,
apparently a consequence of the scale-free properties of the early
tidal torques between neighboring systems responsible for the spin of
individual halos (Peebles 1969; White 1984; Barnes \& Efstathiou 1987;
Heavens \& Peacock 1988; Steinmetz \& Bartelmann 1995; Eisenstein \&
Loeb 1995a; Cole \& Lacey 1996; Catelan \& Theuns 1996). This
distribution approximately follows a lognormal distribution in the
vicinity of the peak, \beq p(\lambda) d\lambda= \frac{1}
{\sigma_{\lambda} \sqrt{2 \pi}} \exp \left [-\frac{ \ln^2(\lambda/
\bar{\lambda})}{2\sigma_{\lambda}^2} \right] \frac{d\lambda}{\lambda}\
, \eeq with $\bar{\lambda}=0.05$ and $\sigma_{\lambda}=0.5$ following
Mo et al.\ (1998), who determined these values based on the N-body
simulations of Warren et al.\ (1992). Although Mo et al.\ (1998)
suggest a lower cutoff on $\lambda$ due to disk instability, it is
unclear if halos with low $\lambda$ indeed cannot contain disks. If a
dense bulge exists, it can prevent bar instabilities, or if a bar
forms it may be weakened or destroyed when a bulge subsequently forms
(Sellwood \& Moore 1999).

\subsection{Phenomenological Prescription for Star Formation}
\label{sec5.2}

Schmidt (1959) put forth the hypothesis that the rate of star
formation in a given region varies as a power of the gas density
within that region. Thus, the star formation rate can be parameterized
as \beq \frac{d \rho_*}{dt} \propto \rho_g^N\ , \label{schmidt}\eeq
where $\rho_*$ is the mass density of stars, and $\rho_g$ is the mass
density of gas. Although Schmidt originally focused on different
regions within our own Galaxy, this relation has since been used to
interpret observations of the global star formation rates in different
galaxies.

One particular value of $N$ is theoretically favored for
self-gravitating disks (e.g., Larson 1992; Elmegreen 1994). The star
formation rate can be written in the form \beq \frac{d \rho_*}{dt}
=\epsilon \frac{\rho_g}{t_c}\ , \eeq where $\epsilon$ is an efficiency
coefficient, and $t_c$ is a characteristic time for star formation. If
$t_c$ is proportional to the dynamical free-fall time, i.e., $t_c
\propto \rho_g^{-1/2}$, then $N=1.5$. However, observations yield
estimates of surface densities $\Sigma$, not volume densities
$\rho$. If the average gas scale height is roughly constant in
different environments, then the same relation as
equation~(\ref{schmidt}) should hold between the surface densities of
stars and gas, with $N=1.5$.

Such a relation has, indeed, been observed to hold over a large range
of physical conditions in galaxies. Synthetic models which include
stellar evolution tracks and stellar atmosphere models are used to
infer star formation rates using spectral observations of stellar
populations. Star formation rates have been inferred in this way in
the disks of normal spiral and irregular galaxies, most often using
H$_{\alpha}$ luminosities. Star formation also occurs in much denser
environments in the nuclear regions of galaxies, where far-infrared
luminosities are most useful for determining star formation
rates. Thus, the relation between star formation and gas density has
been measured over conditions ranging from the outskirts of normal
disks to the central engines of infrared-luminous starburst
galaxies. The result is a tight correlation in accordance with the
Schmidt law, empirically given by (Kennicutt 1998) \beq \Sigma_{SFR} =
(2.5 \pm 0.7) \times 10^{-4} \left[\frac{ \Sigma_{gas}} {1 M_{\sun}
{\rm pc}^{-2}} \right]^{1.4 \pm 0.15} M_{\sun}\, {\rm yr}^{-1}\, {\rm
kpc}^{-2}\ , \eeq where $\Sigma_{SFR}$ and $\Sigma_{gas}$ are the
disk-averaged star formation rate and gas surface densities,
respectively. This relation is observed to hold over almost five
orders of magnitude in gas surface density.


\section{\bf Radiative Feedback from the First Sources of Light}
\label{sec6}

\subsection{Escape of Ionizing Radiation from Galaxies}
\label{sec6.1}

The intergalactic ionizing radiation field, a key ingredient in the
development of reionization, is determined by the amount of ionizing
radiation escaping from the host galaxies of stars and quasars.  The
value of the escape fraction as a function of redshift and galaxy mass
remains a major uncertainty in all current studies, and could affect
the cumulative radiation intensity by orders of magnitude at any given
redshift. Gas within halos is far denser than the typical density of
the IGM, and in general each halo is itself embedded within an
overdense region, so the transfer of the ionizing radiation must be
followed in the densest regions in the universe. Numerical simulations
of reionization are limited in their resolution of the densest regions
and in the accuracy of their treatment of radiative transfer.

The escape of ionizing radiation ($h\nu > 13.6$eV, $\lambda < 912$
{\AA}) from the disks of present-day galaxies has been studied in
recent years in the context of explaining the extensive diffuse
ionized gas layers observed above the disk in the Milky Way (Reynolds
et al.\ 1995) and other galaxies (e.g., Rand 1996; Hoopes, Walterbos,
\& Rand 1999). Theoretical models predict that of order 3--14\% of the
ionizing luminosity from O and B stars escapes the Milky Way disk
(Dove \& Shull 1994; Dove, Shull, \& Ferrara 2000).  A similar escape
fraction of $f_{\rm esc}=6$\% was determined by Bland-Hawthorn \&
Maloney (1999) based on H$\alpha$ measurements of the Magellanic
Stream.  From {\it Hopkins Ultraviolet Telescope} observations of four
nearby starburst galaxies (Leitherer et al.\ 1995; Hurwitz, Jelinsky,
\& Dixon 1997), the escape fraction was estimated to be in the range
3\%$<f_{\rm esc} < 57$\%.  If similar escape fractions characterize
high-redshift galaxies, then stars could have provided a major
fraction of the background radiation that reionized the IGM (e.g.,
Madau \& Shull 1996; Madau 1999).  However, the escape fraction from
high-redshift galaxies, which formed when the universe was much denser
($\rho\propto (1+z)^3$), may be significantly lower than that
predicted by models meant to describe present-day galaxies.  Current
reionization calculations assume that galaxies are isotropic point
sources of ionizing radiation and adopt escape fractions in the range
$5\% < f_{\rm esc} < 60\%$ (see, e.g., Gnedin 2000a, Miralda-Escud\'e
et al.\ 2000).

Clumping is known to have a significant effect on the penetration and
escape of radiation from an inhomogeneous medium (e.g., Boiss\'e 1990;
Witt \& Gordon 1996, 2000; Neufeld 1991; Haiman \& Spaans 1999;
Bianchi et al.\ 2000).  The inclusion of clumpiness introduces several
unknown parameters into the calculation, such as the number and
overdensity of the clumps, and the spatial correlation between the
clumps and the ionizing sources.  An additional complication may arise
from hydrodynamic feedback, whereby part of the gas mass is expelled
from the disk by stellar winds and supernovae (\S \ref{sec7}).

\begin{figure}[htbp] 
\includegraphics{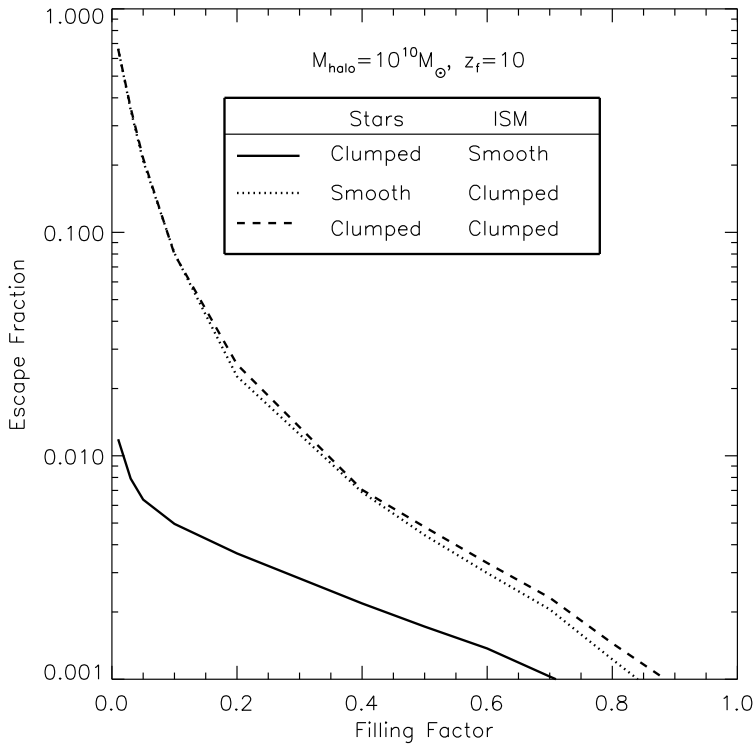}
\vspace{3.65in}
\caption{Escape fractions of stellar ionizing photons from a gaseous
disk embedded within a $10^{10}M_\odot$ halo which formed at $z=10$
(from Wood \& Loeb 2000). The curves show three different cases of
clumpiness within the disk. The volume filling factor refers to either
the ionizing emissivity, the gas clumps, or both, depending on the
case. The escape fraction is substantial ($\ga 1\%$) only if the gas
distribution is highly clumped. (Note: ISM is interstellar medium)}
\label{fig6a}
\end{figure}

Wood \& Loeb (2000) used a three-dimensional radiation transfer code
to calculate the steady-state escape fraction of ionizing photons from
disk galaxies as a function of redshift and galaxy mass. The gaseous
disks were assumed to be isothermal, with a sound speed $c_s\sim
10~{\rm km~s^{-1}}$, and radially exponential, with a scale-length
based on the characteristic spin parameter and virial radius of their
host halos. The corresponding temperature of $\sim 10^4$ K is typical
for a gas which is continuously heated by photo-ionization from stars.
The sources of radiation were taken to be either stars embedded in the
disk, or a central quasar. For stellar sources, the predicted increase
in the disk density with redshift resulted in a strong decline of the
escape fraction with increasing redshift. The situation is different
for a central quasar. Due to its higher luminosity and central
location, the quasar tends to produce an ionization channel in the
surrounding disk through which much of its ionizing radiation escapes
from the host. In a steady state, only recombinations in this
ionization channel must be balanced by ionizations, while for stars
there are many ionization channels produced by individual star-forming
regions and the total recombination rate in these channels is very
high. Escape fractions $\ga 10\%$ were achieved for stars at $z\sim
10$ only if $\sim 90\%$ of the gas was expelled from the disks or if
dense clumps removed the gas from the vast majority ($\ga 80\%$) of
the disk volume (see Figure~\ref{fig6a}). This analysis applies only
to halos with virial temperatures $\ga 10^4$ K. Ricotti \& Shull
(2000) reached similar conclusions but for a quasi-spherical
configuration of stars and gas.  They demonstrated that the escape
fraction is substantially higher in low-mass halos with a virial
temperature $\la 10^4$ K.  However, the formation of stars in such
halos depends on their uncertain ability to cool via the efficient
production of molecular hydrogen (see \S \ref{sec3.3}).

\noindent
  
\vspace{-0.21in} The main uncertainty in the above predictions
involves the distribution of the gas inside the host galaxy, as the
gas is exposed to the radiation released by stars and the mechanical
energy deposited by supernovae.  Given the fundamental role played by
the escape fraction, it is desirable to calibrate its value
observationally.  Recently, Steidel, Pettini, \& Adelberger (2001)
reported a preliminary detection of significant Lyman continuum flux
in the composite spectrum of 29 Lyman break galaxies (LBG) with
redshifts in the range $z = 3.40\pm 0.09$. They co-added the spectra
of these galaxies in order to be able to measure the low flux. Another
difficulty in the measurement comes from the need to separate the
Lyman-limit break caused by the interstellar medium from that already
produced in the stellar atmospheres. After correcting for
intergalactic absorption, Steidel et al.\ (2001) inferred a ratio
between the emergent flux density at 1500\AA~ and 900\AA~ (rest frame)
of $4.6 \pm 1.0$. Taking into account the fact that the stellar
spectrum should already have an intrinsic Lyman discontinuity of a
factor of $\sim 3$--5, but that only $\sim 15$--$20\%$ of the 1500\AA~
photons escape from typical LBGs without being absorbed by dust
(Pettini et al.\ 1998a; Adelberger \& Steidel 2000), the inferred
900\AA~ escape fraction is $f_{\rm esc} \sim 10$--$20\%$. However, the
observed blue spectrum suggests that these 29 particular LBGs may have
a very low dust content, and the escape fraction in these galaxies may
be $50\%$ or higher (Haehnelt \etal 2001). Thus, although the galaxies
in this sample were drawn from the bluest quartile of the LBG spectral
energy distributions, the measurement implies that this quartile may
itself dominate the hydrogen-ionizing background relative to quasars
at $z\sim 3$.

\subsection{Propagation of Ionization Fronts in the IGM}
\label{sec6.2}

The radiation output from the first stars ionizes hydrogen in a
growing volume, eventually encompassing almost the entire IGM within a
single \ion{H}{2} bubble. In the early stages of this process, each
galaxy produces a distinct \ion{H}{2} region, and only when the
overall \ion{H}{2} filling factor becomes significant do neighboring
bubbles begin to overlap in large numbers, ushering in the ``overlap
phase'' of reionization. Thus, the first goal of a model of
reionization is to describe the initial stage, when each source
produces an isolated expanding \ion{H}{2} region.

We assume a spherical ionized volume $V$, separated from the
surrounding neutral gas by a sharp ionization front. In the case of a
stellar ionizing spectrum, most ionizing photons are just above the
hydrogen ionization threshold of 13.6 eV, where the absorption
cross-section is high and a very thin layer of neutral hydrogen is
sufficient to absorb all the ionizing photons. On the other hand, an
ionizing source such as a quasar produces significant numbers of
higher energy photons and results in a thicker transition region.

In the absence of recombinations, each hydrogen atom in the IGM would
only have to be ionized once, and the ionized proper volume $V_p$
would simply be determined by \beq \nb_H V_p=\Ng\ , \eeq where $\nb_H$
is the mean number density of hydrogen and $\Ng$ is the total number
of ionizing photons produced by the source. However, the increased
density of the IGM at high redshift implies that recombinations cannot
be neglected. Indeed, in the case of a steady ionizing source (and
neglecting the cosmological expansion), a steady-state volume would be
reached corresponding to the Str\"{o}mgren sphere, with recombinations
balancing ionizations: \beq \alpha_B \nb_H^2 V_p=\frac{d\, \Ng}{dt}\ ,
\eeq where the recombination rate depends on the square of the density
and on the case B recombination coefficient $\alpha_B=2.6\times
10^{-13}$ cm$^3$ s$^{-1}$ for hydrogen at $T=10^4$ K. The exact
evolution for an expanding \ion{H}{2} region, including a non-steady
ionizing source, recombinations, and cosmological expansion, is given
by (Shapiro \& Giroux 1987) \beq \nb_H\left( \frac{dV_p}{dt}-3 H
V_p\right)= \frac{d\, \Ng}{dt} - \alpha_B \left<n_H^2\right> V_p\
. \label{front} \eeq In this equation, the mean density $\nb_H$ varies
with time as $1/a^3(t)$. A critical feature of the physics of
reionization is the dependence of recombination on the square of the
density. This means that if the IGM is not uniform, but instead the
gas which is being ionized is mostly distributed in high-density
clumps, then the recombination time is very short. This is often dealt
with by introducing a volume-averaged clumping factor $C$ (in general
time-dependent), defined by\footnote{The recombination rate depends on
the number density of electrons, and in using equation~(\ref{clump})
we are neglecting the small contribution caused by partially or fully
ionized helium.} \beq C=\left<n_H^2\right>/\nb_H^2 \label{clump}\
. \eeq

If the ionized volume is large compared to the typical scale of
clumping, so that many clumps are averaged over, then $C$ can be
assumed to be approximately spatially uniform. In general,
equation~(\ref{front}) can be solved by supplementing it with
equation~(\ref{clump}) and specifying $C$. Switching to the comoving
volume $V$, the resulting equation is \beq \frac{dV}{dt}=
\frac{1}{\nb_H^0} \frac{d\, \Ng}{dt}- \alpha_B \frac{C}{a^3} \nb_H^0
V\ , \label{HIIreg} \eeq where the present number density of hydrogen
is \beq \nb_H^0=1.88\times 10^{-7} \left(\frac{\Omega_b
h^2}{0.022}\right)\ {\rm cm}^{-3}\ . \eeq This number density is lower
than the total number density of baryons $\nb_b^0$ by a factor of
$\sim 0.76$, corresponding to the primordial mass fraction of
hydrogen. The solution for $V(t)$ (generalized from Shapiro \& Giroux
1987) around a source which turns on at $t=t_i$ is \beq
V(t)=\int_{t_i}^t \frac{1}{\nb_H^0} \frac{d\, \Ng}{dt'} \,e^{F(t',t)}
dt'\ ,\label{HIIsoln} \eeq where \beq F(t',t)=-\alpha_B \nb_H^0
\int_{t'}^t \frac{C(t'')} {a^3(t'')}\, dt''\
\label{Fgen}. \eeq At high redshift (equations~(\ref{highz1}) and
(\ref{highz2}) in \S \ref{sec2.1}), and with the additional assumption
of a constant $C$, the function $F$ simplifies as follows. Defining
\beq f(t)=a(t)^{-3/2}
\label{foft}\ , \eeq we derive \beq F(t',t)=-\frac{2}{3}\frac{ \alpha_B
\nb_H^0} {\sqrt{\Omm} H_0}\,C \left[f(t')-f(t)\right]=-0.262
\left[f(t') -f(t)\right] \ , \eeq where the last equality assumes
$C=10$ and our standard choice of cosmological parameters: $\Omm=0.3$,
$\Oml=0.7$, and $\Omega_b=0.045$ (see the end of \S \ref{sec1}).
Although this expression for $F(t',t)$ is in general an accurate
approximation at high redshift, in the particular case of the
$\Lambda$CDM model (where $\Omm+\Oml=1$) we get the exact result by
replacing equation~(\ref{foft}) with \beq
f(t)=\sqrt{\frac{1}{a^3}+\frac{1-\Omm}{\Omm}}\ . \label{fLCDM} \eeq

\begin{figure}[htbp]
\epsscale{0.7}
\plotone{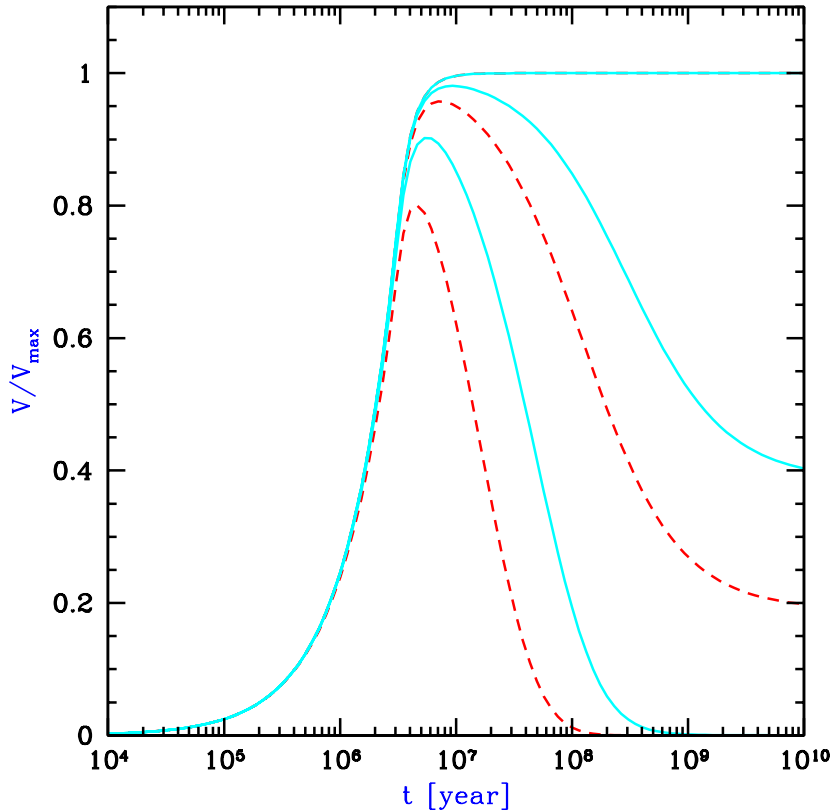}
\caption{Expanding \ion{H}{2} region around an isolated ionizing
source. The comoving ionized volume $V$ is expressed in units of the
maximum possible volume, $V_{\rm max}=4\pi r_{\rm max}^3/3$ [with
$r_{\rm max}$ given in equation~(\ref{rmax})], and the time is
measured after an instantaneous starburst which produces ionizing photons
according to equation~(\ref{zoltan}). We consider a source turning on at
$z=10$ (solid curves) or $z=15$ (dashed curves), with three cases for each:
no recombinations, $C=1$, and $C=10$, in order from top to bottom. The
no-recombination curve is identical for the different source redshifts.}
\label{fig6b}
\end{figure}

The size of the resulting \ion{H}{2} region depends on the halo which
produces it. Consider a halo of total mass $M$ and baryon fraction
$\Omega_b/\Omm$. To derive a rough estimate, we assume that baryons
are incorporated into stars with an efficiency of $f_{\rm star}=10\%$,
and that the escape fraction for the resulting ionizing radiation is
also $f_{\rm esc}=10\%$. If the stellar IMF is similar to the one
measured locally [Scalo 1998; equation (\ref{imf})], then $N_{\gamma}
\approx 4000$ ionizing photons are produced per baryon in stars (for a
metallicity equal to $1/20$ of the solar value). We define a parameter
which gives the overall number of ionizations per baryon, \beq \Ni
\equiv N_{\gamma} \, f_{\rm star}\, f_{\rm esc}\ . \eeq If we neglect
recombinations then we obtain the maximum comoving radius of the
region which the halo of mass $M$ can ionize, \beq r_{\rm max}=
\left(\frac{3}{4\pi}\, \frac{\Ng} {\nb_H^0} \right)^{1/3} =
\left(\frac{3}{4\pi}\, \frac{\Ni} {\nb_H^0}\, \frac{\Omega_b}{\Omm}\,
\frac{M}{m_p} \right)^{1/3}= 675\, {\rm kpc} \left( \frac{\Ni}{40}\,
\frac{M} {10^9 M_{\sun}}\right)^{1/3}\ , \label{rmax} \eeq for our
standard set of parameters. Note that this radius is larger than the
halo virial radius [equation (\ref{rvir})] by a factor of $\sim 20$,
essentially independent of redshift and halo mass. The actual radius
never reaches this size if the recombination time is shorter than the
lifetime of the ionizing source. For an instantaneous starburst with
the Scalo (1998) IMF [equation (\ref{imf})], the production rate of
ionizing photons can be approximated as (Haiman, personal
communication) \beq \frac{d\, \Ng}{dt}=\frac{\alpha-1}{\alpha}
\frac{\Ng} {t_s}\times \left\{
\begin{array}{ll} \ \ \, 1 & \mbox{if
$t<t_s$,} \\ \left(\frac{t}{t_s}\right)^{-\alpha}\ & \mbox{otherwise,}
\end{array} \right. \label{zoltan} \eeq
where $\Ng=4000$, $\alpha=4.5$, and the most massive stars fade away
with the characteristic time-scale $t_s=3\times 10^6$ yr. In
Figure~\ref{fig6b} we show the time evolution of the volume ionized by
such a source, with the volume shown in units of the maximum volume
$V_{\rm max}$ which corresponds to $r_{\rm max}$ in
equation~(\ref{rmax}). We consider a source turning on at $z=10$
(solid curves) or $z=15$ (dashed curves), with three cases for each:
no recombinations, $C=1$, and $C=10$, in order from top to bottom
(Note that the result is independent of redshift in the case of no
recombinations). When recombinations are included, the volume rises
and reaches close to $V_{\rm max}$ before dropping after the source
turns off. At large $t$ recombinations stop due to the dropping
density, and the volume approaches a constant value (although $V \ll
V_{\rm max}$ at large $t$ if $C=10$).

We obtain a similar result for the size of the \ion{H}{2} region
around a galaxy if we consider a mini-quasar rather than stars. For
the typical quasar spectrum (Elvis et al.\ 1994), if we assume a
radiative efficiency of $\sim 6\%$ then roughly 11,000 ionizing
photons are produced per baryon incorporated into the black hole
(Haiman, personal communication). The efficiency of incorporating the
baryons in a galaxy into a central black hole is low ($\la 0.6\%$ in
the local universe, e.g.\ Magorrian et al.\ 1998; see also \S
\ref{sec8.2.2}), but the escape fraction for quasars is likely to be
close to unity, i.e., an order of magnitude higher than for stars (see
\S \ref{sec6.1}). Thus, for every baryon in galaxies, up to $\sim 65$
ionizing photons may be produced by a central black hole and $\sim 40$
by stars, although both of these numbers for $\Ni$ are highly
uncertain. These numbers suggest that in either case the typical size
of \ion{H}{2} regions before reionization may be $\la 1$ Mpc or $\sim
10$ Mpc, depending on whether $10^8 M_{\sun}$ halos or $10^{12}
M_{\sun}$ halos dominate.

\subsection{Reionization of the IGM}
\label{sec6.3}

\subsubsection{Hydrogen Reionization}
\label{sec6.3.1}

In this section we summarize recent progress, both analytic and
numerical, made toward elucidating the basic physics of reionization
and the way in which the characteristics of reionization depend on the
nature of the ionizing sources and on other input parameters of
cosmological models.

The process of the reionization of hydrogen involves several distinct
stages. The initial, ``pre-overlap'' stage (using the terminology of
Gnedin 2000a) consists of individual ionizing sources turning on and
ionizing their surroundings. The first galaxies form in the most
massive halos at high redshift, and these halos are biased and are
preferentially located in the highest-density regions. Thus the
ionizing photons which escape from the galaxy itself (see \S
\ref{sec6.1}) must then make their way through the surrounding
high-density regions, which are characterized by a high recombination
rate. Once they emerge, the ionization fronts propagate more easily
into the low-density voids, leaving behind pockets of neutral,
high-density gas. During this period the IGM is a two-phase medium
characterized by highly ionized regions separated from neutral regions
by ionization fronts.  Furthermore, the ionizing intensity is very
inhomogeneous even within the ionized regions, with the intensity
determined by the distance from the nearest source and by the ionizing
luminosity of this source.

The central, relatively rapid ``overlap'' phase of reionization begins when
neighboring \ion{H}{2} regions begin to overlap. Whenever two ionized
bubbles are joined, each point inside their common boundary becomes exposed
to ionizing photons from both sources. Therefore, the ionizing intensity
inside \ion{H}{2} regions rises rapidly, allowing those regions to expand
into high-density gas which had previously recombined fast enough to remain
neutral when the ionizing intensity had been low. Since each bubble
coalescence accelerates the process of reionization, the overlap phase has
the character of a phase transition and is expected to occur rapidly, over
less than a Hubble time at the overlap redshift. By the end of this stage
most regions in the IGM are able to see several unobscured sources, and
therefore the ionizing intensity is much higher than before overlap and it
is also much more homogeneous. An additional ingredient in the rapid
overlap phase results from the fact that hierarchical structure formation
models predict a galaxy formation rate that rises rapidly with time at the
relevant redshift range. This process leads to a state in which the
low-density IGM has been highly ionized and ionizing radiation reaches
everywhere except for gas located inside self-shielded, high-density
clouds. This marks the end of the overlap phase, and this important
landmark is most often referred to as the 'moment of reionization'.

Some neutral gas does, however, remain in high-density structures
which correspond to Lyman Limit systems and damped Ly$\alpha$ systems
seen in absorption at lower redshifts. The high-density regions are
gradually ionized as galaxy formation proceeds, and the mean ionizing
intensity also grows with time. The ionizing intensity continues to
grow and to become more uniform as an increasing number of ionizing
sources is visible to every point in the IGM. This ``post-overlap''
phase continues indefinitely, since collapsed objects retain neutral
gas even in the present universe. The IGM does, however, reach another
milestone (of limited significance) at $z \sim 1.6$, the breakthrough
redshift (which is determined by the probability of intersecting Lyman
limit systems; Madau, Haardt, \& Rees 1999). Below this redshift, all
ionizing sources are visible to each other, while above this redshift
absorption by the Ly$\alpha$ forest clouds implies that only sources
in a small redshift range are visible to a typical point in the IGM.

Semi-analytic models of the pre-overlap stage focus on the evolution
of the \ion{H}{2} filling factor, i.e., the fraction of the volume of
the universe which is filled by \ion{H}{2} regions. We distinguish
between the naive filling factor $F_{\rm H\ II}$ and the actual
filling factor or porosity $Q_{\rm H\ II}$. The naive filling factor
equals the number density of bubbles times the average volume of each,
and it may exceed unity since when bubbles begin to overlap the
overlapping volume is counted multiple times. However, as explained
below, in the case of reionization the linearity of the physics means
that $F_{\rm H\ II}$ is a very good approximation to $Q_{\rm H\ II}$
up to the end of the overlap phase of reionization.

The model of individual \ion{H}{2} regions presented in the previous
section can be used to understand the development of the total filling
factor. Starting with equation~(\ref{HIIreg}), if we assume a common
clumping factor $C$ for all \ion{H}{2} regions then we can sum each
term of the equation over all bubbles in a given large volume of the
universe, and then divide by this volume. Then $V$ is replaced by the
filling factor and $\Ng$ by the total number of ionizing photons
produced up to some time $t$, per unit volume. The latter quantity
equals the mean number of ionizing photons per baryon times the mean
density of baryons $\nb_b$. Following the arguments leading to
equation~(\ref{rmax}), we find that if we include only stars then \beq
\frac {\nb_\gamma} {\nb_b}= \Ni F_{\rm col}\ , \label{ngnb} \eeq where
the collapse fraction $F_{\rm col}$ is the fraction of all the baryons
in the universe which are in galaxies, i.e., the fraction of gas which
settles into halos and cools efficiently inside them. In writing
equation~(\ref{ngnb}) we are assuming instantaneous production of
photons, i.e., that the time-scale for the formation and evolution of
the massive stars in a galaxy is short compared to the Hubble time at
the formation redshift of the galaxy. In a model based on
equation~(\ref{HIIreg}), the near-equality between $F_{\rm H\ II}$ and
$Q_{\rm H\ II}$ results from the linearity of this equation. First,
the total number of ionizations equals the total number of ionizing
photons produced by stars, i.e., all ionizing photons contribute
regardless of the spatial distribution of sources; and second, the
total recombination rate is proportional to the total ionized volume,
regardless of its topology. Thus, even if two or more bubbles overlap
the model remains an accurate approximation for $Q_{\rm H\ II}$ (at
least until $Q_{\rm H\ II}$ becomes nearly equal to 1). Note, however,
that there still are a number of important simplifications in the
model, including the assumption of a homogeneous (though possibly
time-dependent) clumping factor, and the neglect of feedback whereby
the formation of one galaxy may suppress further galaxy formation in
neighboring regions. These complications are discussed in detail below
and in \S \ref{sec6.5} and \S \ref{sec7}.

Under these assumptions we convert equation~(\ref{HIIreg}), which
describes individual \ion{H}{2} regions, to an equation which
statistically describes the transition from a neutral universe to a
fully ionized one (compare Madau et al.\ 1999 and Haiman \& Loeb
1997): \beq \frac{dQ_{\rm H\ II}}{dt}=\frac{\Ni}{0.76} \frac{dF_{\rm
col}}{dt}- \alpha_B \frac{C}{a^3} \nb_H^0 Q_{\rm H\ II}
\label{QIIeqn}\ , \eeq where we assumed a primordial mass fraction of 
hydrogen of 0.76. The solution (in analogy with
equation~(\ref{HIIsoln})) is \beq Q_{\rm H\ II}(t) =\int_{0}^t
\frac{\Ni} {0.76} \frac{dF_{\rm col}}{dt'}\,e^{F(t',t)} dt'\ , \eeq
where $F(t',t)$ is determined by
equations~(\ref{Fgen})--(\ref{fLCDM}).
  
\begin{figure}[htbp]
\epsscale{0.7}
\plotone{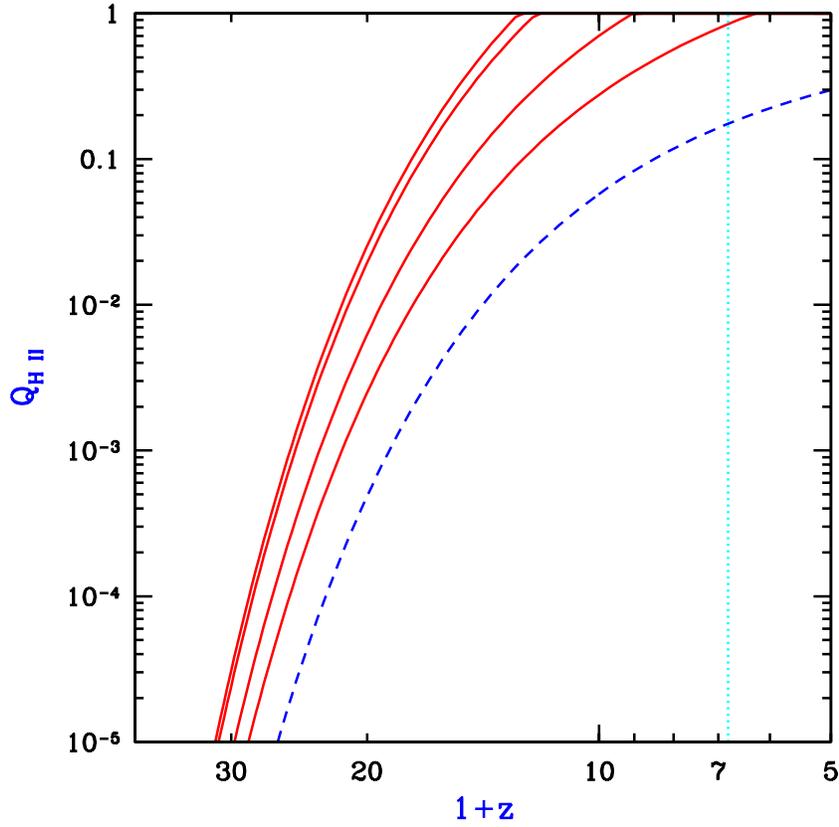}
\caption{Semi-analytic calculation of the reionization of the IGM (for
$\Ni=40$), showing the redshift evolution of the filling factor
$Q_{\rm H\ II}$. Solid curves show $Q_{\rm H\ II}$ for a clumping
factor $C=0$ (no recombinations), $C=1$, $C=10$, and $C=30$, in order
from left to right. The dashed curve shows the collapse fraction
$F_{\rm col}$, and the vertical dotted line shows the $z=5.8$
observational lower limit (Fan et al.\ 2000) on the reionization
redshift.}
\label{fig6c}
\end{figure}

A simple estimate of the collapse fraction at high redshift is the
mass fraction (given by equation~(\ref{PSerfc}) in the Press-Schechter
model) in halos above the cooling threshold, which is the minimum mass
of halos in which gas can cool efficiently. Assuming that only atomic
cooling is effective during the redshift range of reionization (\S
\ref{sec3.3}), the minimum mass corresponds roughly to a halo of
virial temperature $T_{\rm vir}=10^4$ K, which can be converted to a
mass using equation~(\ref{tvir}). With this prescription we derive
(for $\Ni=40$) the reionization history shown in Figure~\ref{fig6c}
for the case of a constant clumping factor $C$. The solid curves show
$Q_{\rm H\ II}$ as a function of redshift for a clumping factor $C=0$
(no recombinations), $C=1$, $C=10$, and $C=30$, in order from left to
right. Note that if $C \sim 1$ then recombinations are unimportant,
but if $C \ga 10$ then recombinations significantly delay the
reionization redshift (for a fixed star-formation history). The dashed
curve shows the collapse fraction $F_{\rm col}$ in this model. For
comparison, the vertical dotted line shows the $z=5.8$ observational
lower limit (Fan et al.\ 2000) on the reionization redshift.

Clearly, star-forming galaxies in CDM hierarchical models are capable
of ionizing the universe at $z\sim 6$--15 with reasonable parameter
choices. This has been shown by a number of theoretical, semi-analytic
calculations (Fukugita \& Kawasaki 1994; Shapiro, Giroux, \& Babul
1994; Kamionkowski, Spergel, \& Sugiyama 1994; Tegmark, Silk, \&
Blanchard 1994; Haiman \& Loeb 1997; Valageas \& Silk 1999; Chiu \&
Ostriker 2000; Ciardi et al.\ 2000) as well as numerical simulations
(Cen \& Ostriker 1993; Gnedin \& Ostriker 1997; Gnedin
2000a). Similarly, if a small fraction ($\la 1\%$) of the gas in each
galaxy accretes onto a central black hole, then the resulting
mini-quasars are also able to reionize the universe, as has also been
shown using semi-analytic models (Fukugita \& Kawasaki 1994; Haiman \&
Loeb 1998; Valageas \& Silk 1999). Note that the prescription whereby
a constant fraction of the galactic mass accretes onto a central black
hole is based on local observations (see \S \ref{sec8.2.2}) which
indicate that $z=0$ galaxies harbor central black holes of mass equal
to $\sim 0.2$--$0.6\%$ of their bulge mass. Although the bulge
constitutes only a fraction of the total baryonic mass of each galaxy,
the higher gas-to-stellar mass ratio in high-redshift galaxies, as
well as their high merger rates compared to their low-redshift
counterparts, suggest that a fraction of a percent of the total gas
mass in high-redshift galaxies may have contributed to the formation
of quasar black holes.

Although many models yield a reionization redshift around 7--12, the
exact value depends on a number of uncertain parameters affecting both
the source term and the recombination term in
equation~(\ref{QIIeqn}). The source parameters include the formation
efficiency of stars and quasars and the escape fraction of ionizing
photons produced by these sources. The formation efficiency of
low-mass galaxies may also be reduced by feedback from galactic
outflows. These parameters affecting the sources are discussed
elsewhere in this review (see \S \ref{sec5.2}, \ref{sec8.2.2},
\ref{sec6.1}, and \ref{sec7}). Even when the clumping is
inhomogeneous, the recombination term in equation~(\ref{QIIeqn}) is
generally valid if $C$ is defined as in equation~(\ref{clump}), where
we take a global volume average of the square of the density inside
ionized regions (since neutral regions do not contribute to the
recombination rate). The resulting mean clumping factor depends on the
density and clustering of sources, and on the distribution and
topology of density fluctuations in the IGM. Furthermore, the source
halos should tend to form in overdense regions, and the clumping
factor is affected by this cross-correlation between the sources and
the IGM density.

Valageas \& Silk (1999) and Chiu \& Ostriker (2000) calculated the
clumping factor semi-analytically by averaging over the IGM on the one
hand and virialized halos on the other hand, with the average weighed
according to the gas fraction in halos. The semi-analytic methods used
in these two detailed calculations of reionization have different
advantages: Valageas \& Silk (1999) included a model for clumping and
absorption by Ly$\alpha$ clouds, but Chiu \& Ostriker (2000) used a
generally more realistic two-phase model with separate ionized and
neutral regions. Miralda-Escud\'e, Haehnelt, \& Rees (2000) went
further in their modeling of the clumping factor by attempting to
account for the geometry of ionized regions. They presented a simple
model for the distribution of density fluctuations, and more generally
they discussed the implications of inhomogeneous clumping during
reionization. They noted that as ionized regions grow, they more
easily extend into low-density regions, and they tend to leave behind
high-density concentrations, with these neutral islands being ionized
only at a later time. They therefore argued that, since at
high-redshift the collapse fraction is low, most of the high-density
regions, which would dominate the clumping factor if they were
ionized, will in fact remain neutral and occupy only a tiny fraction
of the total volume. Thus, the development of reionization through the
end of the overlap phase should occur almost exclusively in the
low-density IGM, and the effective clumping factor during this time
should be $\sim 1$, making recombinations relatively unimportant (see
Figure~\ref{fig6c}). Only in the post-reionization phase,
Miralda-Escud\'e et al.\ (2000) argued, do the high density clouds and
filaments become gradually ionized as the mean ionizing intensity
further increases.

\begin{figure}[htbp]
\epsscale{0.7} 
\plotone{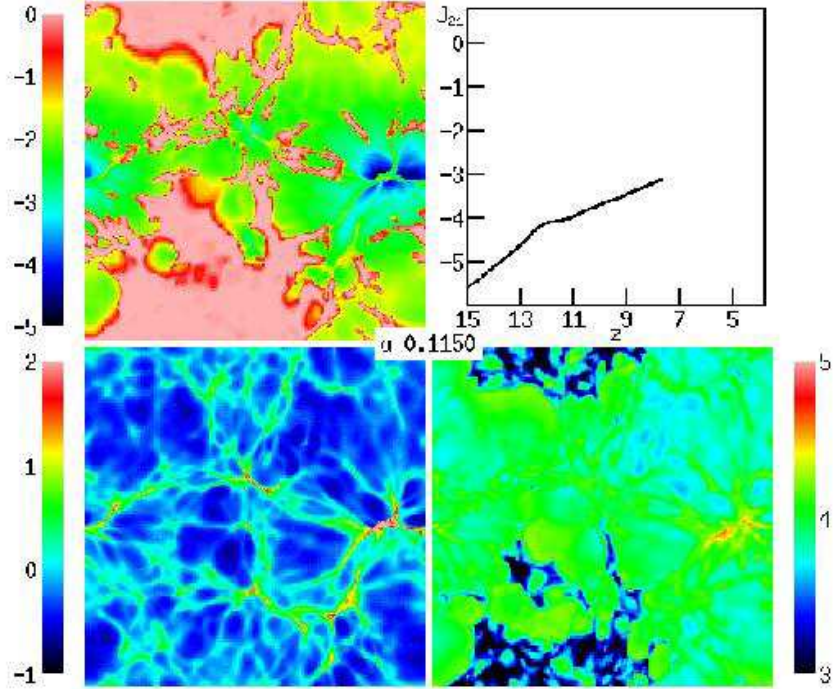}
\caption{Visualization at $z=7.7$ of a numerical simulation of
reionization, adopted from Figure~3c of Gnedin (2000a). The panels
display the logarithm of the neutral hydrogen fraction (upper left),
the gas density in units of the cosmological mean (lower left), and
the gas temperature in Kelvin (lower right). These panels show a
two-dimensional slice of the simulation (not a two-dimensional
projection). Also shown is the redshift evolution of the logarithm of
the ionizing intensity averaged over the entire simulation volume
(upper right). Note the periodic boundary conditions.}
\label{fig6d}
\end{figure}

\begin{figure}[htbp]
\epsscale{0.7}
\plotone{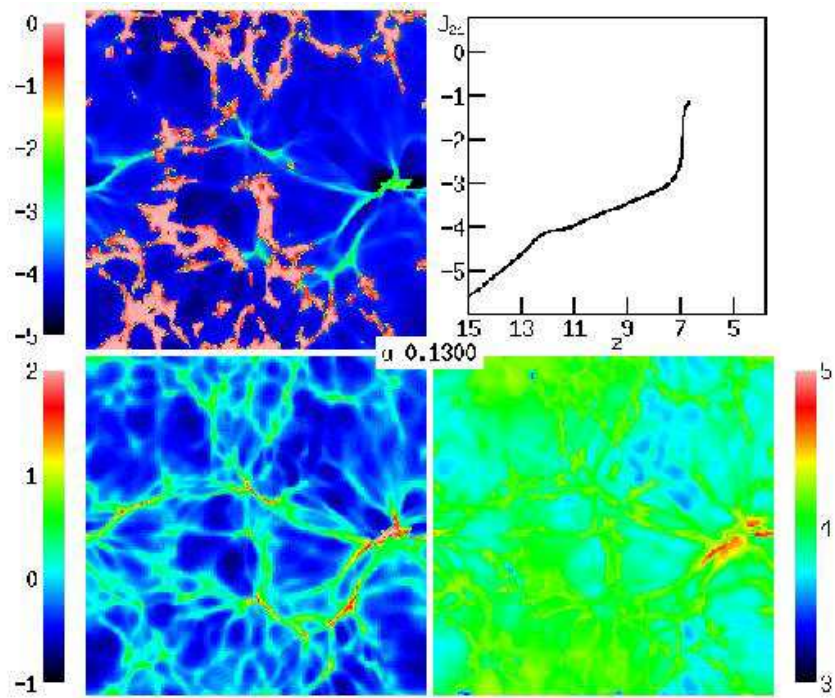}
\caption{Visualization at $z=6.7$ of a numerical simulation of
reionization, adopted from Figure~3c of Gnedin (2000a). The panels
display the logarithm of the neutral hydrogen fraction (upper left),
the gas density in units of the cosmological mean (lower left), and
the gas temperature in Kelvin (lower right). These panels show a
two-dimensional slice of the simulation (not a two-dimensional
projection). Also shown is the redshift evolution of the logarithm of
the ionizing intensity averaged over the entire simulation volume
(upper right). Note the periodic boundary conditions.}
\label{fig6e}
\end{figure}

The complexity of the process of reionization is illustrated by the
recent numerical simulation by Gnedin (2000a) of stellar reionization
(in $\Lambda$CDM with $\Omm=0.3$). This simulation uses a formulation
of radiative transfer which relies on several rough approximations;
although it does not include the effect of shadowing behind
optically-thick clumps, it does include for each point in the IGM the
effects of an estimated local optical depth around that point, plus a
local optical depth around each ionizing source. This simulation helps
to understand the advantages of the various theoretical approaches,
while pointing to the complications which are not included in the
simple models. Figures~\ref{fig6d} and \ref{fig6e}, taken from
Figure~3 in Gnedin (2000a), show the state of the simulated universe
just before and just after the overlap phase, respectively. They show
a thin (15 $h^{-1}$ comoving kpc) slice through the box, which is 4
$h^{-1}$ Mpc on a side. The simulation achieves a spatial resolution
of $1 h^{-1}$ kpc, and uses $128^3$ each of dark matter particles and
baryonic particles (with each baryonic particle having a mass of
$5\times 10^5 M_{\sun}$). The figures show the redshift evolution of
the ionizing intensity averaged over the entire volume, $J_{21}$
(upper right panel), and, visually, the spatial distribution of three
quantities: the neutral hydrogen fraction (upper left panel), the gas
density (lower left panel), and the gas temperature (lower right
panel). Note the obvious features around the edges, resulting from the
periodic boundary conditions assumed in the simulation (e.g., the left
and right edges match identically). Also note that the intensity
$J_{21}$ is defined as the radiation intensity at the Lyman limit,
expressed in units of $10^{-21}\, \mbox{ erg cm}^{-2} \mbox{ s}^{-1}
\mbox{ sr} ^{-1}\mbox{Hz}^{-1}$. For a given source emission, the
intensity inside \ion{H}{2} regions depends on absorption and
radiative transfer through the IGM (e.g., Haardt \& Madau 1996; Abel
\& Haehnelt 1999)

Figure~\ref{fig6d} shows the two-phase IGM at $z=7.7$, with ionized
bubbles emanating from many independent sources, although there is one
main concentration (located at the right edge of the image, vertically
near the center; note the periodic boundary conditions). The bubbles
are shown expanding into low density regions and beginning to overlap
at the center of the image. The topology of ionized regions is clearly
complex: While the ionized regions are analogous to islands in an
ocean of neutral hydrogen, the islands themselves contain small lakes
of dense neutral gas. One aspect which has not been included in
theoretical models of clumping is clear from the figure. The sources
themselves are located in the highest density regions (these being the
sites where the earliest galaxies form) and must therefore ionize the
gas in their immediate vicinity before the radiation can escape into
the low density IGM. For this reason, the effective clumping factor is
of order 100 in the simulation and also, by the overlap redshift,
roughly ten ionizing photons have been produced per baryon. As
emphasized by Gnedin (2000a), some of these numbers are resolution
dependent, since the clumping factor accounts only for absorption by
gas at the highest resolvable density. A higher-resolution simulation
would have higher-density gas clumps and --- depending on the geometry
of those clumps --- a higher or possibly lower clumping factor than
the low-resolution simulation. Figure~\ref{fig6e} shows that by
$z=6.7$ the low density regions have all become highly ionized along
with a rapid increase in the ionizing intensity. The only neutral
islands left are the highest density regions which are not near the
sources (compare the two panels on the left). However, we emphasize
that the quantitative results of this simulation must be considered
preliminary, since the effects of increased resolution and a more
accurate treatment of radiative transfer are yet to be
explored. Methods are being developed for incorporating a more
complete treatment of radiative transfer into three dimensional
cosmological simulations (e.g., Abel, Norman, \& Madau 1999; Razoumov
\& Scott 1999).

Gnedin, Ferrara, \& Zweibel (2000) investigated an additional effect
of reionization. They showed that the Biermann battery in cosmological
ionization fronts inevitably generates coherent magnetic fields of an
amplitude $\sim 10^{-19}$ Gauss. These fields form as a result of the
breakout of the ionization fronts from galaxies and their propagation
through the \ion{H}{1} filaments in the IGM. Although the fields are
too small to directly affect galaxy formation, they could be the seeds
for the magnetic fields observed in galaxies and X-ray clusters today.

If quasars contribute substantially to the ionizing intensity during
reionization then several aspects of reionization are modified compared to
the case of pure stellar reionization. First, the ionizing radiation
emanates from a single, bright point-source inside each host galaxy, and
can establish an escape route (\ion{H}{2} funnel) more easily than in the case
of stars which are smoothly distributed throughout the galaxy (\S
\ref{sec6.1}). Second, the hard photons produced by a quasar penetrate
deeper into the surrounding neutral gas, yielding a thicker ionization
front.  Finally, the quasar X-rays catalyze the formation of H$_2$
molecules and allow stars to keep forming in very small halos (\S
\ref{sec3.3}).

Oh (2000) showed that star-forming regions may also produce
significant X-rays at high redshift. The emission is due to inverse
Compton scattering of CMB photons off relativistic electrons in the
ejecta, as well as thermal emission by the hot supernova remnant. The
spectrum expected from this process is even harder than for typical
quasars, and the hard photons photoionize the IGM efficiently by
repeated secondary ionizations. The radiation, characterized by
roughly equal energy per logarithmic frequency interval, would produce
a uniform ionizing intensity and lead to gradual ionization and
heating of the entire IGM. Thus, if this source of emission is indeed
effective at high redshift, it may have a crucial impact in changing
the topology of reionization. Even if stars dominate the emission, the
hardness of the ionizing spectrum depends on the initial mass
function. At high redshift it may be biased toward massive,
efficiently ionizing stars (see \S \ref{sec4.1.1}), but this remains
very much uncertain.

Semi-analytic as well as numerical models of reionization depend on an
extrapolation of hierarchical models to higher redshifts and
lower-mass halos than the regime where the models have been compared
to observations. These models have the advantage that they are based
on the current CDM paradigm which is supported by a variety of
observations of large-scale structure, galaxy clustering, and the
CMB. The disadvantage is that the properties of high-redshift galaxies
are derived from those of their host halos by prescriptions which are
based on low-redshift observations, and these prescriptions will only
be tested once abundant data is available on galaxies which formed
during the reionization era. An alternative approach to analyzing the
possible ionizing sources which brought about reionization is to
extrapolate from the observed populations of galaxies and quasars at
currently accessible redshifts. This has been attempted, e.g., by
Madau et al.\ (1999) and Miralda-Escud\'e et al.\ (2000). The general
conclusion is that a high-redshift source population similar to the
one observed at $z=3$--4 would produce roughly the needed ionizing
intensity for reionization. A precise conclusion, however, remains
elusive because of the same kinds of uncertainties as those found in
the models based on CDM: The typical escape fraction, and the faint
end of the luminosity function, are both not well determined even at
$z=3$--4, and in addition the clumping factor at high redshift must be
known in order to determine the importance of recombinations. Future
direct observations of the source population at redshifts approaching
reionization may help resolve some of these questions.

\subsubsection{Helium Reionization}
\label{sec6.3.2}

The sources that reionized hydrogen very likely caused the single
reionization of helium from \ion{He}{1} to \ion{He}{2}. Neutral helium is
ionized by photons of 24.6 eV or higher energy, and its recombination rate
is roughly equal to that of hydrogen. On the other hand, the ionization
threshold of \ion{He}{2} is 54.4 eV, and fully ionized helium recombines
$\ga 5$ times faster than hydrogen. This means that for both quasars and
galaxies, the reionization of \ion{He}{2} should occur later than the
reionization of hydrogen, even though the number of helium atoms is smaller
than hydrogen by a factor of 13.  The lower redshift of \ion{He}{2}
reionization makes it more accessible to observations and allows it to
serve in some ways as an observational preview of hydrogen reionization.

\begin{figure}
\epsscale{0.7}
\plotone{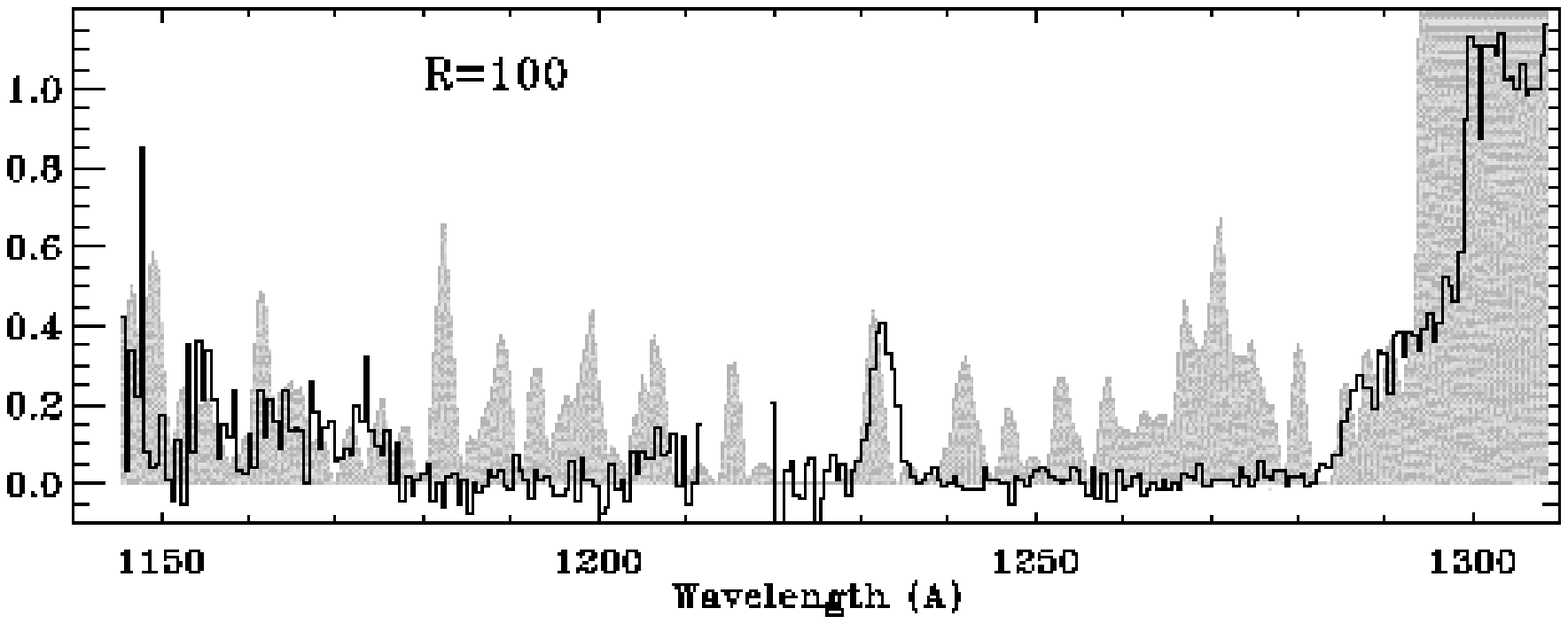}
\caption{Ly$\alpha$ absorption by intergalactic \ion{He}{2}. This
spectrum of the $z=3.286$ quasar $Q\ 0302-003$, adopted from Figure~4
of Heap et al.\ (2000), was obtained using the Space Telescope Imaging
Spectrograph. The observed spectrum (solid line) is compared to a
simulated spectrum (gray shading) based on the \ion{H}{1} Ly$\alpha$
forest observed in the same quasar. In deriving the simulated
spectrum, Heap et al.\ assumed a ratio of \ion{He}{2} to \ion{H}{1}
column densities of 100, and pure turbulent line broadening.}
\label{VI42fig1}
\end{figure}

The Ly$\alpha$ absorption by intergalactic \ion{He}{2} (at wavelength
$304$\AA) has been observed in four quasars at redshifts $2.4 < z <
3.2$ (Jakobsen et al.\ 1994; Davidsen et al.\ 1996; Hogan et al.\
1997; Reimers et al.\ 1997; Anderson et al.\ 1999; Heap et al.\
2000). The results are consistent among the different quasars, and we
illustrate them here with one particular spectrum. In
Figure~\ref{VI42fig1}, adopted from Figure~4 of Heap et al.\ (2000),
we show a portion of the spectrum of the $z=3.286$ quasar $Q\
0302-003$, obtained with the Space Telescope Imaging Spectrograph
on-board the {\it Hubble Space Telescope}.\, The observed spectrum
(solid line) is compared to a simulated spectrum (gray shading) based
on the \ion{H}{1} Ly$\alpha$ forest observed in the same quasar. In
deriving the simulated spectrum, Heap et al.\ assumed a ratio of
\ion{He}{2} to \ion{H}{1} column densities of 100, and pure turbulent
line broadening. The wavelength range shown in the figure corresponds
to \ion{He}{2} Ly$\alpha$ in the redshift range 2.8--3.3.

The observed flux shows a clear break shortward of the quasar emission
line at an observed $\lambda=1300$ \AA. Relatively near the quasar, at
$\lambda=$1285--1300\AA, a shelf of relatively high transmission is
likely evidence of the 'proximity effect', in which the emission from
the quasar itself creates a highly ionized local region with a reduced
abundance of absorbing ions. In the region at $\lambda= $1240--1280
\AA\ ($z=3.08$--3.21), on the other hand, the very low flux level
implies an average optical depth of $\tau\sim 4.5$--5 due to
\ion{He}{2}. Another large region with average $\tau \sim 4$, a region
spanning $\sim 100$ comoving Mpc along the line of sight, is evident
at $\lambda= $1180--1210\AA\ ($z =2.88$--2.98). The strong continuous
absorption in these large regions, and the lack of correlation with
the observed \ion{H}{1} Ly$\alpha$ forest, is evidence for a
\ion{He}{2} Gunn-Peterson absorption trough due to the diffuse IGM. It
also suggests a rather soft UV background with a significant stellar
contribution, i.e., a background that ionizes the diffuse hydrogen
much more thoroughly than \ion{He}{2}. Significant emission is
observed in between the two regions of constant high absorption. A
small region around 1216\AA\ is contaminated by geo-coronal
Ly$\alpha$, but the emission at 1230--1235\AA\ apparently corresponds
to a real, distinct gap in the \ion{He}{2} abundance, which could be
caused by a local source photo-ionizing a region of radius $\sim 10$
comoving Mpc. The region at $\lambda=$1150--1175\AA\, ($z=2.78$--2.86)
shows a much higher overall transmission level than the regions at
slightly higher redshift. Heap et al.\ measure an average $\tau=1.9$
in this region, and note that the significant correlation of the
observed spectrum with the simulated one suggests that much of the
absorption is due to a \ion{He}{2} Ly$\alpha$ forest while the
low-density IGM provides a relatively low opacity in this region. The
authors conclude that the observed data suggest a sharp opacity break
occurring between $z=3.0$ and 2.9, accompanied by a hardening of the
UV ionizing background.  However, even the relatively high opacity at
$z
\ga 3$ only requires $\sim 0.1\%$ of helium atoms not to be fully
ionized, in a region at the mean baryon density. Thus, the overlap
phase of full helium reionization may have occurred significantly
earlier, with the ionizing intensity already fairly uniform but still
increasing with time at $z \sim 3$.

The properties of helium reionization have been investigated
numerically by a number of authors. Zheng \& Davidsen (1995) modeled
the \ion{He}{2} proximity effect, and a number of authors
(Miralda-Escud\'e et al.\ 1996; Croft et al.\ 1997; Zhang et al.\ 
1998) used numerical simulations to show that the observations
generally agree with cold dark matter models. They also found that
helium absorption particularly tests the properties of under-dense
voids which produce much of the \ion{He}{2} opacity but little opacity
in \ion{H}{1}. According to the semi-analytic model of inhomogeneous
reionization of Miralda-Escud\'e, Haehnelt, \& Rees (2000; see also \S
\ref{sec6.3.1}), the total emissivity of observed quasars at redshift
3 suffices to completely reionize helium before $z=3$. They find that
the observations at $z\sim 3$ can be reproduced if a population of
low-luminosity sources, perhaps galaxies, has ionized the low-density
IGM up to an overdensity of around 12 relative to the cosmological
mean, with luminous quasars creating the observed gaps of transmitted
flux.

The conclusion that an evolution of the ionization state of helium has
been observed is also strengthened by several indirect lines of
evidence. Songaila \& Cowie (1996) and Songaila (1998) found a rapid
increase in the \ion{Si}{4}/\ion{C}{4} ratio with decreasing redshift
at $z=3$, for intermediate column density hydrogen Ly$\alpha$
absorption lines. They interpreted this evolution as a sudden
hardening below $z=3$ of the spectrum of the ionizing background.
Boksenberg et al.\ (1998) also found an increase in the
\ion{Si}{4}/\ion{C}{4} ratio, but their data implied a much more
gradual increase from $z=3.8$ to $z=2.2$. 

The full reionization of helium due to a hard ionizing spectrum should
also heat the IGM to 20,000 K or higher, while the IGM can only reach
$\sim$10,000 K during a reionization of hydrogen alone (although a
temperature of $\sim$15,000 K may be reached due to Compton heating by
the hard X-ray background: Madau \& Efstathiou 1999). This increase in
temperature can serve as an observational probe of helium
reionization, and it should also increase the suppression of dwarf
galaxy formation (\S \ref{sec6.5}). The temperature of the IGM can be
measured by searching for the smallest line-widths among hydrogen
Ly$\alpha$ absorption lines (Schaye et al.\ 1999). In general, bulk
velocity gradients contribute to the line width on top of thermal
velocities, but a lower bound on the width is set by thermal
broadening, and the narrowest lines can be used to measure the
temperature. Several different measurements (Ricotti et al.\ 2000;
Schaye et al.\ 2000; Bryan \& Machacek 2000; McDonald et al.\ 2000)
have found a nearly isothermal IGM at a temperature of $\sim$20,000 K
at $z=3$, higher than expected in ionization equilibrium and
suggestive of photo-heating due to ongoing reionization of
helium. However, the measurement errors remain too large for a firm
conclusion about the redshift evolution of the IGM temperature or its
equation of state.

Clearly, the reionization of helium is already a rich phenomenological
subject. Our knowledge will benefit from measurements of increasing
accuracy, made toward many more lines of sight, and extended to higher
redshift. New ways to probe helium will also be useful. For example,
Miralda-Escud\'e (2000) has suggested that continuum \ion{He}{2}
absorption in soft X-rays can be used to determine the \ion{He}{2}
fraction along the line of sight, although the measurement requires an
accurate subtraction of the Galactic contribution to the absorption,
based on the Galactic \ion{H}{1} column density as determined by 21 cm
maps.

\subsection{Photo-evaporation of Gaseous Halos After Reionization}
\label{sec6.4}

The end of the reionization phase transition resulted in the emergence
of an intense UV background that filled the universe and heated the
IGM to temperatures of $\sim 1$--$2\times 10^4$K (see the previous
section). After ionizing the rarefied IGM in the voids and filaments
on large scales, the cosmic UV background penetrated the denser
regions associated with the virialized gaseous halos of the first
generation of objects. A major fraction of the collapsed gas had been
incorporated by that time into halos with a virial temperature $\la
10^4$K, where the lack of atomic cooling prevented the formation of
galactic disks and stars or quasars. Photoionization heating by the
cosmic UV background could then evaporate much of this gas back into
the IGM. The photo-evaporating halos, as well as those halos which did
retain their gas, may have had a number of important consequences just
after reionization as well as at lower redshifts.

In this section we focus on the process by which gas that had already
settled into virialized halos by the time of reionization was
evaporated back into the IGM due to the cosmic UV background. This
process was investigated by Barkana \& Loeb (1999) using semi-analytic
methods and idealized numerical calculations. They first considered an
isolated spherical, centrally-concentrated dark matter halo containing
gas. Since most of the photo-evaporation occurs at the end of overlap,
when the ionizing intensity builds up almost instantaneously, a sudden
illumination by an external ionizing background may be assumed.
Self-shielding of the gas implies that the halo interior sees a
reduced intensity and a harder spectrum, since the outer gas layers
preferentially block photons with energies just above the Lyman limit.
It is useful to parameterize the external radiation field by a
specific intensity per unit frequency, $\nu$, \beq J_{\nu}=10^{-21}\,
J_{21}\, \left(\frac{\nu}{\nu_L}\right)^ {-\alpha}\mbox{ erg
cm}^{-2}\mbox{ s}^{-1}\mbox{ sr} ^{-1}\mbox{Hz}^{-1}\ , \label{Inu}
\eeq where $\nu_L$ is the Lyman limit frequency, and $J_{21}$ is the
intensity at $\nu_L$ expressed in units of $10^{-21}\, \mbox{ erg
cm}^{-2} \mbox{ s}^{-1} \mbox{ sr} ^{-1}\mbox{Hz}^{-1}$. The intensity
is normalized to an expected post-reionization value of around unity
for the ratio of ionizing photon density to the baryon density.
Different power laws can be used to represent either quasar spectra
($\alpha \sim 1.8$) or stellar spectra ($\alpha \sim 5$).

Once the gas is heated throughout the halo, some fraction of it
acquires a sufficiently high temperature that it becomes unbound. This
gas expands due to the resulting pressure gradient and eventually
evaporates back into the IGM. The pressure gradient force (per unit
volume) 
$\kB \nabla (T \rho/\mu m_p)$ competes with the gravitational force of
$\rho\, G M/r^2$. Due to the density gradient, the ratio between the
pressure force and the gravitational force is roughly equal to the ratio
between the thermal energy $\sim \kB T$ and the gravitational binding
energy $\sim \mu m_p G M/r$ (which is $\sim \kB T_{\rm vir}$ at the virial
radius $r_{\rm vir}$) per particle. Thus, if the kinetic energy exceeds the
potential energy (or roughly if $T>T_{\rm vir}$), the repulsive pressure
gradient force exceeds the attractive gravitational force and expels the
gas on a dynamical time (or faster for halos with $T\gg T_{\rm vir}$).

\begin{figure}[htbp]
\epsscale{0.7}
\plottwo{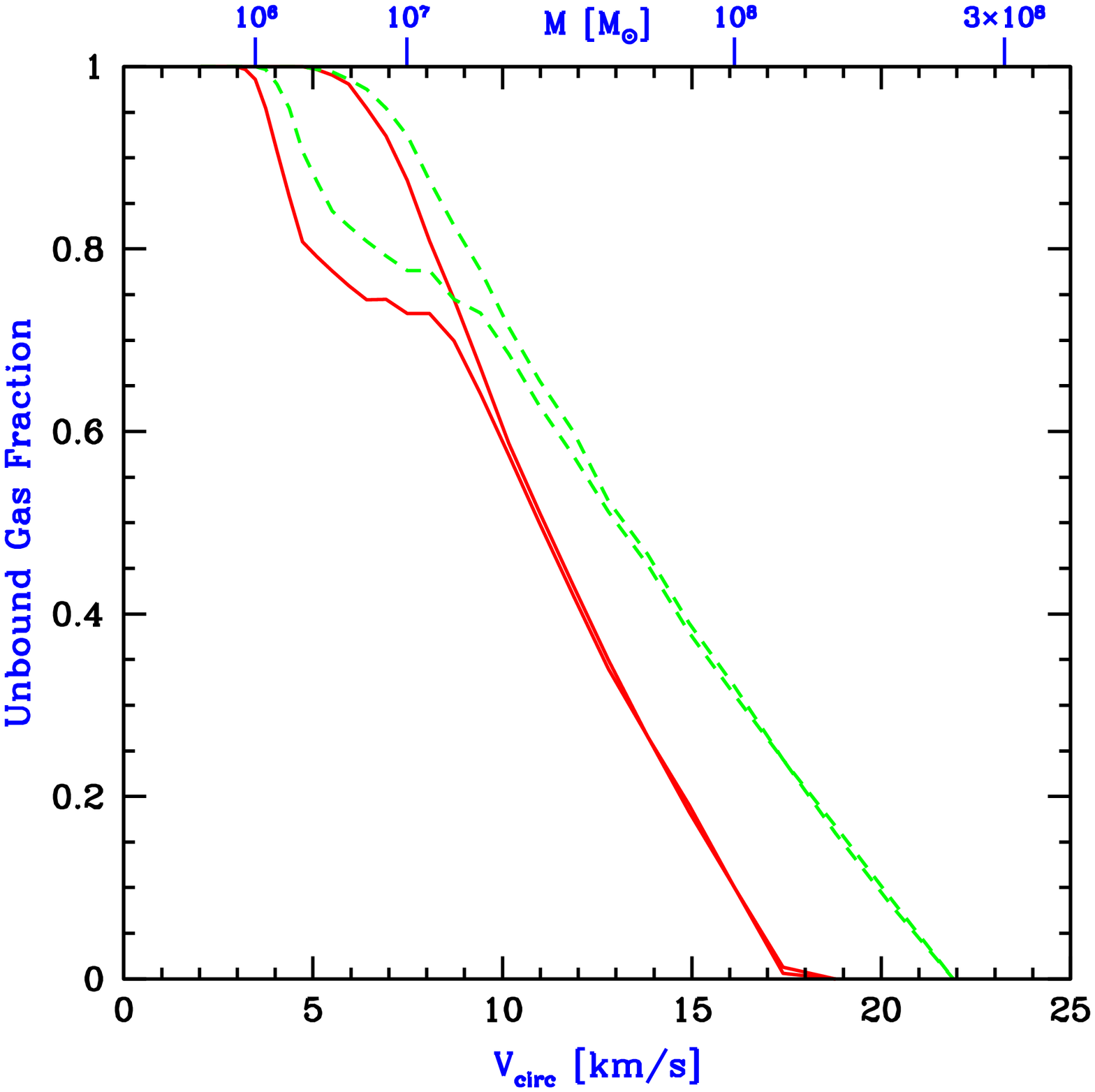}{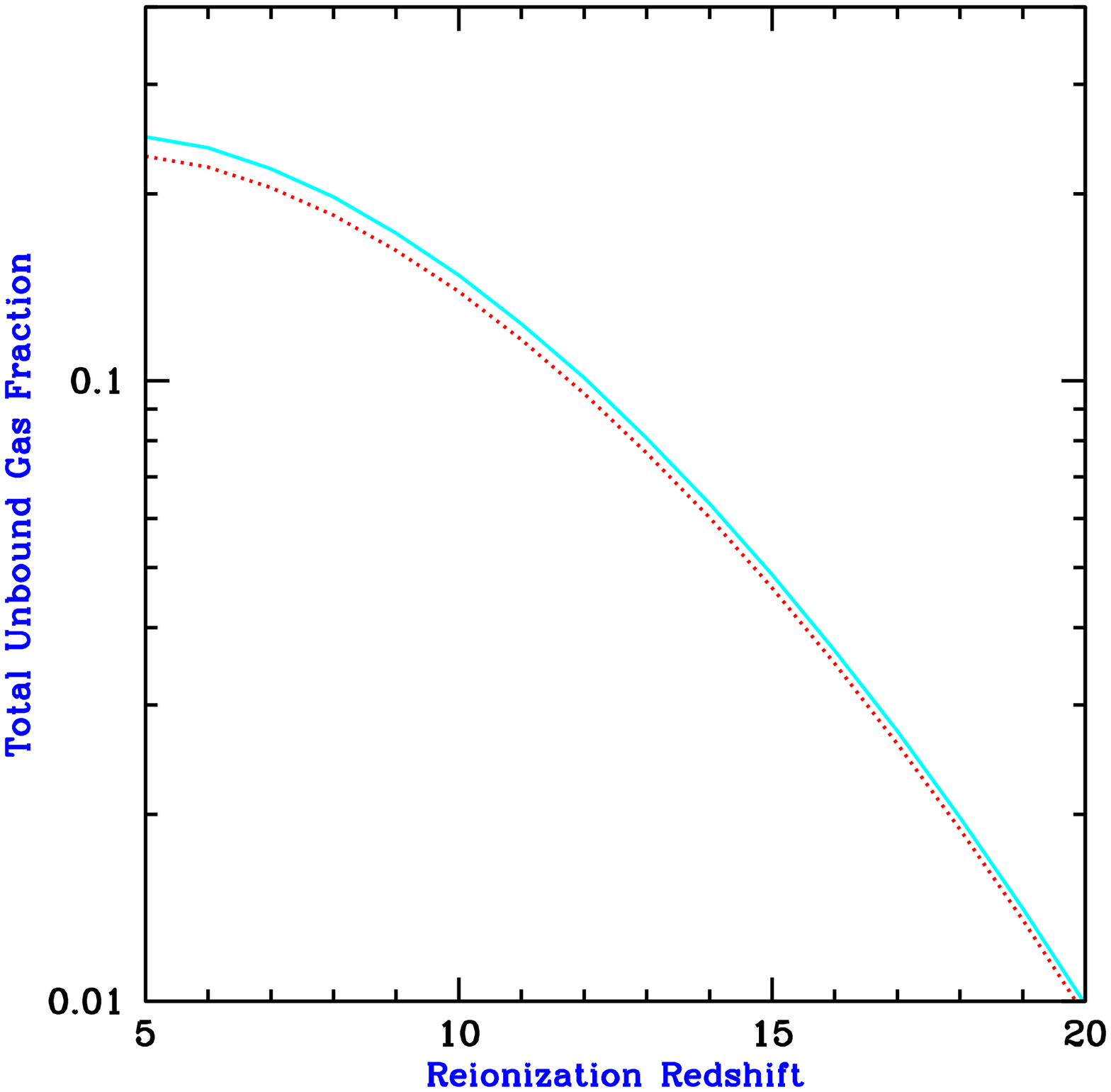}
\caption{Effect of photo-evaporation on individual halos and on the
overall halo population. The left panel shows the unbound gas fraction
(within the virial radius) versus total halo circular velocity or
mass, adopted from Figure~3 of Barkana \& Loeb (1999). The two pairs
of curves correspond to spectral index $\alpha=5$ (solid) or
$\alpha=1.8$ (dashed), in each case at $z=8$. In each pair, assuming
an optically-thin halo leads to the upper curve, while the lower curve
shows the result of including radiative transfer and self
shielding. The right panel shows the total fraction of gas in the
universe which evaporates from halos at reionization, versus the
reionization redshift, adopted from Figure~7 of Barkana \& Loeb
(1999). The solid line assumes a spectral index $\alpha=1.8$, and the
dotted line assumes $\alpha=5$.}
\label{fig6.7}
\end{figure}

The left panel of Figure~\ref{fig6.7} (adopted from Figure~3 of
Barkana \& Loeb 1999) shows the fraction of gas within the virial
radius which becomes unbound after reionization, as a function of the
total halo circular velocity, with halo masses at $z=8$ indicated at
the top. The two pairs of curves correspond to spectral index
$\alpha=5$ (solid) or $\alpha=1.8$ (dashed). In each pair, a
calculation which assumes an optically-thin halo leads to the upper
curve, but including radiative transfer and self-shielding modifies
the result to the one shown by the lower curve. In each case
self-shielding lowers the unbound fraction, but it mostly affects only
a neutral core containing $\sim 30\%$ of the gas. Since high energy
photons above the Lyman limit penetrate deep into the halo and heat
the gas efficiently, a flattening of the spectral slope from
$\alpha=5$ to $\alpha=1.8$ raises the unbound gas fraction. This
Figure is essentially independent of redshift if plotted in terms of
circular velocity, but the conversion to a corresponding mass does
vary with redshift. The characteristic circular velocity where most of
the gas is lost is $\sim 10$--$15~{\rm km~s^{-1}}$, but clearly the
effect of photo-evaporation is gradual, going from total gas removal
down to no effect over a range of a factor of $\sim 100$ in halo mass.

Given the values of the unbound gas fraction in halos of different
masses, the Press-Schechter mass function (\S \ref{sec2.4}) can be
used to calculate the total fraction of the IGM which goes through the
process of accreting onto a halo and then being recycled into the IGM
at reionization. The low-mass cutoff in this sum over halos is given
by the lowest mass halo in which gas has assembled by the reionization
redshift. This mass can be estimated by the linear Jeans mass
$\mjeans$ in equation (\ref{eq:m_j}) in \S \ref{sec3.1}. The Jeans
mass does not in general precisely equal the limiting mass for
accretion (see the discussion in the next section). Indeed, at a given
redshift some gas can continue to fall into halos of lower mass than
the Jeans mass at that redshift. On the other hand, the larger Jeans
mass at higher redshifts means that a time-averaged Jeans mass may be
more appropriate, as indicated by the filtering mass. In practice, the
Jeans mass is sufficiently accurate since at $z\sim 10$--20 it agrees
well with the values found in the numerical spherical collapse
calculations of Haiman, Thoul, \& Loeb (1996).

The right panel of Figure~\ref{fig6.7} (adopted from Figure~7 of
Barkana \& Loeb 1999) shows the total fraction of gas in the universe
which evaporates from halos at reionization, versus the reionization
redshift. The solid line assumes a spectral index $\alpha=1.8$, and
the dotted line assumes $\alpha=5$, showing that the result is
insensitive to the spectrum. Even at high redshift, the amount of gas
which participates in photo-evaporation is significant, which suggests
a number of possible implications as discussed below. The gas fraction
shown in the figure represents most ($\sim 60$--$80\%$ depending on
the redshift) of the collapsed fraction before reionization, although
some gas does remain in more massive halos.

The photo-evaporation of gas out of large numbers of halos may have
interesting implications. First, gas which falls into halos and is
expelled at reionization attains a different entropy than if it had
stayed in the low-density IGM. The resulting overall reduction in the
entropy is expected to be small --- the same as would be produced by
reducing the temperature of the entire IGM by a factor of $\sim 1.5$
--- but localized effects near photo-evaporating halos may be more
significant. Furthermore, the resulting $\sim 20~{\rm km~s^{-1}}$
outflows induce small-scale fluctuations in peculiar velocity and
temperature. These outflows are usually well below the resolution
limit of most numerical simulations, but some outflows were resolved
in the simulation of Bryan et al.\ (1998). The evaporating halos may
consume a significant number of ionizing photons in the post-overlap
stage of reionization (e.g., Haiman, Abel, \& Madau 2000), but a
definitive determination requires detailed simulations which include
the three-dimensional geometry of source halos and sink halos.

Although gas is quickly expelled out of the smallest halos,
photo-evaporation occurs more gradually in larger halos which retain
some of their gas. These surviving halos initially expand but they
continue to accrete dark matter and to merge with other halos. These
evaporating gas halos could contribute to the high column density end
of the Ly$\alpha$ forest (Bond, Szalay, \& Silk 1988). Abel \& Mo
(1998) suggested that, based on the expected number of surviving
halos, a large fraction of the Lyman limit systems at $z\sim 3$ may
correspond to mini-halos that survived reionization. Surviving halos
may even have identifiable remnants in the present universe, as
discussed in \S \ref{sec9.3}. These ideas thus offer the possibility
that a population of halos which originally formed prior to
reionization may correspond almost directly to several populations
that are observed much later in the history of the universe. However,
the detailed dynamics of photo-evaporating halos are complex, and
detailed simulations are required to confirm these ideas.
Photo-evaporation of a gas cloud has been followed in a two dimensional
simulation with radiative transfer, by Shapiro \& Raga (2000). They
found that an evaporating halo would indeed appear in absorption as a
damped Ly$\alpha$ system initially, and as a weaker absorption system
subsequently. Future simulations will clarify the contribution to
quasar absorption lines of the entire population of photo-evaporating
halos.

\subsection{Suppression of the Formation of Low Mass Galaxies}
\label{sec6.5}

At the end of overlap, the cosmic ionizing background increased
sharply, and the IGM was heated by the ionizing radiation to a
temperature $\ga 10^4$ K. Due to the substantial increase in the IGM
temperature, the intergalactic Jeans mass increased dramatically,
changing the minimum mass of forming galaxies (Rees 1986; Efstathiou
1992; Gnedin \& Ostriker 1997; Miralda-Escud\'e \& Rees 1998).

Gas infall depends sensitively on the Jeans mass. When a halo more
massive than the Jeans mass begins to form, the gravity of its dark
matter overcomes the gas pressure. Even in halos below the Jeans mass,
although the gas is initially held up by pressure, once the dark
matter collapses its increased gravity pulls in some gas (Haiman,
Thoul, \& Loeb 1996). Thus, the Jeans mass is generally higher than
the actual limiting mass for accretion. Before reionization, the IGM
is cold and neutral, and the Jeans mass plays a secondary role in
limiting galaxy formation compared to cooling. After reionization, the
Jeans mass is increased by several orders of magnitude due to the
photoionization heating of the IGM, and hence begins to play a
dominant role in limiting the formation of stars. Gas infall in a
reionized and heated universe has been investigated in a number of
numerical simulations. Thoul \& Weinberg (1996) inferred, based on a
spherically-symmetric collapse simulation, a reduction of $\sim 50\%$
in the collapsed gas mass due to heating, for a halo of circular
velocity $V_c\sim 50\ {\rm km\ s}^{-1}$ at $z=2$, and a complete
suppression of infall below $V_c \sim 30\ {\rm km\ s}^{-1}$. Kitayama
\& Ikeuchi (2000) also performed spherically-symmetric simulations but
included self-shielding of the gas, and found that it lowers the
circular velocity thresholds by $\sim 5\ {\rm km\ s}^{-1}$. Three
dimensional numerical simulations (Quinn, Katz, \& Efstathiou 1996;
Weinberg, Hernquist, \& Katz 1997; Navarro \& Steinmetz 1997) found a
significant suppression of gas infall in even larger halos ($V_c \sim
75\ {\rm km\ s}^{-1}$), but this was mostly due to a suppression of
late infall at $z\la 2$.

When a volume of the IGM is ionized by stars, the gas is heated to a
temperature $T_{\rm IGM}\sim 10^4$ K. If quasars dominate the UV
background at reionization, their harder photon spectrum leads to
$T_{\rm IGM}\sim 2\times 10^4$ K. Including the effects of dark
matter, a given temperature results in a linear Jeans mass (see \S
\ref{sec3.1}) corresponding to a halo circular velocity of \beq V_J=81
\left(\frac{T_{\rm IGM}}{1.5\times 10^4 {\rm K}}\right)^{1/2}\
\left[\frac{1}{\Ommz}\ \frac{\Delta_c}{18 \pi^2}\right]^{1/6}\ {\rm
km\ s}^{-1}, \eeq where we used equation~(\ref{Vceqn}) and assumed
$\mu=0.6$. In halos with $V_c>V_J$, the gas fraction in infalling gas
equals the universal mean of $\Omega_b/\Omm$, but gas infall is
suppressed in smaller halos. Even for a small dark matter halo, once
it collapses to a virial overdensity of $\Delta_c/\Ommz$ relative to
the mean, it can pull in additional gas. A simple estimate of the
limiting circular velocity, below which halos have essentially no gas
infall, is obtained by substituting the virial overdensity for the
mean density in the definition of the Jeans mass. The resulting
estimate is \beq V_{\rm lim}=34 \left(\frac{T_{\rm IGM}}{1.5\times
10^4 {\rm K}}\right)^{1/2}\ {\rm km\ s}^{-1}. \eeq This value is in
rough agreement with the numerical simulations mentioned above (see
also the related discussion in \S \ref{sec3.2}).

Although the Jeans mass is closely related to the rate of gas infall
at a given time, it does not directly yield the total gas residing in
halos at a given time. The latter quantity depends on the entire
history of gas accretion onto halos, as well as on the merger
histories of halos, and an accurate description must involve a
time-averaged Jeans mass. Gnedin (2000b) showed that the gas content
of halos in simulations is well fit by an expression which depends on
the filtering mass, a particular time-averaged Jeans mass (Gnedin \&
Hui 1998; see also \S \ref{sec3.1}). Gnedin (2000b) calculated the
Jeans and filtering masses using the mean temperature in the
simulation to define the sound speed, and found the following fit to
the simulation results: \beq \bar{M_g}=\frac{f_b M}{\left [1+
\left(2^{1/3}-1\right) M_C/M \right]^3}\ , \eeq where $\bar{M_g}$ is
the average gas mass of all objects with a total mass $M$,
$f_b=\Omega_b/\Omm$ is the universal baryon fraction, and the
characteristic mass $M_C$ is the total mass of objects which on
average retain $50\%$ of their gas mass. The characteristic mass was
well fit by the filtering mass at a range of redshifts from $z=4$ up
to $z\sim 15$.


\section{\bf Feedback from Galactic Outflows}
\label{sec7}

\subsection{Propagation of Supernova Outflows in the IGM}
\label{sec7.1}

Star formation is accompanied by the violent death of massive stars in
supernova explosions. In general, if each halo has a fixed baryon
fraction and a fixed fraction of the baryons turns into massive stars,
then the total energy in supernova outflows is proportional to the
halo mass. The binding energies of both the supernova ejecta and of
all the gas in the halo are proportional to the halo mass
squared. Thus, outflows are expected to escape more easily out of
low-mass galaxies, and to expel a greater fraction of the gas from
dwarf galaxies. At high redshifts, most galaxies form in relatively
low-mass halos, and the high halo merger rate leads to vigorous star
formation. Thus, outflows may have had a great impact on the earliest
generations of galaxies, with consequences that may include metal
enrichment of the IGM and the disruption of dwarf galaxies. In this
subsection we present a simple model for the propagation of individual
supernova shock fronts in the IGM. We discuss some implications of
this model, but we defer to the following subsection the brunt of the
discussion of the cosmological consequences of outflows.

For a galaxy forming in a given halo, the supernova rate is related to
the star formation rate. In particular, for a Scalo (1998) initial
stellar mass function, if we assume that a supernova is produced by
each $M>8 M_{\odot}$ star, then on average one supernova explodes for
every 126 $M_{\odot}$ of star formation, expelling an ejecta mass of
$\sim 3\, M_{\odot}$ including $\sim 1\, M_{\odot}$ of heavy
elements. We assume that the individual supernovae produce expanding
hot bubbles which merge into a single overall region delineated by an
outwardly moving shock front. We assume that most of the baryons in
the outflow lie in a thin shell, while most of the thermal energy is
carried by the hot interior. The total ejected mass, which is lifted
out of the halo by the outflow, equals a fraction $\fg$ of the total
halo gas mass. The ejected mass includes some of the supernova ejecta
itself. We let $\fe$ denote the fraction of the supernova ejecta that
winds up in the outflow (with $\fe \le 1$ since some metals may be
deposited in the disk and not ejected). Since at high redshift most of
the halo gas is likely to have cooled onto a disk, we assume that the
mass carried by the outflow remains constant until the shock front
reaches the halo virial radius. We assume an average supernova energy
of $10^{51}E_{51}$ erg, a fraction $\fw$ of which remains in the
outflow after the outflow escapes from the disk. The outflow must
overcome the gravitational potential of the halo, which we assume to
have a Navarro, Frenk, \& White (1997) density profile [NFW; see
equation (\ref{NFW}) in \S (\ref{sec2.3})]. Since the entire shell
mass must be lifted out of the halo, we include the total shell mass
as well as the total injected energy at the outset. This assumption is
consistent with the fact that the burst of star formation in a halo is
typically short compared to the total time for which the corresponding
outflow expands.

The escape of an outflow from an NFW halo depends on the concentration
parameter $\cN$ of the halo. Simulations by Bullock et al.\ (2000)
indicate that the concentration parameter decreases with redshift, and
their results may be extrapolated to our regime of interest (i.e., to
smaller halo masses and higher redshifts) by assuming that \beq
\cN=\left(\frac{M}{10^9 M_{\sun}}\right) ^{-0.1}\, \frac{25}{(1+z)}\
. \eeq Although we calculate below the dynamics of each outflow in
detail, it is also useful to estimate which halos can generate
large-scale outflows by comparing the kinetic energy of the outflow to
the potential energy needed to completely escape (i.e., to infinite
distance) from an NFW halo. We thus find that the outflow can escape
from its originating halo if the circular velocity is below a critical
value given by \beq V_{\rm crit} = 200 \sqrt{\frac{E_{51}\fw
(\eta/0.1)} {\fg\ g(\cN)}}\ {\rm km\ s}^{-1} \ , \label{Vcrit} \eeq
where the efficiency $\eta$ is the fraction of baryons incorporated in
stars, and \beq g(x)=\frac{x^2}{(1+x)\ln(1+x)-x} \ .  \eeq Note that
the contribution to $\fg$ of the supernova ejecta itself is $0.024
\eta \fe$, so the ejecta mass is usually negligible unless $\fg \la
1\%$. Equation (\ref{Vcrit}) can also be used to yield the maximum gas
fraction $\fg$ which can be ejected from halos, as a function of their
circular velocity. Although this equation is most general, if we
assume that the parameters $\fg$ and $\fw$ are independent of $M$ and
$z$ then we can normalize them based on low-redshift observations. If
we specify $\cN \sim 10$ (with $g(10)=6.1$) at $z=0$, then setting
$E_{51}=1$ and $\eta=10\%$ yields the required energy efficiency as a
function of the ejected halo gas fraction: \beq \fw = 1.5 \fg
\left[\frac{V_{\rm crit}}{100\ {\rm km\ s}^{-1}} \right]^2\ . \eeq A
value of $V_{\rm crit} \sim 100\ {\rm km\ s}^{-1}$ is suggested by
several theoretical and observational arguments which are discussed in
the next subsection. However, these arguments are not conclusive, and
$V_{\rm crit}$ may differ from this value by a large factor,
especially at high redshift (where outflows are observationally
unconstrained at present). Note the degeneracy between $\fg$ and $\fw$
which remains even if $V_{\rm crit}$ is specified. Thus, if $V_{\rm
crit} \sim 100\ {\rm km\ s}^{-1}$ then a high efficiency $\fw \sim 1$
is required to eject most of the gas from all halos with $V_c < V_{\rm
crit}$, but only $\fw \sim 10\%$ is required to eject 5--10$\%$ of the
gas. The evolution of the outflow does depend on the value of $\fw$
and not just the ratio $\fw/\fg$, since the shell accumulates material
from the IGM which eventually dominates over the initial mass carried
by the outflow.

We solve numerically for the spherical expansion of a galactic
outflow, elaborating on the basic approach of Tegmark, Silk, \& Evrard
(1993). We assume that most of the mass $m$ carried along by the
outflow lies in a thin, dense, relatively cool shell of proper radius
$R$. The interior volume, while containing only a fraction $\fin \ll
1$ of the mass $m$, carries most of the thermal energy in a hot,
isothermal plasma of pressure $p_{\rm int}$ and temperature $T$. We
assume a uniform exterior gas, at the mean density of the universe (at
each redshift), which may be neutral or ionized, and may exert a
pressure $p_{\rm ext}$ as indicated below. We also assume that the
dark matter distribution follows the NFW profile out to the virial
radius, and is at the mean density of the universe outside the halo
virial radius. Note that in reality an overdense distribution of gas
as well as dark matter may surround each halo due to secondary infall.

The shell radius $R$ in general evolves as follows: \beq m \frac{d^2R}
{dt^2}= 4 \pi R^2 \delta p-\left (\frac{dR}{dt} - H R\right) \frac{dm}
{dt}- \frac{G m} {R^2} \left(M(R)+ \frac{1} {2}m \right) + \frac{8}{3}
\pi G R m \rho_{\Lambda} \ , \eeq where the right-hand-side includes
forces due to pressure, sweeping up of additional mass, gravity, and a
cosmological constant, respectively\footnote{The last term, which is
due to the cosmological constant, is an effective repulsion which
arises in the Newtonian limit of the full equations of General
Relativity.}. The shell is accelerated by internal pressure and
decelerated by external pressure, i.e., $\delta p=p_{\rm int}-p_{\rm
ext}$. In the gravitational force, $M(R)$ is the total enclosed mass,
not including matter in the shell, and $\frac{1} {2}m$ is the
effective contribution of the shell mass in the thin-shell
approximation (Ostriker \& McKee 1988). The interior pressure is
determined by energy conservation, and evolves according to (Tegmark
et al.\ 1993): \beq \frac{d p_{\rm int}} {dt}=\frac{L}{2 \pi R^3}-5\,
\frac{p_{\rm int}} {R}\, \frac{d R}{dt} \ , \eeq where the luminosity
$L$ incorporates heating and cooling terms. We include in $L$ the
supernova luminosity $L_{\rm sn}$ (during a brief initial period of
energy injection), cooling terms $L_{\rm cool}$, ionization $L_{\rm
ion}$, and dissipation $L_{\rm diss}$. For simplicity, we assume
ionization equilibrium for the interior plasma, and a primordial
abundance of hydrogen and helium. We include in $L_{\rm cool}$ all
relevant atomic cooling processes in hydrogen and helium, i.e.,
collisional processes, Bremsstrahlung emission, and Compton cooling
off the CMB. Compton scattering is the dominant cooling process for
high-redshift outflows. We include in $L_{\rm ion}$ only the power
required to ionize the incoming hydrogen upstream, at the energy cost
of 13.6 eV per hydrogen atom. The interaction between the expanding
shell and the swept-up mass dissipates kinetic energy. The fraction
$f_d$ of this energy which is re-injected into the interior depends on
complex processes occurring near the shock front, including
turbulence, non-equilibrium ionization and cooling, and so (following
Tegmark et al.\ 1993) we let \beq L_{\rm diss}=\frac{1}{2} f_d
\frac{dm}{dt} \left( \frac{dR}{dt} - H R \right)^2\ , \eeq where we
set $f_d=1$ and compare below to the other extreme of $f_d=0$.

In an expanding universe, it is preferable to describe the propagation
of outflows in terms of comoving coordinates since, e.g., the critical
result is the maximum {\it comoving}\, size of each outflow, since
this size yields directly the total IGM mass which is displaced by the
outflow and injected with metals. Specifically, we apply the following
transformation (Shandarin 1980): \beq d\hat{t}=a^{-2} dt,\ \
\hat{R}=a^{-1}R,\ \ \hat{p}=a^5 p,\ \hat{\rho}= a^3 \rho\ . \eeq For
$\Oml=0$, Voit (1996) obtained (with the time origin $\hat{t}=0$ at
redshift $z_1$): \beq \hat{t}=\frac{2}{\Omm H_0} \left[ \sqrt{1+\Omm
z_1}-\sqrt{1+\Omm z}\, \right]\ , \eeq while for $\Omm+\Oml=1$ there
is no simple analytic expression. We set $\beta=\hat{R}/\hat{r}_{\rm
vir}$, in terms of the virial radius $r_{\rm vir}$ [equation
(\ref{rvir})] of the source halo. We define $\alpha_S^1$ as the ratio
of the shell mass $m$ to $\frac{4}{3} \pi \hat{\rho}_b\, \hat{r}_{\rm
vir}^3$, where $\hat{\rho}_b=\rho_b(z=0)$ is the mean baryon density
of the Universe at $z=0$. More generally, we define \beq
\alpha_S(\beta) \equiv \frac{m}{\frac{4}{3} \pi \hat{\rho}_b \,
\hat{R}^3} = \left\{ \begin{array}{ll} \alpha_S^1 / \beta^3 & \mbox{if
$\beta<1$} \\ 1+\left(\alpha_S^1-1 \right)/\beta^3 & \mbox{otherwise.}
\end{array} \right. \eeq Here we assumed, as noted above, that the
shell mass is constant until the halo virial radius is reached, at
which point the outflow begins to sweep up material from the IGM. We
thus derive the following equations: \beq \frac{d^2 \hat{R}} {d
\hat{t}^2}= \left\{ \begin{array}{ll} \frac{3}{\alpha_S(\beta)}
\frac{\hat{p}}{\hat{\rho}_b\, \hat{R} }-\frac{a}{2} \hat{R} H_0^2 \Omm
\bar{\delta} (\beta) & \mbox{if $\beta<1$} \vspace{.1in} \\
\frac{3}{\alpha_S(\beta) \hat{R}}
\left[\frac{\hat{p}}{\hat{\rho}_b}-\left( \frac{d\hat{R}} {d\hat{t}}
\right)^2 \right]-\frac{a}{2} \hat{R} H_0^2 \Omm \bar{\delta} (\beta)
+ \frac{a}{4} \hat{R} H_0^2 \Omega_b \alpha_S(\beta) &
\mbox{otherwise,} \end{array} \right. \eeq along with \beq
\frac{d}{d\hat{t}} \left( \hat{R}^5 \hat{p}_{\rm int} \right)=
\frac{a^4}{2 \pi} L \hat{R}^2\ . \eeq In the evolution equation for
$\hat{R}$, for $\beta < 1$ we assume for simplicity that the baryons
are distributed in the same way as the dark matter, since in any case
the dark matter halo dominates the overall gravitational potential.
For $\beta>1$, however, we correct (via the last term on the
right-hand side) for the presence of mass in the shell, since at
$\beta \gg 1$ this term may become important. The $\beta > 1$ equation
also includes the braking force due to the swept-up IGM mass. The
enclosed mean overdensity for the NFW profile [equation (\ref{NFW}) in
\S (\ref{sec2.3})] surrounded by matter at the mean density is \beq
\bar\delta(\beta)= \left\{
\begin{array}{ll} \frac{ \Delta_c}{\Ommz \beta^3} \frac{\ln (1+\cN
\beta) - \cN \beta/(1+ \cN \beta)} {\ln (1+\cN )-\cN/ (1+\cN)}&
\mbox{if $\beta<1$} \vspace{.1in} \\ \left( \frac{ \Delta_c}{\Ommz}-1
\right)\frac{1} {\beta^3} & \mbox{otherwise.}
\end{array} \right. \eeq

The physics of supernova shells is discussed in Ostriker \& McKee
(1988) along with a number of analytical solutions. The propagation of
cosmological blast waves has also been computed by Ostriker \& Cowie
(1981), Bertschinger (1985) and Carr \& Ikeuchi (1985). Voit (1996)
derived an exact analytic solution to the fluid equations which,
although of limited validity, is nonetheless useful for understanding
roughly how the outflow size depends on several of the parameters. The
solution requires an idealized case of an outflow which at all times
expands into a homogeneous IGM. Peculiar gravitational forces, and the
energy lost in escaping from the host halo, are neglected, cooling and
ionization losses are also assumed to be negligible, and the external
pressure is not included. The dissipated energy is assumed to be
retained, i.e., $f_d$ is set equal to unity. Under these conditions,
the standard Sedov-Taylor self-similar solution (Sedov 1946, 1959;
Taylor 1950) generalizes to the cosmological case as follows (Voit
1996): \beq \hat{R}=\left( \frac{\xi \hat{E}_0}{ \hat{\rho}_b} \right)
^{1/5} \hat{t}^{\,2/5}\ , \label{Voit} \eeq where $\xi=2.026$ and
$\hat{E}_0= E_0/(1+z_1)^2$ in terms of the initial (i.e., at
$t=\hat{t}=0$ and $z=z_1$) energy $E_0$. Numerically, the comoving
radius is \beq \hat{R}= 280 \left( \frac{0.022}{\Omega_b h^2}\,
\frac{E_0}{10^{56}{\rm erg}} \right) ^{1/5} \left(\frac{10}{1+z_1} \,
\frac{\hat{t}}{10^{10}{\rm yr}} \right) ^{2/5}\ {\rm kpc}\ . \eeq

In solving the equations described above, we assume that the shock
front expands into a pre-ionized region which then recombines after a
time determined by the recombination rate. Thus, the external pressure
is included initially, it is turned off after the pre-ionized region
recombines, and it is then switched back on at a lower redshift when
the universe is reionized. When the ambient IGM is neutral and the
pressure is off, the shock loses energy to ionization. In practice we
find that the external pressure is unimportant during the initial
expansion, although it {\it is}\, generally important after
reionization. Also, at high redshift ionization losses are much
smaller than losses due to Compton cooling. In the results shown
below, we assume an instantaneous reionization at $z=9$.

\begin{figure}[htbp]
\epsscale{0.7}
\plotone{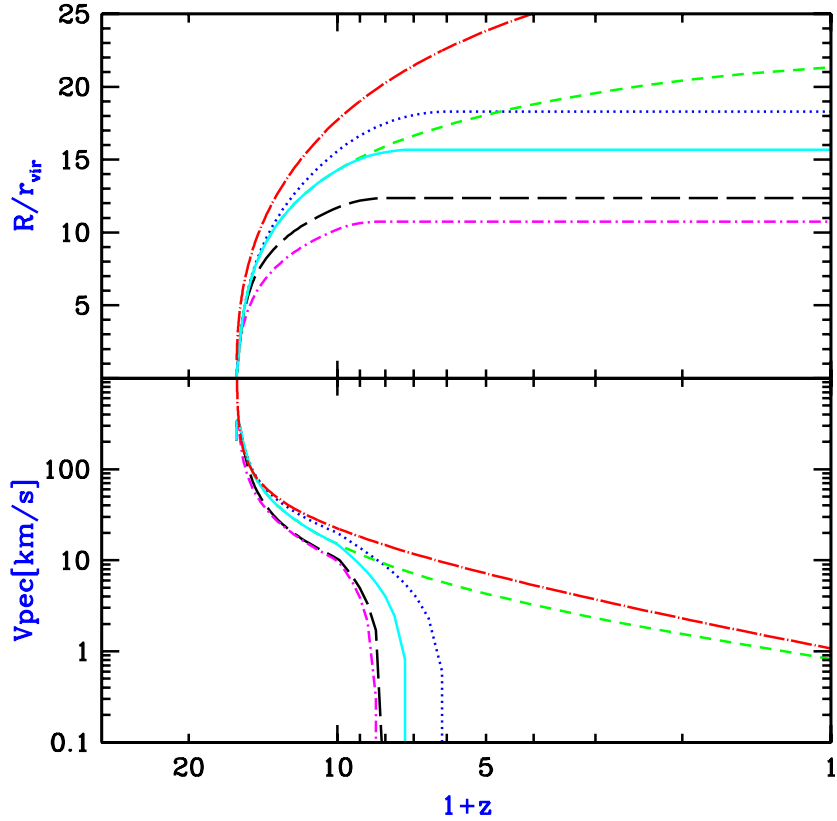}
\caption{Evolution of a supernova outflow from a $z=15$ halo of
circular velocity $V_c=20$ km/s. Plotted are the shell comoving radius
in units of the virial radius of the source halo (top panel), and the
physical peculiar velocity of the shock front (bottom panel). Results
are shown for the standard parameters $\fin=0.1$, $f_d=1$, $\fw=75\%$,
and $\fg=50\%$ (solid curve). Also shown for comparison are the cases
of no cooling (dotted curve), no reionization (short-dashed curve),
$f_d=0$ (long-dashed curve), or $\fw=15\%$ and $\fg=10\%$ (dot-short
dashed curve), as well as the simple Voit (1996) solution of equation
(\ref{Voit}) for the standard parameter set (dot-long dashed
curve). In cases where the outflow halts, we freeze the radius at the
point of maximum expansion.}
\label{figVII1}
\end{figure}

Figure~\ref{figVII1} shows the results for a starting redshift $z=15$,
for a halo of mass $5.4 \times 10^7 M_{\sun}$, stellar mass $8.0
\times 10^5 M_{\sun}$, comoving $\hat{r}_{\rm vir}=12$ kpc, and
circular velocity $V_c=20$ km/s. We show the shell comoving radius in
units of the virial radius of the source halo (top panel), and the
physical peculiar velocity of the shock front (bottom panel). Results
are shown (solid curve) for the standard set of parameters $\fin=0.1$,
$f_d=1$, $\fw=75\%$, and $\fg=50\%$. For comparison, we show several
cases which adopt the standard parameters except for no cooling
(dotted curve), no reionization (short-dashed curve), $f_d=0$
(long-dashed curve), or $\fw=15\%$ and $\fg=10\%$ (dot-short dashed
curve). When reionization is included, the external pressure halts the
expanding bubble. We freeze the radius at the point of maximum
expansion (where $d \hat{R}/d\hat{t}=0$), since in reality the shell
will at that point begin to spread and fill out the interior volume
due to small-scale velocities in the IGM. For the chosen parameters,
the bubble easily escapes from the halo, but when $\fw$ and $\fg$ are
decreased the accumulated IGM mass slows down the outflow more
effectively. In all cases the outflow reaches a size of 10--20 times
$\hat{r}_{\rm vir}$, i.e., 100--200 comoving kpc. If all the metals
are ejected (i.e., $\fe=1$), then this translates to an average
metallicity in the shell of $\sim 1$--5$\times 10^{-3}$ in units of
the solar metallicity (which is $2\%$ by mass). The asymptotic size of
the outflow varies roughly as $\fw^{1/5}$, as predicted by the simple
solution in equation (\ref{Voit}), but the asymptotic size is rather
insensitive to $\fg$ (at a fixed $\fw$) since the outflow mass becomes
dominated by the swept-up IGM mass once $\hat{R} \ga 4 \hat{r}_{\rm
vir}$. With the standard parameter values (i.e., those corresponding
to the solid curve), Figure~\ref{figVII1} also shows (dot-long dashed
curve) the Voit (1996) solution of equation (\ref{Voit}). The Voit
solution behaves similarly to the no-reionization curve at low
redshift, although it overestimates the shock radius by $\sim 30\%$,
and the overestimate is greater compared to the more realistic case
which does include reionization.

\begin{figure}[htbp]
\epsscale{0.7}
\plotone{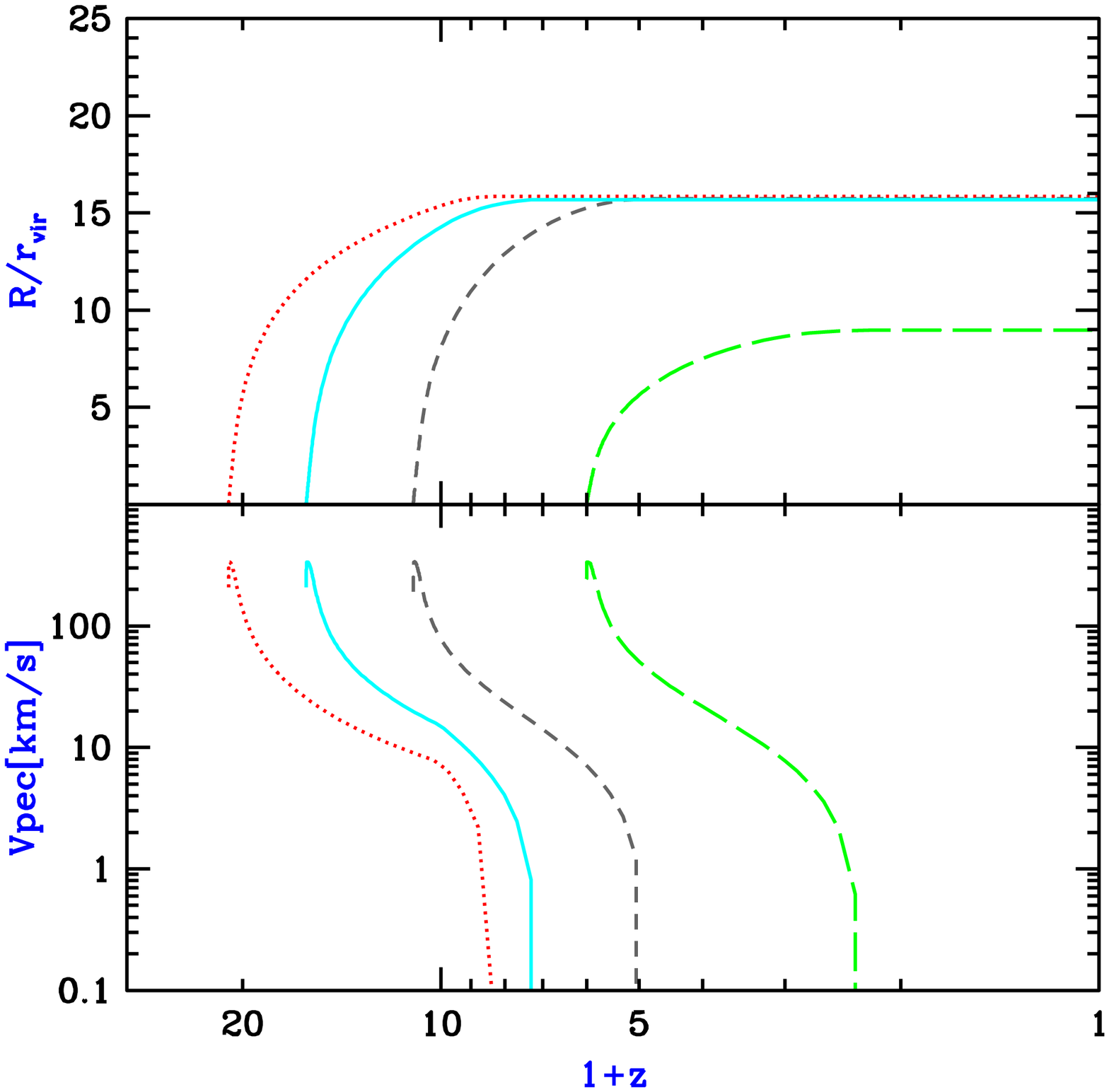}
\caption{Evolution of supernova outflows at different redshifts. The
top and bottom panels are arranged similarly to Figure~\ref{figVII1}.
The $z=15$ outflow (solid curve) is repeated from
Figure~\ref{figVII1}, and it is compared here to outflows with the
same parameters but starting at $z=20$ (dotted curve), $z=10$
(short-dashed curve), and $z=5$ (long-dashed curve). A $V_c=20$ km/s
halo is assumed except for $z=5$, in which case a $V_c=42$ km/s halo
is assumed to produce the outflow (see text).}
\label{figVII2}
\end{figure}

Figure~\ref{figVII2} shows different curves than Figure~\ref{figVII1}
but on an identical layout. A single curve starting at $z=15$ (solid
curve) is repeated from Figure~\ref{figVII1}, and it is compared here
to outflows with the same parameters but starting at $z=20$ (dotted
curve), $z=10$ (short-dashed curve), and $z=5$ (long-dashed curve). A
$V_c=20$ km/s halo, with a stellar mass equal to $1.5\%$ of the total
halo mass, is chosen at the three higher redshifts, but at $z=5$ a
$V_c=42$ km/s halo is assumed. Because of the suppression of gas
infall after reionization (\S \ref{sec6.5}), we assume that the $z=5$
outflow is produced by supernovae from a stellar mass equal to only
$0.3\%$ of the total halo mass (with a similarly reduced initial shell
mass), thus leading to a relatively small final shell radius. The main
conclusion from both Figures is the following: In all cases, the
outflow undergoes a rapid initial expansion over a fractional redshift
interval $\delta z/z \sim 0.2$, at which point the shell has slowed
down to $\sim 10$ km/s from an initial 300 km/s. The rapid
deceleration is due to the accumulating IGM mass. External pressure
from the reionized IGM completely halts all high-redshift outflows,
and even without this effect most outflows would only move at $\sim
10$ km/s after the brief initial expansion. Thus, it may be possible
for high-redshift outflows to pollute the Lyman alpha forest with
metals without affecting the forest hydrodynamically at $z \la
4$. While the bulk velocities of these outflows may dissipate quickly,
the outflows do sweep away the IGM and create empty bubbles. The
resulting effects on observations of the Lyman alpha forest should be
studied in detail (some observational signatures of feedback have been
suggested recently by Theuns, Mo, \& Schaye 2001).

Barkana \& Loeb (2001, in preparation) derive the overall filling
factor of supernova bubbles based on this formalism. In the following
subsection we survey previous analytic and numerical work on the
collective astrophysical effects of galactic outflows. 


\subsection{Effect of Outflows on Dwarf Galaxies and on the IGM}
\label{sec7.2}

Galactic outflows represent a complex feedback process which affects
the evolution of cosmic gas through a variety of phenomena. Outflows
inject hydrodynamic energy into the interstellar medium of their host
galaxy. As shown in the previous subsection, even a small fraction of
this energy suffices to eject most of the gas from a dwarf galaxy,
perhaps quenching further star formation after the initial burst. At
the same time, the enriched gas in outflows can mix with the
interstellar medium and with the surrounding IGM, allowing later
generations of stars to form more easily because of metal-enhanced
cooling. On the other hand, the expanding shock waves may also strip
gas in surrounding galaxies and suppress star formation.

Dekel \& Silk (1986) attempted to explain the different properties of
diffuse dwarf galaxies in terms of the effect of galactic outflows
(see also Larson 1974; Vader 1986, 1987). They noted the observed
trends whereby lower-mass dwarf galaxies have a lower surface
brightness and metallicity, but a higher mass-to-light ratio, than
higher mass galaxies. They argued that these trends are most naturally
explained by substantial gas removal from an underlying dark matter
potential. Galaxies lying in small halos can eject their remaining gas
after only a tiny fraction of the gas has turned into stars, while
larger galaxies require more substantial star formation before the
resulting outflows can expel the rest of the gas. Assuming a wind
efficiency $\fw \sim 100\%$, Dekel \& Silk showed that outflows in
halos below a circular velocity threshold of $V_{\rm crit} \sim 100$
km/s have sufficient energy to expel most of the halo
gas. Furthermore, cooling is very efficient for the characteristic gas
temperatures associated with $V_{\rm crit} \la 100$ km/s halos, but it
becomes less efficient in more massive halos. As a result, this
critical velocity is expected to signify a dividing line between
bright galaxies and diffuse dwarf galaxies. Although these simple
considerations may explain a number of observed trends, many details
are still not conclusively determined. For instance, even in galaxies
with sufficient energy to expel the gas, it is possible that this
energy gets deposited in only a small fraction of the gas, leaving the
rest almost unaffected.

Since supernova explosions in an inhomogeneous interstellar medium
lead to complicated hydrodynamics, in principle the best way to
determine the basic parameters discussed in the previous subsection
($\fw$, $\fg$, and $\fe$) is through detailed numerical simulations of
individual galaxies. Mac Low \& Ferrara (1999) simulated a gas disk
within a $z=0$ dark matter halo. The disk was assumed to be azimuthal
and initially smooth. They represented supernovae by a central source
of energy and mass, assuming a constant luminosity which is maintained
for 50 million years. They found that the hot, metal-enriched ejecta
can in general escape from the halo much more easily than the colder
gas within the disk, since the hot gas is ejected in a tube
perpendicular to the disk without displacing most of the gas in the
disk. In particular, most of the metals were expelled except for the
case with the most massive halo considered (with $10^9 M_{\sun}$ in
gas) and the lowest luminosity ($10^{37}$ erg/s, or a total injection
of $2 \times 10^{52}$ erg). On the other hand, only a small fraction
of the total gas mass was ejected except for the least massive halo
(with $10^6 M_{\sun}$ in gas), where a luminosity of $10^{38}$ erg/s
or more expelled most of the gas. We note that beyond the standard
issues of numerical resolution and convergence, there are several
difficulties in applying these results to high-redshift dwarf
galaxies. Clumping within the expanding shells or the ambient
interstellar medium may strongly affect both the cooling and the
hydrodynamics. Also, the effect of distributing the star formation
throughout the disk is unclear since in that case several
characteristics of the problem will change; many small explosions will
distribute the same energy over a larger gas volume than a single
large explosion [as in the Sedov-Taylor solution (Sedov 1946, 1959;
Taylor 1950); see, e.g., equation (\ref{Voit})], and the geometry will
be different as each bubble tries to dig its own escape route through
the disk. Also, high-redshift disks should be denser by orders of
magnitude than $z=0$ disks, due to the higher mean density of the
universe at early times.  Thus, further numerical simulations of this
process are required in order to assess its significance during the
reionization epoch.

Some input on these issues also comes from observations. Martin (1999)
showed that the hottest extended X-ray emission in galaxies is
characterized by a temperature of $\sim 10^{6.7}$ K. This hot gas,
which is lifted out of the disk at a rate comparable to the rate at
which gas goes into new stars, could escape from galaxies with
rotation speeds of $\la 130$ km/s. However, these results are based on
a small sample that includes only the most vigorous star-forming local
galaxies, and the mass-loss rate depends on assumptions about the
poorly understood transfer of mass and energy among the various phases
of the interstellar medium.

Many authors have attempted to estimate the overall cosmological
effects of outflows by combining simple models of individual outflows
with the formation rate of galaxies, obtained via semi-analytic
methods (Couchman \& Rees 1986; Blanchard, Valls-Gabaud, \& Mamon
1992; Tegmark, et al.\ 1993; Voit 1996; Nath \& Trentham 1997; Prunet
\& Blanchard 2000; Ferrara, Pettini, \& Shchekinov 2000; Scannapieco
\& Broadhurst 2000) or numerical simulations (Gnedin \& Ostriker 1997;
Gnedin 1998; Cen \& Ostriker 1999; Aguirre et al.\ 2000a). The main
goal of these calculations is to explain the characteristic
metallicities of different environments as a function of redshift. For
example, the IGM is observed to be enriched with metals at redshifts
$z\la 5$. Identification of \ion{C}{4}, \ion{Si}{4} and \ion{O}{6}
absorption lines which correspond to Ly$\alpha$ absorption lines in
the spectra of high-redshift quasars has revealed that the low-density
IGM has been enriched to a metal abundance (by mass) of $Z_{\rm
IGM}\sim 10^{-2.5 (\pm 0.5)}Z_\odot$, where $Z_\odot=0.019$ is the
solar metallicity (Meyer \& York 1987; Tytler et al.\ 1995; Songaila
\& Cowie 1996; Lu et al 1998; Cowie \& Songaila 1998; Songaila 1997;
Ellison et al.\ 2000; Prochaska \& Wolfe 2000). The metal enrichment
has been clearly identified down to \ion{H}{1} column densities of
$\sim 10^{14.5}~{\rm cm^{-2}}$. The detailed comparison of
cosmological hydrodynamic simulations with quasar absorption spectra
has established that the forest of Ly$\alpha$ absorption lines is
caused by the smoothly-fluctuating density of the neutral component of
the IGM (Cen et al.\ 1994; Zhang, Anninos \& Norman 1995; Hernquist et
al.\ 1996). The simulations show a strong correlation between the
\ion{H}{1} column density and the gas overdensity $\delta_{\rm gas}$
(e.g., Dav\'{e} et al.\ 1999), implying that metals were dispersed
into regions with an overdensity as low as $\delta_{\rm gas}\sim 3$ or
possibly even lower.

In general, dwarf galaxies are expected to dominate metal enrichment
at high-redshift for several reasons. As noted above and in the
previous subsection, outflows can escape more easily out of the
potential wells of dwarfs. Also, at high redshift, massive halos are
rare and dwarf halos are much more common. Finally, as already noted,
the Sedov-Taylor solution (Sedov 1946, 1959; Taylor 1950) [or equation
(\ref{Voit})] implies that for a given total energy and expansion
time, multiple small outflows (i.e., caused by explosions with a small
individual energy release) fill large volumes more effectively than
would a smaller number of large outflows. Note, however, that the
strong effect of feedback in dwarf galaxies may also quench star
formation rapidly and reduce the efficiency of star formation in
dwarfs below that found in more massive galaxies.

Cen \& Ostriker (1999) showed via numerical simulation that metals
produced by supernovae do not mix uniformly over cosmological
volumes. Instead, at each epoch the highest density regions have much
higher metallicity than the low-density IGM. They noted that early
star formation occurs in the most overdense regions, which therefore
reach a high metallicity (of order a tenth of the solar value) by $z
\sim 3$, when the IGM metallicity is lower by 1--2 orders of
magnitude. At later times, the formation of high-temperature clusters
in the highest-density regions suppresses star formation there, while
lower-density regions continue to increase their metallicity. Note,
however, that the spatial resolution of the hydrodynamic code of Cen
\& Ostriker is a few hundred kpc, and anything occurring on smaller
scales is inserted directly via simple parametrized models.
Scannapieco \& Broadhurst (2000) implemented expanding outflows within
a numerical scheme which, while not a full gravitational simulation,
did include spatial correlations among halos. They showed that winds
from low-mass galaxies may also strip gas from nearby galaxies (see
also Scannapieco, Ferrara, \& Broadhurst 2000), thus suppressing star
formation in a local neighborhood and substantially reducing the
overall abundance of galaxies in halos below a mass of $\sim 10^{10}
M_{\sun}$. Although quasars do not produce metals, they may also
affect galaxy formation in their vicinity via energetic outflows
(Efstathiou \& Rees 1988; Babul \& White 1991; Silk \& Rees 1998;
Natarajan, Sigurdsson, \& Silk 1998).

Gnedin \& Ostriker (1997) and Gnedin (1998) identified another mixing
mechanism which, they argued, may be dominant at high redshift ($z \ga
4$). In a collision between two proto-galaxies, the gas components
collide in a shock and the resulting pressure force can eject a few
percent of the gas out of the merger remnant. This is the merger
mechanism, which is based on gravity and hydrodynamics rather than
direct stellar feedback. Even if supernovae inject most of their
metals in a local region, larger-scale mixing can occur via
mergers. Note, however, that Gnedin's (1998) simulation assumed a
comoving star formation rate at $z \ga 5$ of $\sim 1 M_{\sun}$ per
year per comoving Mpc$^3$, which is 5--10 times larger than the
observed rate at redshift 3--4 (\S \ref{sec8.1}). Aguirre et al.\
(2000a) used outflows implemented in simulations to conclude that
winds of $\sim 300$ km/s at $z \la 6$ can produce the mean metallicity
observed at $z \sim 3$ in the Ly$\alpha$ forest. Aguirre et al.\
(2000b) explored another process, where metals in the form of dust
grains are driven to large distances by radiation pressure, thus
producing large-scale mixing without displacing or heating large
volumes of IGM gas. The success of this mechanism depends on detailed
microphysics such as dust grain destruction and the effect of magnetic
fields. The scenario, though, may be directly testable because it
leads to significant ejection only of elements which solidify as
grains.

Feedback from galactic outflows encompasses a large variety of processes
and influences. The large range of scales involved, from stars or quasars
embedded in the interstellar medium up to the enriched IGM on cosmological
scales, make possible a multitude of different, complementary approaches,
promising to keep galactic feedback an active field of research.

\section{\bf Properties of the Expected Source Population}
\label{sec8}

\subsection{The Cosmic Star Formation History}
\label{sec8.1}

One of the major goals of the study of galaxy formation is to achieve
an observational determination and a theoretical understanding of the
cosmic star formation history. By now, this history has been sketched
out to a redshift $z\sim 4$ (see, e.g., the compilation of Blain et
al.\ 1999a). This is based on a large number of observations in
different wavebands. These include various
ultraviolet/optical/near-infrared observations (Madau et al.\ 1996;
Gallego et al.\ 1996; Lilly et al.\ 1996; Connolly et al.\ 1997;
Treyer et al.\ 1998; Tresse \& Maddox 1998; Pettini et al.\ 1998a,b;
Cowie, Songaila \& Barger 1999; Gronwall 1999; Glazebrook et al.\
1999; Yan et al.\ 1999; Flores et al.\ 1999; Steidel et al.\ 1999). At
the shortest wavelengths, the extinction correction is likely to be
large (a factor of $\sim 5$) and is still highly uncertain. At longer
wavelengths, the star formation history has been reconstructed from
submillimeter observations (Blain et al.\ 1999b; Hughes et al.\ 1998)
and radio observations (Cram 1998). In the submillimeter regime, a
major uncertainty results from the fact that only a minor portion of
the total far infrared emission of galaxies comes out in the observed
bands, and so in order to estimate the star formation rate it is
necessary to assume a spectrum based, e.g., on a model of the dust
emission (see the discussion in Chapman et al.\ 2000). In general,
estimates of the star formation rate (hereafter SFR) apply
locally-calibrated correlations between emission in particular lines
or wavebands and the total SFR. It is often not possible to check
these correlations directly on high-redshift populations, and together
with the other uncertainties (extinction and incompleteness) this
means that current knowledge of the star formation history must be
considered to be a qualitative sketch only.  Despite the relatively
early state of observations, a wealth of new observatories in all
wavelength regions promise to greatly advance the field. In
particular, \NGST\, will be able to detect galaxies and hence
determine the star formation history out to $z\ga 10$.

Hierarchical models have been used in many papers to match observations on
star formation at $z \la 4$ (see, e.g. Baugh et al.\ 1998; Kauffmann \&
Charlot 1998; Somerville \& Primack 1998, and references therein). In
this section we focus on theoretical predictions for the cosmic star
formation rate at higher redshifts. The reheating of the IGM during
reionization suppressed star formation inside the smallest halos (\S
\ref{sec6.5}). Reionization is therefore predicted to cause a drop in
the cosmic SFR. This drop is accompanied by a dramatic fall in the
number counts of faint galaxies. Barkana \& Loeb (2000b) argued that a
detection of this fall in the faint luminosity function could be used
to identify the reionization redshift observationally.

A model for the SFR can be constructed based on the extended
Press-Schechter theory. The starting point is the abundance of dark
matter halos, obtained using the Press-Schechter model. The abundance
of halos evolves with redshift as each halo gains mass through mergers
with other halos. If $dp[M_1,t_1 \rightarrow M,t]$ is the probability
that a halo of mass $M_1$ at time $t_1$ will have merged to form a
halo of mass between $M$ and $M+dM$ at time $t>t_1$, then in the limit
where $t_1$ tends to $t$ we obtain an instantaneous merger rate
$d^2p[M_1\rightarrow M,t]/(dM\ dt)$. This quantity was evaluated by
Lacey \& Cole [1993, their equation~(2.18)], and it is the basis for
modeling the rate of galaxy formation.

Once a dark matter halo has collapsed and virialized, the two
requirements for forming new stars are gas infall and cooling. We
assume that by the time of reionization, photo-dissociation of
molecular hydrogen (see \S \ref{sec3.3}) has left only atomic
transitions as an avenue for efficient cooling. Before reionization,
therefore, galaxies can form in halos down to a circular velocity of
$V_c\sim 17\ {\rm km\ s}^{-1}$, where this limit is set by cooling. On
the other hand, when a volume of the IGM is ionized by stars or
quasars, the gas is heated and the increased pressure suppresses gas
infall into halos with a circular velocity below $V_c\sim 80\ {\rm km\
s}^{-1}$, halting infall below $V_c\sim 30\ {\rm km\ s}^{-1}$ (\S
\ref{sec6.5}). Since the suppression acts only in regions that have
been heated, the reionization feedback on galaxy formation depends on
the fraction of the IGM which is ionized at each redshift. In order to
include a gradual reionization in the model, we take the simulations
of Gnedin (2000a) as a guide for the redshift interval of
reionization.

In general, new star formation in a given galaxy can occur either from
primordial gas or from recycled gas which has already undergone a previous
burst of star formation. The former occurs when a massive halo accretes gas
from the IGM or from a halo which is too small to have formed stars. The
latter occurs when two halos, in which a fraction of the gas has already
turned to stars, merge and trigger star formation in the remaining
gas. Numerical simulations of starbursts in interacting $z=0$ galaxies
(e.g., Mihos \& Hernquist 1994; 1996) found that a merger triggers
significant star formation in a halo even if it merges with a much less
massive partner. Preliminary results (Somerville 2000, private
communication) from simulations of mergers at $z \sim 3$ find that they
remain effective at triggering star formation even when the initial disks
are dominated by gas.  Regardless of the mechanism, we assume that feedback
limits the star formation efficiency, so that only a fraction $\eta$ of
the gas is turned into stars.

Given the SFR and the total number of stars in a halo of mass $M$, the
luminosity and spectrum can be derived from an assumed stellar initial
mass function. We assume an initial mass function which is similar to
the one measured locally. If $n(M)$ is the total number of stars with
masses less than $M$, then the stellar initial mass function,
normalized to a total mass of $1 M_{\sun}$, is (Scalo 1998) \beq
\frac{dn}{d\ln(M)}=\left\{ \begin{array}{ll} 0.396\, M_1^{-0.2}\ , &
0.1 < M_1 < 1.0 \\ 0.396\, M_1^{-1.7}\ , & 1.0 < M_1 < 10 \\
\label{imf} 0.158\, M_1^{-1.3}\ , & 10 < M_1 < 100 \end{array}
\right. \eeq where $M_1=M/M_{\sun}$. We assume a metallicity
$Z=0.001$, and use the stellar population model results of Leitherer
et al.\ (1999) \footnote{Model spectra of star-forming galaxies were
obtained from http://www.stsci.edu/science/starburst99/}. We also
include a Ly$\alpha$ cutoff in the spectrum due to absorption by the
dense Ly$\alpha$ forest. We do not, however, include dust extinction,
which could be significant in some individual galaxies despite the low
mean metallicity expected at high redshift.

Much of the star formation at high redshift is expected to occur in
low mass, faint galaxies, and even \NGST\, may only detect a fraction of
the total SFR. A realistic estimate of this fraction must include the
finite resolution of the instrument as well as its detection limit for
faint sources (Barkana \& Loeb 2000a). We characterize the
instrument's resolution by a minimum circular aperture of angular
diameter $\theta_a$. We describe the sensitivity of \NGST\, by
$F_{\nu}^{\rm ps}$, the minimum spectral flux\footnote{Note that
$F_{\nu}^{\rm ps}$ is the total spectral flux of the source, not just
the portion contained within the aperture.}, averaged over wavelengths
0.6--3.5$\mu$m, required to detect a point source (i.e., a source which
is much smaller than $\theta_a$). For an extended source of diameter
$\theta_s \gg \theta_a$, we assume that the signal-to-noise ratio can
be improved by using a larger aperture, with diameter $\theta_s$. The
noise amplitude scales as the square root of the number of noise (sky)
photons, or the square root of the corresponding sky area. Thus, the
total flux needed for detection of an extended source at a given
signal-to-noise threshold is larger than $F_{\nu}^{\rm ps}$ by a
factor of $\theta_s/ \theta_a$. We adopt a simple interpolation
formula between the regimes of point-like and extended sources, and
assume that a source is detectable if its flux is at least $\sqrt{1+
(\theta_s/ \theta_a)^2}\, F_{\nu}^{\rm ps}$.

We combine this result with a model for the distribution of disk sizes
at each value of halo mass and redshift (\S \ref{sec5.1}). We adopt a
value of $F_{\nu}^{\rm ps}=0.25$ nJy \footnote{We obtained the flux
limit using the \NGST\, calculator at
http://www.ngst.stsci.edu/nms/main/}, assuming a deep 300-hour
integration on an 8-meter \NGST\, and a spectral resolution of
10:1. This resolution should suffice for a $\sim 10\%$ redshift
measurement, based on the Ly$\alpha$ cutoff. We also choose the
aperture diameter to be $\theta_a=0\farcs 06$, close to the expected
\NGST\, resolution at $2\mu$m.

\begin{figure}[htbp]
\epsscale{0.7}
\plotone{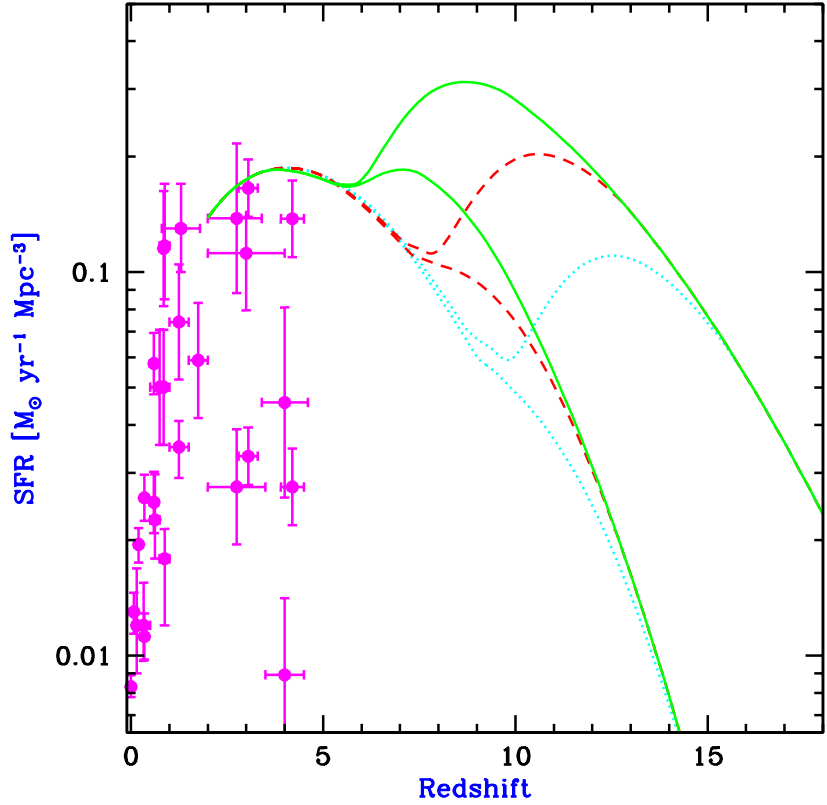}
\caption{Redshift evolution of the SFR (in $M_{\sun}$ per year per
comoving Mpc$^3$), adopted from Figure~1 of Barkana \& Loeb (2000b)
with slight modifications. Points with error bars are observational
estimates (compiled by Blain et al.\ 1999a). Also shown are model
predictions for a reionization redshift $\zr=6$ (solid curves),
$\zr=8$ (dashed curves), and $\zr=10$ (dotted curves), with a star
formation efficiency $\eta=10\%$. In each pair of curves, the upper
one is the total SFR, and the lower one is the fraction detectable
with \NGST\, at a limiting point source flux of 0.25 nJy. We assume
the $\Lambda$CDM model (with parameters given at the end of \S
\ref{sec1}).}
\label{fig8n}
\end{figure}

Figure~\ref{fig8n} shows our predictions for the star formation
history of the universe, adopted from Figure~1 of Barkana \& Loeb
(2000b) with slight modifications (in the initial mass function and
the values of the cosmological parameters). Letting $\zr$ denote the
redshift at the end of overlap, we show the SFR for $\zr=6$ (solid
curves), $\zr=8$ (dashed curves), and $\zr=10$ (dotted curves). In
each pair of curves, the upper one is the total SFR, and the lower one
is the fraction detectable with {\it NGST}.\, The curves assume a star
formation efficiency $\eta=10\%$ and an IGM temperature $T_{\rm IGM}=
2\times 10^4$ K. Although photoionization directly suppresses new gas
infall after reionization, it does not immediately affect mergers
which continue to trigger star formation in gas which had cooled prior
to reionization. Thus, the overall suppression is dominated by the
effect on star formation in primordial (unprocessed) gas. The
contribution from merger-induced star formation is comparable to that
from primordial gas at $z<\zr$, and it is smaller at $z>\zr$. However,
the recycled gas contribution to the {\em detectable} SFR is dominant
at the highest redshifts, since the brightest, highest mass halos form
in mergers of halos which themselves already contain stars. Thus, even
though most stars at $z > \zr$ form out of primordial,
zero-metallicity gas, a majority of stars in detectable galaxies may
form out of the small gas fraction that has already been enriched by
the first generation of stars.

Points with error bars in Figure~\ref{fig8n} are observational
estimates of the cosmic SFR per comoving volume at various redshifts
(as compiled by Blain et al.\ 1999a). We choose $\eta=10\%$ to obtain
a rough agreement between the models and these observations at $z\sim
3$--4. An efficiency of order this value is also suggested by
observations of the metallicity of the Ly$\alpha$ forest at $z=3$
(Haiman \& Loeb 1999b). The SFR curves are roughly proportional to the
value of $\eta$. Note that in reality $\eta$ may depend on the halo
mass, since the effect of supernova feedback may be more pronounced in
small galaxies (\S \ref{sec7}). Figure~\ref{fig8n} shows a sharp
rise in the total SFR at redshifts higher than $\zr$. Although only a
fraction of the total SFR can be detected with {\it NGST},\, the
detectable SFR displays a definite signature of the reionization
redshift. However, current observations at lower redshifts demonstrate
the observational difficulty in measuring the SFR directly. The
redshift evolution of the faint luminosity function provides a
clearer, more direct observational signature. We discuss this topic
next.

\subsection{Number Counts}
\label{sec8.2}

\subsubsection{Galaxies}
\label{sec8.2.1}

As shown in the previous section, the cosmic star formation history should
display a signature of the reionization redshift. Much of the increase in
the star formation rate beyond the reionization redshift is due to star
formation occurring in very small, and thus faint, galaxies. This evolution
in the faint luminosity function constitutes the clearest observational
signature of the suppression of star formation after reionization.
  
\begin{figure}[htbp]
\epsscale{0.7}
\plotone{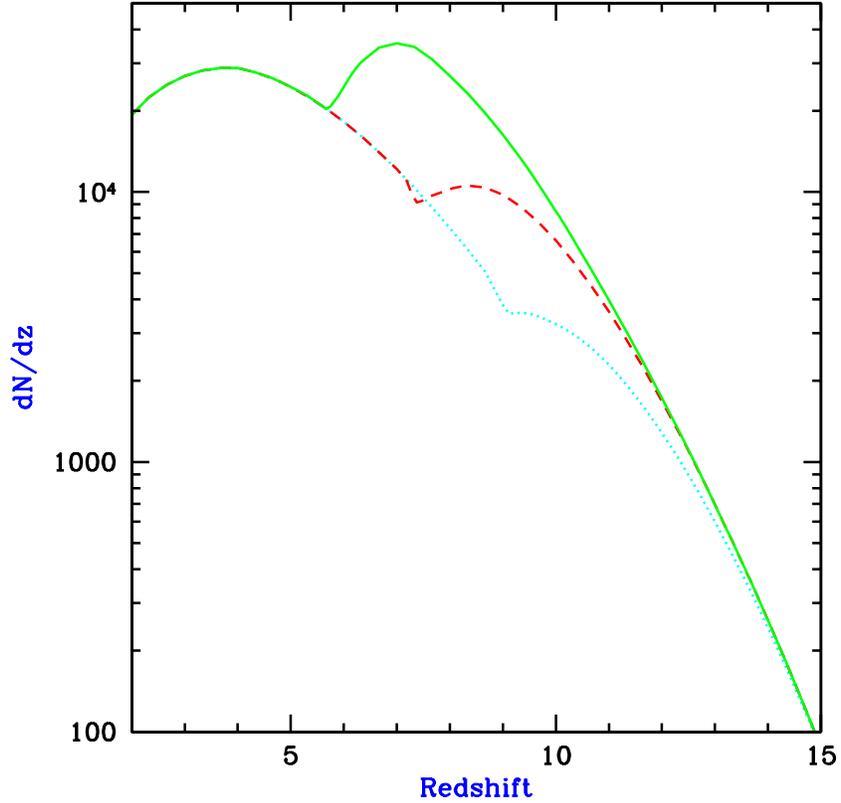}
\caption{Predicted redshift distribution of galaxies observed with
{\it NGST},\, adopted and modified from Figure~7 of Barkana \& Loeb
(2000a). The distribution in the $\Lambda$CDM model (with parameters
given at the end of \S \ref{sec1}), with a star formation efficiency
$\eta=10\%$, is shown for a reionization redshift $\zr=6$ (solid
curve), $\zr=8$ (dashed curve), and $\zr=10$ (dotted curve). The
plotted quantity is $dN/dz$, where $N$ is the number of galaxies per
\NGST\, field of view. All curves assume a limiting point source flux
of 0.25 nJy.}
\label{fig8k}
\end{figure}

Figure~\ref{fig8k} shows the predicted redshift distribution in
$\Lambda$CDM (with parameters given at the end of \S \ref{sec1}) of
galaxies observed with {\it NGST}.\, The plotted quantity is $dN/dz$,
where $N$ is the number of galaxies per \NGST\, field of view
($4\arcmin \times 4\arcmin$). The model predictions are shown for a
reionization redshift $\zr=6$ (solid curve), $\zr=8$ (dashed curve),
and $\zr=10$ (dotted curve), with a star formation efficiency
$\eta=10\%$. All curves assume a point-source detection limit of 0.25
nJy. This plot is updated from Figure~7 of Barkana \& Loeb (2000a) in
that redshifts above $\zr$ are included.


\begin{figure}[htbp]
\epsscale{0.7}
\plotone{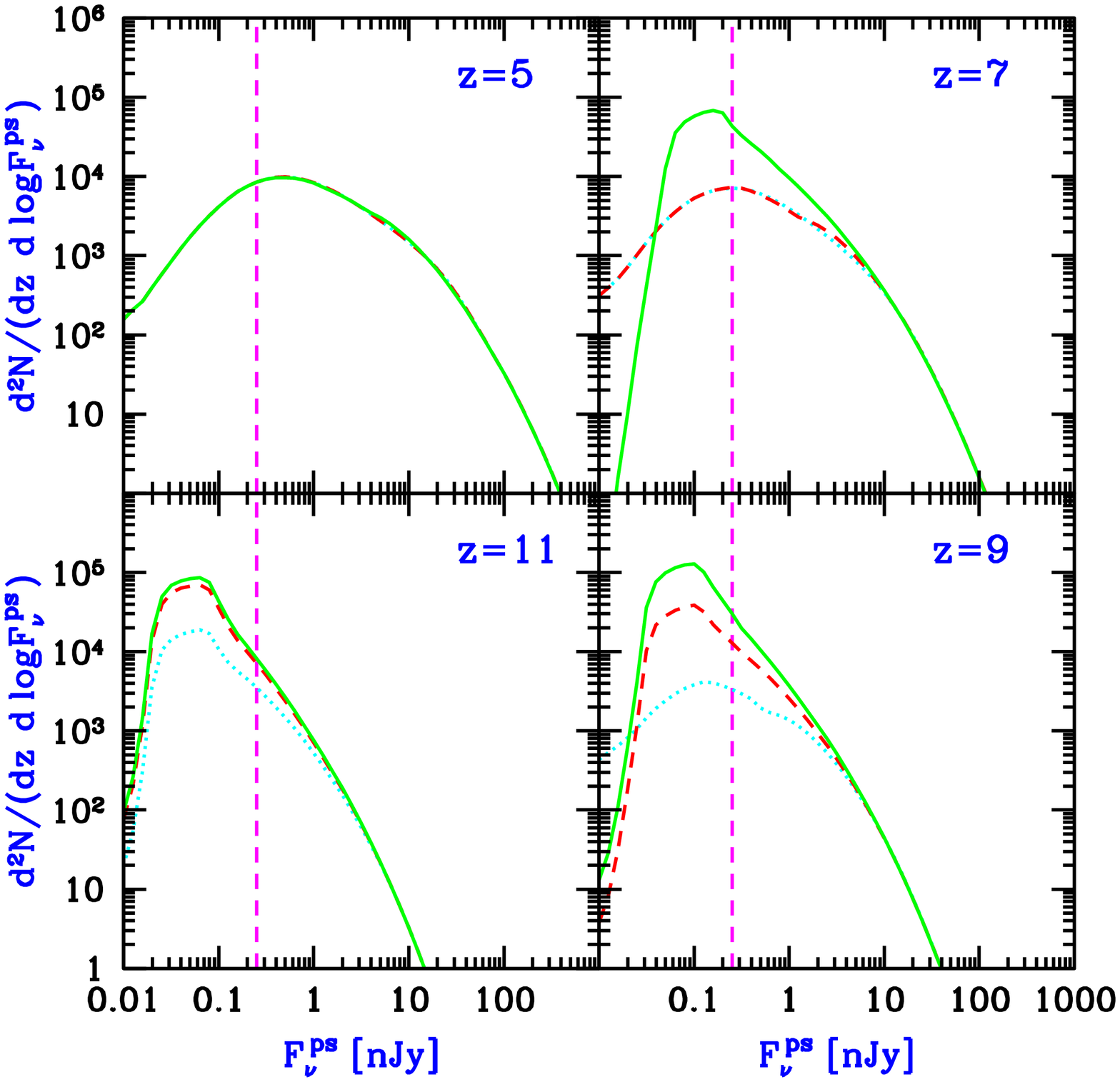}
\caption{Predicted luminosity function of galaxies at a fixed
redshift, adopted and modified from Figure~2 of Barkana \& Loeb
(2000b). With $\eta=10\%$, the curves show $d^2N/(dz\ d\ln
F_{\nu}^{\rm ps})$, where $N$ is the total number of galaxies in a
single field of view of {\it NGST},\, and $F_{\nu}^{\rm ps}$ is the
limiting point source flux averaged over 0.6--3.5$\mu$m for {\it
NGST}.\, The luminosity function is shown at $z=5$, $z=7$, $z=9$, and
$z=11$, with redshift increasing clockwise starting with the upper
left panel. Each case assumes the $\Lambda$CDM model (with parameters
given at the end of \S \ref{sec1}) and a reionization redshift $\zr=6$
(solid curves), $\zr=8$ (dashed curves), or $\zr=10$ (dotted
curves). The expected
\NGST\, detection limit is shown by the vertical dashed line.}
\label{fig8l}
\end{figure}

Clearly, thousands of galaxies are expected to be found at high
redshift. This will allow a determination of the luminosity function
at many redshift intervals, and thus a measurement of its evolution.
As the redshift is increased, the luminosity function is predicted to
gradually change shape during the overlap era of reionization.
Figure~\ref{fig8l} shows the predicted evolution of the luminosity
function for various values of $\zr$. This Figure is adopted from
Figure~2 of Barkana \& Loeb (2000b) with modifications (in the initial
mass function, the values of the cosmological parameters, and the plot
layout). All curves show $d^2N/(dz\ d\ln F_{\nu}^{\rm ps})$, where $N$
is the total number of galaxies in a single field of view of {\it
NGST},\, and $F_{\nu}^{\rm ps}$ is the limiting point source flux at
0.6--3.5$\mu$m for {\it NGST}.\, Each panel shows the result for a
reionization redshift $\zr=6$ (solid curve), $\zr=8$ (dashed curve),
and $\zr=10$ (dotted curve). Figure~\ref{fig8l} shows the luminosity
function as observed at $z=5$ (upper left panel) and (proceeding
clockwise) at $z=7$, $z=9$, and $z=11$.  Although our model assigns a
fixed luminosity to all halos of a given mass and redshift, in reality
such halos would have some dispersion in their merger histories and
thus in their luminosities. We thus include smoothing in the plotted
luminosity functions. Note the enormous increase in the number density
of faint galaxies in a pre-reionization universe. Observing this
dramatic increase toward high redshift would constitute a clear
detection of reionization and of its major effect on galaxy formation.

\subsubsection{Quasars}
\label{sec8.2.2}

Dynamical studies indicate that massive black holes exist in the
centers of most nearby galaxies (Richstone et al.\ 1998; Kormendy \&
Ho 2000; Kormendy 2000, and references therein). This leads to the
profound conclusion that black hole formation is a generic consequence
of galaxy formation.  The suggestion that massive black holes reside
in galaxies and power quasars dates back to the sixties (Zel'dovich
1964; Salpeter 1964; Lynden-Bell 1969).  Efstathiou \& Rees (1988)
pioneered the modeling of quasars in the modern context of galaxy
formation theories. The model was extended by Haehnelt \& Rees (1993)
who added more details concerning the black hole formation efficiency
and lightcurve.  Haiman \& Loeb (1998) and Haehnelt, Natarajan, \&
Rees (1998) extrapolated the model to high redshifts. All of these
discussions used the Press-Schechter theory to describe the abundance
of galaxy halos as a function of mass and redshift. More recently,
Kauffmann \& Haehnelt (2000; also Haehnelt \& Kauffmann 2000) embedded
the description of quasars within semi-analytic modeling of galaxy
formation, which uses the extended Press-Schechter formalism to
describe the merger history of galaxy halos.

In general, the predicted evolution of the luminosity function of
quasars is constrained by the need to match the observed quasar
luminosity function at redshifts $z\la 5$, as well as data from the
Hubble Deep Field (HDF) on faint point-sources.  Prior to
reionization, we may assume that quasars form only in galaxy halos
with a circular velocity $\ga 10~{\rm km~s^{-1}}$ (or equivalently a
virial temperature $\ga 10^4$ K), for which cooling by atomic
transitions is effective. After reionization, quasars form only in
galaxies with a circular velocity $\ga 50\ {\rm km~s^{-1}}$, for which
substantial gas accretion from the warm ($\sim 10^4$ K) IGM is
possible. The limits set by the null detection of quasars in the HDF
are consistent with the number counts of quasars which are implied by
these thresholds (Haiman, Madau, \& Loeb 1999).

For spherical accretion of ionized gas, the bolometric luminosity
emitted by a black hole has a maximum value beyond which radiation
pressure prevents gas accretion. This Eddington luminosity (Eddington
1926) is derived by equating the radiative repulsive force on a free
electron to the gravitational attractive force on an ion in the
plasma,
\begin{equation}
{L \sigma_T\over 4\pi r^2 c} = {GM_{\rm bh}\mu_e m_p \over r^2}\ ,
\end{equation}
where $\sigma_T=6.65\times 10^{-25}~{\rm cm^2}$ is the Thomson
cross-section, $\mu_e m_p$ is the average ion mass per electron, and
$M_{\rm bh}$ is the black hole mass.  Since both forces scale as $r^{-2}$,
the limiting Eddington luminosity is independent of radius $r$ in the
Newtonian regime, and for gas of primordial composition is given by,
\begin{equation}
L_E={4\pi c G M_{\rm bh} \mu_e m_p\over \sigma_T }= 1.45 \times 10^{46}
\left({M_{\rm
bh}\over 10^8 M_\odot}\right)~{\rm erg~s^{-1}}\ .
\end{equation}
Generically, the Eddington limit applies to within a factor of order
unity also to simple accretion flows in a non-spherical geometry
(Frank, King, \& Raine 1992).

The total luminosity of a black hole is related to its mass accretion
rate by the radiative efficiency, $\epsilon$,
\begin{equation}
L=\epsilon \dot{M}_{\rm bh} c^2 .
\end{equation}
For accretion through a thin Keplerian disk onto a Schwarzschild
(non-rotating) black hole, $\epsilon=5.7\%$, while for a maximally rotating
Kerr black hole, $\epsilon=42\%$ (Shapiro \& Teukolsky 1983, p.\ 429).  The
thin disk configuration, for which these high radiative efficiencies are
attainable, exists for $L_{\rm disk}\la 0.5 L_E$ (Laor \& Netzer 1989) .

The accretion time can be defined as
\begin{equation}
\tau={M_{\rm bh}\over {\dot M}_{\rm bh}}= 4\times 10^7~{\rm yr} 
\left(\epsilon \over 0.1\right) \left({L\over L_E}\right)^{-1}\ .
\end{equation}
This time is comparable to the dynamical time inside the central kpc
of a typical galaxy, $t_{\rm dyn}\sim ({\rm 1~kpc}/100~{\rm
km~s^{-1}})=10^7~{\rm yr}$. As long as its fuel supply is not limited
and $\epsilon$ is constant, a black hole radiating at the Eddington
limit will grow its mass exponentially with an $e$-folding time equal
to $\tau$. The fact that $\tau$ is much shorter than the age of the
universe even at high redshift implies that black hole growth is
mainly limited by the feeding rate ${\dot M}_{\rm bh}(t)$, or by the
total fuel reservoir, and not by the Eddington limit.

The ``simplest model'' for quasars involves the following three
assumptions (Haiman \& Loeb 1998):

\noindent
(i) A fixed fraction of the baryons in each ``newly formed'' galaxy
ends up making a central black hole.

\noindent
(ii) Each black hole shines at its maximum (Eddington) luminosity for
a universal amount of time.

\noindent
(iii) All black holes share the same emission spectrum during their
luminous phase.

Note that these assumptions relate only to the most luminous phase of
the black hole accretion process, and they may not be valid during
periods when the radiative efficiency or the mass accretion rate is
very low.  Such periods are not of interest here since they do not
affect the luminosity function of bright quasars, which is the
observable we wish to predict. The first of the above assumptions is
reasonable as long as the fraction of virialized baryons in the
universe is much smaller than unity; it does not include a separate
mechanism for fueling black hole growth during mergers of
previously-formed galaxies, and thus, under this assumption, black
holes would not grow in mass once most of the baryons were virialized.
The second hypothesis is motivated by the fact that for a sufficiently
high fueling rate (which may occur in the early stage of the
collapse/merger of a galaxy), quasars are likely to shine at their
maximum possible luminosity. The resulting luminosity should be close
to the Eddington limit over a period of order $\tau$. The third
assumption can be implemented by incorporating the average quasar
spectrum measured by Elvis et al.\ (1994).

\begin{figure}[htbp] 
\includegraphics{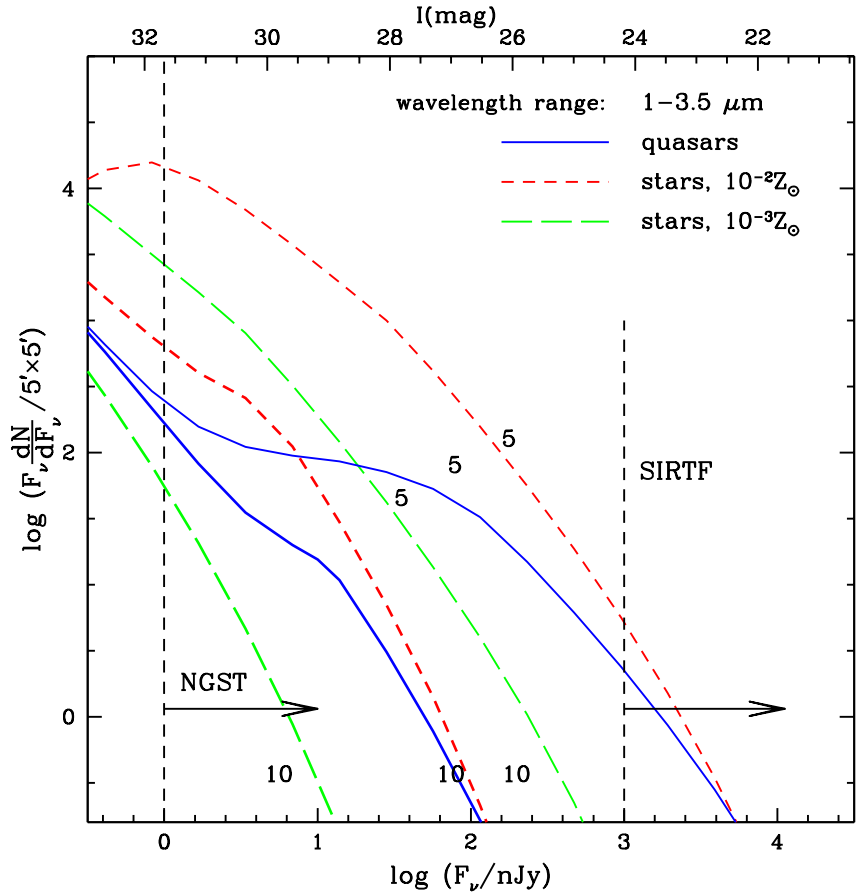}
\vspace{6.2in}
\caption{Infrared number counts of quasars (averaged over the
wavelength interval of $1$--$3.5\mu$m) based on the ``simplest quasar
model'' of Haiman \& Loeb (1999b). The solid curves refer to quasars,
while the long/short dashed curves correspond to star clusters with
low/high normalization for the star formation efficiency.  The curves
labeled ``5'' or ``10'' show the cumulative number of objects with
redshifts above $z=5$ or 10.}
\label{fig8e}
\end{figure}

At high redshifts the number of ``newly formed'' galaxies can be
estimated based on the time-derivative of the Press-Schechter mass
function, since the collapsed fraction of baryons is small and most
galaxies form out of the unvirialized IGM.  Haiman \& Loeb (1998,
1999a) have shown that the above simple prescription provides an
excellent fit to the observed evolution of the luminosity function of
bright quasars between redshifts $2.6<z<4.5$ (see the analytic
description of the existing data in Pei 1995). The observed decline in
the abundance of bright quasars (Schneider, Schmidt, \& Gunn 1991; Pei
1995) results from the deficiency of massive galaxies at high
redshifts. Consequently, the average luminosity of quasars declines
with increasing redshift. The required ratio between the mass of the
black hole and the total baryonic mass inside a halo is $M_{\rm
bh}/M_{\rm gas}=10^{-3.2}\Omm/\Omega_b=5.5\times10^{-3}$, comparable
to the typical value of $\sim 2$--$6\times10^{-3}$ found for the ratio
of black hole mass to spheroid mass in nearby elliptical galaxies
(Magorrian et al.\ 1998; Kormendy 2000). The required lifetime of the
bright phase of quasars is $\sim 10^6$ yr. Figure~\ref{fig8e} shows
the most recent prediction of this model (Haiman \& Loeb 1999a) for the
number counts of high-redshift quasars, taking into account the
above-mentioned thresholds for the circular velocities of galaxies
before and after reionization\footnote{Note that the post-reionization
threshold was not included in the original discussion of Haiman \&
Loeb (1998).}.

We do, however, expect a substantial intrinsic scatter in the ratio
$M_{\rm bh}/M_{\rm gas}$. Observationally, the scatter around the
average value of $\log_{10} (M_{\rm bh}/L)$ is 0.3 (Magorrian et al.\
1998), while the standard deviation in $\log_{10} (M_{\rm bh}/M_{\rm
gas})$ has been found to be $\sigma\sim 0.5$.  Such an intrinsic
scatter would flatten the predicted quasar luminosity function at the
bright end, where the luminosity function is steeply declining.
However, Haiman \& Loeb (1999a) have shown that the flattening
introduced by the scatter can be compensated for through a modest
reduction in the fitted value for the average ratio between the black
hole mass and halo mass by $\sim 50\%$ in the relevant mass range
($10^{8}~{\rm M_\odot}\la M_{\rm bh}\la 10^{10}~{\rm M_\odot}$).

In reality, the relation between the black hole and halo masses may be
more complicated than linear. Models with additional free parameters,
such as a non-linear (mass and redshift dependent) relation between
the black hole and halo mass, can also produce acceptable fits to
the observed quasar luminosity function (Haehnelt et al.\ 1998).  The
nonlinearity in a relation of the type $M_{\rm bh}\propto M_{\rm
halo}^\alpha$ with $\alpha>1$, may be related to the physics of the
formation process of low-luminosity quasars(Haehnelt et al.\ 1998;
Silk \& Rees 1998), and can be tuned so as to reproduce the black hole
reservoir with its scatter in the local universe (Cattaneo, Haehnelt,
\& Rees 1999).  Recently, a tight correlation between the masses of
black holes and the velocity dispersions of the bulges in which they
reside, $\sigma$, was identified in nearby galaxies. Ferrarese \&
Merritt (2000; see also Merritt \& Ferrarese 2001) inferred a
correlation of the type $M_{\rm bh}\propto \sigma^{4.72\pm 0.36}$,
based on a selected sample of a dozen galaxies with reliable $M_{\rm
bh}$ estimates, while Gebhardt et al.\ (2000a,b) have found a somewhat
shallower slope, $M_{\rm bh}\propto \sigma^{3.75(\pm0.3)}$ based on a
significantly larger sample. A non-linear relation of $M_{\rm
bh}\propto \sigma^5 \propto M_{\rm halo}^{5/3}$ has been predicted by
Silk \& Rees (1998) based on feedback considerations, but the observed
relation also follows naturally in the standard semi-analytic models
of galaxy formation (Haehnelt \& Kauffmann 2000).

Figure~\ref{fig8e} shows the predicted number counts in the ``simplest
model'' described above (Haiman \& Loeb 1999a), normalized to a
$5^{\prime}\times5^{\prime}$ field of view.  Figure~\ref{fig8e} shows
separately the number per logarithmic flux interval of all objects
with redshifts $z>5$ (thin lines), and $z>10$ (thick lines). The
number of detectable sources is high; \NGST\, will be able to probe of
order $100$ quasars at $z>10$, and $\sim200$ quasars at $z>5$ per
$5^{\prime}\times5^{\prime}$ field of view.  The bright-end tail of
the number counts approximately follows a power law, with
$dN/dF_\nu\propto F_\nu^{-2.5}$.  The dashed lines show the
corresponding number counts of ``star-clusters'', assuming that each
halo shines due to a starburst that converts a fraction of 2\%
(long-dashed) or 20\% (short-dashed) of the gas into stars.

\begin{figure}[htbp] 
\includegraphics{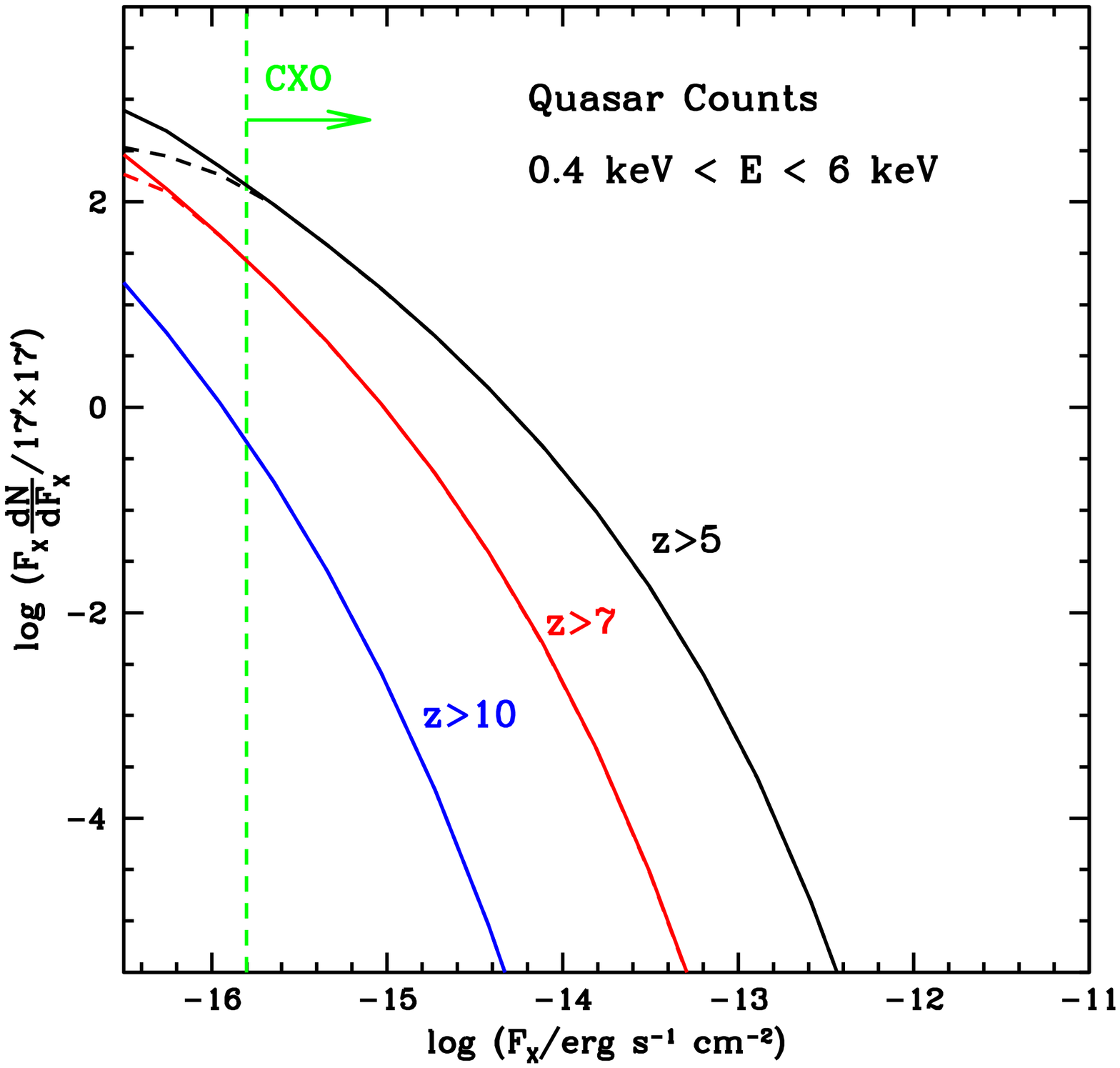}
\vspace{4.9in}
\caption{Total number of quasars with redshift exceeding $z=5$, $z=7$,
and $z=10$ as a function of observed X-ray flux in the {\it CXO}
detection band (from Haiman \& Loeb 1999a). The numbers are normalized
per $17^{\prime}\times 17^{\prime}$ area of the sky. The solid curves
correspond to a cutoff in circular velocity for the host halos of
$v_{\rm circ}\geq 50~{\rm km~s^{-1}}$, the dashed curves to a cutoff
of $v_{\rm circ}\geq 100~{\rm km~s^{-1}}$.  The vertical dashed line
shows the {\it CXO} sensitivity for a 5$\sigma$ detection of a point
source in an integration time of $5\times10^5$ seconds.  }
\label{fig8f}
\end{figure}

Similar predictions can be made in the X-ray regime.
Figure~\ref{fig8f} shows the number counts of high-redshift X-ray
quasars in the above ``simplest model''.  This model fits the X-ray
luminosity function of quasars at $z\sim 3.5$ as observed by ROSAT
(Miyaji, Hasinger, \& Schmidt 2000), using the same parameters
necessary to fit the optical data (Pei 1995).  Deep optical or
infrared follow-ups on deep images taken with the Chandra X-ray
satellite (CXO; see, e.g., Mushotzky et al.\ 2000; Barger et al.\
2001; Giacconi et al.\ 2000) may be used to test these predictions in
the relatively near future.

\noindent
  
\noindent
  
\vspace{-0.61in}
The ``simplest model'' mentioned above predicts that black holes and
stars make comparable contributions to the ionizing background prior
to reionization. Consequently, the reionization of hydrogen and helium
is predicted to occur roughly at the same epoch. A definitive
identification of the \ion{He}{2} reionization redshift will provide
another powerful test of this model. Further constraints on the
lifetime of the active phase of quasars may be provided by future
measurements of the clustering properties of quasars (Haehnelt et al.\ 
1998; Martini \& Weinberg 2001; Haiman \& Hui 2000).

\subsubsection{Supernovae}
\label{sec8.2.3}

The detection of galaxies and quasars becomes increasingly difficult
at a higher redshift. This results both from the increase in the
luminosity distance and the decrease in the average galaxy mass with
increasing redshift. It therefore becomes advantageous to search for
transient sources, such as supernovae or $\gamma$-ray bursts, as
signposts of high-redshift galaxies (Miralda-Escud\'e \& Rees
1997). Prior to or during the epoch of reionization, such sources are
likely to outshine their host galaxies.

The metals detected in the IGM (see \S \ref{sec7.2}) signal the
existence of supernova (SN) explosions at redshifts $z\ga 5$. Since
each SN produces an average of $\sim 1M_\odot$ of heavy elements
(Woosley \& Weaver 1995), the inferred metallicity of the IGM, $Z_{\rm
IGM}$, implies that there should be a supernova at $z\ga 5$ for each
$\sim 1.7 \times 10^4M_\odot \times (Z_{\rm IGM}/10^{-2.5}
Z_\odot)^{-1}$ of baryons in the universe. We can therefore estimate
the total supernova rate, on the entire sky, necessary to produce
these metals at $z\sim 5$. Consider all SNe which are observed over a
time interval $\Delta t$ on the whole sky. Due to the cosmic time
dilation, they correspond to a narrow redshift shell centered at the
observer of proper width $c\Delta t/(1+z)$.  In a flat $\Omm=0.3$
cosmology, the total mass of baryons in a narrow redshift shell of
width $c \Delta t/(1+z)$ around $z=5$ is $\sim [4 \pi
(1+z)^{-3}(1.8c/H_0)^2 c \Delta t]\times [\Omega_b (3H_0^2/8\pi
G)(1+z)^3] = 4.9\, c^3\Omega_b \Delta t/G$.  Hence, for $h=0.7$ the
total supernova rate across the entire sky at $z\ga 5$ is estimated to
be (Miralda-Escud\'e \& Rees 1997),
\begin{equation}
{\dot N}_{SN}\approx 10^8 \left({Z_{\rm IGM}\over
10^{-2.5}Z_\odot}\right)\ {\rm yr}^{-1}\ ,
\end{equation}
or roughly one SN per square arcminute per year.

The actual SN rate at a given observed flux threshold is determined by
the star formation rate per unit comoving volume as a function of
redshift and the initial mass function of stars (Madau, della Valle,
\& Panagia 1998; Woods \& Loeb 1998; Sullivan et al.\ 2000).  To
derive the relevant expression for flux-limited observations, we
consider a general population of transient sources which are standard
candles in peak flux and are characterized by a comoving rate per unit
volume $R(z)$.  The observed number of new events per unit time
brighter than flux $F_\nu$ at observed wavelength $\lambda$ for such a
population is given by
\begin{equation}
\dot{N}(F_\nu;\lambda) = \int_0^{z_{{\rm max}(F_\nu;\lambda)}} R(z)
(1+z)^{-1}
(dV_{\rm c}/dz) dz,
\label{CountsPerYear}
\end{equation}
where $z_{\rm max}(F_\nu,\lambda)$ is the maximum redshift at which a
source will appear brighter than $F_\nu$ at an observed wavelength
$\lambda=c/\nu$, and $dV_{\rm c}$ is the cosmology-dependent comoving
volume element corresponding to a redshift interval $dz$. The above
integrand includes the $(1+z)$ reduction in the apparent rate due to
the cosmic time dilation.

\begin{figure}[htbp] 
\includegraphics{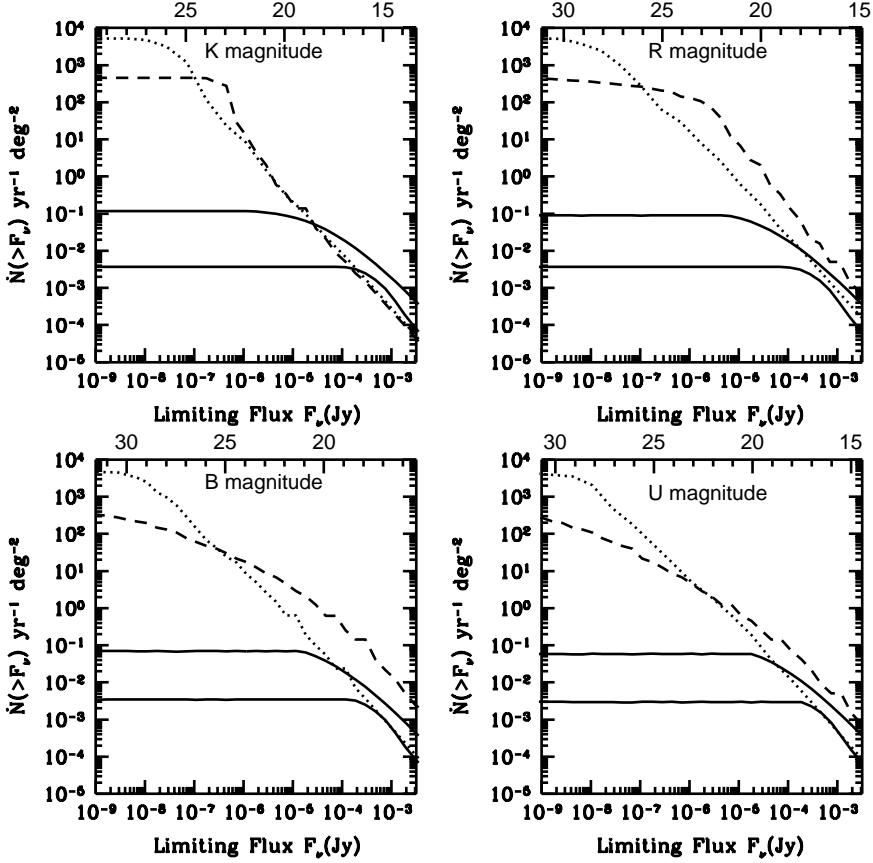}
\vspace{4.5in}
\caption{Predicted cumulative rate $\dot{N}(>F_\nu)$ per year per
square degree of supernovae at four wavelengths, corresponding to the
$K$, $R$, $B$, and $U$ bands (from Woods \& Loeb 1998).  The broken
lines refer to different supernova types, namely SNe Ia ({\it dashed
curves}) and SNe II ({\it dotted curves}). For comparison, the solid
curves show estimates for the rates of gamma-ray burst (GRB)
afterglows; the lower solid curve assumes the best-fit rate and
luminosity for GRB sources which trace the star formation history
(Wijers et al.\ 1998), while the upper solid curve assumes the best-fit
values for a non-evolving GRB population.  }
\label{fig8g}
\end{figure}

Figure~\ref{fig8g} shows the predicted SN rate as a function of
limiting flux in various bands (Woods \& Loeb 1998), based on the
comoving star formation rate as a function of redshift that was
determined empirically by Madau (1997). The actual star formation rate
may be somewhat higher due to corrections for dust extinction (for a
recent compilation of current data, see Blain \& Natarajan 2000).  The
dashed lines correspond to Type Ia SNe and the dotted lines to Type II
SNe. For comparison, the solid lines indicate two crude estimates for
the rate of $\gamma$-ray burst afterglows, which are discussed in
detail in the next section.

\noindent
  
\begin{figure}[htbp] 
\includegraphics{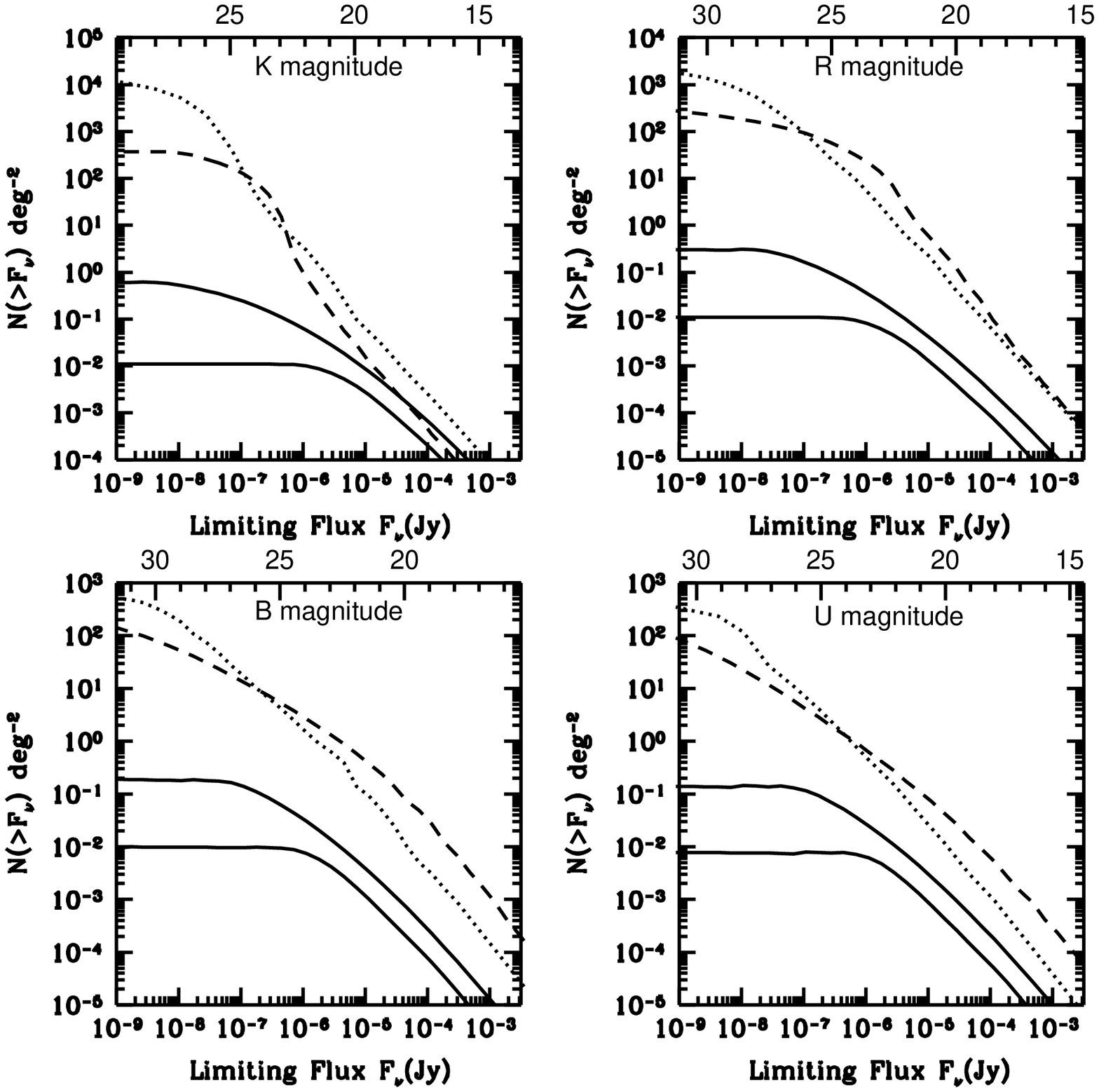}
\vspace{4.7in}
\caption{Cumulative number counts $N(>F_\nu)$ per square degree (from
Woods \& Loeb 1998). The notation is the same as in
Figure~\ref{fig8g}. }
\label{fig8h}
\end{figure}

\vspace{-0.31in}
Equation~(\ref{CountsPerYear}) is appropriate for a threshold
experiment, one which monitors the sky continuously and triggers when
the detected flux exceeds a certain value, and hence identifies the
most distant sources only when they are near their peak flux.  For
search strategies which involve taking a series of ``snapshots'' of a
field and looking for variations in the flux of sources in successive
images, one does not necessarily detect most sources near their peak
flux. In this case, the {\it total}\, number of events (i.e., {\it
not}\, per unit time) brighter than $F_\nu$ at observed wavelength
$\lambda$ is given by
\begin{equation}
N(F_\nu;\lambda) = \int_0^\infty R(z) t_\star (z;F_\nu,\lambda)
(dV_{\rm
c}/dz) dz,
\label{Counts}
\end{equation}
where $t_\star (z;F_\nu,\lambda)$ is the rest-frame duration over
which an event will be brighter than the limiting flux $F_\nu$ at
redshift $z$.  This is a naive estimate of the so-called ``control
time''; in practice, the effective duration over which an event can be
observed is shorter, owing to the image subtraction technique, host
galaxy magnitudes, and a number of other effects which reduce the
detection efficiency (Pain et al.\ 1996). Figure~\ref{fig8h} shows the
predicted number counts of SNe as a function of limiting flux for the
parameters used in Figure~\ref{fig8g} (Woods \& Loeb 1998).

\noindent

\vspace{-0.31in}
Supernovae also produce dust which could process the emission spectrum
of galaxies. Although produced in galaxies, the dust may be expelled
together with the metals out of galaxies by supernova-driven winds.
Loeb \& Haiman (1997) have shown that if each supernova produces $\sim
0.3\, M_\odot$ of galactic dust, and some of the dust is expelled
together with metals out of the shallow potential wells of the early
dwarf galaxies, then the optical depth for extinction by intergalactic
dust may reach a few tenths at $z\sim 10$ for observed wavelengths of
$\sim 0.5$--$1\,\mu$m [see Todini \& Ferrara (2000) for a detailed
discussion on the production of dust in primordial Type II SNe]. The
opacity in fact peaks in this wavelength band since at $z\sim 10$ it
corresponds to rest-frame UV, where normal dust extinction is most
effective. In these estimates, the amplitude of the opacity is
calibrated based on the observed metallicity of the IGM at $z\la 5$.
The intergalactic dust absorbs the UV background at the reionization
epoch and re-radiates it at longer wavelengths. The flux and spectrum
of the infrared background which is produced at each redshift depends
sensitively on the distribution of dust around the ionizing sources,
since the deviation of the dust temperature from the microwave
background temperature depends on the local flux of UV radiation that
it is exposed to. For reasonable choices of parameters, dust could
lead to a significant spectral distortion of the microwave background
spectrum that could be measured by a future spectral mission, going
beyond the upper limit derived by the COBE satellite (Fixsen et
al. 1996).

The metals produced by supernovae may also yield strong molecular line
emission.  Silk \& Spaans (1997) pointed out that the rotational line
emission of CO by starburst galaxies is enhanced at high redshift due
to the increasing temperature of the cosmic microwave background,
which affects the thermal balance and the level populations of the
atomic and molecular species.  They found that the future Millimeter
Array (MMA) could detect a starburst galaxy with a star formation rate
of $\sim 30\, {\rm M_\odot~yr^{-1}}$ equally well at $z=5$ and $z=30$
because of the increasing cosmic microwave background temperature with
redshift. Line emission may therefore be a more powerful probe of the
first bright galaxies than continuum emission by dust.

\subsubsection{Gamma Ray Bursts}
\label{sec8.2.4}

The past decade has seen major observational breakthroughs in the
study of Gamma Ray Burst (GRB) sources. The Burst and Transient Source
Experiment (BATSE) on board the Compton Gamma Ray Observatory (Meegan
et al.\ 1992) showed that the GRB population is distributed
isotropically across the sky, and that there is a deficiency of faint
GRBs relative to a Euclidean distribution.  These were the first
observational clues indicating a cosmological distance scale for GRBs.
The localization of GRBs by X-ray observations with the BeppoSAX
satellite (Costa et al.\ 1997) allowed the detection of afterglow
emission at optical (e.g., van Paradijs et al.\ 1997, 2000) and radio
(e.g., Frail et al.\ 1997) wavelengths up to more than a year
following the events (Fruchter et al.\ 1999; Frail et al.\ 2000). The
afterglow emission is characterized by a broken power-law spectrum
with a peak frequency that declines with time.  The radiation is
well-fitted by a model consisting of synchrotron emission from a
decelerating blast wave (Blandford \& McKee 1976), created by the GRB
explosion in an ambient medium, with a density comparable to that of
the interstellar medium of galaxies (Waxman 1997; Sari, Piran, \&
Narayan 1998; Wijers \& Galama 1999; M\'esz\'aros 1999; but see also
Chevalier \& Li 2000). The detection of spectral features, such as
metal absorption lines in some optical afterglows (Metzger et al.\
1997) and emission lines from host galaxies (Kulkarni et al.\ 2000),
allowed an unambiguous identification of cosmological distances to
these sources.

The nature of the central engine of GRBs is still unknown. Since the
inferred energy release, in cases where it can be securely calibrated
(Freedman \& Waxman 2001; Frail et al.\ 2000), is comparable to that in
a supernova, $\sim 10^{51}~{\rm erg}$, most popular models relate GRBs
to stellar remnants such as neutron stars or black holes (Eichler et
al.\ 1989; Narayan, Paczy\'{n}ski, \& Piran 1992; Paczy\'{n}ski 1991; Usov
1992; Mochkovitch et al.\ 1993; Paczy\'{n}ski 1998; MacFadyen \&
Woosley 1999). Recently it has been claimed that the late evolution of
some rapidly declining optical afterglows shows a component which is
possibly associated with supernova emission (e.g., Bloom et al.\ 1999;
Reichart 1999). If the supernova association is confirmed by detailed
spectra of future afterglows, the GRB phenomenon will be linked to the
terminal evolution of massive stars.

Any association of GRBs with the formation of single compact stars
implies that the GRB rate should trace the star formation history of
the universe (Totani 1997; Sahu et al.\ 1997; Wijers et al.\ 1998; but
see Krumholz, Thorsett \& Harrison 1998).  Owing to their high
brightness, GRB afterglows could in principle be detected out to
exceedingly high redshifts. Just as for quasars, the broad-band
emission of GRB afterglows can be used to probe the absorption
properties of the IGM out to the reionization epoch at redshift $z\sim
10$. Lamb \& Reichart (2000) extrapolated the observed gamma-ray and
afterglow spectra of known GRBs to high redshifts and emphasized the
important role that their detection could play in probing the IGM (see
also Miralda-Escud\'e 1998). In particular, the broad-band afterglow
emission can be used to probe the ionization and metal-enrichment
histories of the intervening IGM during the epoch of reionization.

\begin{figure}[htbp] 
\includegraphics{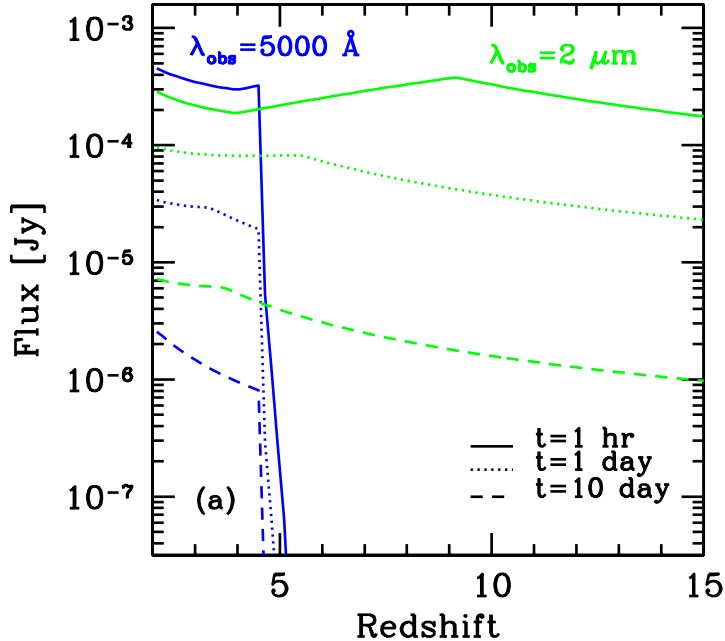}
\vspace{3.3in}
\caption{ Observed flux from a $\gamma$-ray burst afterglow as a
function of redshift (from Ciardi \& Loeb 2000).  The two sets of
curves refer to a photon frequency $\nu=6 \times 10^{14}$ Hz
($\lambda_{obs}=5000$ \AA, thin lines) and $\nu=1.5 \times 10^{14}$ Hz
($\lambda_{obs}=2 \mu$m, thick lines).  Each set shows different
observed times after the GRB trigger; from top to bottom: 1 hour
(solid line), 1 day (dotted) and 10 days (dashed). The sharp
suppression for 5000 \AA\, at $z\ga 4.5$ is due to IGM absorption.}
\label{fig8i}
\end{figure}

\begin{figure}[htbp] 
\includegraphics{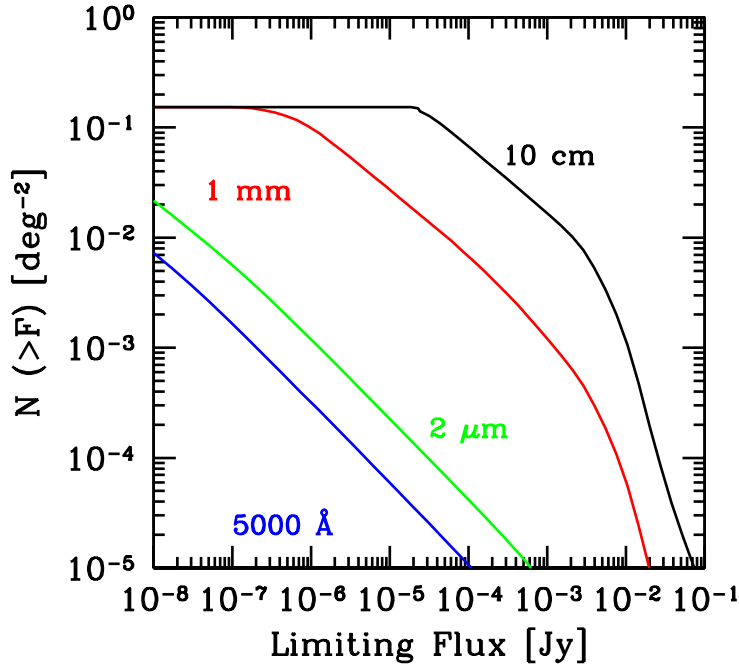}
\vspace{3.4in}
\caption{Predicted number of GRB afterglows per square degree with
observed flux greater than $F$, at several different observed
wavelengths (from Ciardi \& Loeb 2000).  From right to left, the
observed wavelength equals 10 cm, 1 mm, 2 $\mu$m and 5000 \AA.  }
\label{fig8j}
\end{figure}

Ciardi \& Loeb (2000) showed that unlike other sources (such as
galaxies and quasars), which fade rapidly with increasing redshift,
the observed infrared flux from a GRB afterglow at a fixed observed
age is only a weak function of its redshift (Figure~\ref{fig8i}).  A
simple scaling of the long-wavelength spectra and the temporal
evolution of afterglows with redshift implies that at a fixed time-lag
after the GRB in the observer's frame, there is only a mild change in
the {\it observed} flux at infrared or radio wavelengths with
increasing redshift. This results in part from the fact that
afterglows are brighter at earlier times, and that a given observed
time refers to an earlier intrinsic time in the source frame as the
source redshift increases.  The ``apparent brightening'' of GRB
afterglows with redshift could be further enhanced by the expected
increase with redshift of the mean density of the interstellar medium
of galaxies (Wood \& Loeb 2000). Figure~\ref{fig8j} shows the expected
number counts of GRB afterglows, assuming that the GRB rate is
proportional to the star formation rate and that the characteristic
energy output of GRBs is $\sim 10^{52}~{\rm erg}$ and is
isotropic. The figure implies that at any time there should be of
order $\sim 15$ GRBs with redshifts $z\ga 5$ across the sky which are
brighter than $\sim 100$ nJy at an observed wavelength of $\sim
2\mu$m.  The infrared spectrum of these sources could be measured with
{\it NGST}\, as a follow-up on their early X-ray localization with
$\gamma$-ray or X-ray detectors. Prior to reionization, the spectrum
of GRB afterglows could reveal the long sought-after Gunn-Peterson
trough (Gunn \& Peterson 1965) due to absorption by the neutral IGM.

The predicted GRB rate and flux are subject to uncertainties regarding
the beaming of the emission. The beaming angle may vary with observed
time due to the decline with time of the Lorentz factor $\gamma(t)$ of
the emitting material. As long as the Lorentz factor is significantly
larger than the inverse of the beaming angle (i.e., $\gamma \ga
\theta^{-1}$), the afterglow flux behaves as if it were emitted by a
spherically-symmetric fireball with the same explosion energy per unit
solid angle.  However, the lightcurve changes as soon as $\gamma$
declines below $\theta^{-1}$, due to the lateral expansion of the jet
(Rhoads 1997, 1999a,b; Panaitescu \& M\'{e}sz\'{a}ros 1999). Finally,
the isotropization of the energy ends when the expansion becomes
sub-relativistic, at which point the remnant recovers the
spherically-symmetric Sedov-Taylor solution (Sedov 1946, 1959; Taylor
1950) with the total remaining energy. When $\gamma\sim 1$, the
emission occurs from a roughly spherical fireball with the effective
explosion energy per solid angle reduced by a factor of $(2\pi
\theta^2/4\pi)$ relative to that at early times, representing the
fraction of sky around the GRB source which is illuminated by the
initial two (opposing) jets of angular radius $\theta$ (see Ciardi \&
Loeb 2000 for the impact of this effect on the number counts). The
calibration of the GRB event rate per comoving volume, based on the
number counts of GRBs (Wijers et al.\ 1998), is inversely proportional
to this factor.

The main difficulty in using GRBs as probes of the high-redshift
universe is that they are rare, and hence their detection requires
surveys which cover a wide area of the sky.  The simplest strategy for
identifying high-redshift afterglows is through all-sky surveys in the
$\gamma$-ray or X-ray regimes.  In particular, detection of
high-redshift sources will become feasible with the high trigger rate
provided by the forthcoming {\it Swift}\, satellite, to be launched in
2003 (see http://swift.gsfc.nasa.gov/, for more details).  {\it
Swift}\, is expected to localize $\sim$300 GRBs per year, and to
re-point within 20--70 seconds its on-board X-ray and UV-optical
instrumentation for continued afterglow studies. The high-resolution
GRB coordinates obtained by {\it Swift}\, will be transmitted to Earth
within $\sim$50 seconds. Deep follow-up observations will then be
feasible from the ground or using the highly-sensitive infrared
instruments on board {\it NGST}.\, {\it Swift}\, will be sufficiently
sensitive to trigger on the $\gamma$-ray emission from GRBs at
redshifts $z\ga 10$ (Lamb \& Reichart 2000).


\subsection{Distribution of Disk Sizes}
\label{sec8.3}

\begin{figure}[htbp]
\epsscale{0.7}
\plotone{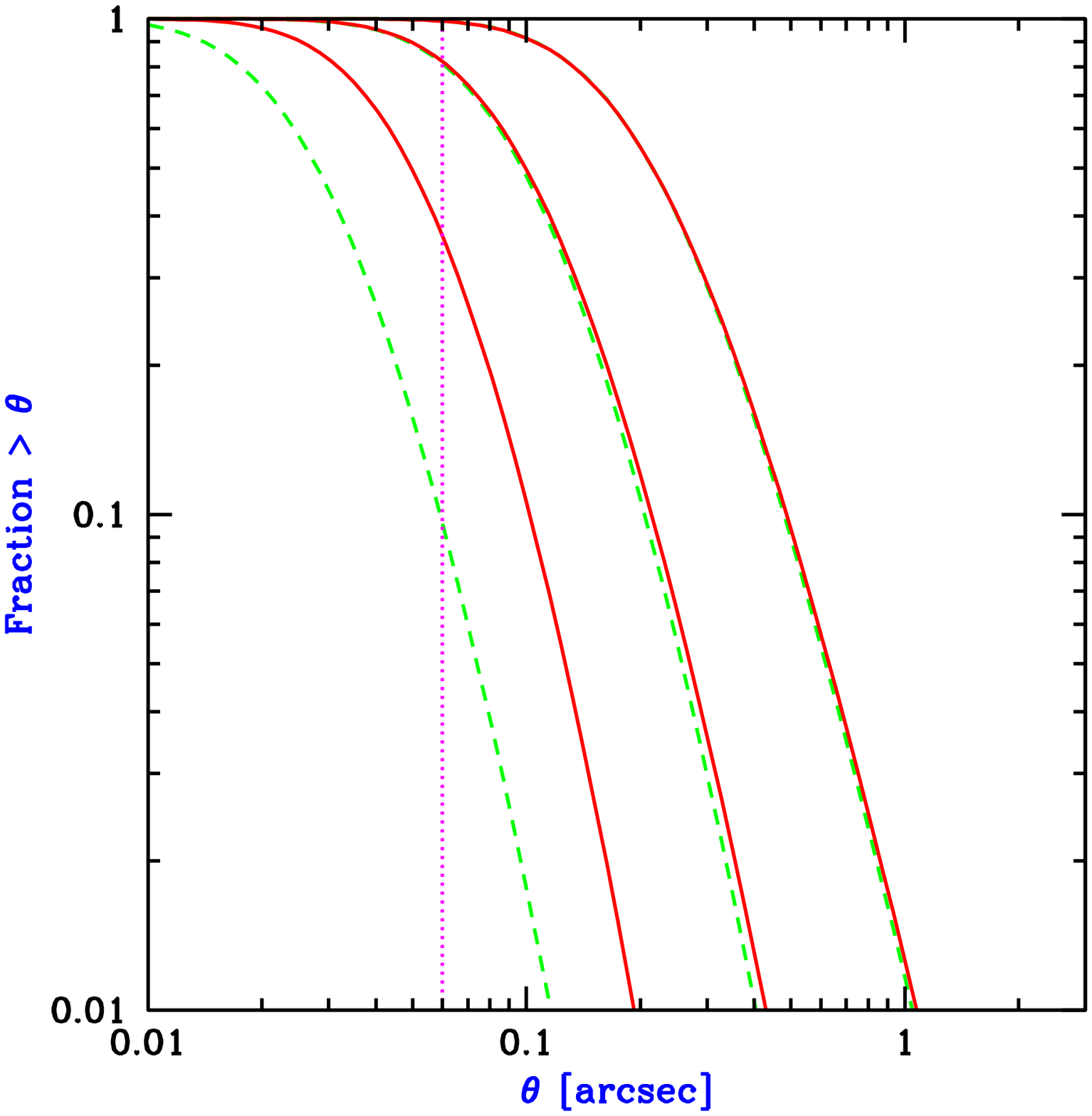}
\caption{Distribution of galactic disk sizes at various redshifts, in
the $\Lambda$CDM model (with parameters given at the end of \S
\ref{sec1}), adopted and modified from Figure~6 of Barkana \& Loeb
(2000a). Given $\theta$ in arcseconds, each curve shows the fraction
of the total number counts contributed by sources larger than
$\theta$. The diameter $\theta$ is measured out to one exponential
scale length. We show three pairs of curves, at $z=2$, $z=5$ and
$z=10$ (from right to left). Each pair includes the distribution for
all galaxies (dashed line), and for galaxies detectable by \NGST\,
(solid line) with a limiting point source flux of 0.25 nJy and with an
efficiency $\eta=10\%$ assumed for the galaxies. The vertical dotted
line indicates the expected \NGST\, resolution of $0\farcs 06$.}
\label{fig8m}
\end{figure}

Given the distribution of disk sizes at each value of halo mass and
redshift (\S \ref{sec5.1}) and the number counts of galaxies (\S
\ref{sec8.2.1}), we derive the predicted size distribution of galactic
disks. Note that although frequent mergers at high redshift may
disrupt these disks and alter the morphologies of galaxies, the
characteristic sizes of galaxies will likely not change
dramatically. We show in Figure~\ref{fig8m} [an updated version of
Figure~6 of Barkana \& Loeb (2000a)] the distribution of galactic disk
sizes at various redshifts, in the $\Lambda$CDM model (with parameters
given at the end of \S \ref{sec1}). Given $\theta$ in arcseconds, each
curve shows the fraction of the total number counts contributed by
sources larger than $\theta$. The diameter $\theta$ is measured out to
one exponential scale length. We show three pairs of curves, at $z=2$,
$z=5$ and $z=10$ (from right to left). Each pair includes the
distribution for all galaxies (dashed line), and for galaxies
detectable by \NGST\, (solid line) with a limiting point source flux
of 0.25 nJy and with an efficiency $\eta=10\%$ assumed for the
galaxies. The vertical dotted line indicates the expected \NGST\,
resolution of $0\farcs 06$.

\noindent
  
\noindent
  
\vspace{-0.41in}
Among detectable galaxies, the typical diameter decreases from
$0\farcs 22$ at $z=2$ to $0\farcs 10$ at $z=5$ and $0\farcs 05$ at
$z=10$. Note that in $\Lambda$CDM (with $\Omm=0.3$) the angular
diameter distance (in units of $c/H_0$) actually decreases from 0.40
at $z=2$ to 0.30 at $z=5$ and 0.20 at $z=10$. Galaxies are still
typically smaller at the higher redshifts because a halo of a given
mass is denser, and thus smaller, at higher redshift, and furthermore
the typical halo mass is larger at low redshift due to the growth of
cosmic structure with time. At $z=10$, the distribution of detectable
galaxies is biased, relative to the distribution of all galaxies,
toward large galaxies, since \NGST\, can only detect the brightest
galaxies. The brightest galaxies tend to lie in the most massive and
therefore largest halos, and this trend dominates over the higher
detection threshold needed for an extended source compared to a point
source (\S \ref{sec8.1}).

Clearly, the angular resolution of \NGST\, will be sufficiently high
to resolve most galaxies. For example, \NGST\, should resolve
approximately $35\%$ of $z=10$ galaxies, $80\%$ of $z=5$ galaxies, and
all but $1\%$ of $z=2$ galaxies. This implies that the shapes of these
high-redshift galaxies can be studied with {\it NGST}.\, It also means
that the high resolution of \NGST\, is crucial in making the majority
of sources on the sky useful for weak lensing studies (although a
mosaic of images is required for good statistics; see also the
following subsection).

\subsection{Gravitational Lensing}
\label{sec8.4}

Detailed studies of gravitational lenses have provided a wealth of
information on galaxies, both through modeling of individual
lens systems (e.g., Schneider, Ehlers, \& Falco 1992; 
Blandford \& Narayan 1992) and from the statistical properties of 
multiply imaged sources (e.g., Turner, Ostriker, \& Gott 1984; 
Maoz \& Rix 1993; Kochanek 1996).

The ability to observe large numbers of high-redshift objects promises
to greatly extend gravitational lensing studies. Due to the increased
path length along the line of sight to the most distant sources, their
probability for being lensed is expected to be the highest among all
possible sources. Sources at $z>10$ will often be lensed by $z>2$
galaxies, whose masses can then be determined with lens modeling.
Similarly, the shape distortions (or weak lensing) caused by
foreground clusters of galaxies will be used to determine the mass
distributions of less massive and higher redshift clusters than
currently feasible. In addition, it will be fruitful to exploit the
magnification of the sources to resolve and study more distant
galaxies than otherwise possible.

These applications have been explored by Schneider \& Kneib (1998),
who investigated weak lensing, and by Barkana \& Loeb (2000a), who
focused on strong lensing. Schneider \& Kneib (1998) noted that the
ability of \NGST\, to take deeper exposures than is possible with
current instruments will increase the observed density of sources on
the sky, particularly of those at high redshifts. The large increase
(by $\sim 2$ orders of magnitude over current surveys) may allow
such applications as a detailed weak lensing mapping of substructure
in clusters. Obviously, the source galaxies must be well resolved to
allow an accurate shape measurement. Barkana \& Loeb (2000a) estimated
the size distribution of galactic disks (see \S \ref{sec8.3}) and
showed that with its expected $\sim 0\farcs06$ resolution, \NGST\,
should resolve most galaxies even at $z \sim 10$.

The probability for strong gravitational lensing depends on the
abundance of lenses, their mass profiles, and the angular diameter
distances among the source, the lens and the observer. The statistics
of existing lens surveys have been used at low redshifts to constrain
the cosmological constant (for the most detailed work see Kochanek
1996, and references therein), although substantial uncertainties
remain regarding the luminosity function of early-type galaxies and
their dark matter content. Given the early stage of observations of
the redshift evolution of galaxies and their dark halos, a theoretical
approach based on the Press-Schechter mass function can be used to
estimate the lensing rate. This approach has been used in the past for
calculating lensing statistics at low redshifts, with an emphasis on
lenses with image separations above $5\arcsec$ (Narayan \& White 1988;
Kochanek 1995; Maoz et al.\ 1997; Nakamura \& Suto 1997; Phillips,
Browne, \& Wilkinson 2001; Ofek et al.\ 2001) or on the lensing rates
of supernovae (Porciani \& Madau 2000; Marri et al.\ 2000).

The probability for producing multiple images of a source at a
redshift $z_S$, due to gravitational lensing by lenses with density
distributed as in a singular isothermal sphere, is obtained
by integrating over lens redshift $z_L$ the differential optical depth
(Turner, Ostriker, \& Gott 1984; Fukugita et al.\ 1992)
\beq d\tau=16 \pi^3 n \left(\frac{\sigma}{c}\right)^4 (1+z_L)^3
\left(\frac{D_{OL} D_{LS}}{D_{OS}}\right)^2 \frac{c dt}{dz_L} d z_L\ ,
\label{dtau}
\eeq in terms of the comoving density of lenses $n$, velocity
dispersion $\sigma$, look-back time $t$, and angular diameter
distances $D$ among the observer, lens and source. More generally we
replace $n \sigma^4$ by \beq 
\label{nsig}
\langle n \sigma^4\rangle = \int \frac{dn(M,z_L)}{dM}\,
\sigma^4(M,z_L)\, dM\ , \eeq where $dn/dM$ is the Press-Schechter halo
mass function. It is assumed that $\sigma(M,z)=V_c(M,z)/\sqrt{2}$\,
and that the circular velocity $V_c(M,z)$ corresponding to a halo of a
given mass is given by equation~(\ref{Vceqn}).

The $\Lambda$CDM model (with $\Omm=0.3$) yields a lensing optical
depth (Barkana \& Loeb 2000a) of $\sim 1\%$ for sources at
$z_S=10$. The fraction of lensed sources in an actual survey is
enhanced, however, by the so-called magnification bias. At a given
observed flux level, unlensed sources compete with lensed sources that
are intrinsically fainter. Since fainter galaxies are more numerous,
the fraction of lenses in an observed sample is larger than the
optical depth value given above. The expected slope of the luminosity
function of the early sources (\S \ref{sec8.2}) suggests an additional
magnification bias of order 5, bringing the fraction of lensed sources
at $z_S=10$ to $\sim 5\%$. The lensed fraction decreases to $\sim 3\%$
at $z=5$. With the magnification bias estimated separately for each
source population, the expected number of detected multiply-imaged
sources per field of view of \NGST\, (which we assume to be $4 \arcmin
\times 4 \arcmin$) is roughly 5 for $z>10$ quasars, 10 for $z>5$
quasars, 10 for $z>10$ galaxies, and 100 for $z>5$ galaxies.

\begin{figure}[htbp]
\epsscale{0.7}
\plotone{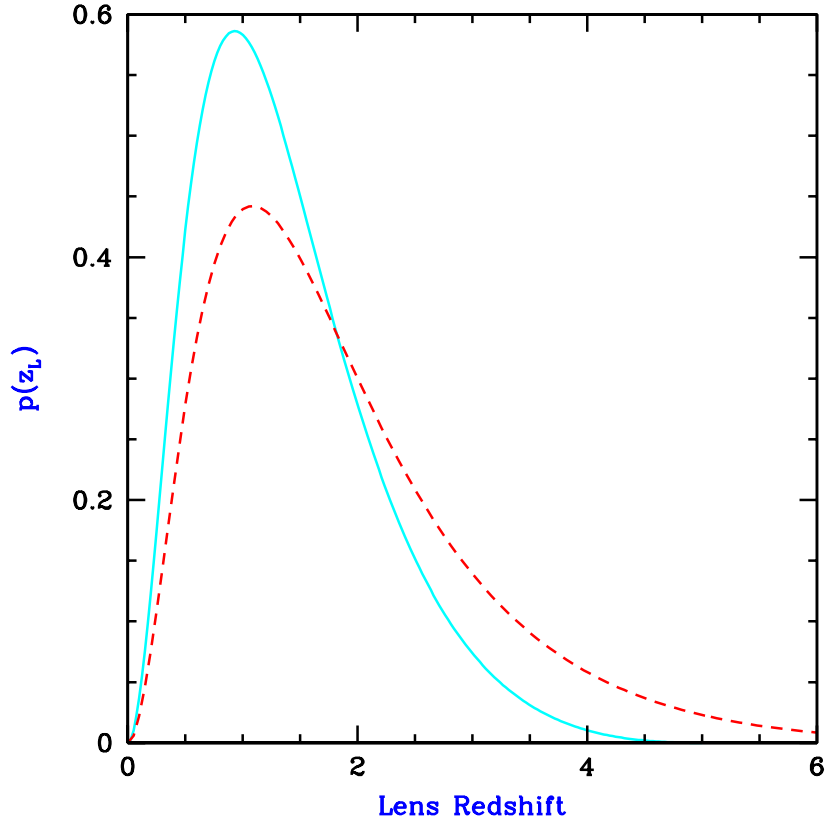}
\caption{Distribution of lens redshifts for a fixed source redshift,
for Press-Schechter halos in $\Lambda$CDM with $\Omm=0.3$ (adopted
from Figure~2 of Barkana \& Loeb 2000a). Shown for a source at $z_S=5$
(solid curve) and for $z_S=10$ (dashed curve). The probability density
$p(z_L)$ is shown, where the fraction of lenses between $z_L$ and
$z_L+dz_L$ is $p(z_L)dz_L$.}
\label{fig_8o}
\end{figure}

High-redshift sources will tend to be lensed by galaxies at relatively
high redshifts. In Figure~\ref{fig_8o} (adopted from Figure~2 of
Barkana \& Loeb 2000a) we show the lens redshift probability density
$p(z_L)$, defined so that the fraction of lenses between $z_L$ and
$z_L+dz_L$ is $p(z_L)dz_L$. We consider a source at $z_S=5$ (solid
curve) or at $z_S=10$ (dashed curve). The curves peak around $z_L=1$,
but in each case a significant fraction of the lenses are above
redshift 2: $20\%$ for $z_S=5$ and $36\%$ for $z_S=10$.

The multiple images of lensed high-redshift sources should be easily
resolvable. Indeed, image separations are typically reduced by a
factor of only 2--3 between $z_S=2$ and $z_S=10$, with the reduction
almost entirely due to redshift evolution in the characteristic mass
of the lenses. With a typical separation of 0.5--1$\arcsec$ for
$z_S=10$, a large majority of lenses should be resolved given the
\NGST\, resolution of $\sim 0\farcs06$.

Lensed sources may be difficult to detect if their images overlap the
lensing galaxy, and if the lensing galaxy has a higher surface
brightness.  Although the surface brightness of a background source
will typically be somewhat lower than that of the foreground lens
(Barkana \& Loeb 2000a), the lensed images should be detectable since
(i) the image center will typically be some distance from the lens
center, of order half the image separation, and (ii) the younger
stellar population and higher redshift of the source will make its
colors different from those of the lens galaxy, permitting an easy
separation of the two in multi-color observations. These helpful
features are evident in the currently known systems which feature
galaxy-galaxy strong lensing. These include two four-image `Einstein
cross' gravitational lenses and other lens candidates discovered by
Ratnatunga et al.\ (1999) in the {\it Hubble Space Telescope}\, Medium
Deep Survey, and a lensed three-image arc detected in the Hubble Deep
Field South and studied in detail by Barkana et al.\ (1999).


\section{\bf Observational Probes of the Epoch of Reionization}
\label{sec9}

\subsection{Spectral Methods of Inferring the Reionization Redshift}
\label{sec9.1}

\subsubsection{Cosmology with \lya Photons}
\label{sec9.1.1}

The scattering cross-section of the \lya resonance line by neutral
hydrogen is given by (\S 23 of Peebles 1993)
\begin{equation}
\sigma_\alpha(\nu) = {3 \lambda_\alpha^2 \Lambda_\alpha^2 \over 8\pi}
{(\nu/\nu_\alpha)^4\over
4\pi^2(\nu-\nu_\alpha)^2+(\Lambda_\alpha^2/4)(\nu/\nu_\alpha)^6},
\label{eq:sig}
\end{equation}
where $\Lambda_\alpha=(8\pi^2 e^2
f_\alpha/3m_ec\lambda_\alpha^2)=6.25\times 10^8~{\rm s^{-1}}$ is the
\lya ($2p\rightarrow 1s$) decay rate, $f_\alpha=0.4162$ is the
oscillator strength, and $\lambda_\alpha=1216$\AA\, and
$\nu_\alpha=(c/\lambda_\alpha)=2.47\times 10^{15}~{\rm Hz}$ are the
wavelength and frequency of the \lya line. The term in the numerator
is responsible for the classical Rayleigh scattering.

We consider a source at a redshift $z_s$ beyond the redshift of
reionization\footnote{We define the reionization redshift to be the
redshift at which the individual \ion{H}{2} regions overlapped and
most of the IGM volume was ionized. In most realistic scenarios, this
transition occurs rapidly on a time-scale much shorter than the age of
the universe (see \S \ref{sec6.3.1}). This is mainly due to the short
distances between neighboring sources.} $\zr$, and the corresponding
scattering optical depth of a uniform, neutral IGM of hydrogen density
$n_{\rm H,0}(1+z)^3$ between the source and the reionization
redshift. The optical depth is a function of the observed wavelength
$\lambda_{\rm obs}$,
\begin{equation}
\tau(\lambda_{\rm obs})=\int_{\zr}^{z_s} dz {cdt\over dz} n_{\rm
H,0} (1+z)^3 \sigma_\alpha\left[\nu_{\rm obs}(1+z)\right],
\end{equation}
where $\nu_{\rm obs}=c/\lambda_{\rm obs}$ and
\begin{equation}
{dt\over dz}=
\left[(1+z)H(z)\right]^{-1}=H_0^{-1}\left[\Omega_m(1+z)^5+
\Omega_\Lambda(1+z)^2+
(1-\Omega_m-\Omega_\Lambda)(1+z)^4\right]^{-1/2}.
\end{equation}

At wavelengths longer than \lya at the source, the optical depth
obtains a small value; these photons redshift away from the line
center along its red wing and never resonate with the line core on
their way to the observer.  Considering only the regime in which
$\vert\nu-\nu_\alpha\vert \gg \Lambda_\alpha$, we may ignore the
second term in the denominator of equation~(\ref{eq:sig}). This leads
to an analytical result for the red damping wing of the Gunn-Peterson
trough (Miralda-Escud\'e 1998)
\begin{equation}
\tau(\lambda_{\rm obs})=\tau_s \left(\Lambda\over
4\pi^2\nu_\alpha\right) {\tilde \lambda}_{\rm obs}^{3/2}\left[
I({\tilde\lambda}_{\rm obs}^{-1}) -
I([(1+\zr)/(1+z_s)]{\tilde\lambda}_{\rm obs}^{-1})\right] ~~~~~~{\rm
for}~~{\tilde\lambda}_{\rm obs}\geq 1~,
\label{eq:shift}
\end{equation}
where $\tau_s$ is given in equation~(\ref{G-P}), and we also define 
\begin{equation}
{\tilde \lambda}_{\rm obs}\equiv {\lambda_{\rm obs}\over
(1+z_s)\lambda_\alpha}
\end{equation}
and
\begin{equation}
I(x)\equiv {x^{9/2}\over 1-x}+{9\over 7}x^{7/2}+{9\over 5}x^{5/2}+ 3
x^{3/2}+9 x^{1/2}-{9\over 2} \ln\left[ {1+x^{1/2}\over 1-x^{1/2}}
\right]\ .
\end{equation}

\begin{figure}[htbp] 
\includegraphics{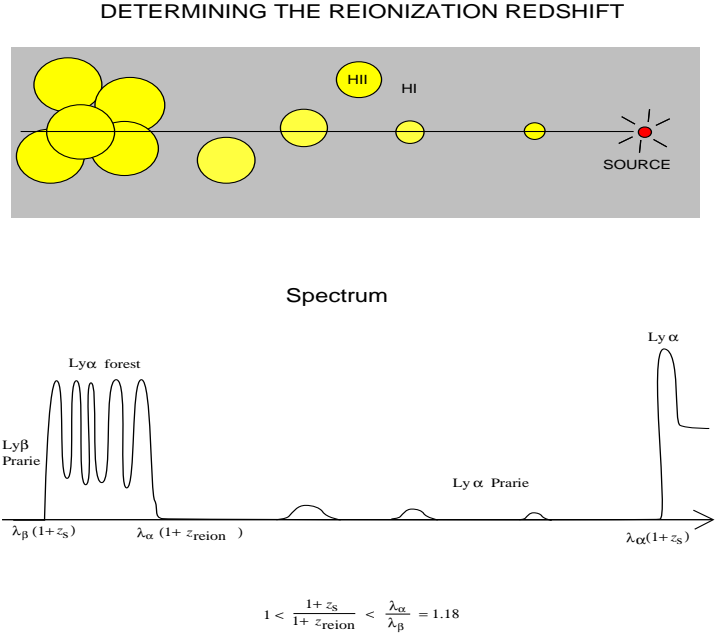}
\vspace{3.18in}
\caption{Sketch of the expected spectrum of a source at a redshift
$z_{s}$ slightly above the reionization redshift $\zr$. The
transmitted flux due to \ion{H}{2} bubbles in the pre-reionization era
and the Ly$\alpha$ forest in the post-reionization era is exaggerated
for illustration.\vspace{.1in}}
\label{fig8a}
\end{figure}

At wavelengths corresponding to the \lya resonance between the source
redshift and the reionization redshift, $(1+\zr)\lambda_\alpha\leq
\lambda_{\rm obs}\leq (1+z_s)\lambda_\alpha$, the optical depth is
given by equation~(\ref{G-P}). Since $\tau_s\sim 10^5$, the flux from
the source is entirely suppressed in this regime. Similarly, the
Ly$\beta$ resonance produces another trough at wavelengths
$(1+\zr)\lambda_\beta\leq \lambda\leq (1+z_s)\lambda_\beta$, where
$\lambda_\beta=(27/32)\lambda_\alpha= 1026\,$\AA, and the same applies
to the higher Lyman series lines.  If $(1+z_s)\geq 1.18(1+\zr)$ then
the \lya and the Ly$\beta$ resonances overlap and no flux is
transmitted in-between the two troughs (see Figure~\ref{fig8a}). The
same holds for the higher Lyman-series resonances down to the Lyman
limit wavelength of $\lambda_c=912$\AA.

At wavelengths shorter than $\lambda_c$, the photons are absorbed when
they photoionize atoms of hydrogen or helium. The bound-free
absorption cross-section from the ground state of a hydrogenic ion
with nuclear charge $Z$ and an ionization threshold $h\nu_0$, is given
by (Osterbrock 1974),
\begin{equation}
\sigma_{bf}(\nu)= {6.30\times 10^{-18}\over Z^2}~{\rm cm^2}\times
\left({\nu_0\over
\nu}\right)^4{e^{4-(4\tan^{-1}\epsilon)/\epsilon}\over 1 -
e^{-2\pi/\epsilon}}~~~~~{\rm for}~~\nu\geq\nu_0,
\end{equation}
where 
\begin{equation}
\epsilon\equiv \sqrt{{\nu\over \nu_0}-1}.
\end{equation}
For neutral hydrogen, $Z=1$ and $\nu_{{\rm H},0}= (c/\lambda_c)
=3.29\times 10^{15}$ Hz ($h\nu_{\rm H,0}=13.60$ eV); for
singly-ionized helium, $Z=2$ and $\nu_{\rm He~II, 0}= 1.31\times
10^{16}~{\rm Hz}$ ($h\nu_{\rm He~II, 0}=54.42$ eV). The cross-section
for neutral helium is more complicated; when averaged over its narrow
resonances it can be fitted to an accuracy of a few percent up to
$h\nu=50$ keV by the fitting function (Verner et al.\ 1996)
\begin{equation}
\sigma_{bf,{\rm He~I}}(\nu)= 9.492\times 10^{-16}~{\rm cm^2}\,\times
\left[(x-1)^2+4.158\right]y^{-1.953}\left(1+ 0.825
y^{1/4}\right)^{-3.188},
\end{equation}
where $x\equiv[(\nu/3.286\times 10^{15}~{\rm Hz}) -0.4434]$, $y\equiv
x^2+4.563$, and the threshold for ionization is $\nu_{\rm He~I,
0}=5.938\times 10^{15}~{\rm Hz}$ ($h\nu_{\rm He~I,0}=24.59$ eV).

For rough estimates, the average photoionization cross-section for a
mixture of hydrogen and helium with cosmic abundances can be
approximated in the range of $54<h\nu \la 10^3$ eV as
$\sigma_{bf}\approx \sigma_0 (\nu/\nu_{\rm H,0})^{-3}$, where
$\sigma_0\approx 6\times 10^{-17}~{\rm cm^2}$ (Miralda-Escud\'e 2000).
The redshift factor in the cross-section then cancels exactly the
redshift evolution of the gas density and the resulting optical depth
depends only on the elapsed cosmic time, $t(\zr)-t(z_s)$. At high
redshifts (equations~(\ref{highz1}) and (\ref{highz2}) in \S
\ref{sec2.1}) this yields,
\begin{eqnarray}
\tau_{bf}(\lambda_{\rm obs})&=&\int_{\zr}^{z_s} dz {cdt\over dz}
n_{0} (1+z)^3 \sigma_{\rm bf}\left[\nu_{\rm obs}(1+z)\right]
\nonumber
\\
&\approx&
1.5\times 10^2 \left({\Omega_bh\over 0.03}\right) \left({\Omega_m\over
0.3}\right)^{-1/2}\left({\lambda\over 100{\rm
\AA}}\right)^{3}\left[{1\over (1+\zr)^{3/2}}-{1\over 
(1+z_s)^{3/2}}\right].
\end{eqnarray}
The bound-free optical depth only becomes of order unity in the
extreme UV to soft X-rays, around $h\nu \sim 0.1$ keV, a regime which
is unfortunately difficult to observe due to Galactic absorption
(Miralda-Escud\'e 2000).

A sketch of the overall spectrum of a source slightly above the
reionization redshift, i.e., with $1<[(1+ z_{\rm s})/(1+\zr)]<1.18$,
is shown in Figure~\ref{fig8a}. The transmitted flux between the
Gunn-Peterson troughs due to Ly$\alpha$ and Ly$\beta$ absorption is
suppressed by the Ly$\alpha$ forest in the post-reionization
epoch. Transmission of flux due to \ion{H}{2} bubbles in the
pre-reionization epoch is expected to be negligible (Miralda-Escud\'e
1998). The redshift of reionization can be inferred in principle from
the spectral shape of the red damping wing (Miralda-Escud\'e \& Rees
1998; Miralda-Escud\'e 1998) or from the transmitted flux between the
Lyman series lines (Haiman \& Loeb 1999a).  However, these signatures
are complicated in reality by damped \lya systems along the
line of sight or by the inhomogeneity or peculiar velocity field of
the IGM in the vicinity of the source.  Moreover, bright sources, such
as quasars, tend to ionize their surrounding environment (Wood \& Loeb
2000) and the resulting \ion{H}{2} region in the IGM could shift the
\lya trough substantially (Cen \& Haiman 2000; Madau \& Rees 2000).

The inference of the \lya transmission properties of the IGM from the
observed spectrum of high-redshift sources suffers from uncertainties
about the precise emission spectrum of these sources, and in
particular the shape of their \lya emission line. The first galaxies
and quasars are expected to have pronounced recombination lines of
hydrogen and helium due to the lack of dust in their interstellar
medium (see \S \ref{sec4.1.3} for more details). Lines such as
$H_\alpha$ or the \ion{He}{2} 1640\,\AA\, line should reach the
observer unaffected by the intervening IGM, since their wavelength is
longer than that of the Ly$\alpha$ transition which dominates the IGM
opacity (Oh 1999).  However, as described above, the situation is
different for the \lya line photons from the source.  As long as
$z_s>\zr$, the intervening neutral IGM acts like a fog and obscures
the view of the \lya line itself [in contrast to the situation with
sources at $z_s<\zr$, where most of the intervening IGM is ionized and
only photons more energetic than \lya are suppressed by the Ly$\alpha$
forest (see Figure~\ref{fig1c})].  Photons which are emitted at the
\lya line center have an initial scattering optical depth of $\sim
10^5$ in the surrounding medium.

\begin{figure}[htbp]
\includegraphics{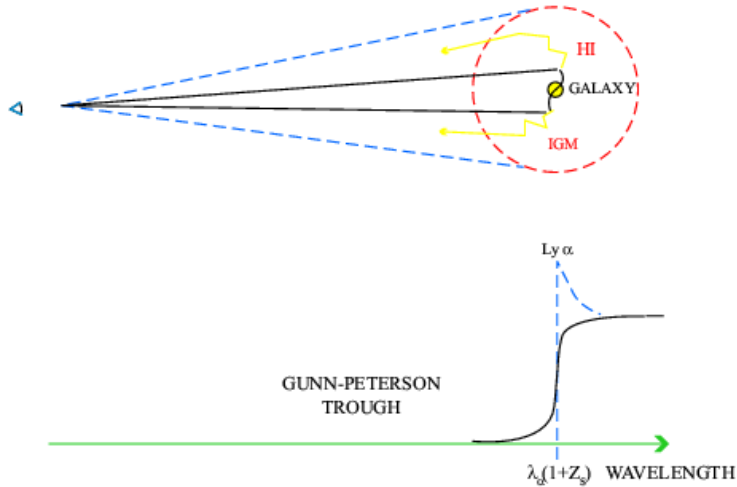}
\vspace{2.85in}
\caption{{\it Loeb-Rybicki halos:}\/ Scattering of \lya line photons
from a galaxy embedded in the neutral IGM prior to reionization. The
line photons diffuse in frequency due to the Hubble expansion of the
surrounding medium and eventually redshift out of resonance and escape
to infinity.  A distant observer sees a \lya halo surrounding the
source, along with a characteristically asymmetric line profile. The
observed line should be broadened and redshifted by about one thousand
${\rm km~s^{-1}}$ relative to other lines (such as H$_\alpha$) emitted
by the galaxy. }
\label{fig8c}
\end{figure}

\begin{figure}[htbp]
\includegraphics{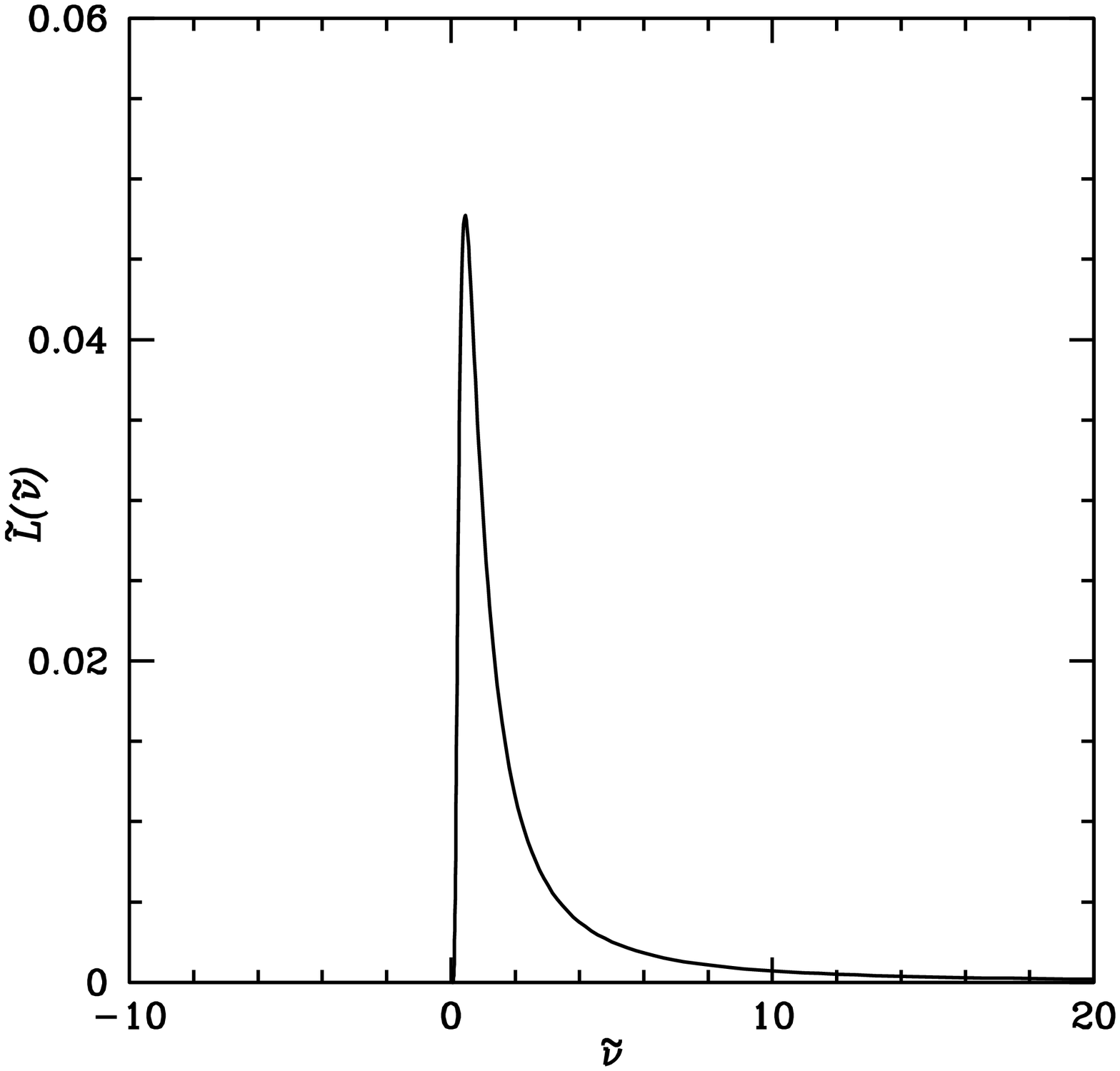}
\vspace{3.3in}
\caption{Monochromatic photon luminosity of a \lya halo as a function
of frequency redshift, ${\tilde \nu}\equiv(\nu_\alpha-\nu)/\nu_\star$.
The observed spectral flux of photons $F(\nu)$ (in photons cm$^{-2}$
s$^{-1}$ Hz$^{-1}$) from the entire \lya halo is $F(\nu) = ({{\tilde
L}({\tilde \nu}) / 4 \pi d_{\rm L}^2})({\lyasource / \nu_\star})
(1+z_{s})^2$ where $\lyasource$ is the production rate of \lya photons
by the source (in ${\rm photons~s^{-1}}$), $\nu={\tilde
\nu}\nu_\star/(1+z_{\rm s})$, and $d_{\rm L}$ is the standard
luminosity distance to the source (from Loeb \& Rybicki 1999). }
\label{fig8b}
\end{figure}

The Ly$\alpha$ line photons are not destroyed but instead are absorbed
and re-emitted\footnote{At the redshifts of interest, $z_s\sim 10$,
the low densities and lack of chemical enrichment of the IGM make the
destruction of \lya photons by two-photon decay or dust absorption
unimportant.}.  Due to the Hubble expansion of the IGM around the
source, the frequency of the photons is slightly shifted by the
Doppler effect in each scattering event. As a result, the photons
diffuse in frequency to the red side of the \lya resonance.
Eventually, when their net frequency redshift is sufficiently large,
they escape and travel freely towards the observer (see
Figure~\ref{fig8c}).  As a result, the source creates a faint
Ly$\alpha$ halo on the sky\footnote{The photons absorbed in the
Gunn-Peterson trough are also re-emitted by the IGM around the
source. However, since these photons originate on the blue side of the
\lya resonance, they travel a longer distance from the source,
compared to the \lya line photons, before they escape to the
observer. The Gunn-Peterson photons are therefore scattered from a
larger and hence dimmer halo around the source.  The Gunn-Peterson
halo is made even dimmer relative to the \lya line halo by the fact
that the luminosity of the source per unit frequency is often much
lower in the continuum than in the \lya line.}.  The well-defined
radiative transfer problem of a point source of \lya photons embedded
in a uniform, expanding neutral IGM was solved by Loeb \& Rybicki
(1999). The \lya halo can be simply characterized by the frequency
redshift relative to the line center, $(\nu-\nu_\alpha)$, which is
required in order to make the optical depth from the source
[equation~(\ref{eq:shift})] equal to unity. At high redshifts, the
leading term in equation~(\ref{eq:shift}) yields
\begin{equation}
\nu_\star=8.85\times 10^{12}~{\rm Hz} \times \left({\Omega_bh\over
0.05\sqrt{\Omega_m}}\right) \left(1+z_s\over 10\right)^{3/2}.
\end{equation}
This is the frequency interval over which the damping wing affects the
source spectrum. A frequency shift of $\nu_\star=8.85\times 10^{12}$
Hz relative to the line center corresponds to a fractional shift of
$(\nu_\star/\nu_\alpha)=(v/c)=3.6\times 10^{-3}$ or a Doppler velocity
of $v\sim 10^3~{\rm km~s^{-1}}$. The halo size is then defined by the
corresponding proper distance from the source at which the Hubble
velocity provides a Doppler shift of this magnitude,
\begin{equation}
r_\star= 1.1 \left({\Omega_b/0.05\over\Omega_m/0.3}\right)~{\rm Mpc} .
\end{equation}
Typically, the \lya halo of a source at $z_{\rm s}\sim 10$ occupies an
angular radius of $\sim 15^{\prime\prime}$ on the sky and yields an
asymmetric line profile as shown in Figures~\ref{fig8c} and
\ref{fig8b}. The scattered photons are highly polarized and so the
shape of the halo would be different if viewed through a polarization
filter (Rybicki \& Loeb 1999).

\noindent
  
\vspace{.55in}

\vspace{-0.31in}
Detection of the diffuse \lya halos around bright high-redshift
sources (which are sufficiently rare so that their halos do not
overlap) would provide a unique tool for probing the distribution and
the velocity field of the neutral IGM before the epoch of
reionization.  The \lya sources serve as lamp posts which illuminate
the surrounding \ion{H}{1} fog. On sufficiently large scales where the
Hubble flow is smooth and the gas is neutral, the Ly$\alpha$
brightness distribution can be used to determine the cosmological mass
densities of baryons and matter. Due to their low surface brightness,
the detection of \lya halos through a narrow-band filter is much more
challenging than direct observation of their sources at somewhat
longer wavelengths. However, \NGST\, might be able to detect the \lya
halos around sources as bright as the quasar discovered by Fan et al.\
(2000) at $z=5.8$ or the galaxy discovered by Hu et al.\ (1999) at
$z=5.74$, even if these sources were moved out to $z\sim 10$ (see \S 4
in Loeb \& Rybicki 1999). The disappearance of
\lya halos below a certain redshift can be used to determine $\zr$.

\subsubsection{21 cm Tomography of the Reionization Epoch}
\label{sec9.1.2}

The ground state of hydrogen exhibits hyperfine splitting involving
the spins of the proton and the electron. The state with parallel
spins (the triplet state) has a slightly higher energy than the state
with anti-parallel spins (the singlet state). The 21 cm line
associated with the spin-flip transition from the triplet to the
singlet state is often used to detect neutral hydrogen in the local
universe. At high redshift, the occurrence of a neutral
pre-reionization IGM offers the prospect of detecting the first
sources of radiation and probing the reionization era by mapping the
21 cm emission from neutral regions. While its energy density is
estimated to be only a $1\%$ correction to that of the CMB, the
redshifted 21 cm emission should display angular structure as well as
frequency structure due to inhomogeneities in the gas density field
(Hogan \& Rees 1979; Scott \& Rees 1990), hydrogen ionized fraction,
and spin temperature (Madau, Meiksin, \& Rees 1997). Some of the
resulting signatures during the pre-overlap phase of reionization (\S
\ref{sec6.3.1}) and during the overlap phase are discussed by Tozzi et
al.\ (2000) and Shaver et al.\ (1999), respectively. Also, the 21 cm
signatures have been explored in a numerical simulation by Gnedin \&
Ostriker (1997). Indeed, a full mapping of the distribution of
\ion{H}{1} as a function of redshift is possible in principle.
Although detecting the presence of the largest \ion{H}{2} regions may
be within the reach of proposed instruments such as the Square
Kilometer Array (hereafter SKA; see Taylor \& Braun 1999), these
instruments may not have sufficient sensitivity at the sub-arcminute
resolution that would be necessary for a detailed mapping. Moreover,
serious technical challenges and problems due to foreground
contamination must be overcome even for an initial detection of the
reionization signal.

The basic physics of the hydrogen spin transition is determined as follows
(for a more detailed treatment, see Madau et al.\ 1997). The ground-state
hyperfine levels of hydrogen tend to thermalize with the CMB background,
making the IGM unobservable. If other processes shift the hyperfine level
populations away from thermal equilibrium, then the gas becomes observable
against the CMB in emission or in absorption. The relative occupancy of the
spin levels is usually described in terms of the hydrogen spin temperature
$T_S$, defined by \beq \frac{n_1}{n_0}=3\,
\exp\left\{-\frac{T_*}{T_S}\right\}\, \eeq where $n_0$ and $n_1$ are the
singlet and triplet hyperfine levels in the atomic ground state ($n=1$),
and $T_*=0.07$ K is defined by $k_B T_*=E_{21}$, where the energy of the 21
cm transition is $E_{21}=5.9 \times 10^{-6}$ eV, corresponding to a
frequency of 1420 MHz. In the presence of the CMB alone, the spin states
reach thermal equilibrium with $T_S=T_{\rm CMB}=2.73 (1+z)$ K on a
time-scale of $T_*/(T_{\rm CMB} A_{10}) \simeq 3 \times 10^5 (1+z)^{-1}$ yr,
where $A_{10}=2.9 \times 10^{-15}$ s$^{-1}$ is the spontaneous decay rate
of the hyperfine transition. This time-scale is much shorter than the age of
the universe at all redshifts after cosmological recombination.

The IGM is observable when the kinetic temperature $T_K$ of the gas differs
from $T_{\rm CMB}$ and an effective mechanism couples $T_S$ to
$T_K$. Although collisional de-excitation of the triplet level (Purcell \&
Field 1956) is a possible mechanism, in the low-density IGM the dominant
mechanism is scattering by Ly$\alpha$ photons (Wouthuysen 1952; Field
1958). Continuum UV photons produced by early radiation sources redshift by
the Hubble expansion into the local Ly$\alpha$ line at a lower
redshift. These photons mix the spin states via the Wouthuysen-Field
process whereby an atom initially in the $n=1$ state absorbs a Ly$\alpha$
photon, and the spontaneous decay which returns it from $n=2$ to $n=1$ can
result in a final spin state which is different from the initial one. Since
the neutral IGM is highly opaque to resonant scattering, the shape of the
radiation spectrum near Ly$\alpha$ is determined by $T_K$ (Field 1959), and
the spin temperature is then a weighed mean of $T_K$ and $T_{\rm CMB}$:
\beq T_S=\frac{T_{\rm CMB}+y_{\alpha} T_K}{1+y_{\alpha}}\ , \eeq where (if
$T_S \gg T_*$) the Ly$\alpha$ pumping efficiency is \beq y_{\alpha} =
\frac{P_{10} T_*} {A_{10} T_K}\ . \eeq Here $P_{10}$ is the indirect
de-excitation rate of the triplet $n=1$ state via the Wouthuysen-Field
process, related to the total scattering rate $P_{\alpha}$ of Ly$\alpha$
photons by $P_{10}=4 P_{\alpha}/27$ (Field 1958). Thus the critical value
of $P_{\alpha}$ is given by the thermalization rate (Madau et al.\ 1997)
\beq P_{\rm th} \equiv \frac{27 A_{10} T_{\rm CMB}}{4 T_*} \simeq 7.6
\times 10^{-12}\, \left(\frac{1+z}{10}\right)\ {\rm s}^{-1}\ . \eeq

A patch of neutral hydrogen at the mean density and with a uniform $T_S$
produces an optical depth at $21 (1+z)$ cm of \beq \tau(z) = 9.0
\times 10^{-3} \left(\frac{T_{\rm CMB}} {T_S} \right) \left (
\frac{\Omega_b h} {0.03} \right) \left(\frac{\Omm}{0.3}\right)^ {-1/2}
\left(\frac{1+z}{10}\right)^{1/2}\ , \eeq assuming a high redshift
$z$. Since the brightness temperature through the IGM is $T_b=T_{\rm CMB}
e^{-\tau}+T_S (1-e^{-\tau})$, the observed differential antenna temperature
of this region relative to the CMB is (Madau et al.\ 1997, with the $\Omm$
dependence added) \beq \delta T_b=(1+z)^{-1} (T_S-T_{\rm CMB})
(1-e^{-\tau}) \simeq 25\, {\rm mK}\, \left( \frac{\Omega_b h} {0.03}
\right) \left(\frac{\Omm}{0.3}\right)^ {-1/2} \left( \frac{1+z} {10}
\right)^{1/2} \left( \frac{T_S-T_{\rm CMB}} {T_S} \right)\ , \eeq where
$\tau \ll 1$ is assumed and $\delta T_b$ has been redshifted to redshift
zero. In overdense regions, the observed $\delta T_b$ is proportional to
the overdensity, and in partially ionized regions $\delta T_b$ is
proportional to the neutral fraction. Thus, if $T_S \gg T_{\rm CMB}$ then
the IGM is observed in emission at a level that is independent of $T_S$. On
the other hand, if $T_S \ll T_{\rm CMB}$ then the IGM is observed in
absorption at a level that is a factor $\sim T_{\rm CMB} / T_S$ larger than
in emission. As a result, a number of cosmic events are expected to leave
observable signatures in the redshifted 21 cm line.

Since the CMB temperature is only $2.73(1+z)$ K, even relatively
inefficient heating mechanisms are expected to heat the IGM above $T_{\rm
CMB}$ well before reionization. Possible preheating sources include soft
X-rays from early quasars or star-forming regions, as well as thermal
bremsstrahlung from ionized gas in collapsing halos. However, even the
radiation from the first stars may suffice for an early preheating. Only
$\sim 10\%$ of the present-day global star formation rate is required
(Madau et al.\ 1997) for a sufficiently strong Ly$\alpha$ background which
produces a scattering rate above the thermalization rate $P_{\rm th}$. Such
a background drives $T_S$ to the kinetic gas temperature, which is
initially lower than $T_{\rm CMB}$ because of adiabatic expansion. Thus,
the entire IGM can be seen in absorption, but the IGM is then heated above
$T_{\rm CMB}$ in $\sim 10^8$ yr (Madau et al.\ 1997) by the atomic recoil
in the repeated resonant Ly$\alpha$ scattering. According to \S
\ref{sec8.1} (also compare Gnedin 2000a), the required level of star
formation is expected to be reached already at $z \sim 20$, with the entire
IGM heated well above the CMB by the time overlap begins at $z \sim
10$. Thus, although the initial absorption signal is in principle
detectable with the SKA (Tozzi et al.\ 2000), it likely occurs at $\la 100$
MHz where Earth-based radio interference is highly problematic.

As individual ionizing sources turn on in the pre-overlap stage of
reionization, the resulting \ion{H}{2} bubbles may be individually
detectable if they are produced by rare and luminous sources such as
quasars. If the \ion{H}{2} region expands into an otherwise unperturbed
IGM, then the expanding shell can be mapped as follows (Tozzi et al.\ 
2000). The \ion{H}{2} region itself, of course, shows neither emission nor
absorption. Outside the ionized bubble, a thin shell of neutral gas is
heated above the CMB temperature and shows up in emission. A much thicker
outer shell is cooler than the CMB due to adiabatic expansion, but
satisfies $T_S=T_K$ and produces absorption. Finally, at large distances
from the quasar, $T_S$ approaches $T_{\rm CMB}$ as the quasar radiation
weakens. For a quasar with an ionizing intensity of $10^{57}$ photons
s$^{-1}$ observed after $\sim 10^7$ yr with $2\arcmin$ resolution and 1 MHz
bandwidth, the signal ranges from -3 to 3 $\mu$Jy per beam (Tozzi et
al.\ 2000). Mapping such regions would convey information on the quasar
number density, ionizing intensity, opening angle, and on the density
distribution in the surrounding IGM. Note, however, that an \ion{H}{2}
region which forms at a redshift approaching overlap expands into a
preheated IGM. In this case, the \ion{H}{2} region itself still appears as
a hole in an otherwise emitting medium, but the quasar-induced heating is
not probed, and there is no surrounding region of absorption to supply an
enhanced contrast.

At redshifts approaching overlap, the IGM should be almost entirely
neutral but with $T_S \gg T_{\rm CMB}$. In this redshift range there
should still be an interesting signal due to density fluctuations. The
same cosmic network of sheets and filaments that gives rise to the
Ly$\alpha$ forest observed at $z \la 5$ should lead to fluctuations in
the 21 cm brightness temperature at higher redshifts. At 150 MHz
($z=8.5$), for observations with a bandwidth of 1 MHz, the root mean
square fluctuation should be $\sim 10$ mK at $1 \arcmin$, decreasing
with scale (Tozzi et al.\ 2000). 

A further signature, observable over the entire sky, should mark the
overlap stage of reionization. During overlap, the IGM is transformed
from being a neutral, preheated and thus emitting gas, to being almost
completely ionized. This disappearance of the emission over a
relatively narrow redshift range can be observed as a drop in the
brightness temperature at the frequencies corresponding to the latter
stages of overlap (Shaver et al.\ 1999). This exciting possibility,
along with those mentioned above, face serious challenges in terms of
signal contamination and calibration. The noise sources include
galactic and extragalactic emission sources, as well as terrestrial
interference, and all of these foregrounds must be modeled and
accurately removed in order to observe the fainter cosmological signal
(see Shaver et al.\ 1999 for a detailed discussion). For the overlap
stage in particular, the sharpness of the spectral feature is the key
to its detectability, but it may be significantly smoothed by
inhomogeneities in the IGM.

\subsection{Effect of Reionization on CMB Anisotropies}
\label{sec9.2}

In standard cosmological models, the early universe was hot and
permeated by a nearly uniform radiation bath. At $z \sim 1200$ the
free protons and electrons recombined to form hydrogen atoms, and most
of the photons last scattered as the scattering cross-section dropped
precipitously. These photons, observed today as the Cosmic Microwave
Background (CMB), thus yield a snapshot of the state of the universe
at that early time. Small fluctuations in the density, velocity, and
gravitational potential lead to observed anisotropies (e.g., Sachs
\& Wolfe 1967; Bennett et al.\ 1996) that can be analyzed to yield a
great wealth of information on the matter content of the universe and
on the values of the cosmological parameters (e.g., Hu 1995; Jungman
et al.\ 1996).

Reionization can alter the anisotropy spectrum, by erasing some of the
primary anisotropy imprinted at recombination, and by generating
additional secondary fluctuations that could be used to probe the era
of reionization itself (see Haiman \& Knox 1999 for a review). The
primary anisotropy is damped since the rescattering leads to a
blending of photons from initially different lines of sight.
Furthermore, not all the photons scatter at the same time, rather the
last scattering surface has a finite thickness. Perturbations on
scales smaller than this thickness are damped since photons scattering
across many wavelengths give canceling redshifts and blueshifts. If
reionization occurs very early, the high electron density produces
efficient scattering, and perturbations are damped on all angular
scales except for the very largest.

The optical depth to scattering over a proper length $dl$ is $ d \tau
= \sigma_T\, n_e\, dl$, where $\sigma_T$ is the Thomson cross-section
and $n_e$ the density of free electrons. If reionization occurs
instantaneously at redshift $z$, then the total scattering optical
depth in $\Lambda$CDM is given by (e.g., \S 7.1.1 of Hu 1995) \beq
\tau = 0.041 \frac{\Omega_b h}{\Omm} \left\{ \left[1-\Omm+\Omm (1 +
z)^3 \right]^{1/2}-1\right\}\ . \eeq With our standard parameters (end
of \S \ref{sec1}) this implies $\tau=0.037$ at the current lower limit
on reionization of $z=5.8$ (Fan et al.\ 2000), with $\tau=0.10$ if
$z=11.6$ and $\tau=0.15$ if $z=15.3$. Recent observations of
small-scale anisotropies (Lange et al.\ 2000; Balbi et al.\ 2000)
revealed a peak in the power spectrum on a $\sim 1^\circ$ scale, as
expected from the primary anisotropies in standard cosmological
models. This indicates that the reionization damping, if present, is
not very large, and the observations set a limit of $\tau < 0.33$ at
$95\%$ confidence (Tegmark \& Zaldarriaga 2000) and, therefore, imply
that reionization must have occurred at $z \la 30$.

However, measuring a small $\tau$ from the temperature anisotropies
alone is expected to be very difficult since the anisotropy spectrum
depends on a large number of other parameters, creating a
near-degeneracy which limits our ability to measure each parameter
separately; the degeneracy of $\tau$ with other cosmological
parameters is due primarily to a degeneracy with the
gravitational-wave background. However, Thomson scattering also
creates net polarization for incident radiation which has a quadrupole
anisotropy. This anisotropy was significant at reionization due to
large-scale structure which had already affected the gas
distribution. The result is a peak in the polarization power spectrum
on large angular scales of order tens of degrees (Zaldarriaga
1997). Although experiments must overcome systematic errors from the
detector itself and from polarized foregrounds (such as galactic dust
emission and synchrotron radiation), parameter estimation models
(Eisenstein, Hu, \& Tegmark 1999; Zaldarriaga, Spergel, \& Seljak
1997) suggest that the peak can be used to measure even very small
values of $\tau$: $2\%$ for the upcoming MAP satellite, and $0.5\%$
for the Planck satellite which will reach smaller angular scales with
higher accuracy.

Reionization should also produce additional temperature anisotropies
on small scales. These result from the Doppler effect. By the time of
reionization, the baryons have begun to follow dark matter potentials
and have acquired a bulk velocity. Since the electrons move with
respect to the radiation background, photons are given a Doppler kick
when they scatter off the electrons. Sunyaev (1978) and Kaiser (1984)
showed, however, that a severe cancellation occurs if the electron
density is homogeneous. Opposite Doppler shifts on crests and troughs
of a velocity perturbation combine to suppress the anisotropy induced
by small-scale velocity perturbations. The cancellation is made more
severe by the irrotational nature of gravitationally-induced flows.
However, if the electron density varies spatially, then the scattering
probability is not equal on the crest and on the trough, and the two
do not completely cancel. Since a non-zero effect requires variation
in both electron density and velocity, it is referred to as a
second-order anisotropy.

The electron density can vary due to a spatial variation in either the
baryon density or the ionized fraction. The former is referred to as
the Ostriker-Vishniac effect (Ostriker \& Vishniac 1986; Vishniac
1987). The latter depends on the inhomogeneous topology of
reionization, in particular on the size of \ion{H}{2} regions due to
individual sources (\S \ref{sec6.2}) and on spatial correlations among
different regions. Simple models have been used to investigate the
character of anisotropies generated during reionization (Gruzinov \&
Hu 1998; Knox et al.\ 1998; Aghanim et al.\ 1996). The
Ostriker-Vishniac effect is expected to dominate all anisotropies at
small angular scales (e.g., Jaffe \& Kamionkowski 1998), below a tenth
of a degree, because the primary anisotropies are damped on such small
scales by diffusion (Silk damping) and by the finite thickness of the
last scattering surface. Anisotropies generated by inhomogeneous
reionization may be comparable to the Ostriker-Vishniac effect, and
could be detected by MAP and Planck, if reionization is caused by
bright quasars with 10 Mpc-size ionized bubbles. However, the smaller
bubbles expected for mini-quasars or for star-forming dwarf galaxies
would produce an anisotropy signal which is weaker and at smaller
angular scales, likely outside the range of the upcoming satellites
(see, e.g., Haiman \& Knox 1999 for discussion). Gnedin \& Jaffe
(2000) used a numerical simulation to show that, in the case of
stellar reionization, the effect on the CMB of patchy reionization is
indeed sub-dominant compared to the contribution of non-linear density
and velocity fluctuations. Nevertheless, a signature of reionization
could still be detected in future measurements of CMB angular
fluctuations on the scale of a few arcseconds (see also Bruscoli \etal
2000, who find a somewhat higher power spectrum due to patchy
reionization).

\subsection{Remnants of High-Redshift Systems in the Local Universe}
\label{sec9.3}

At the end of the reionization epoch, the heating of the IGM resulted
in the photo-evaporation of gas out of halos of circular velocity
$V_c$ above $\sim 10$--$15\ {\rm km~s^{-1}}$ (\S \ref{sec6.4}). The
pressure of the hot gas subsequently shut off gas infall into even
more massive halos, those with $V_c \sim 30\ {\rm km~s^{-1}}$ (\S
\ref{sec6.5}).  Thus, the gas reservoir of photo-evaporating halos
could not be immediately replenished.  Some dwarf galaxies which were
prevented from forming after reionization could have eventually
collected gas at $z=1$--2, when the UV background flux declined
sufficiently (Babul \& Rees 1992; Kepner, Babul, \& Spergel
1997). However, Kepner et al.\ (1997) found that even if the ionizing
intensity $J_{21}$ declines as $(1+z)^4$ below $z=3$, only halos with
$V_c \ga 20\ {\rm km\ s}^{-1}$ can form atomic hydrogen by $z=1$, and
$V_c \ga 25\ {\rm km\ s}^{-1}$ is required to form molecular hydrogen.
While the fact that the IGM was reionized has almost certainly
influenced the abundance and properties of dwarf galaxies observed
today, the exact manifestations of this influence and ways to prove
that they occurred have not been well determined. In this section we
summarize recent work on this topic, which should remain an active
research area.

The suppression of gas infall mentioned above suggests that the
abundance of luminous halos as a function of circular velocity should
show a break, with a significant drop in the abundance below $V_c=30\
{\rm km~s^{-1}}$. Such a drop may in fact be required in order to
reconcile the $\Lambda$CDM model with observations. Klypin et al.\
(1999) and Moore et al.\ (1999) found that the abundance of halos with
$V_c \sim 10$--$30\ {\rm km~s^{-1}}$ in numerical simulations of the
Local Group environment is higher by an order of magnitude than the
observed dwarf galaxy abundance. The predicted and observed abundances
matched well at $V_c > 50\ {\rm km~s^{-1}}$. Bullock et al.\ (2000a)
considered whether photoionization can explain the discrepancy at the
low-mass end by preventing dark matter halos from forming stars. They
assumed that a sub-halo in the Local Group can host an observable
galaxy only if already at reionization its main progenitor contained a
fraction $f$ of the final sub-halo mass. Using semi-analytic modeling,
they found a close match to the observed circular velocity
distribution for $\zr=8$ and $f=0.3$.

These results neglect several complications. As mentioned above, halos
with $V_c \ga 20\ {\rm km\ s}^{-1}$ may be able to accrete gas and
form stars once again at $z \la 1$. Any accreted gas at low redshift
could have been previously enriched with metals and molecules, thus
enabling more efficient cooling. On the other hand, if the progenitors
had a $V_c \la 17\ {\rm km~s^{-1}}$ at $\zr$ then they were not able
to cool and form stars, unless molecular hydrogen had not been
dissociated (\S \ref{sec3.3}). In order to reconcile the
photoionization scenario with the recent episodes of star formation
deduced to have occurred in most dwarf galaxies (e.g., Mateo 1998;
Grebel 1998), a continuous recycling of gas over many generations of
stars must be assumed. This would mean that supernova feedback (\S
\ref{sec7.1}) was unable to shut off star formation even in these
smallest known galaxies. In addition, the existence of a large
abundance of sub-halos may be problematic even if the sub-halos have
no gas, since they would interact with the disk dynamically and tend
to thicken it (Toth \& Ostriker 1992; Moore et al.\ 1999; Velazquez \&
White 1999).

However, the photoionization scenario is useful because it may be testable
through other implications. For example, Bullock et al.\ (2001) used
semi-analytic modeling to show that many subhalos which did form stars
before reionization were tidally disrupted in the Milky Way's gravitational
field, and the resulting stellar streams may be observable. The formation
of the Milky Way's stellar component has also been investigated by White \&
Springel (1999). They combined a scaled-down dark matter cluster simulation
with semi-analytic prescriptions for star formation in halos, and showed
that the oldest stars in the Milky Way should be located mostly in the
inner halo or bulge, but they cannot be easily identified because the
populations of old stars and of low metallicity stars are only weakly
correlated.

Gnedin (2000c) pointed out several observed features of dwarf galaxies
that may be related to their high-redshift histories. First, almost
all Local Group dwarf galaxies with measured star formation histories
show a sharp decline in the star formation rate around 10 Gyr ago.
Gnedin noted that this drop could correspond to the suppression of
star formation due to reionization (\S \ref{sec6.5}) if the measured
ages of the old stellar populations are somewhat underestimated; or
that it could instead correspond to the additional suppression caused
by helium reionization (\S \ref{sec6.3.2}) at $z \ga 3$. He showed
that if only the old stellar population is considered, then the
Schmidt Law (\S \ref{sec5.2}) implies that the luminosity of each
dwarf galaxy, divided by a characteristic volume containing a fixed
fraction of all the old stars, should be proportional to some power of
the central luminosity density. This assumes that the present central
luminosity density of old stars is a good measure of the original gas
density, i.e., that in the core almost all the gas was transformed
into stars, and feedback did not play a role. It also assumes that the
gas density distribution was self-similar in all the dwarf galaxies
during the period when they formed stars, and that the length of this
period was also the same in all galaxies. These assumptions are
required in order for the total stellar content of each galaxy to be
simply related to the central density via the Schmidt Law evaluated at
the center. Taking ten galaxies which have well-measured star
formation histories, and which formed most of their stars more than 10
Gyr ago, Gnedin found a correlation with a power law of $3/2$, as
expected from the Schmidt Law (\S \ref{sec5.2}). Clearly, the
theoretical derivation of this correlation combines many simplistic
assumptions. However, explaining the observed correlation is a
challenge for any competing models, e.g., models where feedback plays
a dominant role in regulating star formation.

Barkana \& Loeb (1999) noted that a particularly acute puzzle is
presented by the very smallest galaxies, the nine dwarf spheroidals in
the Local Group with central velocity dispersions $\sigma \la 10\ {\rm
km\ s}^{-1}$, including five below $7\ {\rm km\ s}^{-1}$ (see recent
reviews by Mateo 1998 \& van den Bergh 2000). These galaxies contain
old stars that must have formed at $z \ga 2$, before the ionizing
background dropped sufficiently to allow them to form. There are
several possible solutions to the puzzle of how these stars formed in
such small halos or in their progenitors which likely had even smaller
velocity dispersions. The solutions are that (i) molecular hydrogen
allowed these stars to form at $z>\zr$, as noted above, (ii) the
measured stellar velocity dispersions of the dwarf galaxies are well
below the velocity dispersions of their dark matter halos, or (iii)
the dwarf galaxies did not form via the usual hierarchical scenario.

One major uncertainty in comparing observations to hierarchical models
is the possibility that the measured velocity dispersion of stars in
the dwarf spheroidals underestimates the velocity dispersion of their
dark halos. Assuming that the stars are in equilibrium, their velocity
dispersion could be lower than that of the halo if the mass profile is
shallower than isothermal beyond the stellar core. The velocity
dispersion and mass-to-light ratio of a dwarf spheroidal could also
appear high if the galaxy is non-spherical or the stellar orbits are
anisotropic. The observed properties of dwarf spheroidals require a
central mass density of order $0.1 M_{\sun}$ pc$^{-3}$ (e.g., Mateo
1998), which is $\sim 7\times 10^5$ times the present critical
density. Thus, only the very inner parts of the halos are sampled by
the central velocity dispersion. Detailed observations of the velocity
dispersion profiles of these galaxies could be used to determine the
circular velocity of the underlying halo more reliably.

A cosmological scenario for the formation of dwarf spheroidal galaxies
is favored by the fact that they are observed to be dark matter
dominated, but this may not rule out all the alternatives. The dwarf
dark halos may have formed at low redshift by the breakup of a much
larger galaxy. Under this scenario, gas forming stars inside the large
parent galaxy would have been unaffected by the photoionizing
background. At low redshift, this galaxy may have collided with the
Milky Way or come close enough to be torn apart, forming at least some
of the dwarf spheroidal systems. Simulations of galaxy encounters
(Barnes \& Hernquist 1992; Elmegreen, Kaufman, \& Thomasson 1993) have
found that shocks in the tidal tails trigger star formation and lead
to the formation of dwarf galaxies, but these galaxies contain only
small amounts of dark matter. However, the initial conditions of these
simulations assumed parent galaxies with a smooth dark matter
distribution rather than clumpy halos with dense sub-halos inside
them. As noted above, simulations (Klypin et al.\ 1999; Moore et al.\
1999) suggest that galaxy halos may have large numbers of dark matter
satellites, and further simulations are needed to test whether these
subhalos can capture stars which form naturally in tidal tails.

A common origin for the Milky Way's dwarf satellites (and a number of
halo globular clusters), as remnants of larger galaxies accreted by
the Milky Way galaxy, has been suggested on independent grounds. These
satellites appear to lie along two (e.g., Majewski 1994) or more
(Lynden-Bell \& Lynden-Bell 1995, Fusi-Pecci et al.\ 1995) polar great
circles. The star formation history of the dwarf galaxies (e.g.,
Grebel 1998) constrains their merger history, and implies that the
fragmentation responsible for their appearance must have occurred early
in order to be consistent with the variation in stellar populations
among the supposed fragments (Unavane, Wyse, \& Gilmore 1996;
Olszewski 1998). Observations of interacting galaxies (outside the
Local Group) also demonstrate that ``tidal dwarf galaxies'' do indeed
form (e.g., Duc \& Mirabel 1997; Hunsberger, Charlton, \& Zaritsky
1996). 


\section{\bf Challenges for the Future}
\label{sec10}

{\it When and how did the first stars and black holes form and when
and how did they ionize most of the gas in the universe?} In this
review we have sketched the first attempts to answer these questions
and the basic physical principles that underlie these attempts. The
coming decade will likely be marked by major advances in our ability
to make theoretical predictions in an attempt to answer these
questions, and will culminate with the launch of {\it NGST},\, a
telescope which is ideally suited for testing these predictions. At
about the same time, the Planck satellite (and perhaps MAP before it)
is expected to directly infer the reionization redshift from
measurements of the CMB polarization power spectrum on large angular
scales. Also in about a decade, next-generation arrays of radio
telescopes may detect the 21 cm emission from the pre-reionization,
neutral warm IGM. The difficult questions just mentioned will receive
their ultimate answers from observations, but it surely is fun to try
to find the answers theoretically in advance, before we can deduce
them by looking through our most technologically-advanced telescopes.


\vspace{0.36in}
\centerline{\bf Acknowledgements}
\vspace{0.06in}

We thank Tom Abel, Steve Furlanetto, Nick Gnedin, Zoltan Haiman, Piero
Madau, Jordi Miralda-Escud\'e and especially the editor Marc
Kamionkowski for providing useful comments after a careful reading of
the manuscript. AL thanks the Institute for Advanced Study at
Princeton for its kind hospitality when the writing of this review
began. RB acknowledges support from Institute Funds; support by the
Smithsonian Institution Visitor Program during a visit to the
Harvard-Smithsonian CfA; and the hospitality of the Weizmann
Institute, Israel, where part of this review was written. This work
was supported in part by NASA grants NAG 5-7039, 5-7768, and NSF
grants AST-9900877, AST-0071019 for AL.



\end{document}